\documentclass[a4paper,11pt]{article}%
\usepackage{graphicx} 
\usepackage[english]{babel}
\usepackage{authblk}
\usepackage{bbm}
\usepackage{tikz}
\usetikzlibrary{trees}
\usetikzlibrary{decorations.pathmorphing}
\usetikzlibrary{decorations.markings}
\usepackage{tikz,ifthen}
\tikzset{beamerprimary/.style={structure.fg, thick}}
\tikzset{beamersecondary/.style={structure.bg, thick}}
\tikzset{boson/.style={draw=structure.fg,decorate, decoration={snake}},
    gauge/.style={decorate, decoration={snake} },
    fermion/.style={postaction={decorate},
        decoration={markings,mark=at position .55 with {\arrow{>}}}},
    fermionloop/.style={postaction={decorate},
        decoration={markings,mark=at position .25 with {\arrow{<}}}}, 
    gluon/.style={decorate, 
        decoration={coil,amplitude=4pt, segment length=5pt}},
    scalar/.style={dashed},
    scalarloop/.style={dashed={decorate},
        decoration={markings,mark=at position .25 with {\arrow{<}}}},
    resonance/.style={double,double distance=1.5pt}	
}
\usetikzlibrary{arrows,shapes,positioning}
\usetikzlibrary{decorations.markings}
\tikzstyle arrowstyle=[scale=1]
\tikzstyle directed=[postaction={decorate,decoration={markings,
    mark=at position .65 with {\arrow[arrowstyle]{stealth}}}}]
\tikzstyle reverse directed=[postaction={decorate,decoration={markings,
    mark=at position .65 with {\arrowreversed[arrowstyle]{stealth};}}}]

\usepackage[utf8]{inputenc}
\RequirePackage[T1]{fontenc}

\RequirePackage{flushend}
\RequirePackage{color}
\usepackage{changes}
\usepackage{multirow}
\usepackage{hyperref} 
\hypersetup{colorlinks,citecolor=blue,linkcolor=blue,urlcolor=blue}

\usepackage{ragged2e} 
\usepackage{subfigure} 
\usepackage{slashed}\usepackage{float}
\usepackage{amsmath,amssymb,lmodern,setspace}
\usepackage{multirow}

\title{New $\tau$-based evaluation of the hadronic contribution to the vacuum polarization piece of the muon anomalous magnetic moment}

\author[1]{J. A. Miranda~\footnote{jmiranda@fis.cinvestav.mx}}
\author[1]{P. Roig~\footnote{proig@fis.cinvestav.mx}}

\affil[1]{Departamento de F\'isica, Centro de Investigaci\'on y de Estudios Avanzados del IPN\\
Apdo. Postal 14-740,07000 Ciudad de M\'exico, M\'exico.}

\date{}

\begin{document}
\maketitle

\abstract{We revisit the isospin-breaking and electromagnetic corrections to the decay $\tau^-\to\pi^-\pi^0\nu_\tau$, which allow its use as input in the two-pion contribution to the (leading order) hadronic vacuum polarization part of the muon anomalous magnetic moment. We extend a previous resonance chiral Lagrangian analysis, which included those operators saturating the next-to-leading order chiral low energy constants, by including the contributions appearing at the next order. As a result, we improve agreement between the two-pion tau decay and $e^+e^-$ data and reduce the discrepancy between experiment and the SM prediction of $a_\mu$ (using $\tau$ input) to the $\sim2\sigma$ level.}

\section*{Introduction}
The anomalous magnetic moment of the (first electron, and then) muon ($a_\mu\equiv(g_\mu-2)/2$) has been crucial for the development of quantum field theory and the understanding of radiative corrections within it. Over the years, it has validated those computed in QED at increasing precision and (in the muon case) started probing the other Standard Model sectors, electroweak and QCD, setting also -and more interestingly- stringent constraints on new physics contributions. In the absence of any direct hint for heavy new particles or interactions at the LHC, clean observables both from experiment and theory -among which $a_\mu$ stands out- are reinforced as a promising gate for the eagerly awaited further (indirect) discoveries in high-energy physics.

With the forthcoming measurement of $a_\mu$ at FNAL \cite{Grange:2015fou} we will finally have an experimental update on the long-standing discrepancy (at 3 to 4 sigmas) between the SM prediction of this observable (recently refined in \cite{Aoyama:2020ynm})~\footnote{The SM prediction \cite{Aoyama:2020ynm} is based on \cite{Davier:2017zfy, Keshavarzi:2018mgv, Colangelo:2018mtw, Hoferichter:2019mqg, Davier:2019can, Keshavarzi:2019abf, Kurz:2014wya, Chakraborty:2017tqp, Borsanyi:2017zdw, Blum:2018mom, Giusti:2019xct, Shintani:2019wai, Davies:2019efs, Gerardin:2019rua, Aubin:2019usy, Giusti:2019hkz, Melnikov:2003xd, Masjuan:2017tvw, Colangelo:2017fiz, Hoferichter:2018kwz, Gerardin:2019vio, Bijnens:2019ghy, Colangelo:2019uex, Pauk:2014rta, Danilkin:2016hnh, Jegerlehner:2017gek, Knecht:2018sci, Eichmann:2019bqf, Roig:2019reh, Colangelo:2014qya, Blum:2019ugy, Aoyama:2012wk, Aoyama:2019ryr, Czarnecki:2002nt, Gnendiger:2013pva} (see also the last developments in refs. \cite{Knecht:2020xyr, Masjuan:2020jsf, Ludtke:2020moa,Hoid:2020xjs, Ananthanarayan:2020vum, Aubin:2020scy, Bijnens:2020xnl}).} and its most accurate measurement, at BNL \cite{Bennett:2006fi}. On the theory side, a tremendous effort driven by the Muon g-2 Theory Initiative \footnote{Its website is https://muon-gm2-theory.illinois.edu/.} has been reducing (and making more robust) the SM errors during the last few years, in order to profit maximally from the new data. In the near future, both the FNAL \cite{Grange:2015fou} and the J-PARC \cite{Abe:2019thb} experiments will shrink the current experimental uncertainty ($63\cdot10^{-11}$) by a factor four. A commensurate improvement on the theory error is essential in maximizing the reach on new physics of these measurements.

The SM uncertainty on $a_\mu$ ($43\cdot10^{-11}$) is saturated by that of the hadronic contributions, where the error of the dominant hadronic vacuum polarization (HVP,LO) part has been reduced to $40\cdot10^{-11}$, versus $17\cdot10^{-11}$ of the light-by-light piece~\cite{Aoyama:2020ynm}. In turn, the HVP,LO contribution is dominated by the $\pi\pi$ cut (yielding $\sim 73\%$ of the overall value), where good-quality data of the corresponding $e^+e^-$ hadronic cross-sections \cite{Aulchenko:2006na, Achasov:2006vp, Akhmetshin:2006wh, Akhmetshin:2006bx, Ambrosino:2008aa, Aubert:2009ad, Ambrosino:2010bv, Lees:2012cj, Ablikim:2015orh, Anastasi:2017eio} enables its computation by dispersive methods \cite{Brodsky:1967sr, Lautrup:1969fr}. Alternatively, one can also use isospin-rotated $\tau \to \pi \pi \nu_\tau$ measurements with that purpose, as was put forward in LEP times \cite{Alemany:1997tn}, despite the required IB corrections cannot be computed in a model-independent way presently. Still, while a lattice QCD computation of these is achieved, the authors find convenient testing the consistency of both extractions of $a_\mu^{HVP,LO_{\pi\pi}}$, in light of the tensions between different sets of $e^+e^-\to\pi^+\pi^-$ data that has not been resolved so far \cite{Aoyama:2020ynm}. 

In addition to the previous data-based determinations of $a_\mu^{HVP,LO}$, lattice QCD is also achieving computations with reduced errors, although not yet competitive with the $e^+e^-$  evaluations \cite{Aoyama:2020ynm}. One notable exception to this being the recent very accurate result ($53\cdot10^{-11}$ error) of the BMW Coll. \cite{Borsanyi:2020mff}, according to which the difference with respect to the SM prediction is at the one sigma level.

Concerning the tau based determination, refs. \cite{Cirigliano:2001er, Cirigliano:2002pv} computed the required isospin violating and electromagnetic corrections using Resonance Chiral Theory ($R\chi T$) \cite{Ecker:1988te, Ecker:1989yg} and refs. \cite{FloresBaez:2006gf, FloresTlalpa:2006gs} using Vector Meson Dominance (VMD). These series of articles were employed by ref. \cite{Davier:2009ag} (updated in refs. \cite{Davier:2010nc, Davier:2013sfa}) which, remarkably, found that the discrepancy of the SM prediction with the measurement is reduced substantially when tau data is employed~\footnote{The difference between the SM prediction of $a_\mu$ and the BNL measurement is $3.7\sigma$ \cite{Aoyama:2020ynm}. If isospin-rotated tau data is employed for $a_\mu^{HVP,LO}$, it amounts to $2.4\sigma$ \cite{Davier:2013sfa}, instead. This difference could in principle be due to new physics effects, hinting at a lepton universality violation in the corresponding non-standard vector and/or tensor couplings at low-energies \cite{Miranda:2018cpf, Cirigliano:2018dyk, Gonzalez-Solis:2019owk}. See the most updated discussions of its connection with $\alpha_{QED}$ in the electroweak fit in refs. \cite{Crivellin:2020zul, Keshavarzi:2020bfy, deRafael:2020uif, Malaescu:2020zuc}.}. Notwithstanding, as precise measurements of $\sigma(e^+e^-\to hadrons)$ became available in the last fifteen years, the $e^+e^-$ based evaluation gained preference over using tau data. Indeed, ref. \cite{Aoyama:2020ynm} concludes that `at the required precision to match the $e^+e^-$ data, the present understanding of the IB (isospin breaking) corrections to $\tau$ data is unfortunately not yet at a level allowing their use for the HVP dispersion integrals', despite ref.~\cite{Jegerlehner:2011ti} claiming that (the model-dependent) $\rho-\gamma$ mixing in the neutral channel makes it agree with the results in the charged current. It is the purpose of this work \footnote{Currently, a lattice evaluation of IB for using tau data in $a_\mu^{HVP,LO_{\pi\pi}}$ is in progress~\cite{Bruno:2018ono}.} to extend previous $R\chi T$ analyses \cite{Cirigliano:2001er, Cirigliano:2002pv} of the required IB corrections to di-pion tau decays so that they can again be useful, when combined with $\sigma(e^+e^-\to\pi^+\pi^-(\gamma))$, to increase the accuracy of the SM prediction of $a_\mu^{HVP,LO}$. In this spirit, we note that F. Jegerlehener \cite{Jegerlehner:2017lbd} indeed combines both sets of data (using the IB corrections of ref.~\cite{Jegerlehner:2011ti}), which reduces the error of $a_\mu^{HVP,LO}$ by $\sim17\%$ \cite{Jegerlehner:2017lbd}.
 
Within the global effort of the Muon g-2 theory initiative, we revisit in this paper the $R\chi T$ computations including operators that -in the chiral limit- start to contribute at $\mathcal{O}(p^6)$. This is possible by the knowledge acquired after the analyses of Cirigliano \textit{et al.} \cite{Cirigliano:2001er, Cirigliano:2002pv} (where operators contributing at $\mathcal{O}(p^4)$ were considered), through a series of works studying operator product expansion (OPE) restrictions on $R\chi T$ couplings on several relevant $3-$point Green functions (and related form factors) \cite{Roig:2019reh, RuizFemenia:2003hm, Cirigliano:2004ue, Cirigliano:2005xn, Cirigliano:2006hb, Guo:2008sh, Dumm:2009kj, Dumm:2009va, Guo:2010dv, Kampf:2011ty, Dumm:2012vb, Chen:2012vw, Colangelo:2012ipa, Dai:2013joa, Guevara:2013wwa, Roig:2013baa, Chen:2013nna, Roig:2014uja, Chen:2017jcw, Guevara:2018rhj, Dai:2019lmj, Kadavy:2020hox,Mateu:2007tr}~\footnote{See also e.g. refs. \cite{Moussallam:1997xx, Peris:1998nj, Knecht:1999gb, Peris:2000tw, Knecht:2001xc, Bijnens:2003rc, Ananthanarayan:2004qk}.}~\footnote{Similar radiative corrections were computed for the $\tau \to \eta \pi \nu_\tau \gamma$ decays in $R\chi T$ \cite{Guevara:2016trs}, even though part of our contributions here were suppressed (and thus neglected) there because of G-parity.}. This procedure will also allow us to evaluate an uncertainty for the results by Cirigliano \textit{et al.} \cite{Cirigliano:2002pv}, which is one of the main outcomes of this work, together with the new results, including operators that start contributing to the $\mathcal{O}(p^6)$ chiral low-energy constants (LECs).
 
 The paper is organized as follows. In section \ref{sec:ampl} we review the main features of the $\tau^-\to\pi^-\pi^0\nu_\tau\gamma$ decays and split the model-independent part from the hadron form factors, computed in $R\chi T$ including new terms, subleading in the chiral expansion. We then recall the short-distance (SD) QCD constraints on the Lagrangian couplings, their phenomenological determinations and explain our estimation of the remaining free couplings, based on chiral counting. After that, in section \ref{GEM} we recap the radiative corrections needed for the tau-based calculation of $a_\mu^{HVP,LO}$ and predict several observables for the processes where the real photon is detected together with the pion pair. Then, in section \ref{sec:IBCorr} we evaluate $a_\mu^{HVP,LO|_{\pi\pi}}$ using tau data, which is the main result of this article. Finally, our conclusions are presented in section \ref{sec:Concl}. Several appendices complement the main material, explaining how the coefficients dominating  uncertainties were fitted, giving a full account of the kinematics, and providing with the complete expressions for the structure-dependent (axial-)vector form factors of the $\tau^-\to\pi^-\pi^0\nu_\tau\gamma$ decays.

\section{$\tau^-\to \pi^-\pi^0\gamma\nu_\tau$ decays}\label{sec:ampl}

\subsection{Amplitude}\label{subsec:ampl}
For the radiative decay $\tau^-\left(P\right)\to \pi^-\left(p_-\right)\pi^0\left(p_0\right)\nu_\tau\left(q\right)\gamma\left(k\right)$, we can split the contribution due to the bremsstrahlung off the initial tau lepton from the one coming from the hadronic part.

We write down the general structure for these processes \cite{Bijnens:1992en, Cirigliano:2002pv}
\begin{equation}\begin{split}
T=e\, G_F V_{ud}^*\epsilon^{\mu}(k)^*&\left\lbrace F_\nu\, \bar{u}\left(q\right)\gamma^\nu\left(1-\gamma_5\right)\left(m_\tau+\slashed P-\slashed k\right)\gamma_\mu u\left(P\right)\right.\\
&\left. +\left(V_{\mu\nu}-A_{\mu\nu}\right)\bar{u}\left(q\right)\gamma^\nu\left(1-\gamma_5\right)u\left(P\right)\right\rbrace,
\end{split}\end{equation}
where $F_\nu=\left(p_0-p_-\right)_\nu f_+\left(s\right)/2P\cdot k$, with the charged pion vector form factor $f_+(s)$ defined through $\left\langle\pi^0\pi^-|\bar{d}\gamma^\mu u|0\right\rangle=\sqrt{2}f_+(s)(p_{-}-p_0)^\mu$ and $s=(p_{-}+p_0)^2$. Gauge invariance ($\epsilon_\mu\to \epsilon_\mu+k_\mu$) implies the Ward identities 

\begin{equation}\label{WardIds}
k_\mu V^{\mu\nu}=\left(p_{-}-p_0\right)^\nu f_+\left(s\right),\;\; k_\mu A^{\mu\nu}=0.
\end{equation}
Imposing eq.~(\ref{WardIds}) and Lorentz invariance, we have the following expression for the vector structure-dependent tensor
\begin{equation}\small\begin{split}
V^{\mu\nu}&=f_+\left[\left(P-q\right)^2\right] \frac{p_-^\mu \left(p_{-}+k -p_0\right)^\nu}{p_-\cdot k}-f_+\left[\left(P-q\right)^2\right]g^{\mu\nu}\\
&+\frac{f_+\left[\left(P-q\right)^2\right]-f_+\left(s\right)}{\left(p_0+p_-\right)\cdot k}\left(p_0+p_-\right)^\mu\left(p_0-p_-\right)^\nu\\
&+v_1\left(g^{\mu\nu}\,p_-\cdot k-p_-^\mu k^\nu\right)+v_2\left(g^{\mu\nu}\,p_0\cdot k-p_0^\mu k^\nu\right)\\
&+v_3\left(p_0\cdot k\, p_-^\mu-p_-\cdot k\,p_0^\mu\right)p_-^\nu+v_4\left(p_0\cdot k\, p_-^\mu-p_-\cdot k\,p_0^\mu\right)\left(p_0+p_-+k\right)^\nu,
\end{split}\end{equation}
and for the axial one
\begin{equation}\begin{split}
A^{\mu\nu} &=ia_1\,\epsilon^{\mu\nu\rho\sigma}\,\left(p_0-p_-\right)_{\rho}k_{\sigma}+ia_2\,W^\nu\,\epsilon^{\mu\lambda\rho\sigma}k_{\lambda}\,p_{-\rho}\,p_{0\sigma}\\
&+ia_3\,\epsilon^{\mu\nu\rho\sigma}k_{\rho}\,W_{\sigma}+ia_4\,\left(p_0+k\right)^\nu\,\epsilon^{\mu\lambda\rho\sigma}\,k_\lambda\,p_{-\rho}\,p_{0\sigma},\\
\end{split}\end{equation}
where $W\equiv P-q=p_-+p_0+k$. We could use the basis given in ref. \cite{Guevara:2016trs} but instead we prefer a modified one that resembles the decomposition in ref. \cite{Cirigliano:2002pv} (see also ref. \cite{Bijnens:1992en}). These tensor structures depend on four vector ($v_i$) and four axial-vector ($a_i$) form factors. For the axial structure, the Schouten's identity has been used.

Taking into account that $\left(P-q\right)^2=s+2\left(p_0+p_-\right)\cdot k$, the Low's theorem \cite{Low:1958sn} is manifestly satisfied
\begin{equation}\begin{split}
V^{\mu\nu}&=f_+\left(s\right) \frac{p_-^\mu }{p_-\cdot k}\left(p_{-}-p_0\right)^\nu +f_+\left(s\right)\left(\frac{p_-^\mu k^\nu}{p_-\cdot k}-g^{\mu\nu} \right)\\
&+2\frac{df_+\left(s\right)}{d\,s}\left(\frac{p_0\cdot k}{p_-\cdot k}p_-^\mu-p_0^\mu\right)\left(p_{-}-p_0\right)^\nu+\mathcal{O}\left(k\right).
\end{split}\end{equation}

\subsection{Theoretical framework}
We will present in the following the model-dependent contributions to the $V_{\mu\nu}$ and $A_{\mu\nu}$ tensors. We will closely 
follow ref. \cite{Cirigliano:2002pv}, extending it to include subleading terms in the chiral expansion. In this reference, a large-$N_C$ \cite{tHooft:1973alw, tHooft:1974pnl, Witten:1979kh} inspired computation was carried out. Specifically, it was restricted to the dominant (for $N_C\to\infty$) tree level diagrams, although the relevant loop corrections for the $\tau^-\to\pi^-\pi^0\nu_\tau\gamma$ decays --giving the $\rho$ (and $a_1$, for completeness) off-shell width~\footnote{We will introduce them following ref. \cite{GomezDumm:2000fz} for the $\rho(770)$ and refs. \cite{Dumm:2009va, Nugent:2013hxa} for the $a_1(1260)$ resonances.}-- were taken into account~\footnote{See refs.~\cite{Rosell:2004mn, Rosell:2005ai, Rosell:2006dt, Portoles:2006nr, Pich:2008jm, SanzCillero:2009pt, Pich:2010sm} for next-to-leading order (NLO) computations in $1/N_C$, allowing to include the scale dependence of the Chiral Pertubation Theory LECs in the low-energy limit of $R\chi T$.}. Also, given the limited phase space of tau decays and the fact that the region $E\lesssim M_\rho+\Gamma_\rho$ is the most important one for the IB corrections needed for $a_\mu^{HVP,LO_{\pi\pi}}$~\cite{Cirigliano:2002pv}, the contribution of the $\rho(1450)$ and other heavier resonances was neglected in this reference (despite the fact that, in the large-$N_C$ limit, there is an infinite tower of resonances per channel), as we will also do~\footnote{Nevertheless, we will include the dominant effect of the $\rho(1450)$ and $\rho(1700)$ resonances in our dispersive pion form factor \cite{Dumm:2013zh, Gonzalez-Solis:2019iod} and check the negligible impact of heavier resonances in the $v_i$ and $a_i$ form factors in our analysis.}. Within this setting, our computation will include all $R\chi T$ operators contributing to the $\mathcal{O}(p^6)$ chiral low-energy constants. Our results agree with those in ref. \cite{Cirigliano:2002pv}, providing the new contributions with resonance operators that are suppressed by one chiral order in the low-energy limit (where possible, our computations have been checked against the results in ref. \cite{Guevara:2016trs}).

As explained in ref. \cite{Cirigliano:2002pv}, this procedure warrants the correct low-energy limit (as given by Chiral Perturbation Theory \cite{Weinberg:1978kz, Gasser:1983yg, Gasser:1984gg, Bijnens:1999sh, Bijnens:2001bb}) and includes consistently the most general pion and photon interactions with the lightest resonances. Demanding the known QCD SD constraints results in relations among the Lagrangian couplings, and chiral counting can be employed to estimate those still unconstrained after using phenomenological information. It should then provide an accurate description of the $\tau^-\to\pi^-\pi^0\nu_\tau\gamma$ decays for $s\lesssim 1\, \mathrm{GeV}^2$, which gives $\sim 99.8\%$ of the whole $a_\mu^{HVP,LO|_{\pi\pi}}$ contribution.

\subsection{Vector Form Factors}\label{ss:VFF}

\begin{figure}[H]
\begin{center}
\begin{tikzpicture}
	\draw[resonance] (-1.5,0)--(0,0);
	\node at (-0.6,-0.2) {\tiny $\rho^-$};
	\draw[fill,white] (-1.5,0) circle [radius=0.12];
	\node at (-1.5,0) {\small$\otimes$};
	\draw[scalar] (0,0)--(0.75,0.75) node[right] {\tiny$\pi^-$};
	\draw[gauge] (0,0)--(0.75,0) node[right] {\tiny $\gamma$};
	\draw[scalar] (0,0)--(0.75,-0.75) node[right] {\tiny $\pi^0$};
	\draw[fill] (0,0) circle [radius=0.04];
\end{tikzpicture}
\begin{tikzpicture}
	\draw[resonance] (2,0)--(3.5,0);
	\node at (2.9,-0.2) {\tiny$\rho^-$};
	\draw[gauge] (2,0)--(2,0.75) node[above] {\tiny$\gamma$};
	\draw[fill,white] (2,0) circle [radius=0.12];
	\node at (2,0) {\small$\otimes$};
	\draw[scalar] (3.5,0)--(4.25,0.75) node[right] {\tiny$\pi^-$};
	\draw[scalar] (3.5,0)--(4.25,-0.75) node[right] {\tiny$\pi^0$};
	\draw[fill] (3.5,0) circle [radius=0.04];
\end{tikzpicture}
\begin{tikzpicture}
	\draw[resonance] (-1.5,0)--(0,0);
	\node at (-0.6,-0.2) {\tiny $\rho^-$};
	\draw[fill,white] (-1.5,0) circle [radius=0.12];
	\node at (-1.5,0) {\small$\otimes$};
	\draw[scalar] (0,0)--(0.75,0.75) node[right] {\tiny$\pi^-$};
	\draw[gauge] (0.375,0.375)--(0.8,-0.05) node[right] {\tiny $\gamma$};
	\draw[scalar] (0,0)--(0.75,-0.75) node[right] {\tiny $\pi^0$};
	\draw[fill] (0,0) circle [radius=0.04];
\end{tikzpicture}
\begin{tikzpicture}

	\draw[resonance] (5.5,0)--(6.7,0);
	\node at (6.2,-0.2) {\tiny$\rho^0$};
	\draw[scalar] (5.5,0)--(6.25,0.75) node[right] {\tiny$\pi^-$};
	\draw[scalar] (5.5,0)--(6.25,-0.75) node[right] {\tiny $\pi^0$};
	\draw[fill,white] (5.5,0) circle [radius=0.12];
	\node at (5.5,0) {\small $\otimes$};
	\draw[gauge] (6.7,0)--(7.5,0) node[right] {\tiny$\gamma$};
	\draw[fill] (6.7,0) circle [radius=0.04];
\end{tikzpicture}
\begin{tikzpicture}
	\draw[resonance] (2,-2)--(3.5,-2);
	\node at (2.8,-2.2) {\tiny $\omega$};
	\draw[scalar] (2,-2)--(2,-1.25) node[above] {\tiny$\pi^-$};
	\draw[fill,white] (2,-2) circle [radius=0.12];
	\node at (2,-2) {\small$\otimes$};
	\draw[scalar] (3.5,-2)--(4.25,-1.25) node[right] {\tiny$\pi^0$};
	\draw[gauge] (3.5,-2)--(4.25,-2.75) node[right] {\tiny$\gamma$};
	\draw[fill] (3.5,-2) circle [radius=0.04];
\end{tikzpicture}
\begin{tikzpicture}
	\draw[resonance] (-1.5,0)--(0,0);
	\node at (-0.6,-0.2) {\tiny $a_1^-$};
	\draw[scalar] (-1.5,0)--(-1.5,0.75) node[above] {\tiny $\pi^0$};
	\draw[fill,white] (-1.5,0) circle [radius=0.12];
	\node at (-1.5,0) {\small$\otimes$};
	\draw[scalar] (0,0)--(0.75,0.75) node[right] {\tiny$\pi^-$};
	\draw[gauge] (0,0)--(0.75,-0.75) node[right] {\tiny $\gamma$};
	\draw[fill] (0,0) circle [radius=0.04];
\end{tikzpicture}
\begin{tikzpicture}
	\draw[scalar] (8.5,-4)--(9.25,-4);
	\node at (8.99,-4.2) {\tiny $\pi^-$};
	\draw[resonance] (9.25,-4)--(10,-4);
	\node at (9.65,-4.2) {\tiny $\rho^0$};
	\draw[scalar] (8.5,-4)--(8.5,-3.25) node[above] {\tiny$\pi^0$};
	\draw[fill,white] (8.5,-4) circle [radius=0.12];
	\node at (8.5,-4) {\small $\otimes$};
	\draw[scalar] (9.25,-4)--(9.25,-3.25) node[above] {\tiny$\pi^-$};
	\draw[fill] (9.25,-4) circle [radius=0.04];
	\draw[gauge] (10,-4)--(10.75,-4) node[right] {\tiny$\gamma$};
	\draw[fill] (10,-4) circle [radius=0.04];
\end{tikzpicture}
\end{center}
\caption{One-resonance exchange contributions from the $R\chi T$ to the vector form factors of the $\tau^-\to\pi^-\pi^0\gamma\nu_\tau$ decays.}\label{VF:fig1}
\end{figure}
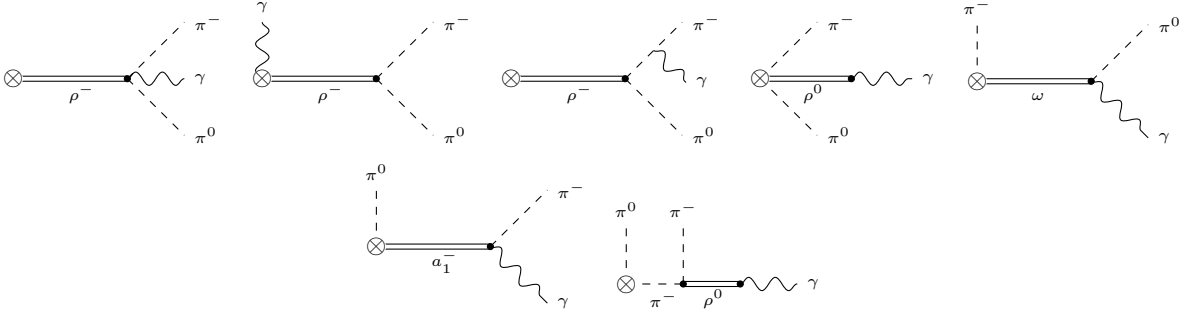

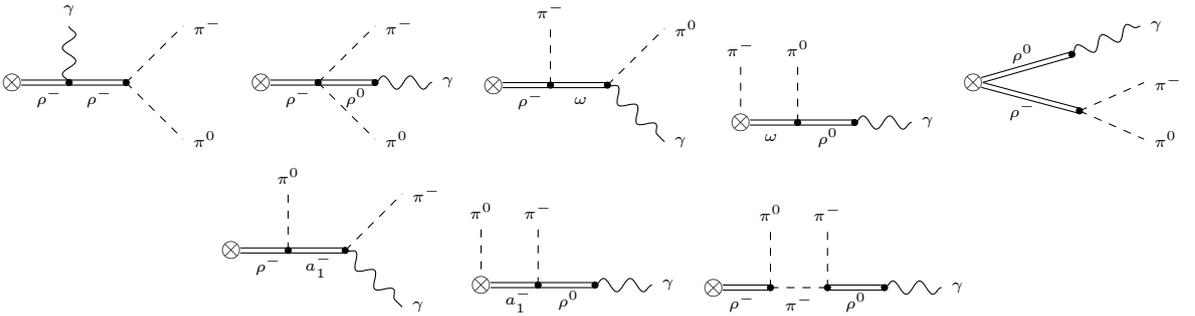
\begin{figure}[H]
\begin{center}
\begin{tikzpicture}
	\draw[resonance] (-1.5,0)--(0,0);
	\node at (-1.01,-0.2) {\tiny $\rho^-$};
	\node at (-0.35,-0.2) {\tiny $\rho^-$};
	\draw[fill,white] (-1.5,0) circle [radius=0.12];
	\node at (-1.5,0) {\small$\otimes$};
	\draw[gauge] (-0.75,0)--(-0.75,0.75) node[above] {\tiny $\gamma$};
	\draw[fill] (-0.75,0) circle [radius=0.04];
	\draw[scalar] (0,0)--(0.75,0.75) node[right] {\tiny$\pi^-$};
	\draw[scalar] (0,0)--(0.75,-0.75) node[right] {\tiny $\pi^0$};
	\draw[fill] (0,0) circle [radius=0.04];
\end{tikzpicture}
\begin{tikzpicture}
	\draw[resonance] (2,0)--(3.5,0);
	\node at (2.49,-0.2) {\tiny$\rho^-$};
	\node at (3.25,-0.2) {\tiny$\rho^0$};
	\draw[fill,white] (2,0) circle [radius=0.12];
	\node at (2,0) {\small$\otimes$};
	\draw[scalar] (2.75,0)--(3.5,0.75) node[right] {\tiny$\pi^-$};
	\draw[scalar] (2.75,0)--(3.5,-0.75) node[right] {\tiny$\pi^0$};
	\draw[fill] (2.75,0) circle [radius=0.04];	
	\draw[gauge] (3.5,0)--(4.25,0) node[right] {\tiny$\gamma$};
	\draw[fill] (3.5,0) circle [radius=0.04];	
\end{tikzpicture}
\begin{tikzpicture}
	\draw[resonance] (5.5,0)--(7,0);
	\node at (5.99,-0.2) {\tiny$\rho^-$};
	\node at (6.65,-0.2) {\tiny$\omega$};
	\draw[fill,white] (5.5,0) circle [radius=0.12];
	\node at (5.5,0) {\small$\otimes$};
	\draw[scalar] (6.25,0)--(6.25,0.75) node[above] {\tiny$\pi^-$};
	\draw[fill] (6.25,0) circle [radius=0.04];
	\draw[scalar] (7,0)--(7.75,0.75) node[right] {\tiny$\pi^0$};
	\draw[gauge] (7,0)--(7.75,-0.75) node[right] {\tiny$\gamma$};
	\draw[fill] (7,0) circle [radius=0.04];
\end{tikzpicture}
\begin{tikzpicture}
	\draw[resonance] (2,-2)--(3.5,-2);
	\node at (3.15,-2.2) {\tiny$\rho^0$};
	\node at (2.4,-2.2) {\tiny$\omega$};
	\draw[scalar] (2,-2)--(2,-1.25) node[above] {\tiny$\pi^-$};
	\draw[fill,white] (2,-2) circle [radius=0.12];
	\node at (2,-2) {\small$\otimes$};
	\draw[scalar] (2.75,-2)--(2.75,-1.25) node[above] {\tiny$\pi^0$};
	\draw[fill] (2.75,-2) circle [radius=0.04];
	\draw[gauge] (3.5,-2)--(4.25,-2) node[right] {\tiny$\gamma$};
	\draw[fill] (3.5,-2) circle [radius=0.04];
\end{tikzpicture}
\begin{tikzpicture}
	\draw[resonance] (5.5,-2)--(6.8,-1.625);
	\node at (6.15,-1.62) {\tiny$\rho^0$};
	\draw[gauge] (6.8,-1.625)--(7.7,-1.25) node[right] {\tiny$\gamma$};
	\draw[fill] (6.8,-1.625) circle [radius=0.04];
	\draw[resonance] (5.5,-2)--(6.9,-2.375);
	\node at (6.15,-2.37) {\tiny$\rho^-$};
	\draw[fill,white] (5.5,-2) circle [radius=0.12];
	\node at (5.5,-2) {\small$\otimes$};
	\draw[scalar] (6.9,-2.375)--(7.75,-2) node[right] {\tiny$\pi^-$};
	\draw[scalar] (6.9,-2.375)--(7.75,-2.75) node[right] {\tiny$\pi^0$};
	\draw[fill] (6.9,-2.375) circle [radius=0.04];
\end{tikzpicture}
\begin{tikzpicture}
	\draw[resonance] (-1.5,0)--(0,0);
	\node at (-1.01,-0.2) {\tiny $\rho^-$};
	\node at (-0.35,-0.2) {\tiny $a_1^-$};
	\draw[fill,white] (-1.5,0) circle [radius=0.12];
	\node at (-1.5,0) {\small$\otimes$};
	\draw[scalar] (-0.75,0)--(-0.75,0.75) node[above] {\tiny $\pi^0$};
	\draw[fill] (-0.75,0) circle [radius=0.04];
	\draw[scalar] (0,0)--(0.75,0.75) node[right] {\tiny$\pi^-$};
	\draw[gauge] (0,0)--(0.75,-0.75) node[right] {\tiny $\gamma$};
	\draw[fill] (0,0) circle [radius=0.04];
\end{tikzpicture}
\begin{tikzpicture}
	\draw[resonance] (1.75,-2)--(3.25,-2);
	\draw[scalar] (1.75,-2)--(1.75,-1.25) node[above] {\tiny$\pi^0$};
	\node at (2.24,-2.2) {\tiny $a_1^-$};
	\node at (2.9,-2.2) {\tiny $\rho^0$};
	\draw[fill,white] (1.75,-2) circle [radius=0.12];
	\node at (1.75,-2) {\small$\otimes$};
	\draw[scalar] (2.5,-2)--(2.5,-1.25) node[above] {\tiny $\pi^-$};
	\draw[fill] (2.5,-2) circle [radius=0.04];
	\draw[gauge] (3.25,-2)--(4,-2) node[right] {\tiny $\gamma$};
	\draw[fill] (3.25,-2) circle [radius=0.04];
\end{tikzpicture}
\begin{tikzpicture}
	\draw[resonance] (7.75,-4)--(8.5,-4);
	\draw[fill,white] (7.75,-4) circle [radius=0.12];
	\node at (7.75,-4) {\small$\otimes$};
	\node at (8.125,-4.2) {\tiny$\rho^-$};
	\draw[scalar] (8.5,-4)--(8.5,-3.25) node[above] {\tiny$\pi^0$};
	\draw[scalar] (8.5,-4)--(9.25,-4);
	\draw[resonance] (9.25,-4)--(10,-4);
	\draw[fill] (8.5,-4) circle [radius=0.04];
	\node at (8.875,-4.2) {\tiny$\pi^-$};
	\node at (9.625,-4.2) {\tiny$\rho^0$};
	\draw[scalar] (9.25,-4)--(9.25,-3.25) node[above] {\tiny$\pi^-$};
	\draw[fill] (9.25,-4) circle [radius=0.04];
	\draw[gauge] (10,-4)--(10.75,-4) node[right] {\tiny$\gamma$};
	\draw[fill] (10,-4) circle [radius=0.04];
\end{tikzpicture}
\end{center}
\caption{Two-resonance exchange contributions from the $R\chi T$ to the vector form factors of the $\tau^-\to\pi^-\pi^0\gamma\nu_\tau$ decays.}\label{VF:fig2}
\end{figure}

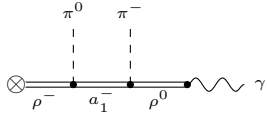
\begin{figure}[H]
\begin{center}
\begin{tikzpicture}
	\draw[resonance] (7.75,-4)--(10,-4);
	\draw[fill,white] (7.75,-4) circle [radius=0.12];
	\node at (7.75,-4) {\small$\otimes$};
	\node at (8.125,-4.2) {\tiny$\rho^-$};
	\draw[scalar] (8.5,-4)--(8.5,-3.25) node[above] {\tiny$\pi^0$};
	\draw[fill] (8.5,-4) circle [radius=0.04];
	\node at (8.875,-4.2) {\tiny$a_1^-$};
	\node at (9.625,-4.2) {\tiny$\rho^0$};
	\draw[scalar] (9.25,-4)--(9.25,-3.25) node[above] {\tiny$\pi^-$};
	\draw[fill] (9.25,-4) circle [radius=0.04];
	\draw[gauge] (10,-4)--(10.75,-4) node[right] {\tiny$\gamma$};
	\draw[fill] (10,-4) circle [radius=0.04];

\end{tikzpicture}
\end{center}
\caption{Three-resonance exchange contributions from the $R\chi T$ to the vector form factors of the $\tau^-\to\pi^-\pi^0\gamma\nu_\tau$ decays.}\label{VF:fig3}
\end{figure}

Within $R\chi T$ \cite{Ecker:1988te, Ecker:1989yg, Cirigliano:2006hb, Kampf:2011ty}, the diagrams contributing to the vector form factors of the $\tau^-\to\pi^-\pi^0\gamma\nu_\tau$ decays including operators that start contributing to the $\mathcal{O}(p^6)$ LECs  are shown in Figs. \ref{VF:fig1}, \ref{VF:fig2} and \ref{VF:fig3} ~\footnote{The contributions involving scalar and pseudoscalar resonances are discussed at the end of section \ref{ss:VFF}.}. The first three diagrams in fig. \ref{VF:fig1} and the first diagram in fig. \ref{VF:fig2} contribute to the pion vector form factor 
entering the structure-independent (SI) piece \footnote{Relevant $R\chi T$ couplings are introduced after eq. (\ref{eqsvs}) and in sec. \ref{subsec:SDC} below.}
\begin{equation}\begin{split}\label{VF:eq06}
f_+\left(s\right)&=1+\frac{G_V F_V}{F^2}\frac{s}{m_\rho^2-s}+\frac{\sqrt{2}F_V\,s}{F^2\left(m_\rho^2-s\right)}\left[2\left(2\lambda_8^V+\lambda_9^V+2\lambda_{10}^V\right)m_\pi^2-s\lambda_{21}^V\right]\\
&+\frac{2\sqrt{2}G_V\,s}{F^2\left(m_\rho^2-s\right)}\left[4\lambda_{6}^Vm_\pi^2-s\lambda_{22}^V\right]\\
&+\frac{4s}{F^2\left(m_\rho^2-s\right)}\left[4\lambda_6^V m_\pi^2-s\lambda_{22}^V\right]\left[2\left(2\lambda_8^V+\lambda_9^V+2\lambda_{10}^V\right)m_\pi^2-s\lambda_{21}^V\right]\,.
\end{split}\end{equation}
The contribution of both the last diagram in fig. \ref{VF:fig1} and the last diagram in fig. \ref{VF:fig2} vanishes for a real photon, as the corresponding ($f_+(0)=1$ part) contribution is already in the SI piece. We note we are using $F\sim 92$ MeV for the pion decay constant and that QCD OPE constraints $\lambda_{21}^V=\lambda_{22}^V=0$ \cite{Cirigliano:2006hb}. In fact, we will see in sec. \ref{subsec:SDC} that all modifications induced by the $\lambda_i^V$ couplings to $f_+(s)$ (\ref{VF:eq06}) vanish once SD QCD constraints are accounted for.\\

For the vector form factors, we get
\begin{subequations}\label{eqsvs}\small\begin{align}
v_1&=v_1^0+v_1^R+v_1^{RR}+v_1^{RRR}+v_{GI1}^{R+RR},\\
v_2&=v_2^0+v_2^R+v_2^{RR}+v_2^{RRR}+v_{GI2}^{R+RR},\\
v_3&=v_3^0+v_3^R+v_3^{RR}+v_3^{RRR}+v_{GI3}^{R+RR},\\
v_4&=v_4^0+v_4^R+v_4^{RR}+v_4^{RRR}+v_{GI4}^{R+RR},
\end{align}\end{subequations}
where $v_i^0$ is the contribution in ref. \cite{Cirigliano:2002pv} ($D^{-1}_R$ stands for the inverse resonance propagator)~\footnote{We recall that $F_R$ gives the coupling of the $R=V,A$ resonance to the $r=v,a$ external current and the $\rho\pi\pi$ vertex receives contributions both from $F_V$ and $G_V$.}
\begin{equation*}\small\begin{split}
v^0_1&\equiv\frac{F_V G_V}{F^2 M_\rho^2}\left(2+2M_\rho^2 D^{-1}_\rho\left[\left(P-q\right)^2\right]+s D^{-1}_\rho\left(s\right)+sM_\rho^2 D^{-1}_\rho\left(s\right) D^{-1}_\rho\left[\left(P-q\right)^2\right]\right)\\
&+\frac{F_V^2}{2F^2M_\rho^2}\left(-1-M_\rho^2 D^{-1}_\rho\left[\left(P-q\right)^2\right]+\left(P-q\right)^2 D^{-1}_\rho\left[\left(P-q\right)^2\right]\right)\\
&+\frac{F_A^2}{F^2 M_{a_1}^2}\left(M_{a_1}^2-m_\pi^2+\frac{1}{2}s\right) D^{-1}_{a_1}\left[\left(p_-+k\right)^2\right],\\
v^0_2&\equiv\frac{F_V G_V \,s}{F^2 M_\rho^2}\left(-D^{-1}_\rho\left(s\right)-M_\rho^2 D^{-1}_\rho\left(s\right) D^{-1}_\rho\left[\left(P-q\right)^2\right]\right)\\
&+\frac{F_V^2}{2F^2M_\rho^2}\left(-1-M_\rho^2 D^{-1}_\rho\left[\left(P-q\right)^2\right]-\left(P-q\right)^2 D^{-1}_\rho\left[\left(P-q\right)^2\right]\right)\\
&+\frac{F_A^2}{F^2 M_{a_1}^2}\left(M_{a_1}^2-m_\pi^2-k\cdot p_-\right) D^{-1}_{a_1}\left[\left(p_-+k\right)^2\right],\\
v^0_3&\equiv\frac{F_A^2}{F^2 M_{a_1}^2} D^{-1}_{a_1}\left[\left(p_-+k\right)^2\right],\\
v^0_4&\equiv-\frac{2F_V G_V}{F^2}D^{-1}_\rho\left(s\right) D^{-1}_\rho\left[\left(P-q\right)^2\right]+\frac{F_V^2}{F^2 M_\rho^2} D^{-1}_\rho\left[\left(P-q\right)^2\right],
\end{split}\end{equation*}
and $v_i^{R}$, $v_i^{RR}$, $v_i^{RRR}$ and $v_{GIi}^{R+RR}$ \footnote{In general, diagrams are gauge-invariant by themselves. Those giving the contribution $v_{GIi}^{R+RR}$ need to be summed to achieve gauge invariance. These are the first three diagrams in fig. \ref{VF:fig1} and the first diagram in fig. \ref{VF:fig2}.} correspond to contributions including operators which do not contribute to the NLO chiral LECs. Due to their length, the expressions for these form factors are in App. \ref{VF}. In writing the new contributions to $v_i$, the basis given in ref. \cite{Cirigliano:2006hb} has been used for the even-intrinsic parity operators (with couplings $\lambda^X_i$) and the basis given in ref. \cite{Kampf:2011ty} has been employed for the odd-intrinsic parity operators ($\kappa^X_i$ couplings). Both sets of $\lambda^X_i$ and $\kappa^X_i$ couplings have dimensions of inverse energy.

Including operators with at most one resonance, only the contribution from the exchange of $\rho$ and $a_1$ resonances on the vector form factor appeared \cite{Cirigliano:2002pv}. Allowing for multi-resonance operators we also have contributions with $\omega$ exchange,  coming from the odd-intrinsic parity sector, for both vector and axial-vector form factors (as well as resonance contributions on the axial form factor, absent in ref. \cite{Cirigliano:2002pv}). Apparently, such $\omega$ contributions were responsible for the larger effect of the IB corrections obtained in refs. \cite{FloresBaez:2006gf, FloresTlalpa:2006gs} with respect to refs. \cite{Cirigliano:2001er, Cirigliano:2002pv}. As a result, ref. \cite{Davier:2009ag} (and later evaluations by this group) ascribed an error to these corrections covering both contradictory evaluations. As we include (among others) contributions with an $\omega-\rho-\pi$ vertex in this work, closer agreement with the VMD evaluation should, in principle, be expected.

We have verified that all diagrams including scalar mesons vanish in the isospin symmetry limit. We point out that all contributions involving pseudoscalar mesons can be obtained from those with an axial-vector resonance by replacing it by a pseudoscalar resonance. Then, at leading chiral order, the saturation of the LECs by spin-one mesons \cite{Ecker:1988te} shows that diagrams including pseudoscalar resonances are suppressed. If we assume that this feature also holds at the next chiral order, then pseudoscalar resonance exchanges could be safely neglected ~\footnote{Since contributions from scalar and pseudoscalar resonances are suppressed,  we will neglect them for the axial form factors in the next section.}.

\subsection{Axial-Vector Form Factors}\label{subsec:AVFFs}
The axial form factors at chiral $\mathcal{O}\left(p^4\right)$ get contibutions from the Wess-Zumino-Witten functional \cite{Wess:1971yu,Witten:1983tw}:
\begin{equation}
a_1^0\equiv\frac{1}{8\pi^2 F^2},\qquad a_2^0\equiv\frac{-1}{4\pi^2 F^2\left[\left(P-q\right)^2-m_\pi^2\right]}.
\end{equation}
The diagrams that receive contributions due to the anomaly are shown in fig. \ref{Ap:fig3}~\footnote{The first diagram, when coupled to a vector current, contributes to the SI piece in $V^{\mu\nu}$.}.

\begin{figure}[H]
\begin{center}
\begin{tikzpicture}
	\draw[scalar] (0,0) -- (1.5,1.5) node [right]{$\pi^-$} ;
	\draw[gauge] (0,0) -- (2.12,0) node[right]{$\gamma$};
	\draw[scalar] (0,0) -- (1.5,-1.5) node [right] {$\pi^0$};
	\draw[fill,white] (0,0) circle [radius=0.12];
	\node at (0,0) {\small$\otimes$};
\end{tikzpicture}
\begin{tikzpicture}
	\draw[scalar] (5.62,0)--(7.74,0);
	\node at (6.68,-0.4) {$\pi^-$};
	\draw[scalar] (7.74,0)--(9.24,1.5) node [right]{$\pi^-$} ;
	\draw[gauge] (7.74,0)--(9.86,0) node[right]{$\gamma$};
	\draw[scalar] (7.74,0)--(9.24,-1.5) node [right] {$\pi^0$};
	\draw[fill,white] (5.62,0) circle [radius=0.12];
	\node at (5.62,0) {\small$\otimes$};
\end{tikzpicture}
\end{center}
\caption{Anomalous diagrams contributing to the axial tensor amplitude $A^{\mu\nu}$ at $\mathcal{O}\left(p^4\right)$.}\label{Ap:fig3}
\end{figure}
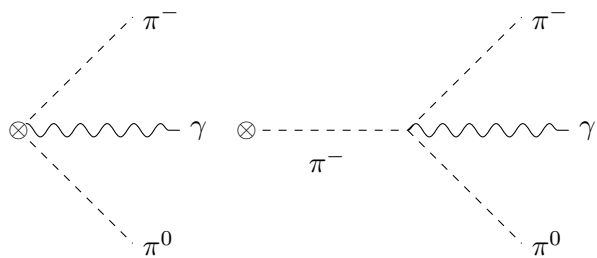
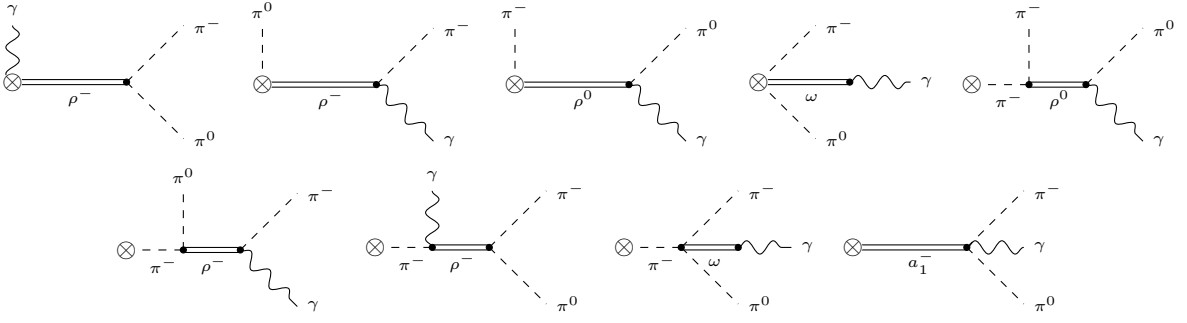
\begin{figure}[H]
\begin{center}
\begin{tikzpicture}
	\draw[resonance] (2,0)--(3.5,0);
	\node at (2.9,-0.2) {\tiny$\rho^-$};
	\draw[gauge] (2,0)--(2,0.75) node[above] {\tiny$\gamma$};
	\draw[fill,white] (2,0) circle [radius=0.12];
	\node at (2,0) {\small$\otimes$};
	\draw[scalar] (3.5,0)--(4.25,0.75) node[right] {\tiny$\pi^-$};
	\draw[scalar] (3.5,0)--(4.25,-0.75) node[right] {\tiny$\pi^0$};
	\draw[fill] (3.5,0) circle [radius=0.04];
\end{tikzpicture}
\begin{tikzpicture}
	\draw[resonance] (8.5,0)--(10,0);
	\node at (9.4,-0.2) {\tiny$\rho^-$};
	\draw[scalar] (8.5,0)--(8.5,0.75) node[above] {\tiny$\pi^0$};
	\draw[fill,white] (8.5,0) circle [radius=0.12];
	\node at (8.5,0) {\small$\otimes$};
	\draw[scalar] (10,0)--(10.75,0.75) node[right] {\tiny$\pi^-$};
	\draw[gauge] (10,0)--(10.75,-0.75) node[right] {\tiny$\gamma$};
	\draw[fill] (10,0) circle [radius=0.04];
\end{tikzpicture}
\begin{tikzpicture}
	\draw[resonance] (-1.5,-2)--(0,-2);
	\node at (-0.6,-2.2) {\tiny $\rho^0$};
	\draw[scalar] (-1.5,-2)--(-1.5,-1.25) node[above] {\tiny$\pi^-$};
	\draw[fill,white] (-1.5,-2) circle [radius=0.12];
	\node at (-1.5,-2) {\small$\otimes$};
	\draw[scalar] (0,-2)--(0.75,-1.25) node[right] {\tiny$\pi^0$};
	\draw[gauge] (0,-2)--(0.75,-2.75) node[right] {\tiny$\gamma$};
	\draw[fill] (0,-2) circle [radius=0.04];
\end{tikzpicture}
\begin{tikzpicture}
	\draw[resonance] (5.5,-2)--(6.7,-2);
	\node at (6.2,-2.2) {\tiny$\omega$};
	\draw[scalar] (5.5,-2)--(6.25,-1.25) node[right] {\tiny$\pi^-$};
	\draw[scalar] (5.5,-2)--(6.25,-2.75) node[right] {\tiny $\pi^0$};
	\draw[fill,white] (5.5,-2) circle [radius=0.12];
	\node at (5.5,-2) {\small $\otimes$};
	\draw[gauge] (6.7,-2)--(7.5,-2) node[right] {\tiny$\gamma$};
	\draw[fill] (6.7,-2) circle [radius=0.04];
\end{tikzpicture}
\begin{tikzpicture}
	\draw[resonance] (9.25,-2)--(10,-2);
	\node at (9.65,-2.2) {\tiny $\rho^0$};
	\draw[scalar] (8.5,-2)--(9.25,-2);
	\node at (8.99,-2.2) {\tiny $\pi^-$};
	\draw[fill] (9.25,-2) circle [radius=0.04];
	\draw[fill,white] (8.5,-2) circle [radius=0.12];
	\node at (8.5,-2) {\small $\otimes$};
	\draw[scalar] (9.25,-2)--(9.25,-1.25) node[above] {\tiny $\pi^-$};
	\draw[scalar] (10,-2)--(10.75,-1.25) node[right] {\tiny $\pi^0$};
	\draw[gauge] (10,-2)--(10.75,-2.75) node[right] {\tiny $\gamma$};
	\draw[fill] (10,-2) circle [radius=0.04];
\end{tikzpicture}
\begin{tikzpicture}
	\draw[scalar] (-1.5,-4)--(-0.75,-4);
	\node at (-1.01,-4.2) {\tiny $\pi^-$};
	\draw[resonance] (-0.75,-4)--(0,-4);
	\node at (-0.35,-4.2) {\tiny $\rho^-$};
	\draw[fill,white] (-1.5,-4) circle [radius=0.12];
	\node at (-1.5,-4) {\small $\otimes$};
	\draw[scalar] (-0.75,-4)--(-0.75,-3.25) node[above] {\tiny$\pi^0$};
	\draw[fill] (-0.75,-4) circle [radius=0.04];
	\draw[scalar] (0,-4)--(0.75,-3.25) node[right] {\tiny$\pi^-$};
	\draw[gauge] (0,-4)--(0.75,-4.75) node[right] {\tiny$\gamma$};
	\draw[fill] (0,-4) circle [radius=0.04];
\end{tikzpicture}
\begin{tikzpicture}
	\draw[scalar] (2,-4)--(2.75,-4);
	\node at (2.49,-4.2) {\tiny $\pi^-$};
	\draw[resonance] (2.75,-4)--(3.5,-4);
	\node at (3.15,-4.2) {\tiny $\rho^-$};
	\draw[fill,white] (2,-4) circle [radius=0.12];
	\node at (2,-4) {\small $\otimes$};
	\draw[gauge] (2.75,-4)--(2.75,-3.25) node[above] {\tiny$\gamma$};
	\draw[fill] (2.75,-4) circle [radius=0.04];
	\draw[scalar] (3.5,-4)--(4.25,-3.25) node[right] {\tiny$\pi^-$};
	\draw[scalar] (3.5,-4)--(4.25,-4.75) node[right] {\tiny$\pi^0$};
	\draw[fill] (3.5,-4) circle [radius=0.04];
\end{tikzpicture}
\begin{tikzpicture}
	\draw[scalar] (5.5,-4)--(6.25,-4);
	\node at (5.99,-4.2) {\tiny $\pi^-$};
	\draw[resonance] (6.25,-4)--(7,-4);
	\node at (6.69,-4.2) {\tiny $\omega$};
	\draw[fill,white] (5.5,-4) circle [radius=0.12];
	\node at (5.5,-4) {\small $\otimes$};
	\draw[scalar] (6.25,-4)--(7,-3.25) node[right] {\tiny$\pi^-$};
	\draw[scalar] (6.25,-4)--(7,-4.75) node[right] {\tiny$\pi^0$};
	\draw[fill] (6.25,-4) circle [radius=0.04];
	\draw[gauge] (7,-4)--(7.7,-4) node[right] {\tiny$\gamma$};
	\draw[fill] (7,-4) circle [radius=0.04];
\end{tikzpicture}
\begin{tikzpicture}
	\draw[resonance] (2,0)--(3.5,0);
	\node at (2.9,-0.2) {\tiny $a_1^-$};
	\draw[fill,white] (2,0) circle [radius=0.12];
	\node at (2,0) {\small$\otimes$};
	\draw[scalar] (3.5,0)--(4.25,0.75) node[right] {\tiny$\pi^-$};
	\draw[gauge] (3.5,0)--(4.25,0) node[right] {\tiny $\gamma$};
	\draw[scalar] (3.5,0)--(4.25,-0.75) node[right] {\tiny $\pi^0$};
	\draw[fill] (3.5,0) circle [radius=0.04];
\end{tikzpicture}
\end{center}
\caption{One-resonance exchange contributions from the $R\chi T$ to the axial-vector form factors of the $\tau^-\to\pi^-\pi^0\gamma\nu_\tau$ decays.}\label{AF:fig1}
\end{figure}

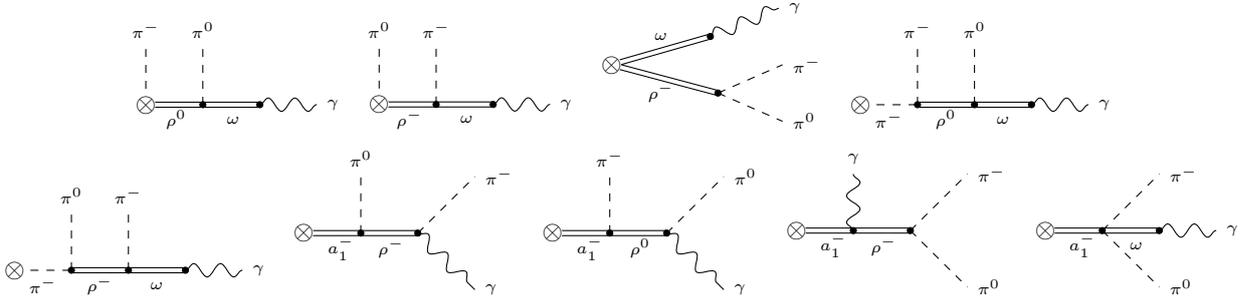
\begin{figure}[H]
\begin{center}
\begin{tikzpicture}
	\draw[resonance] (9,0)--(10.5,0);
	\node at (9.4,-0.2) {\tiny$\rho^0$};
	\node at (10.15,-0.2) {\tiny$\omega$};
	\draw[scalar] (9,0)--(9,0.75) node[above] {\tiny$\pi^-$};
	\draw[fill,white] (9,0) circle [radius=0.12];
	\node at (9,0) {\small$\otimes$};
	\draw[scalar] (9.75,0)--(9.75,0.75) node[above] {\tiny$\pi^0$};
	\draw[fill] (9.75,0) circle [radius=0.04];
	\draw[gauge] (10.5,0)--(11.25,0) node[right] {\tiny$\gamma$};
	\draw[fill] (10.5,0) circle [radius=0.04];
\end{tikzpicture}
\begin{tikzpicture}
	\draw[resonance] (-1.5,-2)--(0,-2);
	\node at (-1.1,-2.2) {\tiny$\rho^-$};
	\node at (-0.35,-2.2) {\tiny$\omega$};
	\draw[scalar] (-1.5,-2)--(-1.5,-1.25) node[above] {\tiny$\pi^0$};
	\draw[fill,white] (-1.5,-2) circle [radius=0.12];
	\node at (-1.5,-2) {\small$\otimes$};
	\draw[scalar] (-0.75,-2)--(-0.75,-1.25) node[above] {\tiny$\pi^-$};
	\draw[fill] (-0.75,-2) circle [radius=0.04];
	\draw[gauge] (0,-2)--(0.75,-2) node[right] {\tiny$\gamma$};
	\draw[fill] (0,-2) circle [radius=0.04];
\end{tikzpicture}
\begin{tikzpicture}
	\draw[resonance] (9,-2)--(10.3,-1.625);
	\node at (9.65,-1.62) {\tiny$\omega$};
	\draw[gauge] (10.3,-1.625)--(11.2,-1.25) node[right] {\tiny$\gamma$};
	\draw[fill] (10.3,-1.625) circle [radius=0.04];
	\draw[resonance] (9,-2)--(10.4,-2.375);
	\node at (9.65,-2.37) {\tiny$\rho^-$};
	\draw[fill,white] (9,-2) circle [radius=0.12];
	\node at (9,-2) {\small$\otimes$};
	\draw[scalar] (10.4,-2.375)--(11.25,-2) node[right] {\tiny$\pi^-$};
	\draw[scalar] (10.4,-2.375)--(11.25,-2.75) node[right] {\tiny$\pi^0$};
	\draw[fill] (10.4,-2.375) circle [radius=0.04];
\end{tikzpicture}
\begin{tikzpicture}
	\draw[scalar] (-0.75,-4)--(0,-4);
	\draw[fill,white] (-0.75,-4) circle [radius=0.12];
	\node at (-0.75,-4) {\small$\otimes$};
	\node at (-0.375,-4.2) {\tiny$\pi^-$};
	\draw[scalar] (0,-4)--(0,-3.25) node[above] {\tiny$\pi^-$};
	\draw[resonance] (0,-4)--(1.5,-4);
	\draw[fill] (0,-4) circle [radius=0.04];
	\node at (0.375,-4.2) {\tiny$\rho^0$};
	\node at (1.125,-4.2) {\tiny$\omega$};
	\draw[scalar] (0.75,-4)--(0.75,-3.25) node[above] {\tiny$\pi^0$};
	\draw[fill] (0.75,-4) circle [radius=0.04];
	\draw[gauge] (1.5,-4)--(2.25,-4) node[right] {\tiny$\gamma$};
	\draw[fill] (1.5,-4) circle [radius=0.04];
\end{tikzpicture}
\begin{tikzpicture}
	\draw[scalar] (3.5,-4)--(4.25,-4);
	\draw[fill,white] (3.5,-4) circle [radius=0.12];
	\node at (3.5,-4) {\small$\otimes$};
	\node at (3.875,-4.2) {\tiny$\pi^-$};
	\draw[scalar] (4.25,-4)--(4.25,-3.25) node[above] {\tiny$\pi^0$};
	\draw[resonance] (4.25,-4)--(5.75,-4);
	\draw[fill] (4.25,-4) circle [radius=0.04];
	\node at (4.625,-4.2) {\tiny$\rho^-$};
	\node at (5.375,-4.2) {\tiny$\omega$};
	\draw[scalar] (5,-4)--(5,-3.25) node[above] {\tiny$\pi^-$};
	\draw[fill] (5,-4) circle [radius=0.04];
	\draw[gauge] (5.75,-4)--(6.5,-4) node[right] {\tiny$\gamma$};
	\draw[fill] (5.75,-4) circle [radius=0.04];
\end{tikzpicture}
\begin{tikzpicture}
	\draw[resonance] (1.75,0)--(3.25,0);
	\node at (2.24,-0.2) {\tiny $a_1^-$};
	\node at (2.9,-0.2) {\tiny $\rho^-$};
	\draw[fill,white] (1.75,0) circle [radius=0.12];
	\node at (1.75,0) {\small$\otimes$};
	\draw[scalar] (2.5,0)--(2.5,0.75) node[above] {\tiny $\pi^0$};
	\draw[fill] (2.5,0) circle [radius=0.04];
	\draw[scalar] (3.25,0)--(4,0.75) node[right] {\tiny$\pi^-$};
	\draw[gauge] (3.25,0)--(4,-0.75) node[right] {\tiny $\gamma$};
	\draw[fill] (3.25,0) circle [radius=0.04];
\end{tikzpicture}
\begin{tikzpicture}
	\draw[resonance] (5,0)--(6.5,0);
	\node at (5.49,-0.2) {\tiny $a_1^-$};
	\node at (6.15,-0.2) {\tiny $\rho^0$};
	\draw[fill,white] (5,0) circle [radius=0.12];
	\node at (5,0) {\small$\otimes$};
	\draw[scalar] (5.75,0)--(5.75,0.75) node[above] {\tiny $\pi^-$};
	\draw[fill] (5.75,0) circle [radius=0.04];
	\draw[scalar] (6.5,0)--(7.25,0.75) node[right] {\tiny$\pi^0$};
	\draw[gauge] (6.5,0)--(7.25,-0.75) node[right] {\tiny $\gamma$};
	\draw[fill] (6.5,0) circle [radius=0.04];
\end{tikzpicture}
\begin{tikzpicture}
	\draw[resonance] (8.25,0)--(9.75,0);
	\node at (8.74,-0.2) {\tiny $a_1^-$};
	\node at (9.4,-0.2) {\tiny $\rho^-$};
	\draw[fill,white] (8.25,0) circle [radius=0.12];
	\node at (8.25,0) {\small$\otimes$};
	\draw[gauge] (9,0)--(9,0.75) node[above] {\tiny $\gamma$};
	\draw[fill] (9,0) circle [radius=0.04];
	\draw[scalar] (9.75,0)--(10.5,0.75) node[right] {\tiny$\pi^-$};
	\draw[scalar] (9.75,0)--(10.5,-0.75) node[right] {\tiny $\pi^0$};
	\draw[fill] (9.75,0) circle [radius=0.04];
\end{tikzpicture}
\begin{tikzpicture}
	\draw[resonance] (5,-2)--(6.5,-2);
	\node at (5.49,-2.2) {\tiny $a_1^-$};
	\node at (6.2,-2.2) {\tiny $\omega$};
	\draw[fill,white] (5,-2) circle [radius=0.12];
	\node at (5,-2) {\small$\otimes$};
	\draw[scalar] (5.75,-2)--(6.5,-1.25) node[right] {\tiny$\pi^-$};
	\draw[scalar] (5.75,-2)--(6.5,-2.75) node[right] {\tiny $\pi^0$};
	\draw[fill] (5.75,-2) circle [radius=0.04];
	\draw[gauge] (6.5,-2)--(7.25,-2) node[right] {\tiny $\gamma$};
	\draw[fill] (6.5,-2) circle [radius=0.04];
\end{tikzpicture}

\end{center}
\caption{Two-resonance exchange contributions from the $R\chi T$ to the axial-vector form factors of the $\tau^-\to\pi^-\pi^0\gamma\nu_\tau$ decays.}\label{AF:fig2}
\end{figure}

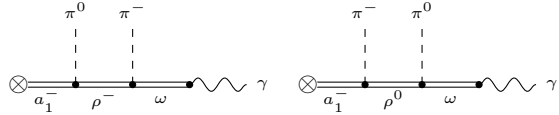
\begin{figure}[H]
\begin{center}
\begin{tikzpicture}
	\draw[resonance] (-0.75,-4)--(1.5,-4);
	\draw[fill,white] (-0.75,-4) circle [radius=0.12];
	\node at (-0.75,-4) {\small$\otimes$};
	\node at (-0.375,-4.2) {\tiny$a_1^-$};
	\draw[scalar] (0,-4)--(0,-3.25) node[above] {\tiny$\pi^0$};
	\draw[fill] (0,-4) circle [radius=0.04];
	\node at (0.375,-4.2) {\tiny$\rho^-$};
	\node at (1.125,-4.2) {\tiny$\omega$};
	\draw[scalar] (0.75,-4)--(0.75,-3.25) node[above] {\tiny$\pi^-$};
	\draw[fill] (0.75,-4) circle [radius=0.04];
	\draw[gauge] (1.5,-4)--(2.25,-4) node[right] {\tiny$\gamma$};
	\draw[fill] (1.5,-4) circle [radius=0.04];
\end{tikzpicture}
\begin{tikzpicture}
	\draw[resonance] (3.5,-4)--(5.75,-4);
	\draw[fill,white] (3.5,-4) circle [radius=0.12];
	\node at (3.5,-4) {\small$\otimes$};
	\node at (3.875,-4.2) {\tiny$a_1^-$};
	\draw[scalar] (4.25,-4)--(4.25,-3.25) node[above] {\tiny$\pi^-$};
	\draw[fill] (4.25,-4) circle [radius=0.04];
	\node at (4.625,-4.2) {\tiny$\rho^0$};
	\node at (5.375,-4.2) {\tiny$\omega$};
	\draw[scalar] (5,-4)--(5,-3.25) node[above] {\tiny$\pi^0$};
	\draw[fill] (5,-4) circle [radius=0.04];
	\draw[gauge] (5.75,-4)--(6.5,-4) node[right] {\tiny$\gamma$};
	\draw[fill] (5.75,-4) circle [radius=0.04];
\end{tikzpicture}
\end{center}
\caption{Three-resonance exchange contributions from the $R\chi T$ to the axial-vector form factors of the $\tau^-\to\pi^-\pi^0\gamma\nu_\tau$ decays.}\label{AF:fig3}
\end{figure}
For the axial form factors, we get
\begin{subequations}\small\begin{align}
a_1&=a_1^0+a_1^R+a_1^{RR}+a_1^{RRR},\\
a_2&=a_2^0+a_2^R+a_2^{RR}+a_2^{RRR},\\
a_3&=a_3^R+a_3^{RR}+a_3^{RRR},\\
a_4&=a_4^R+a_4^{RR}+a_4^{RRR},
\end{align}\end{subequations}
where $a_i^{R}$, $a_i^{RR}$ and $a_i^{RRR}$ include $\mathcal{O}\left(p^6\right)$ vertices. Due to their length, the expressions 
for these form factors appear in App. \ref{AF}.

\subsection{SD constraints}\label{subsec:SDC}
Including operators which start contributing to the $\mathcal{O}\left(p^6\right)$ LECs, we have now so many parameters (see Table \ref{SD:T1}) allowed by the discrete symmetries of QCD and chiral symmetry that, in practice, prevent making phenomenological predictions. It is possible to find relations between these couplings by means of SD properties of QCD and its OPE. We summarize these results in this section.

For the parameters contributing to $2$-point Green Functions (and related form factors), the constraints \cite{Ecker:1988te,Ecker:1989yg,Pich:2002xy,Weinberg:1967kj,Golterman:1999au,Jamin:2000wn,Jamin:2001zq}:
\begin{equation}\label{SD:eq01}\small\begin{split}
F_V G_V=F^2,&\qquad F_V^2-F_A^2=F^2,\\
F_V^2 M_V^2=F_A^2 M_A^2,&\qquad 4c_d c_m=F^2,\\
8\left(c_m^2-d_m^2\right)=F^2,&\qquad c_m=c_d=\sqrt{2}d_m=F/2
\end{split}\end{equation}
are set, respectively, by the known asymptotic behaviour of: the pion vector form factor, the $V-A$ correlator (yielding the Weinberg sum rules), the scalar form factor and the $S-P$ correlator.

We note that the vanishing of the axial pion form factor (giving the $\pi$-to-$\gamma$ matrix element) at infinite momentum transfer demands -if only the original $R\chi T$ Lagrangian \cite{Ecker:1989yg} is used- $2 F_V G_V = F_V^2$. This, together with the two first eqs. in (\ref{SD:eq01}), determine
\begin{equation}\label{SDCEN}
F_V= \sqrt{2} F\,,\quad G_V = \frac{F}{\sqrt{2}}\,,\quad F_A=F\,,
\end{equation}
all in terms of the pion decay constant. These relations were employed in ref. \cite{Cirigliano:2002pv}. We emphasize that -once  operators contributing to the NLO chiral LECs are considered \cite{Cirigliano:2006hb, Kampf:2011ty}-, the relations (\ref{SDCEN}) no longer hold true (see ref. \cite{Roig:2013baa}). Seen from another perspective, consistent sets of SD relations on $n$-point Green functions vary with $n$. For $n=2$ one has the set (\ref{SDCEN}) \cite{Ecker:1988te}. We shall also consider the set obtained for $n=3$ \cite{Cirigliano:2006hb, Kampf:2011ty, Roig:2013baa} (where operators with more than one resonance field start to appear) in the following. We will come back to discussing the actual values of the $F_V,\, G_V\;\text{and}\;F_A$ couplings before closing this section, as they are essential to assess the error associated to the IB corrections computed in ref. \cite{Cirigliano:2002pv}.

Now, we consider $R\chi T$ operators which do not contribute to the NLO chiral LECs. For the even intrinsic parity sector \cite{Cirigliano:2006hb,Guevara:2018rhj}~\footnote{The corresponding coefficients are denoted $\lambda_i^R$, with the upper index showing the resonance fields involved.}:
\begin{equation}\label{SD:eq1}\small\begin{split}
\lambda_{13}^P&=0,\quad  \lambda_{17}^S=\lambda_{18}^S=0,\\
\lambda_{17}^A&=0, \quad \lambda_{6}^V=\lambda_{21}^V=\lambda_{22}^V=0,
\end{split}\end{equation}
using these SD constraints in eq. (\ref{VF:eq06}) and the Brodsky-Lepage behaviour \cite{Brodsky:1973kr,Lepage:1980fj} of $f_+(s)$, we get:
\begin{equation}\small
2\lambda_{8}^V+\lambda_{9}^V+2\lambda_{10}^V=0.
\end{equation}
The study of the $\left\langle VAP\right\rangle$ and $\left\langle SPP\right\rangle$ Green functions yield the following restrictions on the resonance couplings \cite{Cirigliano:2006hb,Cirigliano:2004ue,Cirigliano:2005xn} (the Weinberg sum rules in eq. (\ref{SD:eq01}) were used below):
\begin{equation}\label{SD:eq2}\small\begin{split}
\sqrt{2}\lambda_0 =-4\lambda_1^{VA}-\lambda_2^{VA}-\frac{\lambda_4^{VA}}{2}-\lambda_5^{VA}&=\frac{1}{2\sqrt{2}}\left(\lambda^\prime+\lambda^{\prime\prime}\right),\\
\sqrt{2}\lambda^\prime=\lambda_2^{VA}-\lambda_3^{VA}+\frac{\lambda_4^{VA}}{2}+\lambda_5^{VA}&=\frac{F_V}{2\sqrt{F_V^2-F^2}},\\
\sqrt{2}\lambda^{\prime\prime}=\lambda_2^{VA}-\frac{\lambda_4^{VA}}{2}-\lambda_5^{VA}&=\frac{2F^2-F_V^2}{2F_V\sqrt{F_V^2-F^2}},\\
\lambda_1^{PV}=-4\lambda_2^{PV}=-\frac{F^2}{4\sqrt{2}d_m F_V},\quad \lambda_1^{PA}&=\frac{F^2}{16\sqrt{2}d_m \sqrt{F_V^2-F^2}}.
\end{split}\end{equation}
For the odd-intrinsic parity sector \cite{Kampf:2011ty}~\footnote{The corresponding coefficients are denoted $\kappa_i^R$, with the upper index showing the resonance fields involved.}:
\begin{equation}\label{sd:eq1}\small\begin{split}
\kappa_{14}^V&=\frac{N_C}{256\sqrt{2}\pi^2 F_V},\quad 2\kappa_{12}^V+\kappa_{16}^V=-\frac{N_C}{32\sqrt{2}\pi^2F_V},\quad \kappa_{17}^V=-\frac{N_C}{64\sqrt{2}\pi^2 F_V},\quad \kappa_5^P=0,\\
\kappa^{VV}_2&=\frac{F^2+16\sqrt{2}d_m F_V \kappa_3^{PV}}{32F_V^2}-\frac{N_C M_V^2}{512 \pi^2 F_V^2}, \quad 8\kappa_2^{VV}-\kappa_3^{VV}=\frac{F^2}{8F_V^2}.
\end{split}\end{equation}
The analysis of the $\left\langle VAS \right\rangle$ Green's function yields \cite{Kampf:2011ty}:
\begin{equation}\label{SD:eq3}\small\begin{split}
\kappa_2^S=\kappa_{14}^A=0,\quad \kappa_4^V&=2\kappa_{15}^V,\quad \kappa_6^{VA}=\frac{F^2}{32F_AF_V},\\
F_V\left(2\kappa_1^{SV}+\kappa_2^{SV}\right)&=2F_A\kappa_1^{SA}=\frac{F^2}{16\sqrt{2}c_m}\,,
\end{split}\end{equation}
and through the study of the $\left\langle VVA \right\rangle$ Green's function in ref. \cite{Roig:2019reh}:
\begin{equation}\small\label{SD:eq015}
F_V \kappa_5^{VA}=-\frac{N_C M_V^2}{64\pi^2F_A}.
\end{equation}
A comparison between two basis for the odd-intrinsic operators \cite{RuizFemenia:2003hm,Kampf:2011ty} was given in ref. \cite{Roig:2013baa}, which is consistent with those in eq. (\ref{sd:eq1})~\footnote{We note, particularly, the last of these eqs., which is at odds with (\ref{SDCEN}).}
\begin{equation}\label{SD:eq4}\small\begin{array}{rcl}
M_V\left(2\kappa_{12}^V+4\kappa_{14}^V+\kappa_{16}^V-\kappa_{17}^V\right)=&4c_3+c_1 &=0,\\
M_V\left(2\kappa_{12}^V+\kappa_{16}^V-2\kappa_{17}^V\right)=&c_1-c_2+c_5&=0,\\
-M_V\kappa_{17}^V=&c_5-c_6&=\frac{N_C M_V}{64\sqrt{2}\pi^2F_V},\\
M_V\, \kappa_{15}^V=&c_{4} &,\\
8\kappa_2^{VV}=&d_1+8d_2&=\frac{F^2}{8F_V^2}-\frac{N_C M_V^2}{64\pi^2F_V^2},\\
\kappa_3^{VV}=& d_3&=-\frac{N_C}{64\pi^2}\frac{M_V^2}{F_V^2},\\
& 1+\frac{32\sqrt{2}F_V d_m\kappa_3^{PV}}{F^2}&=0,\\
& F_V^2 &=3F^2,\\
\end{array}\end{equation}
For the even- and odd-intrinsic parity sectors, there are 115 (EIP)+67 (OIP)=182 operators saturating the $\mathcal{O}\left(p^6\right)$ LECs but only a few of them contribute to a given process. The form factors of the $\tau^-\to\pi^-\pi^0\gamma\nu_\tau$ decays at $\mathcal{O}\left(p^6\right)$ are given by 32 (EIP)+23(OIP)=55 operators (Table \ref{SD:T1}). Taking into account the relations in eqs. (\ref{SD:eq1})-(\ref{SD:eq4}) we get 24 (EIP)+17 (OIP)=41 undetermined couplings.\\
\begin{table}[htbp]
\begin{center}
\begin{tabular}{|c|c|}
\hline
\multicolumn{2}{|c|}{Even-intrinsic parity (EIP)\cite{Cirigliano:2006hb}}\\
\hline
$\hat{O}_i^V$ & 6,7,8,9,10,12,13,14,15,16,17,18,19,20,21,22\\
$\hat{O}_i^A$ & 4,12,13,15,16,17\\
$\hat{O}_i^{VV}$ & 2,3,4,5,7\\
$\hat{O}_i^{VA}$ & 1,2,3,4,5\\
\hline
\multicolumn{2}{|c|}{Odd-intrinsic parity (OIP)\cite{Kampf:2011ty}}\\
\hline
$\hat{O}_{i\mu\nu\alpha\beta}^V$ & 1,2,3,6,7,8,9,10,11,12,14,16,17\\
$\hat{O}_{i\mu\nu\alpha\beta}^A$ & 5,6,7\\
$\hat{O}_{i\mu\nu\alpha\beta}^{VV}$ & 2,3,4\\
$\hat{O}_{i\mu\nu\alpha\beta}^{VA}$ & 2,3,4,5\\
\hline
\end{tabular}
\caption{Operators contributing at $\mathcal{O}\left(p^6\right)$ to the vector and axial-vector form factors.}
\label{SD:T1}
\end{center}
\end{table}
In order to estimate the unknown parameters, we basically followed (but for the results in appendix \ref{Fit}) the strategy devised in ref. \cite{Guevara:2016trs}. We will restore to the available phenomenological information on these couplings and estimate -based on chiral counting- those for which we lack it.

Eq. (\ref{SD:eq2}) leaves two $\lambda_i^{VA}$ couplings undetermined, the numerical values of the restricted combinations (see their definitions in terms of the $\lambda_i^{VA}$ in \cite{Cirigliano:2004ue}) are:
\begin{equation}\label{SD:eq6}\small
\lambda^\prime\sim 0.4,\quad \lambda^{\prime\prime}\sim -0.14, \quad \lambda_0\sim 0.07.
\end{equation}
Since the same linear combination of $\lambda_4^{VA}$ and $\lambda_5^{VA}$ is in all couplings in eq. (\ref{SD:eq6}), we choose $\lambda_4^{VA}$ as independent. By similar reasons we take $\lambda_2^{VA}$ as the other independent coupling. Based on eq. (\ref{SD:eq6}), we  conservatively estimate $\left\vert \lambda_2^{VA}\right\vert \sim \left\vert \lambda_4^{VA}\right\vert \leq0.4$.

According to ref. \cite{Cirigliano:2006hb} the $\lambda_i^X$ couplings can be estimated from low energy couplings $C_i^R$ of the $\mathcal{O}\left(p^6\right)$ $\chi PT$ Lagrangian as~\footnote{Couplings of operators with two resonance fields are dimensionless \cite{Cirigliano:2006hb, Kampf:2011ty}.} 
\begin{equation}\small\begin{split}
\vert\lambda_i^V\vert&\sim \frac{3M_V^2}{2F}C_i^R\sim 0.025\,\mathrm{GeV}^{-1},\\
\vert\lambda_i^{VV}\vert&\sim \frac{M_V^4}{2F^2}C_i^R\sim 0.1,
\end{split}\end{equation}
where we take the relation $\vert C_i^R\vert\sim\frac{1}{F^2\left(4\pi\right)^4}$ linked to $\vert L_i^R\vert\sim\frac{1}{\left(4\pi\right)^2}\sim5\cdot10^{-3}$ which corresponds to the typical size of the $\mathcal{O}\left(p^4\right)$ LECs. This sets a reasonable upper bound on $\vert \lambda_i^V\vert\sim\vert \lambda_i^A\vert\lesssim 0.025\,\mathrm{GeV}^{-1}$ and $\vert\lambda_i^{VV}\vert\sim\vert\lambda_i^{VA}\vert\lesssim 0.1$.

For the anomalous sector, we have the following predictions from the eq. (\ref{SD:eq4}): $-M_V\kappa_{17}^V=c_5-c_6\sim 0.016$, $8\kappa^{VV}_2=d_1+8d_2\sim -0.070$ and $\kappa_3^{VV}=d_3\sim -0.112$. There is a sign ambiguity on the determination of $c_3$ from $\tau^-\to\eta\pi^-\pi^0\nu_\tau$ decays \cite{Dumm:2012vb}. We will take $c_3=0.007^{+0.020}_{-0.012}$ according to the determinations by Y. H. Chen {\it et al.} in refs. \cite{Chen:2012vw,Chen:2013nna,Chen:2014yta} (which is also in agreement with the most elaborated $e^+e^-\to(\eta/\pi^0)\pi^+\pi^-$ fit \cite{Dai:2013joa}). Although $c_4$ was first evaluated by studying $\sigma(e^+ e^-\to KK\pi)$ in ref. \cite{Dumm:2009va}, this yielded an inconsistent result for $\tau^-\to K^-\gamma\nu_\tau$ branching ratio \cite{Guo:2010dv}, so we will use $c_4=-0.0024\pm0.0006$ \cite{Chen:2013nna} as the most reliable estimation. In view of all these results, we will take $\vert c_i\vert \lesssim 0.015$ as a reasonable estimate, which is translated to $\vert\kappa_i^V\vert \lesssim 0.025\, \mathrm{GeV}^{-1}$. Since there is not enough information on $\kappa_i^A$, we will take $\vert\kappa_i^A\vert\sim \vert\kappa_i^V\vert\lesssim 0.025\, \mathrm{GeV}^{-1}$. We will see in the following sections that the observables that we consider and the IB corrections for $a_\mu^{HVP,LO|_{\pi\pi}}$ depend mostly on the $\kappa_i^V$ couplings (besides $F_V$, $G_V$ and $F_A$) for this reason we perform a global fit to better bind these couplings (see App. \ref{Fit})~\footnote{The results obtained assuming $\vert\kappa_i^V\vert \lesssim 0.025\, \mathrm{GeV}^{-1}$ can be found in https://arxiv.org/abs/2007.11019v1. While both results agree remarkably, the errors are reduced in the current procedure.}.

We turn now to the remaining couplings. We will employ $d_2=0.08\pm0.08$, which has been determined simultaneously with $c_3$ \cite{Chen:2012vw,Chen:2013nna,Chen:2014yta,Dai:2013joa}. For $d_4$ we will assume $\vert d_4\vert <0.15$, or in terms of $\kappa_i^{VV}$, we get $\vert \kappa_i^{VV}\vert\lesssim0.1$. Again we will adopt $\vert\kappa_i^{VA}\vert\sim\vert \kappa_i^{VV}\vert\lesssim0.1$, which agrees with the prediction $\kappa_{5}^{VA}\sim-0.14$ in eq. (\ref{SD:eq015}).

Using only operators contributing to the $\mathcal{O}\left(p^4\right)$ LECs we have the consistent set for $2$-point Green functions (\ref{SDCEN}). However, including operators which start contributing at $\mathcal{O}\left(p^6\right)$, we shall use the relations for $2$ and $3$-point Green functions (eq. (\ref{SD:eq01}) and eqs. (\ref{SD:eq1}) to (\ref{SD:eq4})). In particular, $F_V=\sqrt{3}F$, which implies (via (\ref{SD:eq01})) $G_V=F/\sqrt{3}$ and $F_A=\sqrt{2}F$. Therefore, we will also be showing the Cirigliano \textit{et al.} results \cite{Cirigliano:2002pv} with the latter set of constraints (inconsistent for $2$-point Green functions) so that the impact of the change of {$F_V,\,F_A \text{ and } G_V$}  between these two cases is appreciated.

\textbf{We will refer to the original \cite{Cirigliano:2002pv} constraints (\ref{SDCEN}) as `$F_V=\sqrt{2}F$' and by `$F_V=\sqrt{3}F$' to their consistent set of values ($F_V=\sqrt{3}F,\, G_V=F/\sqrt{3},\, F_A=\sqrt{2}F$) up to $3$-point Green functions}. In this last way, we stress that the consistent set of SD constraints in both parity sectors  \cite{Cirigliano:2004ue,Cirigliano:2006hb,Kampf:2011ty,Roig:2013baa} determines the $F_V=\sqrt{3}F$ relations (among many others, reviewed in this section).

\section{Radiative corrections for hadronic vacuum polarization\label{GEM}}

The four-body differential decay width is given by \cite{Cirigliano:2002pv}~\footnote{Although the analytical results in this section were presented in the quoted reference, we include them here given their importance in the evaluation of the relevant IB corrections, and take advantage to add a few explanations to previous discussions of this subject \cite{Cirigliano:2002pv, FloresTlalpa:2006gs}.}
\begin{equation}
 d\Gamma = \frac{\left(2\pi\right)^4}{2m_\tau}\overline{\left\vert\mathcal{M}\right\vert^2} \delta^{4}\left(P-p_--p_0-k-q\right)\frac{d^3 p_-}{\left(2\pi\right)^3 2E_-}\frac{d^3 p_0}{\left(2\pi\right)^3 2E_0}\frac{d^3 q}{\left(2\pi\right)^3 2E_\nu}\frac{d^3 k}{\left(2\pi\right)^3 2E_\gamma},
\end{equation}
using the relation $\frac{d^3 p_-}{2E_-}\frac{d^3 p_0}{2E_0}=\frac{\pi^2}{4 m_\tau^2}\,ds\,du\,dx$ and integrating over the three-momentum of the photon and neutrino ~\footnote{The kinematics for these decays are in App. \ref{kin}.}, we get
\begin{equation}\label{Appx4:eq20}
 d\Gamma=\frac{1}{32\left(2\pi\right)^6 m_\tau^3}\left[\int\frac{d^3 q}{2E_\nu}\frac{d^3 k}{2E_\gamma}\, \overline{\left\vert\mathcal{M}\right\vert^2} \delta^{4}\left(P-p_--p_0-k-q\right)\right]ds\,du\,dx,
\end{equation}
working at leading order in the Low expansion and in the isospin limit $m_u=m_d$, we have
\begin{equation}
 \mathcal{M}=e\,\epsilon^{*\mu}\left(k\right) \mathcal{M}^{(0)}_{\pi\pi}\left(\frac{p_{-\mu}}{p_-\cdot k}-\frac{P_{\mu}}{P\cdot k}\right)+\mathcal{O}\left(k^0\right),
\end{equation}
where $\mathcal{M}^{(0)}_{\pi\pi}=G_F V_{ud}^{*}\sqrt{S_{EW}} f_+\left(s\right)\left(p_--p_0\right)_\nu \bar{u}\left(q\right)\gamma^\nu \left(1-\gamma_5\right)u\left(P\right)$ is the amplitude at leading order for the non-radiative decay that includes the SD electroweak radiative corrections ($S_{EW}$). At $\mathcal{O}\left(k^{-1}\right)$, the amplitude for the radiative decay is proportional to the amplitude of the non-radiative decay according to the Low's theorem \cite{Low:1958sn}.

The unpolarized spin-averaged squared amplitude is given by
\begin{equation}\label{Appx4:eq22}\begin{split}
 \overline{\vert \mathcal{M}\vert^2}=&4\pi\alpha \overline{\vert\mathcal{M}^{(0)}_{\pi\pi}\vert^2}\sum_{\gamma}\epsilon^{*\mu}\left(k\right)\epsilon^\nu\left(k\right)\left(\frac{p_{-\mu}}{p_-\cdot k+\frac{1}{2}M_\gamma^2}-\frac{P_{\mu}}{P\cdot k-\frac{1}{2}M_\gamma^2}\right)\\
 &\times\left(\frac{p_{-\nu}}{p_-\cdot k+\frac{1}{2}M_\gamma^2}-\frac{P_{\nu}}{P\cdot k-\frac{1}{2}M_\gamma^2}\right)+\mathcal{O}\left(k^{-1}\right),
\end{split}\end{equation}
using the relation $\sum_{\gamma}\epsilon^{*\mu}\left(k\right)\epsilon^\nu\left(k\right)=-g^{\mu\nu}$ and massive photons ($k^\mu k_\mu=M_\gamma^2$). The sum over photon polarizations should include the longitudinal part, since our photon has mass and the amplitude is no longer gauge invariant. We do not take into account this contribution because it will vanish in the limit $M_\gamma \to 0$.

Thus, eq. (\ref{Appx4:eq22}) becomes
\begin{equation}\label{Appx4:eq23}\begin{split}
  \overline{\vert \mathcal{M}\vert^2}=&4\pi\alpha \overline{\vert\mathcal{M}^{(0)}_{\pi\pi}\vert^2}\left(\frac{2P\cdot p_-}{\left(p_-\cdot k+\frac{1}{2}M_\gamma^2\right)\left(P\cdot k-\frac{1}{2}M_\gamma^2\right)}-\frac{m_\pi^2}{\left(p_-\cdot k+\frac{1}{2}M_\gamma^2\right)^2}\right.\\
  &-\left.\frac{m_\tau^2}{\left(P\cdot k-\frac{1}{2}M_\gamma^2\right)^2}\right)+\mathcal{O}\left(k^{-1}\right),
\end{split}\end{equation}
where
\begin{equation}\label{Appx4:eq24}
 \overline{\vert\mathcal{M}^{(0)}_{\pi\pi}\vert^2}=4G_F^2 \left\vert V_{ud}\right\vert^2 S_{EW} \left\vert f_+\left(s\right)\right\vert^2 \left(D\left(s,u\right)+\mathcal{O}\left(k\right)\right),
\end{equation}
with $D(s,u)=\frac{1}{2}m_\tau^2\left(m_\tau^2-s\right)+2m_\pi^4-2u(m_\tau^2-s+2m_\pi^2)+2u^2$. Eq. (\ref{Appx4:eq23}) does not  contribute at $\mathcal{O}\left(k^{-1}\right)$, these terms are canceled out by those in eq. (\ref{Appx4:eq24}) according to the Burnett-Kroll theorem \cite{Burnett:1967km}.

Replacing eqs. (\ref{Appx4:eq23}) and (\ref{Appx4:eq24}) in eq. (\ref{Appx4:eq20}), we get
\begin{equation}\begin{split}
 d\Gamma=&\frac{\alpha G_F^2 \vert V_{ud}\vert^2 S_{EW}}{4(2\pi)^4m_\tau^3}\left\vert f_+\left(s\right)\right\vert^2 D\left(s,u\right)\left(2 P\cdot p_-\, I_{11}\left(s,u,x\right)-m_\pi^2\,I_{02}\left(s,u,x\right)\right.\\
 &\qquad\left.-m_\tau^2\,I_{20}\left(s,u,x\right)\right)ds\,du\,dx+\mathcal{O}\left(k^0\right),
\end{split}\end{equation}
the $I_{mn}\left(s,u,x\right)$ is defined as
\begin{equation}\label{Appx4:eq31}
 I_{mn}\left(s,u,x\right)=\frac{1}{2\pi}\int\frac{d^3 q}{2E_\nu}\frac{d^3 k}{2E_\gamma}\frac{\delta^{4}\left(P-p_--p_0-k-q\right)}{\left(P\cdot k-\frac{1}{2}M_\gamma^2\right)^m \left(p_-\cdot k+\frac{1}{2}M_\gamma^2\right)^n} ,
\end{equation}
performing an integration over $x$, we can split the decay width according to the integration region 
\begin{equation}
 \frac{d^2\Gamma}{ds\,du}=\left.\frac{d^2\Gamma}{ds\,du}\right\vert_{\mathcal{D}^{III}}+\left.\frac{d^2\Gamma}{ds\,du}\right\vert_{\mathcal{D}^{IV/III}}+\mathcal{O}\left(k^0\right),
\end{equation}
where
\begin{equation}\begin{split}
 \left.\frac{d^2\Gamma}{ds\,du}\right\vert_{\mathcal{D}^{III}}=&\frac{\alpha G_F^2 \vert V_{ud}\vert^2 S_{EW}}{4(2\pi)^4m_\tau^3}\left\vert f_+\left(s\right)\right\vert^2 D\left(s,u\right)\times\\
 &\left(J_{11}\left(s,u,M_\gamma\right)+J_{02}\left(s,u,M_\gamma\right)+J_{20}\left(s,u,M_\gamma\right)\right),
\end{split}\end{equation}
and
\begin{equation}\begin{split}
 \left.\frac{d^2\Gamma}{ds\,du}\right\vert_{\mathcal{D}^{IV/III}}=&\frac{\alpha G_F^2 \vert V_{ud}\vert^2 S_{EW}}{4(2\pi)^4m_\tau^3}\left\vert f_+\left(s\right)\right\vert^2 D\left(s,u\right)\times\\
 &\left(K_{11}\left(s,u\right)+K_{02}\left(s,u\right)+K_{20}\left(s,u\right)\right),
\end{split}\end{equation}
with
\begin{equation}
 J_{mn}\left(s,u,M_\gamma\right)=c_{mn}\int^{x_+(s,u)}_{M_\gamma^2}dx\,I_{mn}\left(s,u,x\right),
\end{equation}
\begin{equation}
 K_{mn}\left(s,u\right)=c_{mn}\int^{x_+(s,u)}_{x_-(s,u)}dx\,I_{mn}\left(s,u,x\right),
\end{equation}
and 
\begin{equation}
 c_{mn}=\left\lbrace \begin{array}{ll}
                       2P\cdot p_- & m=n=1,\\
                       -m_\tau^2 & m=2,\, n=0,\\
                       -m_{\pi^-}^2 & m=0,\, n=2.\\
                      \end{array}\right. 
\end{equation}

Eq. (\ref{Appx4:eq31}) is an invariant, so we can evaluate it in any reference frame in order to simplify the integration, working in the $\gamma-\nu_\tau$ center of mass, we have
\begin{equation}
 I_{mn}\left(s,u\right)=\frac{1}{2^3(2\pi)}\int \frac{x-M_\gamma^2}{x\left(P\cdot k-\frac{1}{2}M_\gamma^2\right)^m \left(p_-\cdot k+\frac{1}{2}M_\gamma^2\right)^n} d\cos\theta_\nu\, d\phi_-.
\end{equation}
Integrating this equation over $x$ in $\mathcal{D}_{IV/III}$ and $\mathcal{D}_{III}$, as in refs. \cite{Cirigliano:2002pv,FloresTlalpa:2008zz} we get ($\mathrm{Li}_2(x)=-\int^1_0 \frac{dt}{t}\log(1-xt)$)
\begin{equation}\label{Appx4:eq39}\begin{split}
 J_{11}(s,u)&=\log\left(\frac{2x_+(s,u)\bar{\gamma}}{M_\gamma}\right)\frac{1}{\bar{\beta}}\log\left(\frac{1+\bar{\beta}}{1-\bar{\beta}}\right)\\
 &+\frac{1}{\bar{\beta}}\left(Li_2(1/Y_2)-Li_2(Y_1)+\log^2(-1/Y_2)/4-\log^2(-1/Y_1)/4\right),
\end{split}\end{equation}
\begin{equation}
J_{20}\left(s,u\right)= \log\left(\frac{M_\gamma(m_\tau^2-s)}{m_\tau\,x_+(s,u)}\right), 
\end{equation}
\begin{equation}
 J_{02}\left(s,u\right)=\log\left(\frac{M_\gamma(m_\tau^2+m_{\pi^0}^2-s-u)}{m_\pi^-\,x_+(s,u)}\right),
\end{equation}
\begin{equation}
 K_{20}\left(s,u\right)=K_{0,2}\left(s,u\right)=\log\left(\frac{x_-(s,u)}{x_+(s,u)}\right)\,,
\end{equation}
where the expressions in eq. (\ref{Appx4:eq39}) are given by
\begin{equation}
Y_{1,2}=\frac{1-2\bar{\alpha}\pm\sqrt{(1-2\bar{\alpha})^2-(1-\bar{\beta}^2)}}{1+\bar{\beta}},
\end{equation}
with 
\begin{subequations}\begin{align*}
\bar{\alpha}&=\frac{(m_\tau^2-s)(m_\tau^2+m_{\pi^0}^2-s-u)}{(m_{\pi^-}^2+m_\tau^2-u)}\cdot \frac{\lambda(u,m_{\pi^-}^2,m_\tau^2)}{2\bar{\delta}},\\
\bar{\beta}&=-\frac{\sqrt{\lambda(u,m_{\pi^-}^2,m_\tau^2)}}{m_{\pi^-}^2+m_\tau^2-u},\\
\bar{\gamma}&=\frac{\sqrt{\lambda(u,m_{\pi^-}^2,m_\tau^2)}}{2\sqrt{\bar{\delta}}},\\
\bar{\delta}&=-m_{\pi^0}^4m_\tau^2+m_{\pi^-}^2(m_\tau^2-s)(m_{\pi^0}^2-u)-su(-m_\tau^2+s+u)\\
&\quad +m_{\pi^0}^2(-m_\tau^4+su+m_\tau^2 s+m_\tau^2 u).
\end{align*}\end{subequations}
Experimentally, it is impossible to measure the full photon spectrum because of acceptances, efficiencies and cuts. For this reason, we need to calculate the inclusive decay width, since we can not distinguish the radiative decay from the non-radiative decay for low-energy (or collinear) photons.

For the non-radiative decay, we have 
\begin{equation}
 \frac{d^2\Gamma}{ds\,du}=\frac{G_F^2\vert V_{ud}\vert^2 S_{EW}}{64\pi^3 m_\tau^3}\left\vert f_+(s) \right\vert^2\left(1+f^{elm}_{loop}\left(u,M_\gamma\right)\right)^2 D\left(s,u\right),
\end{equation}
that includes isospin violation and photonic corrections according to ref. \cite{Cirigliano:2001er}, where $f^{elm}_{loop}(u,M_\gamma)$ is given by
\begin{equation}\begin{split}
f^{elm}_{loop}\left(u,M_\gamma\right)=&\frac{\alpha}{4\pi}\left((u-m_{\pi}^2)\mathcal{A}(u)+(u-m_\pi^2-m_\tau^2)\mathcal{B}(u)\right.\\
&\qquad\left. +2(m_\pi^2+m_\tau^2-u)\mathcal{C}\left(u,M_\gamma\right)+2\log\frac{m_\pi m_\tau}{M_\gamma^2}\right),
\end{split}\end{equation}
with
\begin{subequations}\begin{align*}
\mathcal{A}(u)&=\frac{1}{u}\left(-\frac{1}{2}\log r_\tau +\frac{2-y_\tau}{\sqrt{r_\tau}}\frac{x_\tau}{1-x_\tau^2}\log x_\tau\right),\\
\mathcal{B}(u)&=\frac{1}{u}\left(\frac{1}{2}\log r_\tau +\frac{2r_\tau-y_\tau}{\sqrt{r_\tau}}\frac{x_\tau}{1-x_\tau^2}\log x_\tau\right),\\
\mathcal{C}(u,M_\gamma)&=\frac{1}{m_\tau m_\pi}\frac{x_\tau}{1-x_\tau^2}\left(-\frac{1}{2}\log^2x_\tau +2\log x_\tau\log\left(1-x_\tau^2\right)-\frac{\pi^2}{6}+\frac{1}{8}\log^2 r_\tau\right.\\
&\left.+Li_2\left(x_\tau^2\right)+Li_2\left(1-\frac{x_\tau}{\sqrt{r_\tau}}\right)+Li_2\left(1-x_\tau\sqrt{r_\tau}\right)-\log x_\tau\log\frac{M_\gamma^2}{m_\tau m_\pi}\right),
\end{align*}\end{subequations}
in terms of the variables
\[r_\tau=\frac{m_\tau^2}{m_\pi^2},\quad y_\tau=1+r_\tau-\frac{u}{m_\pi^2},\quad x_\tau=\frac{1}{2\sqrt{r_\tau}}\left(y_\tau-\sqrt{y_\tau^2-4r_\tau}\right),\]
Thus, the inclusive decay width is
\begin{equation}\label{Appx4:eq45}
\left.\frac{d^2\Gamma}{ds\,du}\right\vert_{\pi\pi(\gamma)}=\frac{G_F^2 \vert V_{ud}\vert^2 S_{EW}}{64\pi^3m_\tau^3}\left\vert f_+(s)\right\vert^2 D\left(s,u\right)\Delta\left(s,u\right),
\end{equation}
where
\begin{equation}\begin{split}
 \Delta\left(s,u\right)&=1+2f_{loop}^{elm}\left(u,M_\gamma\right)+g_{rad}\left(s,u,M_\gamma\right).
\end{split}\end{equation}
In the previous expression we neglected the quadratic term for $f^{elm}_{loop}\left(u,M_\gamma\right)$, and
\begin{equation}
 g_{rad}\left(s,u,M_\gamma\right)=g_{brems}\left(s,u,M_\gamma\right)+g_{rest}\left(s,u\right),
\end{equation}
with
\begin{subequations}\begin{align}
 g_{brems}\left(s,u,M_\gamma\right)&=\frac{\alpha}{\pi}\left(J_{11}(s,u,M_\gamma)+J_{20}(s,u,M_\gamma)+J_{02}(s,u,M_\gamma)\right),\\
 g_{rest}\left(s,u\right)&=\frac{\alpha}{\pi}\left(K_{11}(s,u)+K_{20}(s,u)+K_{02}(s,u)\right).
\end{align}\end{subequations}
Integrating eq. (\ref{Appx4:eq45}) over $u$, and using 
\[\int_{u_-(s)}^{u_+(s)}D\left(s,u\right)du=\frac{m_\tau^6}{6}\left(1-\frac{s}{m_\tau^2}\right)^2\left(1-\frac{4m_\pi^2}{s}\right)^{3/2}\left(1+\frac{2s}{m_\tau^2}\right),\]
we have
\begin{equation}\label{Appx4:eq49}\begin{split}
\left.\frac{d\Gamma}{ds}\right\vert_{\pi\pi(\gamma)}=&\frac{G_F^2 \vert V_{ud}\vert^2 m_\tau^3 S_{EW}}{384\pi^3}\left\vert f_+(s)\right\vert^2 \left(1-\frac{s}{m_\tau^2}\right)^2\left(1-\frac{4m_\pi^2}{s}\right)^{3/2}\times\\
&\qquad\left(1+\frac{2s}{m_\tau^2}\right)G_{EM}(s),
\end{split}\end{equation}
for this we follow the same notation as in ref. \cite{Cirigliano:2002pv},
\begin{equation}
 G_{EM}(s)=\frac{\int_{\mathcal{R}^{IV}}D\left(s,u\right)\Delta\left(s,u\right)du}{\int_{u_-(s)}^{u_+(s)}D\left(s,u\right)du}.
\end{equation}
We can split the electromagnetic correction factor ($G_{EM}(s)$) in two parts, $G_{EM}^{(0)}(s)$ and $G_{EM}^{rest}(s)$, the first one corresponds to taking $g_{rest}(s,u)\to 0$ and the second one is the remainder of $G_{EM}(s)$, 
\begin{subequations}\begin{align}\label{Appx4:eq51a}
 G^{(0)}_{EM}(s)=&\frac{\int_{\mathcal{R}^{III}}D\left(s,u\right)\left(1+2f_{loop}^{elm}\left(u,M_\gamma\right)+g_{brems}\left(s,u,M_\gamma\right)\right)du}{\int_{u_-(s)}^{u_+(s)}D\left(s,u\right)du},\\
 G_{EM}^{rest}(s)=&\frac{\int_{\mathcal{R}^{IV/III}}D\left(s,u\right)g^{rest}\left(s,u\right)du}{\int_{u_-(s)}^{u_+(s)}D\left(s,u\right)du}.
\end{align}\end{subequations}
In eq. (\ref{Appx4:eq51a}), the term $2f_{loop}^{elm}(u,M_\gamma)+g_{brems}(s,u,M_\gamma)$ is finite when $M_\gamma\to 0$, 
\begin{equation}\begin{split}
 2f_{loop}^{elm}(u,M_\gamma)+g_{brems}(s,u,M_\gamma)=&\frac{\alpha}{4\pi}\left((u-m_{\pi}^2)\mathcal{A}(u)+(u-m_\pi^2-m_\tau^2)\mathcal{B}(u)\right.\\
&\qquad\left. +2(m_\pi^2+m_\tau^2-u)\mathcal{C}\left(u\right)\right)\\
&+\frac{\alpha}{\pi}\left(J_{11}(s,u)+J_{20}(s,u)+J_{02}(s,u)\right).
\end{split}\end{equation}
In this limit, we have
\begin{equation}\begin{split}
 \mathcal{C}(u)&=\frac{1}{m_\tau m_\pi}\frac{x_\tau}{1-x_\tau^2}\left(-\frac{1}{2}\log^2x_\tau +2\log x_\tau\log\left(1-x_\tau^2\right)-\frac{\pi^2}{6}+\frac{1}{8}\log^2 r_\tau\right.\\
&\left.+Li_2\left(x_\tau^2\right)+Li_2\left(1-\frac{x_\tau}{\sqrt{r_\tau}}\right)+Li_2\left(1-x_\tau\sqrt{r_\tau}\right)\right),
\end{split}\end{equation}
\begin{equation}\begin{split}
 J_{11}(s,u)&=\frac{1}{2}\log\left(\frac{4x_+^2(s,u)\bar{\gamma}^2}{m_\pi m_\tau}\right)\frac{1}{\bar{\beta}}\log\left(\frac{1+\bar{\beta}}{1-\bar{\beta}}\right)\\
 &+\frac{1}{\bar{\beta}}\left(Li_2(1/Y_2)-Li_2(Y_1)+\log^2(-1/Y_2)/4-\log^2(-1/Y_1)/4\right),
\end{split}\end{equation}
\begin{equation}
 J_{20}(s,u)=\log\left(\frac{m_\tau^2-s}{x_+(s,u)}\right),
\end{equation}
\begin{equation}
 J_{02}(s,u)=\log\left(\frac{m_\tau^2+m_\pi^2-s-u}{x_+(s,u)}\right),
\end{equation}
where $x_+\left(s,u\right)$ is defined in eq. (\ref{Appx4:eq18}).

\begin{figure}[H]
\includegraphics[width=7.4cm]{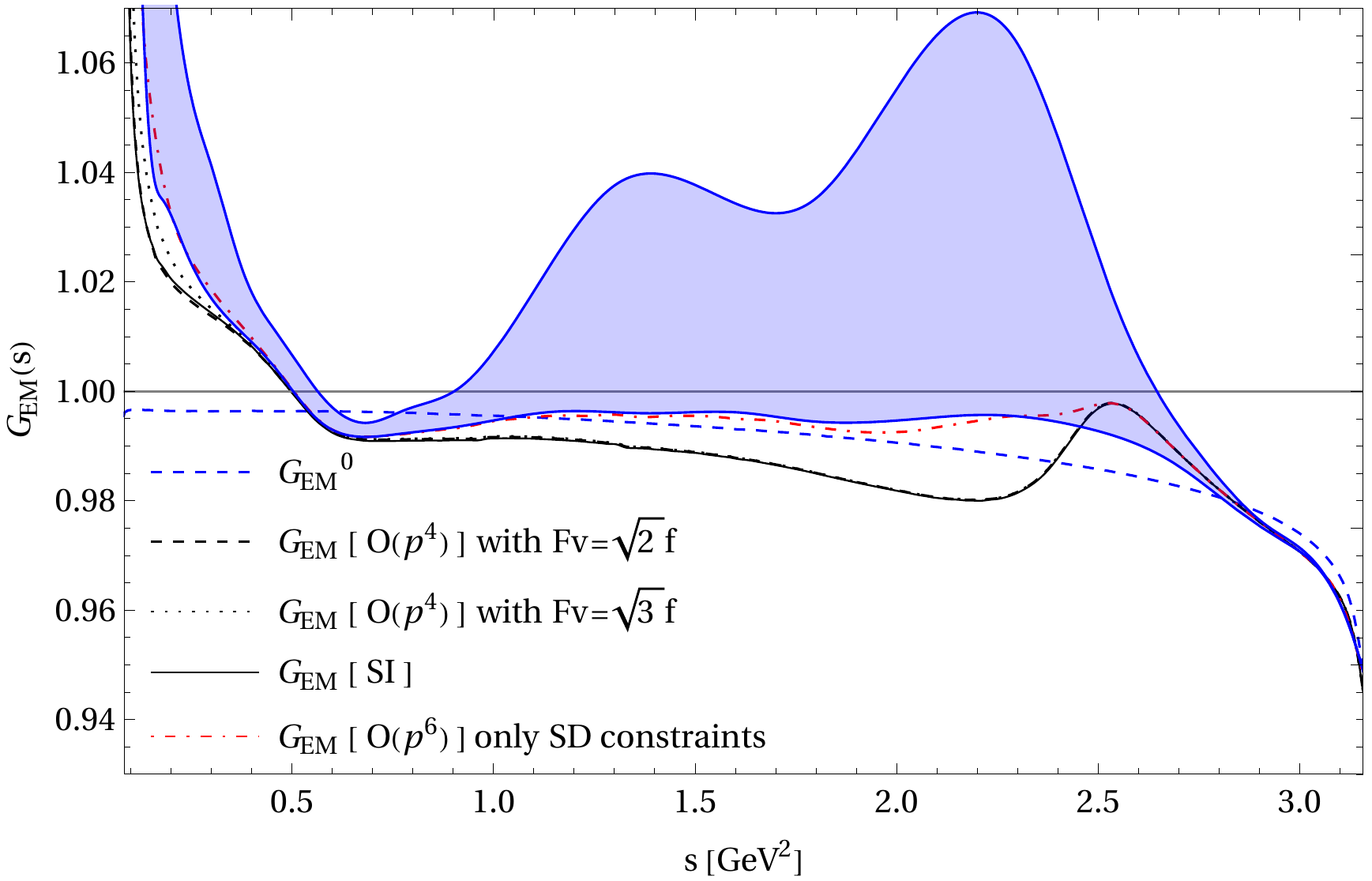}	
\includegraphics[width=7.4cm]{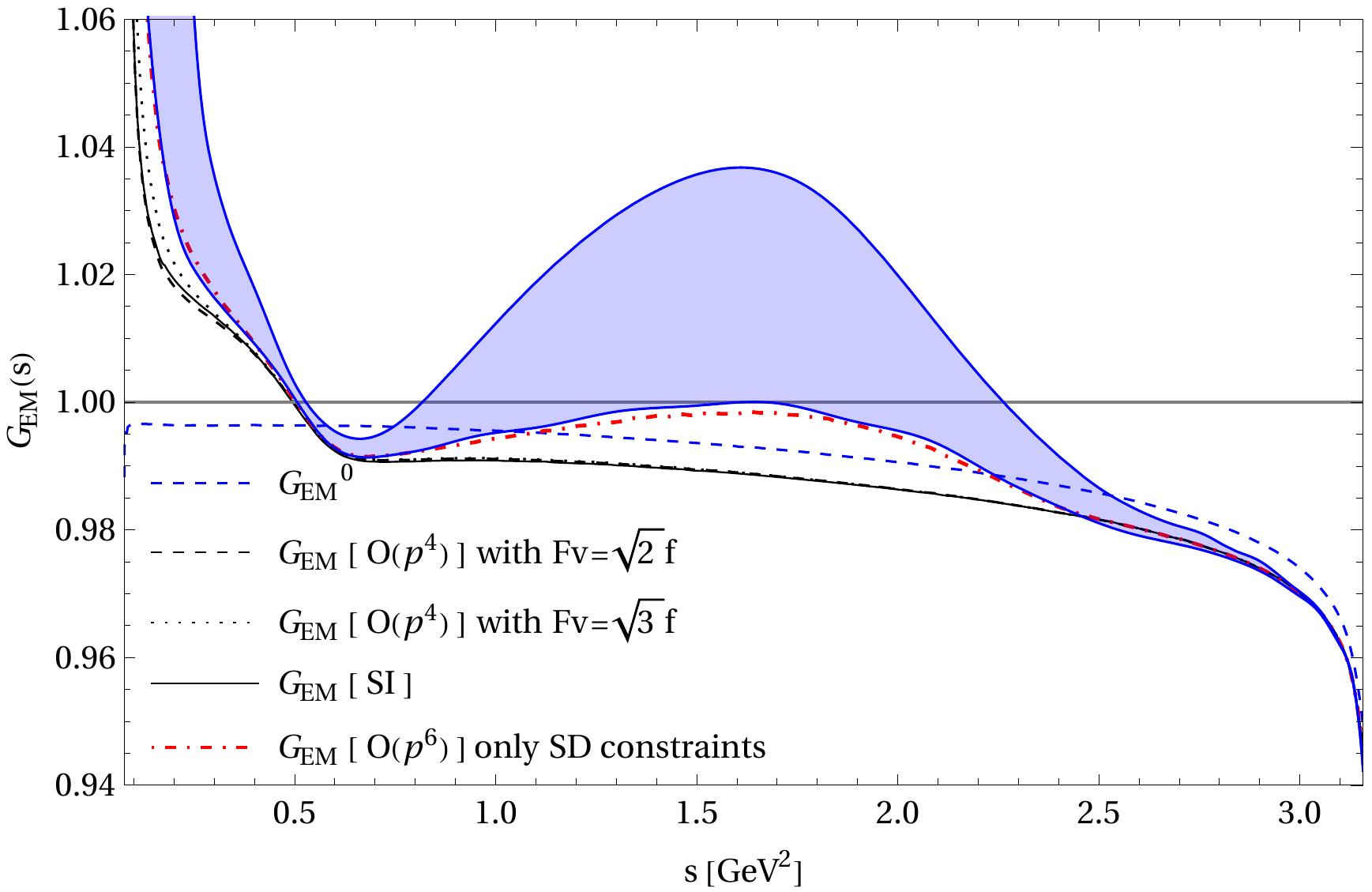}	
\centering			
\caption{Correction function $G_{EM}^{(0)}\left(s\right)$ in eq. (\ref{Appx4:eq51a}) (blue dashed line). The solid line shows the $G_{EM}(s)$ function neglecting the structure-dependent part (SI), i.e. by taking $v_1=v_2=v_3=v_4=a_1=a_2=a_3=a_4=0$, the dashed and dotted lines are the $\mathcal{O}\left(p^4\right)$ $G_{EM}(s)$ function  (with either $F_V=\sqrt{2}F$ or $F_V=\sqrt{3}F$ constraints). The blue shaded region is the full $\mathcal{O}\left(p^6\right)$ contribution, including (overestimated) uncertainties. The left-hand side plot corresponds to the dispersive parametrization \cite{Dumm:2013zh} while the right-hand side corresponds to the Guerrero-Pich parametrization \cite{Guerrero:1997ku} of the form factor (the latter was used in ref. \cite{Cirigliano:2002pv}).}\label{Appx4:fig3}
\end{figure}

The leading Low approximation for $G^{(0)}_{EM}\left(s\right)$ is plotted in fig. \ref{Appx4:fig3}. This function has two poles, one at $s=4m_\pi^2$ and the other at $s=m_\tau^2$.

We will use the same conventions as ref. \cite{Cirigliano:2002pv}, so we denote as `complete Bremsstrahlung' the amplitude where the structure-dependent (`\textit{SD}') part vanishes, i.e. $v_1=v_2=v_3=v_4=a_1=a_2=a_3=a_4=0$. \textbf{For convenience, we will refer in the following simply as $\mathcal{O}\left(p^4\right)$ and $\mathcal{O}\left(p^6\right)$ to the contributions from $R\chi T$ including operators that contribute up to $\mathcal{O}\left(p^4\right)$ and up to $\mathcal{O}\left(p^6\right)$ chiral LECs, respectively}~\footnote{The different SD constraints applying in each case were discussed at length in section \ref{subsec:SDC}.}.

In $G_{EM}(s)$, the difference between using the $F_V=\sqrt{2}F$ or $F_V=\sqrt{3}F$ constraints at $\mathcal{O}\left(p^4\right)$ is only appreciated for $s\lesssim 0.35$ GeV$^2$, with the latter set producing the largest deviation with respect to the SI result (fig. \ref{Appx4:fig3}). It is important to note that -as put forward in ref.~\cite{Cirigliano:2002pv}- with $F_V=\sqrt{2}F$ constraints (those consistent for $2$-point Green functions) the impact of the `\textit{SD}' corrections on $G_{EM}(s)$ is negligible and the evaluation with \textit{SI} gives already an excellent approximation. On the contrary, we find that using the $F_V=\sqrt{3}F$ set this is no longer true, which will increase the $G_{EM}(s)$ correction in $a_\mu^{HVP,LO|_{\pi\pi}}$ using $\tau$ data (even before adding the $\mathcal{O}(p^6)$ contributions).

In fig. \ref{Appx4:fig3} several contributions to the $G_{EM}(s)$ function are shown: the $G_{EM}^{(0)}$ part by a dashed blue line and the complete Bremsstrahlung (SI) contribution with a solid black line. The full amplitude including all $R\chi T$ operators which contribute at $\mathcal{O}\left(p^4\right)$ ($\mathcal{O}\left(p^6\right)$) are represented by black dashed/dotted (red dashed-dotted) lines in fig. \ref{Appx4:fig3}. For the $\mathcal{O}\left(p^4\right)$ contribution we distinguish between using $F_V=\sqrt{2}F$ ($F_V=\sqrt{3}F$), represented by dashed (dotted) lines. Compared to previous results \cite{Cirigliano:2001er, Cirigliano:2002pv, FloresBaez:2006gf, FloresTlalpa:2006gs}, we note the appearance of a bump near the end of the phase space on $G_{EM}\left(s\right)$  due to the inclusion of the $\rho(1450)$ and the $\rho(1700)$ resonances in the dispersive representation of the vector form factor \cite{Dumm:2013zh, Gonzalez-Solis:2019iod}. The blue band in fig. \ref{Appx4:fig3} shows the uncertainty of the $\mathcal{O}(p^6)$ contribution, evaluated according to that on the couplings which were determined phenomenologically or estimated from chiral counting in section \ref{subsec:SDC} (see also appendix \ref{Fit}) \footnote{These were varied assuming Gaussian errors, and the band was generated so as to cover all data points obtained in 100 spectrum simulations. Results were stable upon increasing statistics. The corresponding blue bands were obtained similarly in Figs. \ref{Appx4:fig5} to \ref{Appx4:fig6}.}. While the central values of the $\mathcal{O}\left(p^6\right)$ corrections change mildly the results obtained at $\mathcal{O}\left(p^4\right)$~\footnote{This is reasonable, since $SI$ is basically unchanged by the $\mathcal{O}\left(p^4\right)$ contributions.}, their huge uncertainty band suggests that our estimate of the $R\chi T$ couplings which start contributing at $\mathcal{O}\left(p^6\right)$ was very conservative (one naively expects a $\sim 1/N_C$ uncertainty for a large-$N_C$ expansion~\footnote{This rough estimate of the parametric uncertainty is supported by the computation of $\chi PT$ LECs including such corrections (see e. g. refs. \cite{Rosell:2006dt, Pich:2008jm, Pich:2010sm}). We note that in this work resonance widths (dominant next-to-leading order effect in the large-$N_C$ expansion for the considered decays) are included. Also the uncertainty corresponding to including excited resonances (an infinite number of them appears for $N_C\to\infty$) was checked to be negligible.}). Lacking a better way for this estimation, we consider this uncertainty band as a conservative upper limit on the corresponding uncertainties. Therefore, our error bands at $\mathcal{O}\left(p^6\right)$ should be regarded accordingly in the following. On the contrary, the small modification induced by those $\mathcal{O}\left(p^6\right)$ couplings fixed by SD constraints (with all remaining ones vanishing) with respect to the $\mathcal{O}\left(p^4\right)$ \cite{Cirigliano:2002pv} results, suggests that the difference between those is a realistic estimate of the missing subdominant terms in ref. \cite{Cirigliano:2002pv} \footnote{These were not estimated in ref. \cite{Cirigliano:2002pv} as SI was already an excellent approximation to the result up to $\mathcal{O}\left(p^4\right)$ (using the $F_V=\sqrt{2}F$ set).} and will be given as such in the remainder of the paper.

\subsection{Radiative decay}\label{subsec:RadDec}
The differential decay width \cite{FloresTlalpa:2008zz} is given by 
\begin{equation}\label{Appx4:eq63}
d\Gamma =\frac{\lambda^{1/2}\left(s,m_{\pi^0}^2,m_{\pi^-}^2\right)}{2\left(4\pi\right)^6m_\tau^2 s}\overline{\left\vert\mathcal{M}\right\vert^2}\,dE_\gamma\,dx\,ds\,d\cos\theta_-\,d\phi_-,
\end{equation}
where $\overline{\left\vert\mathcal{M}\right\vert^2}$ is the unpolarized spin-averaged squared amplitude that corresponds to the $\tau^-\to\pi^-\pi^0\gamma\,\nu_\tau$ decays, and $E_\gamma$ is the photon energy in the $\tau$ rest frame. It is not worth to quote here the full analytical expression for $\overline{\left\vert\mathcal{M}\right\vert^2}$.

For these decays, we have the following integration region
\begin{equation}\begin{split}
\mathcal{D}=&\left\lbrace E_\gamma^{min}\leq E_\gamma\leq E_\gamma^{max},\,x_{min}\leq x\leq x_{max},\,s_{min}\leq s \leq s_{max},\right.\\
&\left.-1\leq\cos\theta_-\leq+1,\, 0\leq\phi_-\leq2\pi\right\rbrace,
\end{split}\end{equation}
with boundaries
\begin{equation}\begin{array}{rcl}
\frac{(m_\tau^2-s+x)}{4m_\tau^2}-\frac{\lambda^{1/2}\left(s,x,m_\tau^2\right)}{4m_\tau}\leq& E_\gamma\left(s,x\right)&\leq\frac{(m_\tau^2-s+x)}{4m_\tau}+\frac{\lambda^{1/2}\left(s,x,m_\tau^2\right)}{4m_\tau},\\
4m_\pi^2\leq&s\left(x\right)&\leq\left(m_\tau-\sqrt{x}\right)^2,\\
0\leq&x&\leq\left(m_\tau-2m_\pi\right)^2,\\
\end{array}
\end{equation}
or interchanging the last two limits,
\begin{equation}\begin{array}{rcl}
0\leq& x\left(s\right)&\leq \left(m_\tau-\sqrt{s}\right)^2,\\
4m_\pi^2\leq& s&\leq m_\tau^2.\\
\end{array}\end{equation}
There are other ways to write these,
\begin{equation}\begin{array}{rcl}
4m_\pi^2\leq&s\left(x,E_\gamma\right)&\leq \frac{(m_\tau-2E_\gamma)(2m_\tau E_\gamma -x)}{2E_\gamma}\\
0\leq& x\left(E_\gamma\right)&\leq \frac{2E_\gamma(m_\tau^2-4m_\pi^2-2m_\tau E_\gamma)}{m_\tau-2E_\gamma},\\
E_\gamma^{cut}\leq&E_\gamma&\leq \frac{m_\tau^2-4m_\pi^2}{2m_\tau},\\
\end{array}\end{equation}
or exchanging $x\leftrightarrow E_\gamma$, 
\begin{equation}\begin{array}{rcl}
\frac{(m_\tau^2+x-4m_\pi^2)}{4m_\tau}-\frac{\lambda^{1/2}\left(x,m_\tau^2,4m_\pi^2\right)}{4m_\tau}\leq& E_\gamma\left(s\right)&\leq \frac{(m_\tau^2+x-4m_\pi^2)}{4m_\tau}+\frac{\lambda^{1/2}\left(x,m_\tau^2,4m_\pi^2\right)}{4m_\tau},\\
0\leq& x &\leq (m_\tau-2m_\pi)^2,\\
\end{array}\end{equation}
and
\begin{equation}\begin{array}{rcl}
0\leq&x\left(s,E_\gamma\right)&\leq \frac{2E_\gamma(m_\tau^2-s-2E_\gamma m_\tau)}{m_\tau-2E_\gamma}\\
4m_\pi^2\leq& s\left(E_\gamma\right)&\leq m_\tau(m_\tau-2E_\gamma),\\
E_\gamma^{cut}\leq&E_\gamma&\leq \frac{m_\tau^2-4m_\pi^2}{2m_\tau}.\\
\end{array}\end{equation}
Further, interchanging $s\leftrightarrow E_\gamma$, we get
\begin{equation}\label{Appx4:eq70}\begin{array}{rcl}
E_\gamma^{cut}\leq&  E_\gamma\left(s\right)&\leq \frac{m_\tau^2-s}{2m_\tau},\\
4m_\pi^2\leq& s &\leq m_\tau(m_\tau-2E_\gamma^{cut}).\\
\end{array}\end{equation}

We recall that this amplitude has IR divergences due to soft photons, i.e. $E_\gamma\to 0$, which is the same problem with $M_\gamma\to 0$ outlined in the previous section. Correspondingly, the experiment is not able to measure photons with energies smaller than some $E_\gamma^{cut}$ (which is related with the experimental resolution).

Concerning the $\mathcal{O}\left(p^6\right)$ contributions, once we employ the relations obtained from the SD behaviour of QCD and its OPE, it is seen that observables are basically insensitive (at the percent level of precision) to $\mathcal{O}(1)$ changes of all the couplings but $\kappa_i^V$ (the $\rho-\omega-\pi$ vertex is described by these couplings), which will saturate the (overestimated) uncertainty of our predictions at this order.

If we integrate eq. (\ref{Appx4:eq63}) using the limits in eq. (\ref{Appx4:eq70}) and the dispersive vector form factor  \cite{Dumm:2013zh, Gonzalez-Solis:2019iod}, we get the $\pi^-\pi^0$ invariant mass distribution, the photon energy distribution and the branching ratios as a function of $E_\gamma^{cut}$, shown in Figs. \ref{Appx4:fig4}, \ref{Appx4:fig05}, \ref{Appx4:fig5}, \ref{Appx4:fig5.5} and \ref{Appx4:fig6} and summarized in Table \ref{Appx4:tab1}. In these figures, the dotdashed red line corresponds to taking the limit where all the couplings at $\mathcal{O}\left(p^6\right)$ vanish except for those constrained by SD and the band overestimates the corresponding uncertainties.\\

\begin{table}[htbp]
\begin{center}
\small{\begin{tabular}{|c|c|c|c|}
\hline
$E_{\gamma}^{cut}$ & BR(Brems) & BR($F_V=\sqrt{2}F$) $\left[\mathcal{O}\left(p^4\right)\right]$ &  BR($F_V=\sqrt{3}F$) $\left[\mathcal{O}\left(p^4\right)\right]$ \\
\hline
\hline
$100\,\mathrm{MeV}$ & $8.6\times 10^{-4}$ & $9.0\times 10^{-4}$  & $9.5\times 10^{-4}$  \\
$300\,\mathrm{MeV}$ & $1.7\times 10^{-4}$ & $1.9\times 10^{-4}$ & $2.3\times 10^{-4}$  \\
$500\,\mathrm{MeV}$ & $2.8\times 10^{-5}$ & $3.9\times 10^{-5}$  & $5.4\times 10^{-5}$\\
\hline
\end{tabular}}
\caption{Branching ratios Br($\tau^-\to\pi^-\pi^0\gamma\nu_\tau$) for different values of $E_\gamma^{cut}$. The second column corresponds to the complete Bremsstrahlung and the third and fourth to the $\mathcal{O}\left(p^4\right)$ contributions. }
\label{Appx4:tab1}
\end{center}
\end{table}

\begin{table}[htbp]
\begin{center}
\small{\begin{tabular}{|c|c|c|}
\hline
$E_{\gamma}^{cut}$ & BR(SD) $\left[\mathcal{O}\left(p^6\right)\right]$ & BR $\left[\mathcal{O}\left(p^6\right)\right]$
\\
\hline
\hline
$100\,\mathrm{MeV}$ & $1.3\times 10^{-3}$ & $(1.9\pm0.3)\times 10^{-3}$ \\
$300\,\mathrm{MeV}$ & $5.1\times 10^{-4}$ & $(1.1\pm0.3)\times 10^{-3}$ \\
$500\,\mathrm{MeV}$ & $2.4\times 10^{-4}$ & $(0.6\pm0.2)\times 10^{-3}$ \\
\hline
\end{tabular}}
\caption{Branching ratios Br($\tau^-\to\pi^-\pi^0\gamma\nu_\tau$) for different $E_\gamma^{cut}$ values at $\mathcal{O}\left(p^6\right)$.}
\label{Appx4:tab2}
\end{center}
\end{table}

\begin{figure}[H]
\includegraphics[width=6.5cm]{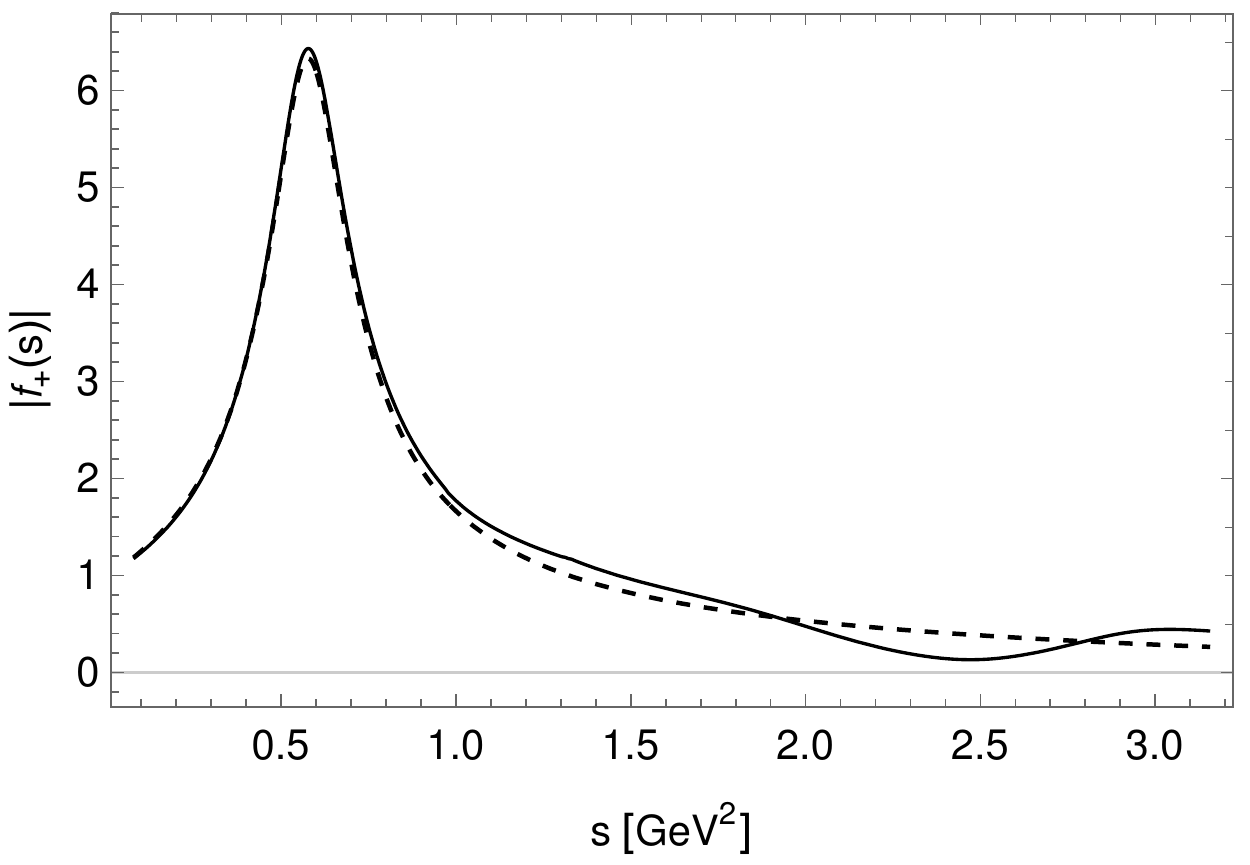}	
\includegraphics[width=6.5cm]{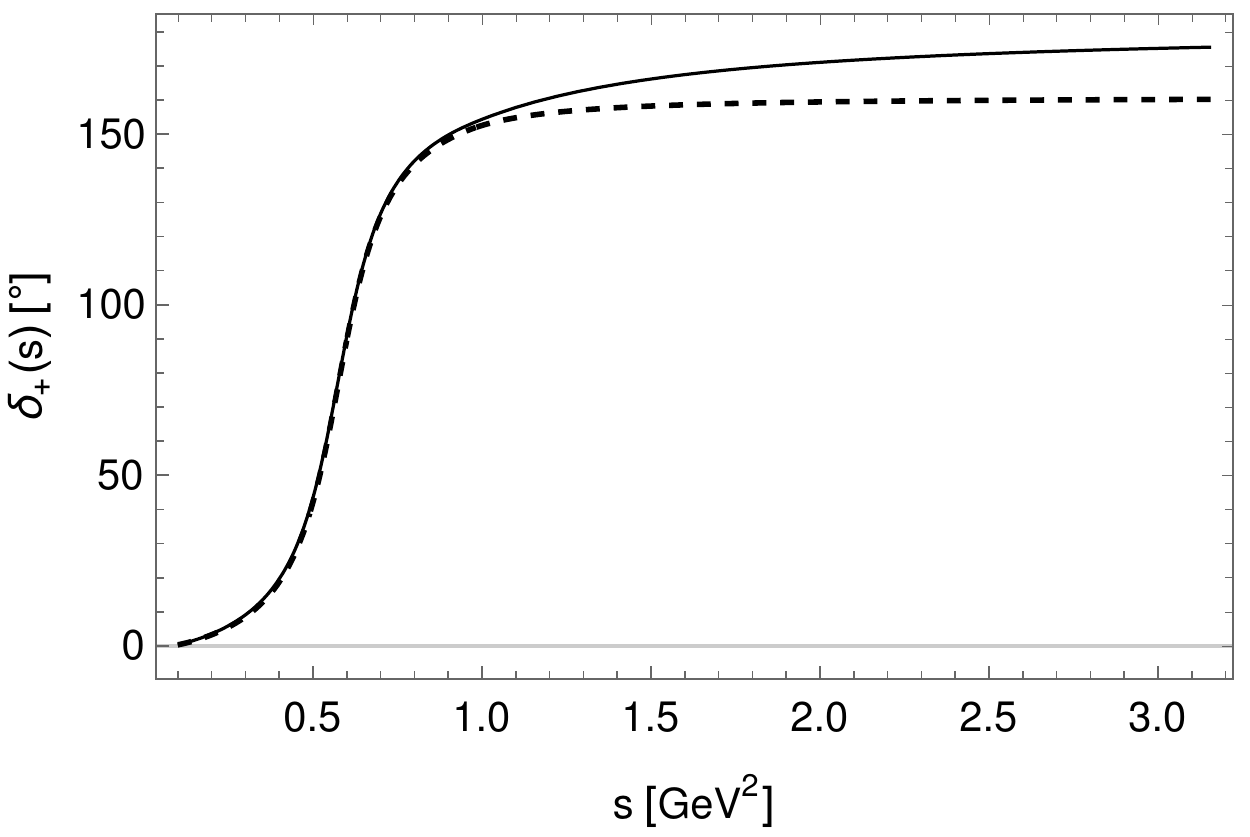}
\centering			
\caption{Modulus and phase of the pion vector form factor, $f_+(s)$. The solid line corresponds to the dispersive representation used in ref. \cite{Dumm:2013zh} while the dashed line corresponds to the Guerrero-Pich parametrization \cite{Guerrero:1997ku} employed by ref. \cite{Cirigliano:2002pv}.}\label{Appx4:fig3.5}
\end{figure}

\begin{figure}[H]
\includegraphics[width=9cm]{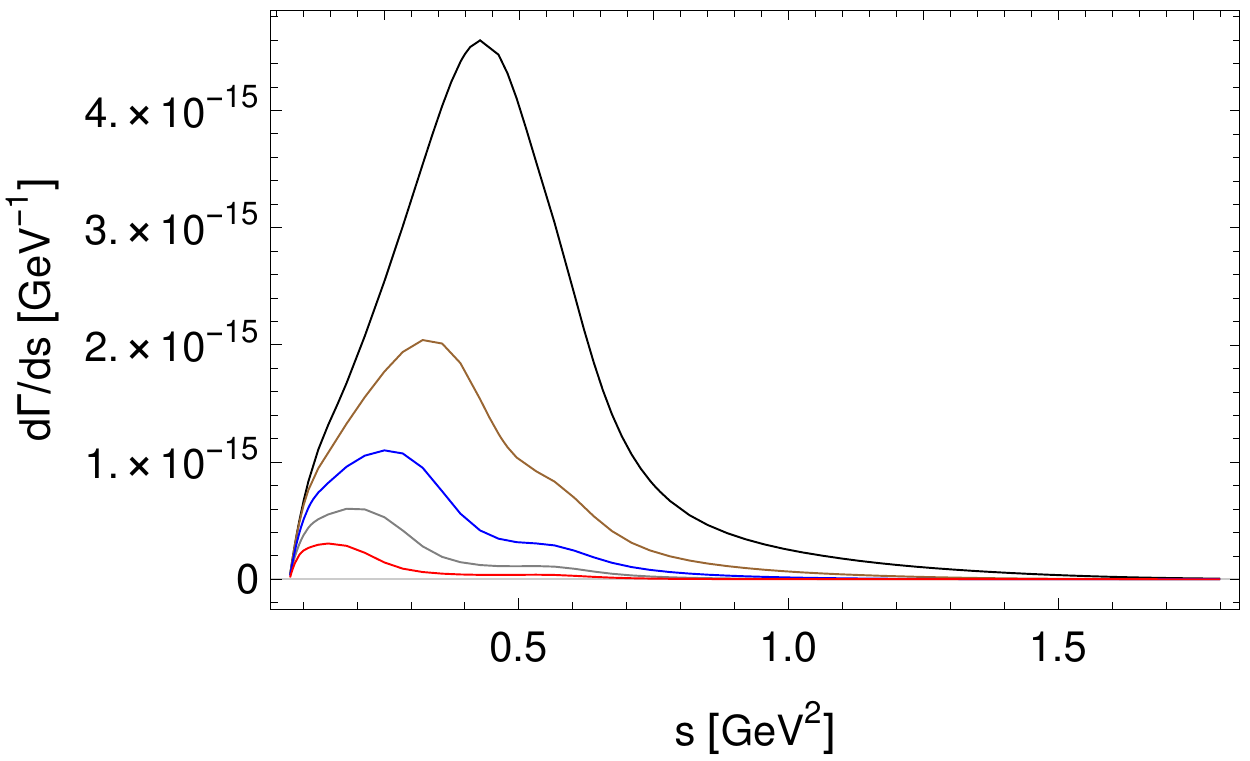}	
\centering			
\caption{The $\pi^-\pi^0$ hadronic invariant mass distribution for the $\tau^-\to\pi^-\pi^0\gamma\nu_\tau$ decays for $E_{\gamma}^{cut}=100\,\mathrm{MeV}$ (black), $E_{\gamma}^{cut}=200\,\mathrm{MeV}$ (brown), $E_{\gamma}^{cut}=300\,\mathrm{MeV}$ (blue), $E_{\gamma}^{cut}=400\,\mathrm{MeV}$ (gray) and $E_{\gamma}^{cut}=500\,\mathrm{MeV}$ (red) using only the Bremsstrahlung (SI) contribution.}\label{Appx4:fig4}
\end{figure}

\begin{figure}[H]
\includegraphics[width=10cm]{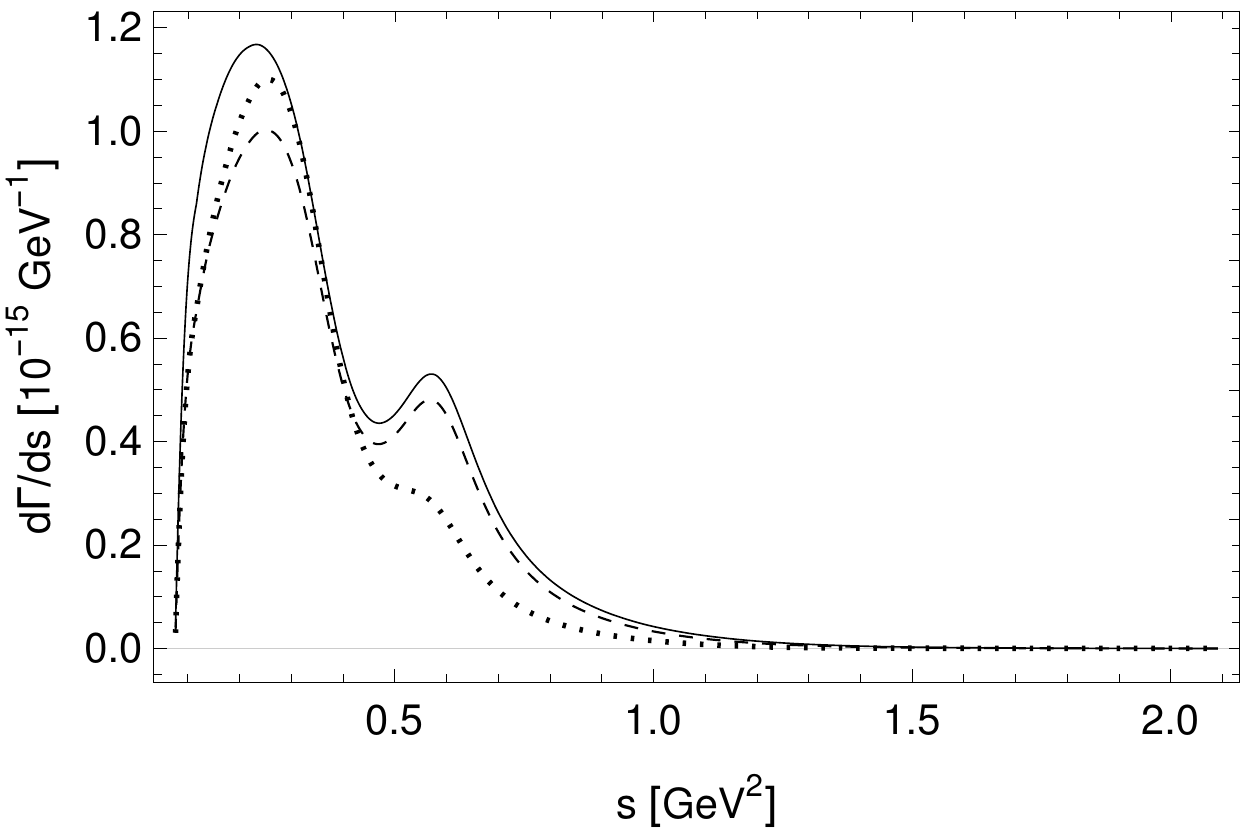}		
\centering			
\caption{The $\pi^-\pi^0$ hadronic invariant mass distributions for $E_\gamma^{cut}=300\,\mathrm{MeV}$. The solid and dashed lines represent the $\mathcal{O}\left(p^4\right)$ corrections using $F_V=\sqrt{3}F$ and $F_V=\sqrt{2}F$, respectively. The dotted line stands for the Bremsstrahlung contribution (SI).}\label{Appx4:fig05}
\end{figure}

\begin{figure}[H]
\includegraphics[width=10cm]{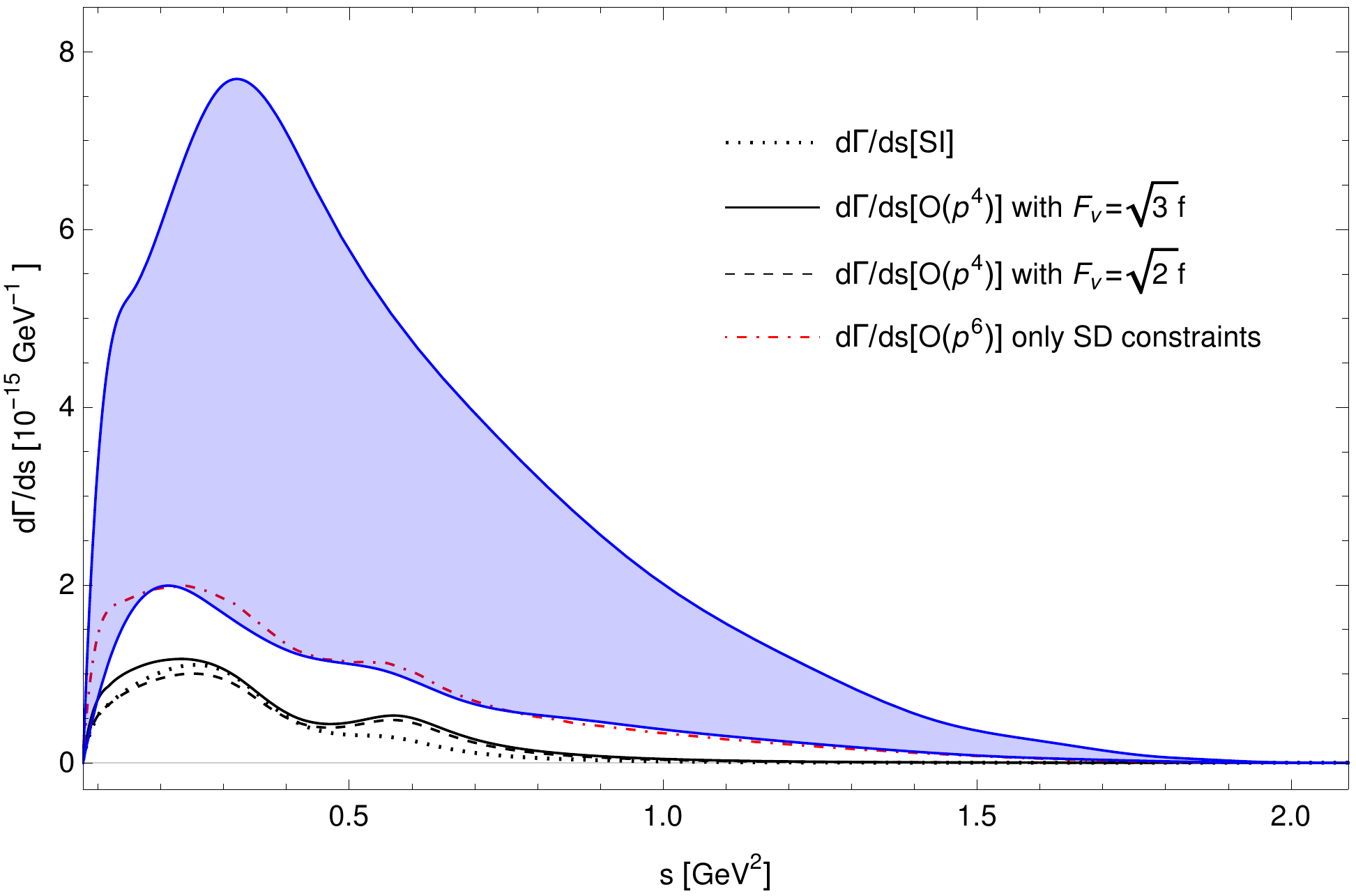}		
\centering			
\caption{The $\pi^-\pi^0$ hadronic invariant mass distributions for $E_\gamma^{cut}=300\,\mathrm{MeV}$. The solid and dashed line represent the $\mathcal{O}\left(p^4\right)$ corrections using $F_V=\sqrt{3}F$ and $F_V=\sqrt{2}F$, respectively. The dotted line represents the Bremsstrahlung contribution (SI). The dotdashed red line corresponds to using only SD constraints at $\mathcal{O}\left(p^6\right)$ and the blue shaded region overestimates the corresponding uncertainties.}\label{Appx4:fig5}
\end{figure}

\begin{figure}[H]
\includegraphics[width=10cm]{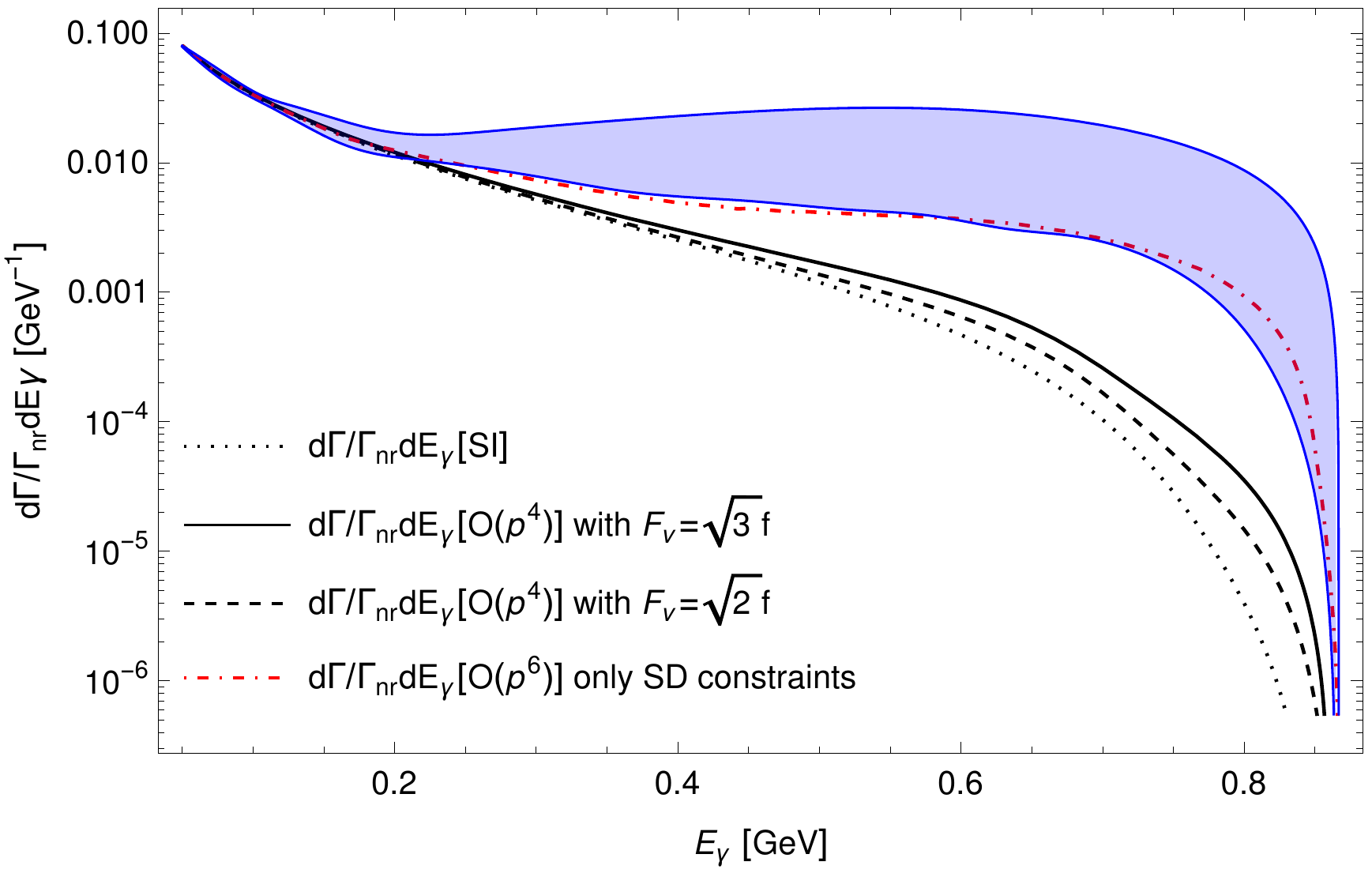}
\centering			
\caption{Photon energy distribution for the $\tau^-\to\pi^-\pi^0\gamma\nu_\tau$ decays normalized with the non-radiative decay width. The dotted line represents the Bremsstrahlung contribution. The solid and dashed lines represent the $\mathcal{O}\left(p^4\right)$ corrections using $F_V=\sqrt{3}F$ and $F_V=\sqrt{2}F$, respectively. The dotdashed red line corresponds to using only SD constraints at $\mathcal{O}\left(p^6\right)$ (with overestimated uncertainties in the blue shaded area).}\label{Appx4:fig5.5}
\end{figure}

\begin{figure}[H]
\includegraphics[width=10cm]{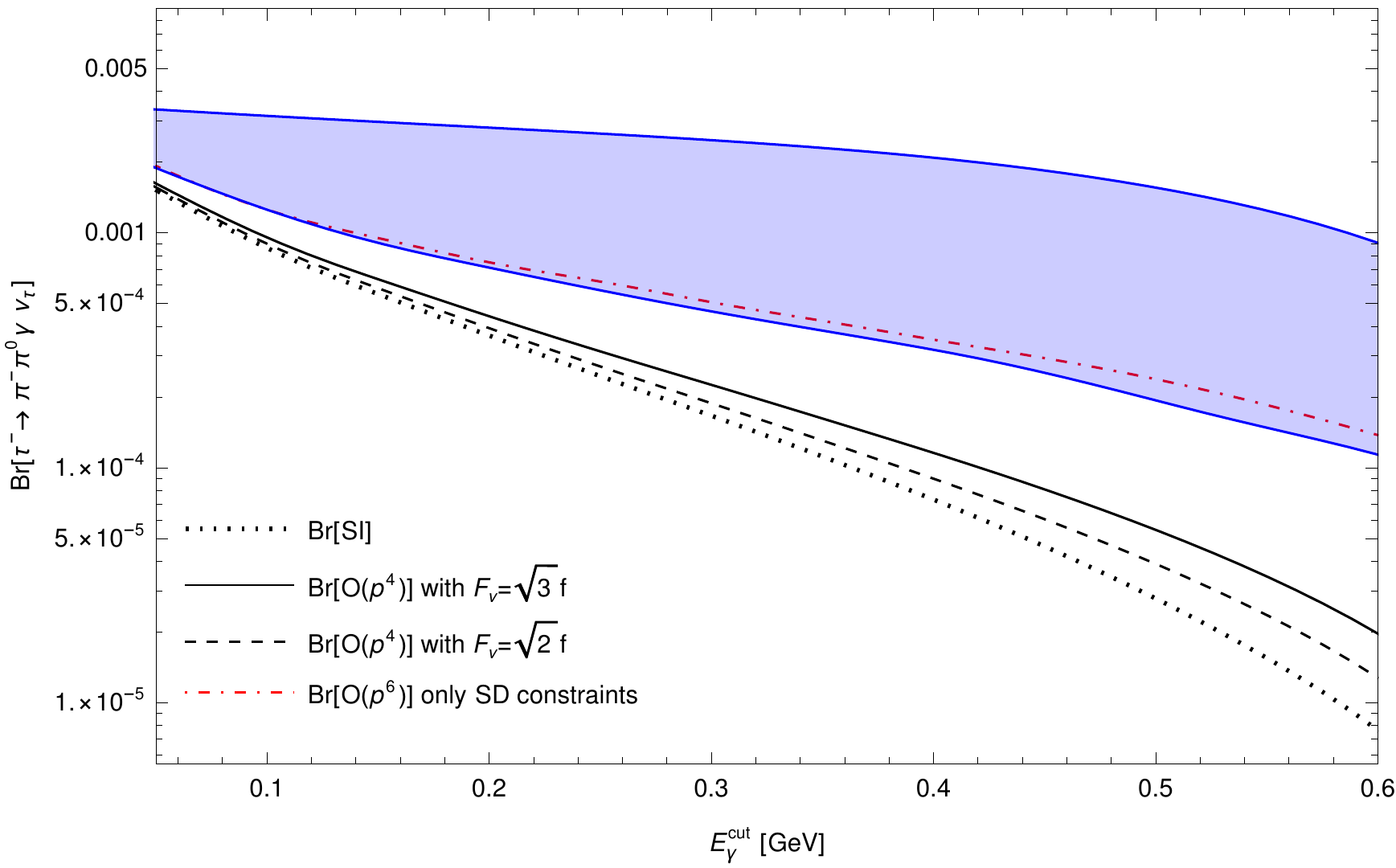}	
\centering			
\caption{Branching ratio for the $\tau^-\to\pi^-\pi^0\gamma\nu_\tau$ decays as a function of $E^{cut}_\gamma$.  The dotted line represents the Bremsstrahlung contribution, the solid line and dashed line represent the $\mathcal{O}\left(p^4\right)$ corrections using $F_V=\sqrt{3}F$ and $F_V=\sqrt{2}F$, respectively. The dotdashed red line is the $\mathcal{O}\left(p^6\right)$ contribution using only SD constraints and neglecting all other couplings. The blue shaded region overestimates the $\mathcal{O}\left(p^6\right)$ uncertainties.}\label{Appx4:fig6}
\end{figure}

As it can be observed from Table \ref{Appx4:tab1} and fig. \ref{Appx4:fig6}, the main contribution at $\mathcal{O}\left(p^4\right)$ corresponds to the complete Bremsstrahlung (SI) amplitude (in agreement with ref. \cite{Cirigliano:2002pv}), and the value for the branching ratio becomes smaller with larger values of $E_\gamma^{cut}$. The values in Table \ref{Appx4:tab1} are slightly different from those reported in ref. \cite{Cirigliano:2002pv}, this effect is mainly due to the parametrization of the pion vector form factor (see fig. \ref{Appx4:fig3.5}). The form factor obtained from the dispersion relation \cite{Dumm:2013zh} is above the one obtained using the Guerrero-Pich parametrization \cite{Guerrero:1997ku} at $s\simeq M_\rho^2$, and also the former includes the $\rho\left(1450\right)$ and $\rho\left(1700\right)$ resonances.\\

According to our discussion on error estimation of the $\mathcal{O}\left(p^4\right)$ result (including the uncertainty coming from missing higher-order terms from the result at $\mathcal{O}\left(p^6\right)$ when only SD constraints are used), we have -for $E_\gamma^{cut}=300$ MeV- $BR(\tau^-\to\pi^-\pi^0\nu_\tau)\,=\,(1.9^{+3.2}_{-0.0})\cdot 10^{-4}$.

The spectrum for these decays with $v_i=a_i=0$ is plotted in fig. \ref{Appx4:fig4}, the dominant peak corresponds to bremsstrahlung off the $\pi^-$, and the secondary receives two contributions: one from bremsstrahlung off the $\tau$ lepton and another from a resonance exchange in $V_{\mu\nu}$ (for $E_\gamma^{cut}\leq 100\,\mathrm{MeV}$, these two are merged into one single peak). The rate and spectrum are dominated by the complete bremsstrahlung (SI) contribution.\\

In fig. \ref{Appx4:fig05}, we show the distribution for $E_\gamma^{cut}=300\,\mathrm{MeV}$ taking into account the SI contribution (dotted line) and the $\mathcal{O}\left(p^4\right)$ amplitude obtained using $F_V=\sqrt{2}F$ (dashed line) and $F_V=\sqrt{3}F$ (solid line), the most important contribution corresponds to the $\rho$ resonance exchange at $s\sim0.6\,\mathrm{GeV}^2$. The main difference between these two approaches is seen in fig. \ref{Appx4:fig05}, where up to $s\sim 0.4\,\mathrm{GeV}^2$ the dashed line is below and the solid line is above the bremmstrahlung (SI) contribution (dotted line). The dashed line is quite similar to the distribution in fig. 2 of ref. \cite{Cirigliano:2002pv} while the solid line resembles closely the distribution in fig. 4.6 of ref. \cite{FloresTlalpa:2008zz} obtained from the vector meson dominance (VMD) model \cite{Sakurai:1960ju} neglecting the $\omega$-resonance contribution.\\

In fig. \ref{Appx4:fig5} we show a comparison between the di-pion distribution at different orders. As we can see, the inclusion of the corrections at $\mathcal{O}\left(p^6\right)$ gives a noticeable enhancement at low $s$.

For the photon energy distribution, fig. \ref{Appx4:fig5.5}, we can differentiate between the full amplitude (solid, dashed lines up to $\mathcal{O}\left(p^4\right)$ and dotdashed red line up to $\mathcal{O}\left(p^6\right)$) and the bremsstrahlung contribution (dotted line) but, as in the case of the branching fraction, the distribution decreases for high-energies. In the case of the $\mathcal{O}\left(p^6\right)$ distribution there is an enhancement at middle and high photon energies.\\

According to Figs. \ref{Appx4:fig05} to \ref{Appx4:fig6}, measurements of the $\pi\pi$ invariant mass, of the photon spectrum and the partial decay width, for a reasonable cut on $E_\gamma$ (at low enough energies the inner bremmstrahlung contribution hides completely any structure-dependent effect), could decrease substantially the uncertainty of the $\mathcal{O}\left(p^6\right)$ computation. This was already emphasized in ref. \cite{Cirigliano:2002pv} but remained unmeasured at BaBar and Belle. We hope these data can finally be acquired and analyzed at Belle-II.

\begin{figure}[H]
\includegraphics[width=7.5cm]{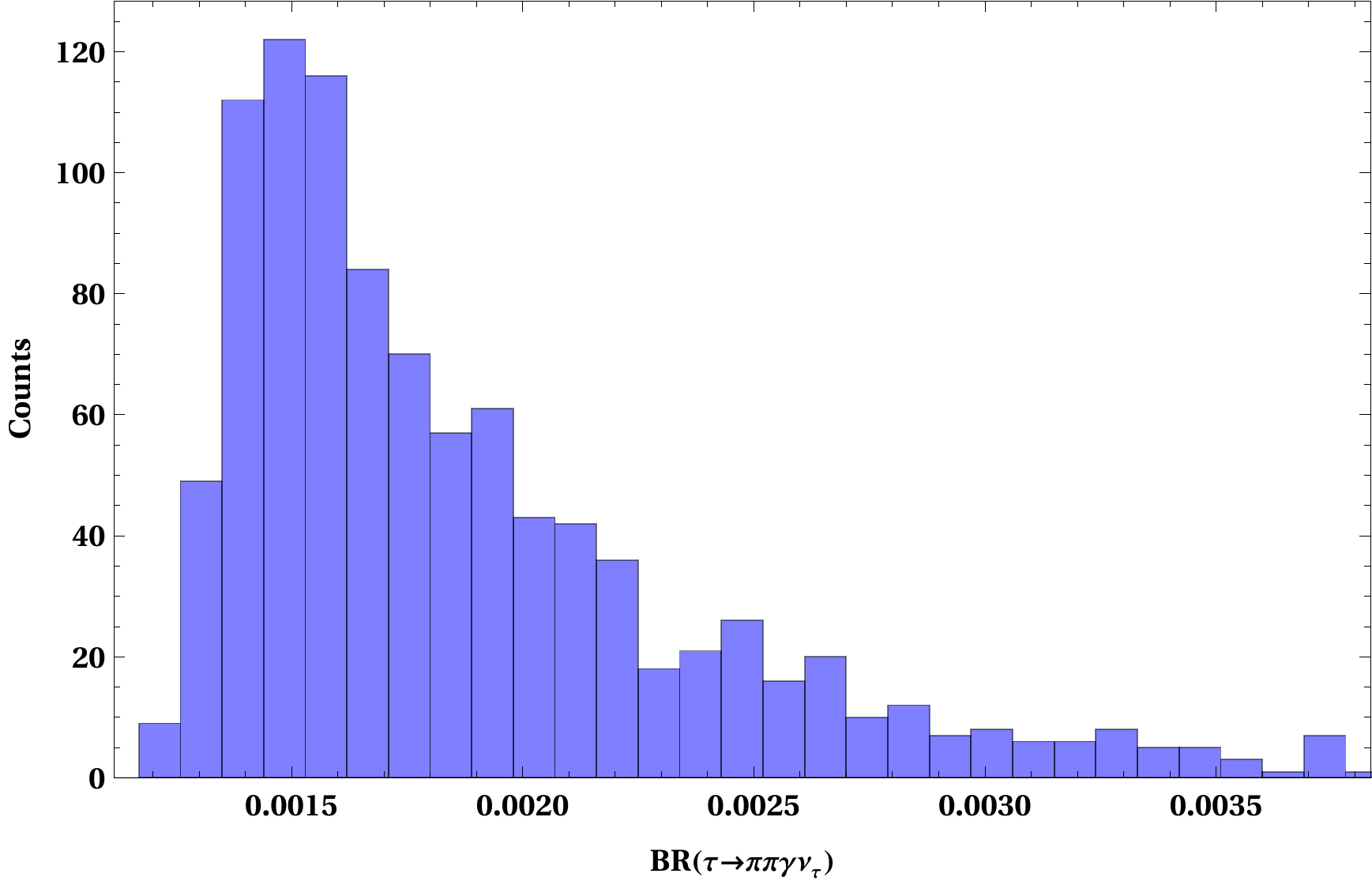}	
\includegraphics[width=7.5cm]{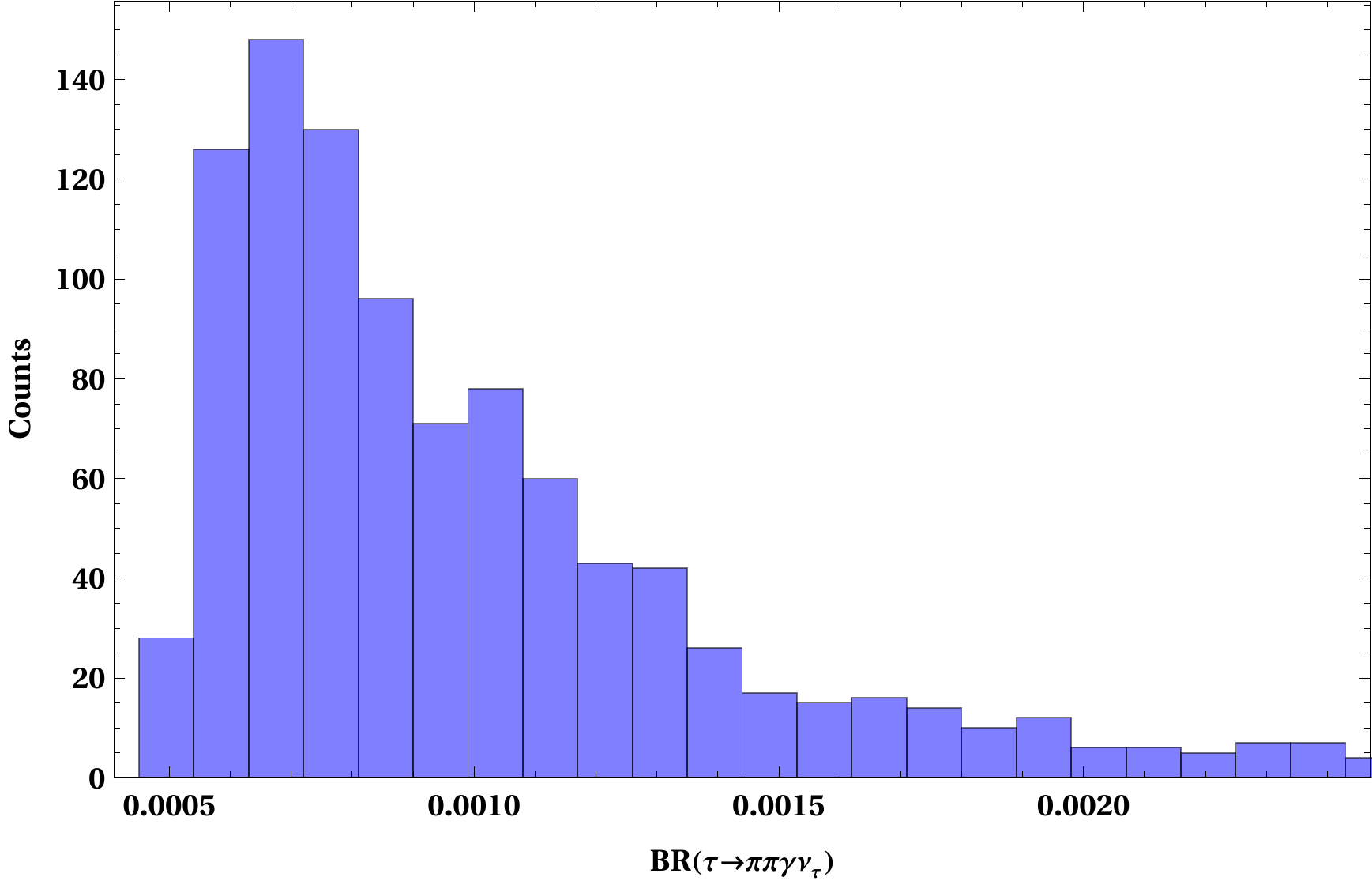}	
\includegraphics[width=7.5cm]{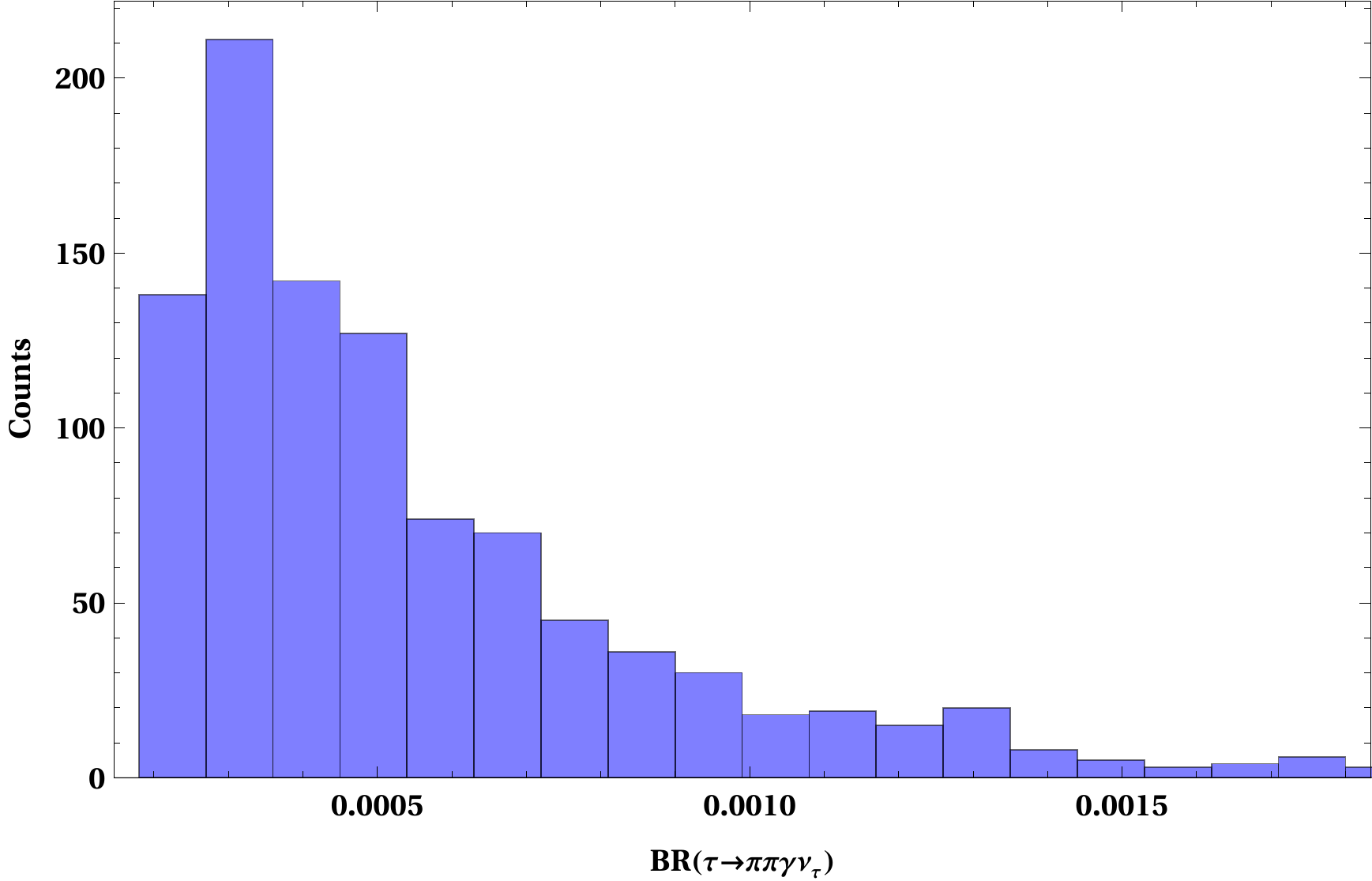}	
\centering			
\caption{Predictions for the branching ratio at $\mathcal{O}\left(p^6\right)$ for a sample of 1000 points, with $E_{cut}=100,\,300,\:\mathrm{and}\:500\:\mathrm{MeV}$ from top to bottom.}\label{Appx4:fig15.1}
\end{figure}

In fig. \ref{Appx4:fig15.1}, we show the branching ratio for $E^{cut}_\gamma=100,\,300,\,\mathrm{and}\,500\,\mathrm{MeV}$ from top to bottom. The outcomes were summarized in Table \ref{Appx4:tab2}.

\section{IB corrections to $a_\mu^{HVP,LO\, \pi\pi}$}\label{sec:IBCorr}
We can evaluate the leading contributions to the hadronic vacuum polarization (HVP) by means of the dispersion relation \cite{Gourdin:1969dm},
\begin{equation}
a^{HVP,LO}_\mu=\frac{1}{4\pi^3}\int^{\infty}_{s_{thr}} ds \,K(s) \sigma^0_{e^- e^+\to hadrons}(s),
\end{equation}
where $K(s)$ is a smooth QED kernel concentrated at low energies, which increases the $E\lesssim M_\rho$ contribution,
\begin{equation}
K(s)=\frac{x^2}{2}(2-x^2)+\frac{(1+x^2)(1+x)^2}{x^2}\left(\ln (1+x)-x+\frac{x^2}{2}\right)+\frac{(1+x)}{(1-x)}x^2\ln (x),
\end{equation}
with
\[x=\frac{1-\beta_\mu}{1+\beta_\mu},\qquad \beta_\mu=\sqrt{1-4m_\mu^2/s},\]
and $ \sigma_{e^- e^+\to hadrons}^0(s)$ is the bare hadronic cross section \footnote{Although final state radiation would belong to HVP,NLO it is always included in HVP,LO (and not in HVP,NLO) as eliminating this radiation from the measured data is unfeasible. Thus, a final state radiation (FSR) factor is also needed in the radiative corrections discussed below.}. We can relate the hadronic spectral function from $\tau$ decays to the $e^+e^-$ hadronic cross section by including the radiative corrections and the IB  effects. For the $\pi\pi$ final state, we have \cite{Cirigliano:2001er, Cirigliano:2002pv}:
\begin{equation}
\sigma_{\pi\pi}^0=\left[\frac{K_\sigma(s)}{K_\Gamma (s)}\frac{d\Gamma_{\pi\pi[\gamma]}}{ds}\right]\frac{R_{IB}(s)}{S_{EW}},
\end{equation}
where
\begin{equation}\begin{split}
K_\Gamma(s)&=\frac{G_F^2\vert V_{ud}\vert^2 m_\tau^3}{384 \pi^3}\left(1-\frac{s}{m_\tau^2}\right)^2\left(1+\frac{2s}{m_\tau^2}\right),\\
K_\sigma(s)&=\frac{\pi \alpha^2}{3s},
\end{split}\end{equation}
and the IB corrections 
\begin{equation}
R_{IB}(s)=\frac{FSR(s)}{G_{EM}(s)}\frac{\beta^3_{\pi^+\pi^-}}{\beta^3_{\pi^+\pi^0}}\left\vert\frac{F_V(s)}{f_+(s)}\right\vert^2.
\end{equation}
The $S_{EW}$ term encodes the SD electroweak corrections \cite{Sirlin:1974ni,Sirlin:1977sv,Sirlin:1981ie,Marciano:1985pd,Marciano:1988vm,Braaten:1990ef,Marciano:1993sh,Erler:2002mv} and $FSR(s)$ accounts for the radiation from the final-state pions \cite{Schwinger:1989ix, Drees:1990te}. The $G_{EM}(s)$ term was already discussed at length in section \ref{GEM}, the $\beta^3_{\pi^+\pi^-}/\beta^3_{\pi^+\pi^0}$ term is a phase space factor and the last term in $R_{IB}(s)$ is a ratio between the neutral ($F_V(s)$) and the charged ($f_+(s)$) pion form factor.

In order to study the effect of the radiative correction $G_{EM}(s)$ on $a_\mu^{HVP,LO}[\pi\pi]$, we have evaluated the following expression \cite{Cirigliano:2002pv}
\begin{equation}\label{IB:eq69}
\Delta a_\mu^{HVP,LO}=\frac{1}{4\pi^3}\int_{s_1}^{s_2}ds\,K(s)\left[\frac{K_\sigma(s)}{K_\Gamma (s)}\frac{d\Gamma_{\pi\pi[\gamma]}}{ds}\right]\left(\frac{R_{IB}(s)}{S_{EW}}-1\right),
\end{equation}
taking $S_{EW}=1$, $\frac{\beta^3_{\pi^+\pi^-}}{\beta^3_{\pi^+\pi^0}}=1$ and $\left\vert\frac{F_V(s)}{f_+(s)}\right\vert^2=1$. The results are summarized in Table \ref{HVP:tab1} using DR form factor. The results obtained for the $G_{EM}^{(0)}$ and the complete $\mathcal{O}\left(p^4\right)$ contribution (with $F_V=\sqrt{2}F$) agree with those in \cite{Cirigliano:2002pv}, which are $+16\cdot10^{-11}$ and $-10\cdot10^{-11}$, respectively (for the whole integral).
In Table \ref{HVP:tab2}, we summarized the results obtained using the Guerrero-Pich \cite{Guerrero:1997ku} parametrization of the form factor (which only accounts for the completely dominant $\rho$ exchange), which are in nice agreement with those found with the dispersive form factor (that also includes the $\rho(1450)$ and $\rho(1700)$ effects). This checks, a posteriori, that excited resonance contributions make a negligible effect in the $G_{EM}(s)$ corrections to $a_\mu^{HVP,LO}$~\footnote{By replacing $D_\rho^{-1}(x)$ by $(1+\beta_{\rho'})^{-1}(D_\rho^{-1}(x)+\beta_{\rho'} D_{\rho'}^{-1}(x))$, with $\beta_{\rho'}\in[0.12,0.15]$ \cite{Gonzalez-Solis:2019iod} throughout the $v_i$ and $a_i$ form factors, we have verified that the impact of the $\rho'$ on the $G_{EM}(s)$ correction to $a_\mu^{HVP,LO|_{\pi\pi,\tau}}$ is negligible. Similarly, the error induced by other excited resonances shall also be irrelevant.}.

The values in the last column of Tables \ref{HVP:tab1} and \ref{HVP:tab2} were obtained evaluating the eq. (\ref{IB:eq69}) according to the couplings discussed in section \ref{subsec:SDC} for a sample of $200$ points for each interval of integration (results were stable under increasing this number).

\begin{table}[htbp]
\begin{center}
\small{\begin{tabular}{|c|c|c|c|c|c|c|}
\hline
$\left[s_1,s_2\right]$ & $\Delta a_{\mu,\,\mathrm{G_{EM}^{(0)}}}^{\mathrm{HVP,LO}}$ & $\Delta a_{\mu,\,\mathrm{SI}}^{\mathrm{HVP,LO}}$ & $\Delta a_{\mu,\,\left[\mathcal{O}\left(p^4\right)\right]}^{\mathrm{HVP,LO}}$ &  $\Delta a_{\mu,\,\left[\mathcal{O}\left(p^4\right)\right]}^{\mathrm{HVP,LO}}$ & $\Delta a_{\mu,\,\left[SD\right]}^{\mathrm{HVP,LO}}$ & $\Delta a_{\mu,\,\left[\mathcal{O}\left(p^6\right)\right]}^{\mathrm{HVP,LO}}$ \\
\hline
\hline
$\left[4m_\pi^2,1\,\mathrm{GeV}^2\right]$ & $+17.8$ & $-11.0$ & $-11.3$  & $-17.0$ & $-32.4$ & $-74.8\pm 44.0$\\
$\left[4m_\pi^2,2\,\mathrm{GeV}^2\right]$ & $+18.3$ & $-10.1$ & $-10.3$  & $-16.0$ & $-31.9$ & $-75.9\pm 45.5$\\
$\left[4m_\pi^2,3\,\mathrm{GeV}^2\right]$ & $+18.4$ & $-10.0$  & $-10.2$ & $-15.9$ & $-31.9$ & $-75.9\pm 45.6$\\
$\left[4m_\pi^2,m_\tau^2\right]$ & $+18.4$ & $-10.0$  & $-10.2$ & $-15.9$ & $-31.9$ & $-75.9\pm 45.6$\\
\hline
\end{tabular}}
\caption{Contributions to $\Delta a_\mu^{HVP,LO}$ in units of $10^{-11}$ using the dispersive representation of the form factor. From the two evaluations labelled $\mathcal{O}\left(p^4\right)$, the left(right) one corresponds to $F_V=\sqrt{2}F$($F_V=\sqrt{3}F$).}
\label{HVP:tab1}
\end{center}
\end{table}

\begin{table}[htbp]
\begin{center}
\small{\begin{tabular}{|c|c|c|c|c|c|c|}
\hline
$\left[s_1,s_2\right]$ & $\Delta a_{\mu,\,\mathrm{G_{EM}^{(0)}}}^{\mathrm{HVP,LO}}$ & $\Delta a_{\mu,\,\mathrm{SI}}^{\mathrm{HVP,LO}}$ & $\Delta a_{\mu,\,\left[\mathcal{O}\left(p^4\right)\right]}^{\mathrm{HVP,LO}}$ &  $\Delta a_{\mu,\,\left[\mathcal{O}\left(p^4\right)\right]}^{\mathrm{HVP,LO}}$ & $\Delta a_{\mu,\,\left[SD\right]}^{\mathrm{HVP,LO}}$ & $\Delta a_{\mu,\,\left[\mathcal{O}\left(p^6\right)\right]}^{\mathrm{HVP,LO}}$\\
\hline
\hline
$\left[4m_\pi^2,1\,\mathrm{GeV}^2\right]$ & $+17.3$ & $-10.2$ & $-10.4$  & $-15.9$ & $-28.3$ & $-63.2\pm 16.5$\\
$\left[4m_\pi^2,2\,\mathrm{GeV}^2\right]$ & $+17.7$ & $-9.4$ & $-9.6$  & $-15.2$ & $-28.1$ &  $-58.1\pm 12.2$\\
$\left[4m_\pi^2,3\,\mathrm{GeV}^2\right]$ & $+17.8$ & $-9.3$  & $-9.5$ & $-15.1$ & $-28.0$ & $-67.8\pm 17.5$\\
$\left[4m_\pi^2,m_\tau^2\right]$ & $+17.8$ & $-9.3$  & $-9.5$ & $-15.1$ & $-28.0$ & $-64.9\pm 13.4$\\
\hline
\end{tabular}}
\caption{Contributions to $\Delta a_\mu^{HVP,LO}$ in units of $10^{-11}$ using the GP parametrization of the form factor. From the two evaluations labelled $\mathcal{O}\left(p^4\right)$, the left(right) one corresponds to $F_V=\sqrt{2}F$($F_V=\sqrt{3}F$).}
\label{HVP:tab2}
\end{center}
\end{table}

\begin{figure}[ht]
\begin{center}
\includegraphics[width=11cm]{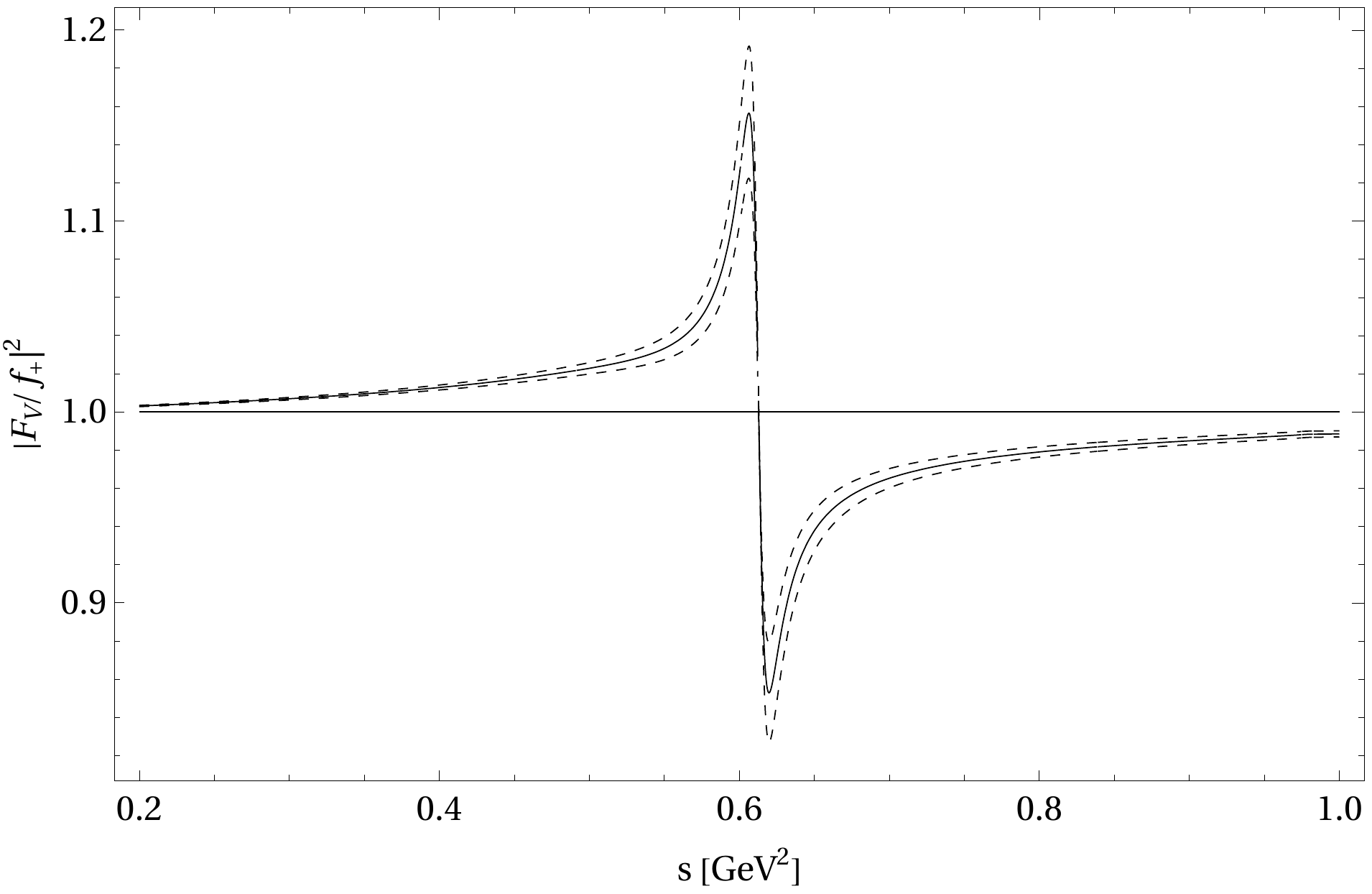}
\caption{Ratio of the form factors (FF1) for $\theta_{\rho\omega}=(-3.5\pm0.7)\times 10^{-3}\:\mathrm{GeV}^2$. The solid line represents the mean value.}\label{RFF}
\end{center}
\end{figure}

The other contributions are summarized in Table \ref{HVP:tab3}.
\begin{itemize}
\item The $S_{EW}$ contribution $S_{EW}=1.0201$ gives $\Delta a_\mu^{HVP,LO}=-103.1\times 10^{-11}$, consistent with earlier determinations (using slightly different values of $S_{EW}$) and with a negligible error.
\item The phase space (PS) correction induces $\Delta a_\mu^{HVP,LO}=-74.5\times 10^{-11}$ (trivially in agreement with previous computations), again with tiny uncertainties.
\item The final state radiation (FSR, which is formally $NLO$) yields $\Delta a_\mu^{HVP,LO}=+45.5(4.6)\times 10^{-11}$, in accord with ref. \cite{Davier:2009ag} (its value was not quoted in ref. \cite{Cirigliano:2002pv}).
\item The correction due to the ratio of the form factors (fig. \ref{RFF}) is harder to evaluate. We have considered two alternatives, labelled FF1 and FF2, that we explain next. We use the following numerical inputs for the $\rho-\omega$ mixing parameter $\theta_{\rho\omega}=(-3.5\pm0.7)\times10^{-3}\:\mathrm{GeV}^2$ \cite{Cirigliano:2002pv} and $\Gamma_{\rho^0}-\Gamma_{\rho^+}=0.3\pm1.3$ MeV, $m_{\rho^\pm}-m_{\rho^0}=0.7\pm0.8$ MeV and $m_{\rho^0}=775.26\pm0.25$ MeV from PDG \cite{Tanabashi:2018oca}.

In FF1, as in ref. \cite{Cirigliano:2002pv}, we include the measurement of the $\pi\pi\gamma$ channel of the $\rho^0$ $\Gamma_{\rho^0\to\pi^+\pi^-\gamma}=1.5\pm0.2$ MeV, and the measurement of $\Gamma_{\rho^0\to\pi^0\gamma}$ and $\Gamma_{\rho^+\to\pi^+\gamma}$ which are approximately $0.1$ MeV \cite{Zyla:2020oca}. Thus, we estimate $\Gamma_{\rho^0\to\pi^+\pi^-\gamma}-\Gamma_{\rho^\pm\to\pi^\pm\pi^0\gamma}=1.5\pm1.3$ MeV. In this way, we get a positive correction of $\Delta a_\mu^{HVP,LO}=+40.9(48.9)\times 10^{-11}$. The uncertainty on the third column of Table \ref{HVP:tab3} (FF1) corresponds to sum the errors due to uncertainties of $\rho-\omega$ mixing ($8.5$), the $\rho^+-\rho^0$ mass difference ($15.9$), and the $\rho^+-\rho^0$ width difference ($45.5$) in quadrature (in units of $10^{-11}$).

On the other hand, in FF2 we use the same numerical inputs for $\Gamma_{\rho^0\to\pi^+\pi^-\gamma}-\Gamma_{\rho^\pm\to\pi^\pm\pi^0\gamma}=0.45\pm0.45$ MeV as in ref. \cite{Cirigliano:2002pv} (and all the others as we did before), we obtain a positive correction of $\Delta a_\mu^{HVP,LO}=+77.6(24.0)\times 10^{-11}$. The uncertainty on the fourth column Table \ref{HVP:tab3} (FF2) corresponds to sum the errors due to uncertainties of $\rho-\omega$ mixing ($8.6$), the $\rho^+-\rho^0$ mass difference ($15.9$), and the $\rho^+-\rho^0$ width difference ($15.8$) in quadrature (in units of $10^{-11}$).

This correction was $+(61\pm26\pm3)\cdot10^{-11}$ in \cite{Cirigliano:2002pv} and $+(86\pm32\pm7)\cdot10^{-11}$ in \cite{Davier:2009ag}, in agreement (despite the big errors) with our FF2 and FF1 determinations, respectively.
\item Finally, we get $(-15.9^{+5.7}_{-16.0})\cdot10^{-11}$ ($(-76\pm46)\cdot10^{-11}$) for the $G_{EM}(s)$ correction at $\mathcal{O}(p^4)$ ($\mathcal{O}(p^6)$), versus $-10\cdot10^{-11}$ in \cite{Cirigliano:2002pv} and $-37\cdot10^{-11}$ in \cite{FloresBaez:2006gf} (from the last two results, $(-19.2\pm9.0)\cdot10^{-11}$ was used in \cite{Davier:2009ag}). As explained before, the previous uncertainty on the $\mathcal{O}(p^6)$ can only be taken as an upper bound on it. Also interesting is the $G_{EM}(s)$ correction when only the couplings restricted by SD are used (with all others at this order set to zero), which allows us to estimate the effect of missing higher-order terms on the $\mathcal{O}(p^4)$ result quoted above. This $\mathcal{O}(p^4)$ result, which is our reference value, is consistent with both the earlier $R\chi T$ \cite{Cirigliano:2002pv} and the VMD \cite{Davier:2009ag} evaluations, albeit with a larger (asymmetric) error.
\end{itemize}

\begin{table}[htbp]
\begin{center}
\small{\begin{tabular}{|c|c|c|c|c|c|c|}
\hline
$\left[s_1,s_2\right]$ & $\mathrm{S_{EW}}$ & $\mathrm{PS}$ & $ \mathrm{FSR}$ & $\mathrm{FF1}$ &  $\mathrm{FF2}$ & $\mathrm{EM}$ \\
\hline
\hline
$\left[4m_\pi^2,1\,\mathrm{GeV}^2\right]$ & $-101.1$ & $-74.1$ & $+44.7$ & $+41.8\pm49.0 $  & $+78.4\pm24.5$ & $-17.0^{+5.7}_{-15.4}$ \\
$\left[4m_\pi^2,2\,\mathrm{GeV}^2\right]$ & $-103.1$ & $-74.4$ & $+45.5$ & $+40.9\pm48.9$  & $+77.6\pm24.0$ & $-16.0^{+5.7}_{-15.9}$ \\
$\left[4m_\pi^2,3\,\mathrm{GeV}^2\right]$ & $-103.1$ & $-74.5$  & $+45.5$ & $+40.9\pm48.9$ & $+77.6\pm24.0$ & $-15.9^{+5.7}_{-16.0}$ \\
$\left[4m_\pi^2,m_\tau^2\right]$ & $-103.1$ & $-74.5$  & $+45.5$ & $+40.9\pm48.9$ & $+77.6\pm24.0$ & $-15.9^{+5.7}_{-16.0}$ \\
\hline
\end{tabular}
\begin{tabular}{|c|c|c|}
\hline
$\left[s_1,s_2\right]$ &  $\Delta a_\mu (\mathrm{FF1})$ & $\Delta a_\mu (\mathrm{FF2})$\\
\hline\hline
$\left[4m_\pi^2,1\,\mathrm{GeV}^2\right]$ & $-105.7^{+49.5}_{-51.6}$ & $-69.1^{+25.6}_{-29.3}$ \\
$\left[4m_\pi^2,2\,\mathrm{GeV}^2\right]$ & $-107.1^{+49.4}_{-51.6}$ & $-70.4^{+25.1}_{-29.2}$ \\
$\left[4m_\pi^2,3\,\mathrm{GeV}^2\right]$ & $-107.1^{+49.4}_{-51.7}$ & $-70.4^{+25.1}_{-29.2}$ \\
$\left[4m_\pi^2,m_\tau^2\right]$ & $-107.1^{+49.4}_{-51.7}$ & $-70.4^{+25.1}_{-29.2}$ \\
\hline
\end{tabular}}
\caption{Contributions to $\Delta a_\mu^{HVP,LO}$ in units of $10^{-11}$ using the DR form factor as the reference one.}
\label{HVP:tab3}
\end{center}
\end{table}

\begin{figure}[ht]
\begin{center}
\includegraphics[width=7.4cm]{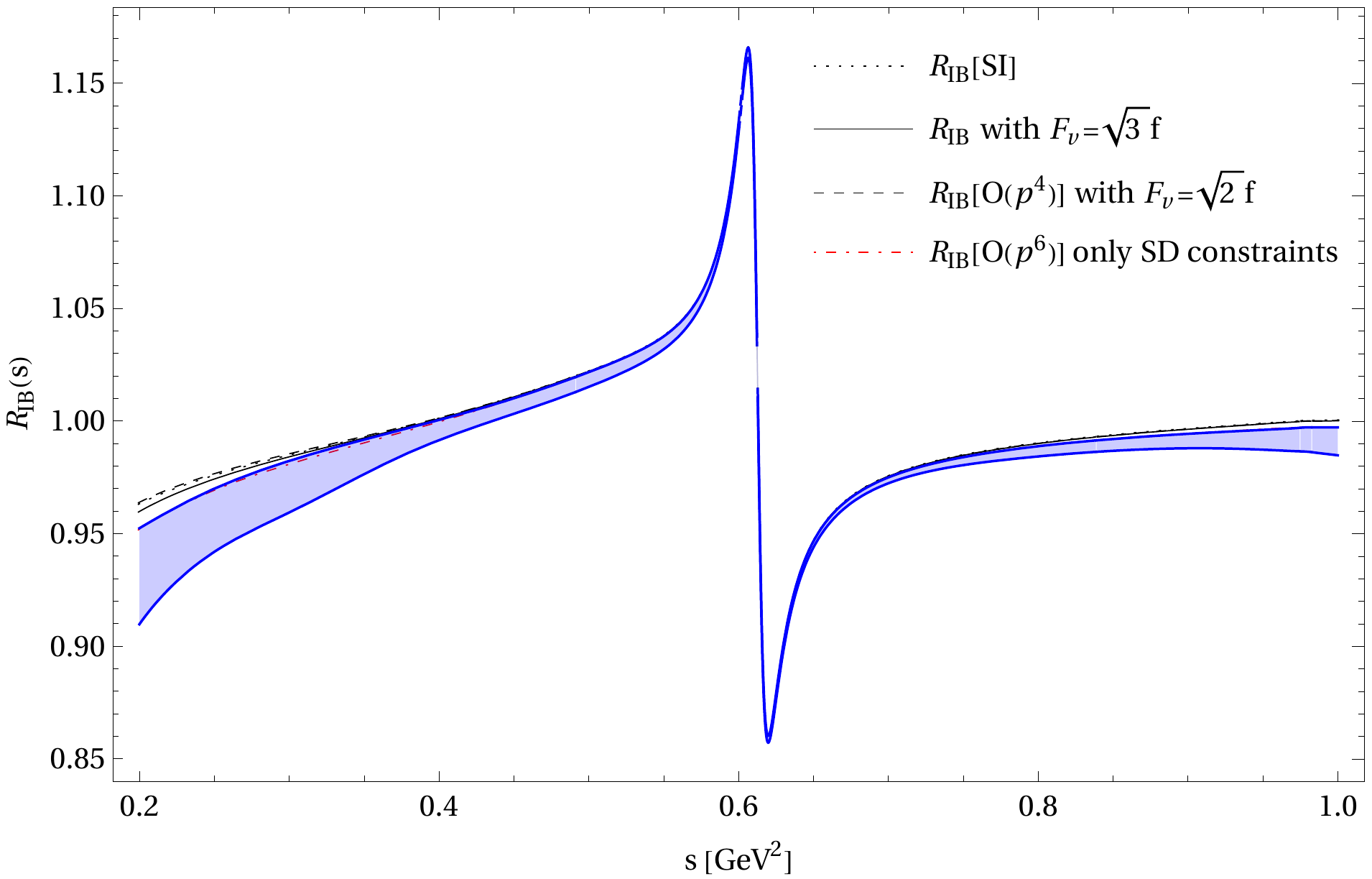}
\includegraphics[width=7.4cm]{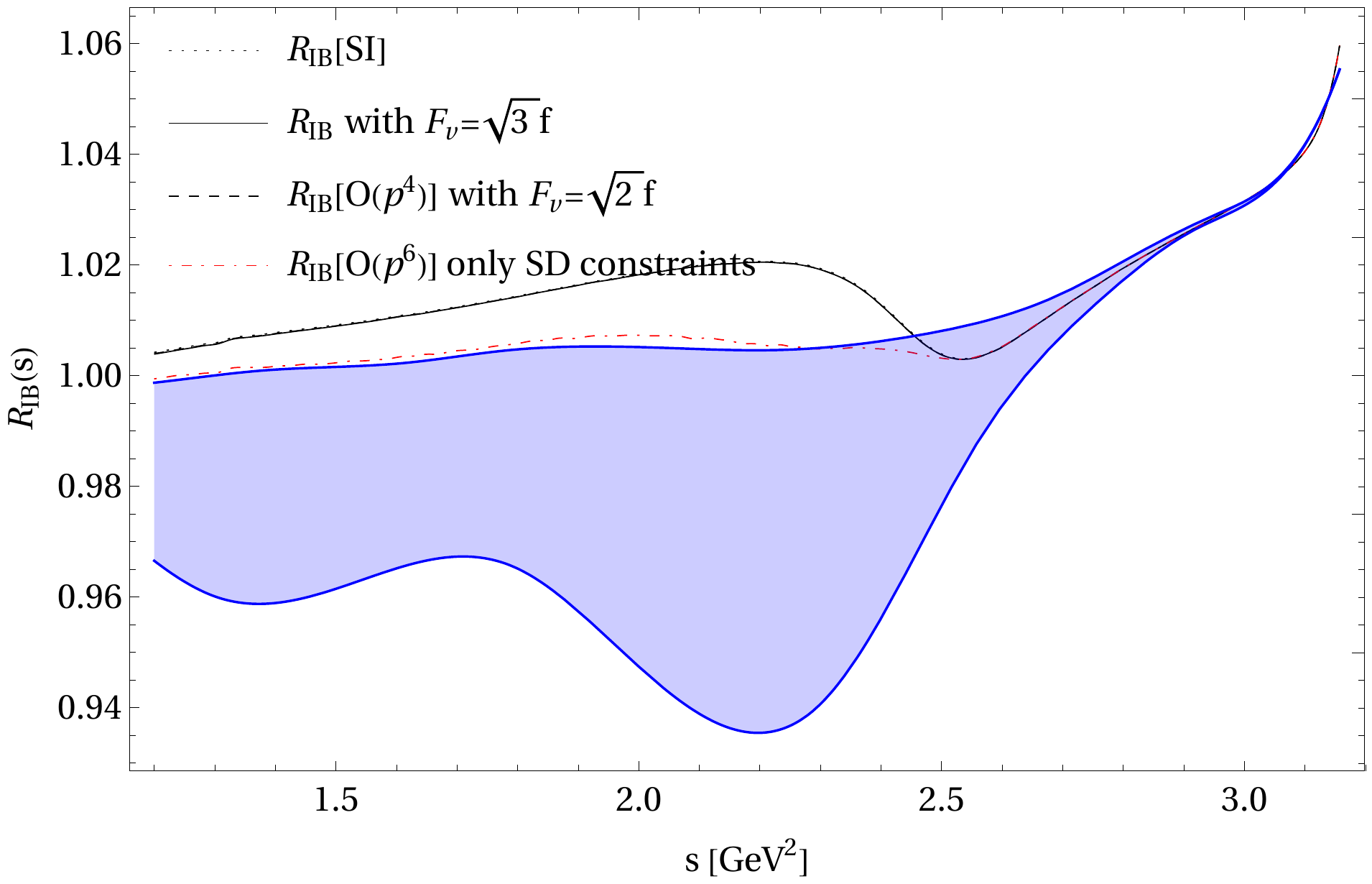}
\caption{Full IB correction factor $R_{IB}(s)$ for the different orders of approximation in $G_{EM}(s)$ using the central values given in (FF1). The blue region corresponds to the (overestimated) corrections at $\mathcal{O}(p^6)$ in $G_{EM}(s)$.}\label{RIB}
\end{center}
\end{figure}

In fig. \ref{RIB}, we show the full IB correction factor $R_{IB}(s)$ for the different orders of approximation in the $G_{EM}(s)$ factor using the DR parametrization of the form factor. As we can see, there is a difference between the contributions at $\mathcal{O}(p^4)$ and those at $\mathcal{O}(p^6)$ for energies below $\sim 0.5\mathrm{GeV}^2$ and above $\sim 0.7\mathrm{GeV}^2$. 

An important cross-check is the branching fraction $B_{\pi\pi^0}=\Gamma(\tau\to\pi\pi^0\nu_\tau)/\Gamma_\tau$ which is a directly measured quantity. It can also be evaluated from the $I=1$ component of the $e^+e^-\to\pi^+\pi^-(\gamma)$ cross section after taking into account the IB corrections. The branching fraction is given by
\begin{equation}
B_{\pi\pi^0}^{CVC}= B_{e}\int_{4m_\pi^2}^{m_\tau^2}ds\,\sigma_{\pi^+\pi^-(\gamma)}(s)\mathcal{N}(s)\frac{S_{EW}}{R_{IB}(s)},
\end{equation}
where 
\begin{equation}
\mathcal{N}(s)=\frac{3\left\vert V_{ud}\right\vert^2}{2\pi\alpha_0^2 m_\tau^2}s\left(1-\frac{s}{m_\tau^2}\right)^2\left(1+\frac{2s}{m_\tau^2}\right).
\end{equation}
Using the most recent data obtained from BaBar \cite{Lees:2012cj}~\footnote{We thank to Alex Keshavarzi and Bogdan Malaescu for providing us tables with the measurement of the $e^+e^-\to\pi^+\pi^-(\gamma)$ cross section.} for the $e^+e^-\to\pi^+\pi^-(\gamma)$ cross section and taking the same numerical inputs as we did for FF1, we get
\begin{equation}
B_{\pi\pi^0}^{CVC}=\left\lbrace \begin{array}{ll}
(24.76  \pm0.11    \pm0.25  \pm0.01  \pm 0.01 \pm 0.02)\%, & \text{SI},\\
(24.77  \pm0.11    \pm0.25  \pm0.01  \pm 0.01 \pm 0.02)\%, & F_V=\sqrt{2}F,\\
(24.77  \pm0.11    \pm0.25  \pm0.01  \pm 0.01 \pm 0.02)\%, & F_V=\sqrt{3}F,\\
(24.80  \pm0.11    \pm0.25  \pm0.01  \pm 0.01 \pm 0.02)\%, & \text{SD},\\
\end{array}\right.
\end{equation}
where `SI', `$F_V=\sqrt{2}F$', `$F_V=\sqrt{2}F$' and `SD' correspond to the different approximations of the $G_{EM}(s)$ factor. The result for $F_V=\sqrt{2}F$ is our reference one, with a negligible uncertainty from the missing higher-order terms starting at $\mathcal{O}(p^6)$.

On the other hand, when we use the same numerical inputs as in the case of FF2, we get (again our reference result is the $F_V=\sqrt{2}F$ one, with the uncertainties quoted below)
\begin{equation}
B_{\pi\pi^0}^{CVC}=\left\lbrace \begin{array}{ll}
(24.57  \pm0.11    \pm0.08  \pm0.01  \pm 0.01 \pm 0.02)\%, & \text{SI},\\
(24.57  \pm0.11    \pm0.08  \pm0.01  \pm 0.01 \pm 0.02)\%, & F_V=\sqrt{2}F,\\
(24.58  \pm0.11    \pm0.08  \pm0.01  \pm 0.01 \pm 0.02)\%, & F_V=\sqrt{3}F,\\
(24.61  \pm0.11    \pm0.08  \pm0.01  \pm 0.01 \pm 0.02)\%, & \text{SD}.\\
\end{array}\right.
\end{equation}

In both cases, the first error corresponds to the statistical experimental uncertainty on $\sigma_{\pi\pi(\gamma)}$, the second is related to uncertainty on the $\rho^+-\rho^0$ width difference, the third to the uncertainty in the $\rho^+-\rho^0$ mass difference and the fourth to the uncertainty of the $\rho-\omega$ mixing. The last error corresponds to the corrections induced by FSR on $B_{\pi\pi^0}^{CVC}$, which reduces $\sim-0.20(2)\%$ the $\pi\pi$ branching fraction.

If we include all the couplings contributing to $G_{EM}(s)$ at $\mathcal{O}(p^6)$ according to section \ref{subsec:SDC} we have an additional error associated to the EM contributions. Thus, we get
\begin{equation}
B_{\pi\pi^0}^{CVC}=(24.80  \pm0.11  \pm0.25  \pm0.01  \pm 0.01 \pm 0.02  ^{+0.21}_{-0.01})\%,
\end{equation}
for FF1, and 
\begin{equation}
B_{\pi\pi^0}^{CVC}=(24.61  \pm0.11  \pm0.08 \pm0.01  \pm 0.01 \pm 0.02 ^{+0.21}_{-0.01})\%.
\end{equation}
for FF2. Both previous results match perfectly our reference determinations obtained with $F_V=\sqrt{2}F$.

These results are in good agreement (though better for FF1) with the value reported by the Belle \cite{Fujikawa:2008ma} collaboration, 
\begin{equation}
B_{\pi\pi^0}^{\tau}=(25.24\pm0.01\pm0.39)\%,
\end{equation}
where the first uncertainty is statistical and the second is systematic. Nonetheless, they are in some tension with the very precise ALEPH measurement $(25.471\pm0.097\pm0.085)\%$ \cite{Schael:2005am}.

\begin{figure}[htbp]
\includegraphics[width=7.4cm]{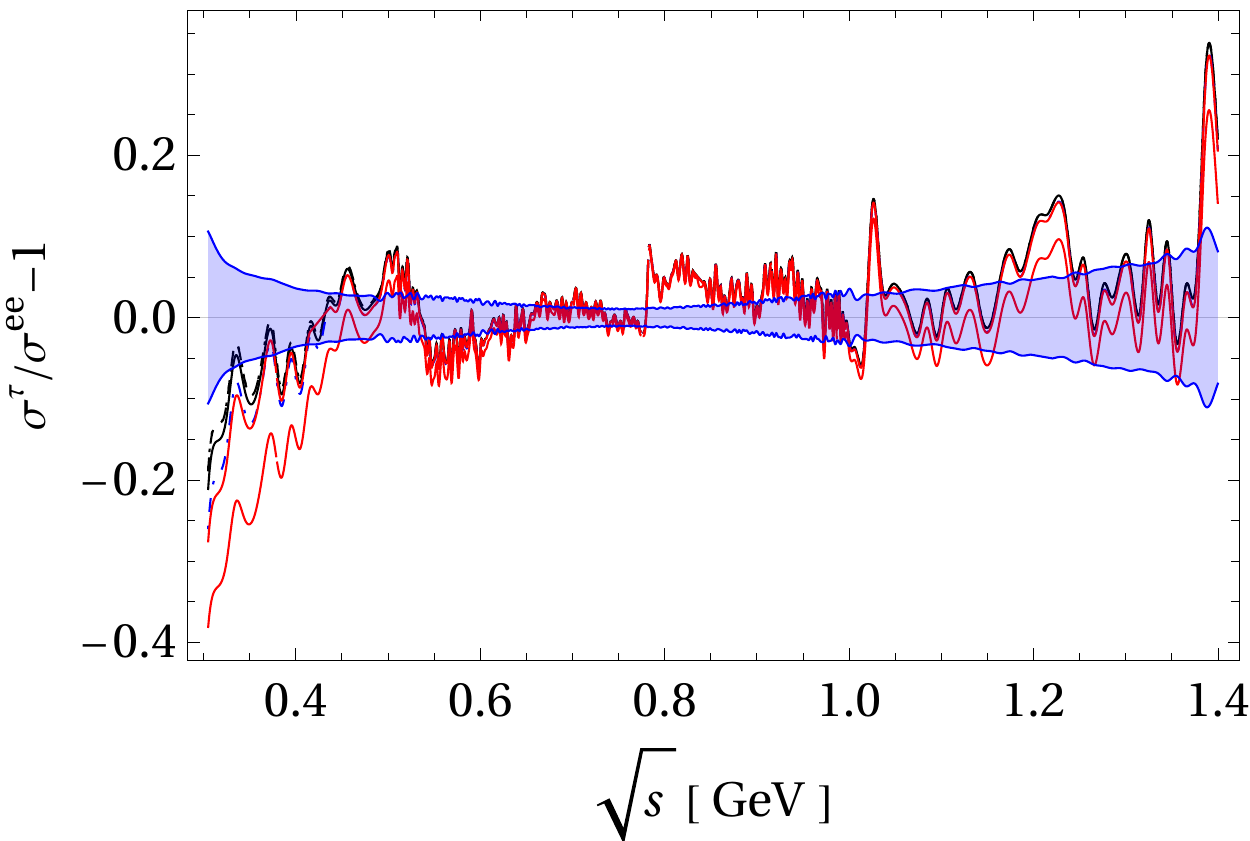}	
\includegraphics[width=7.4cm]{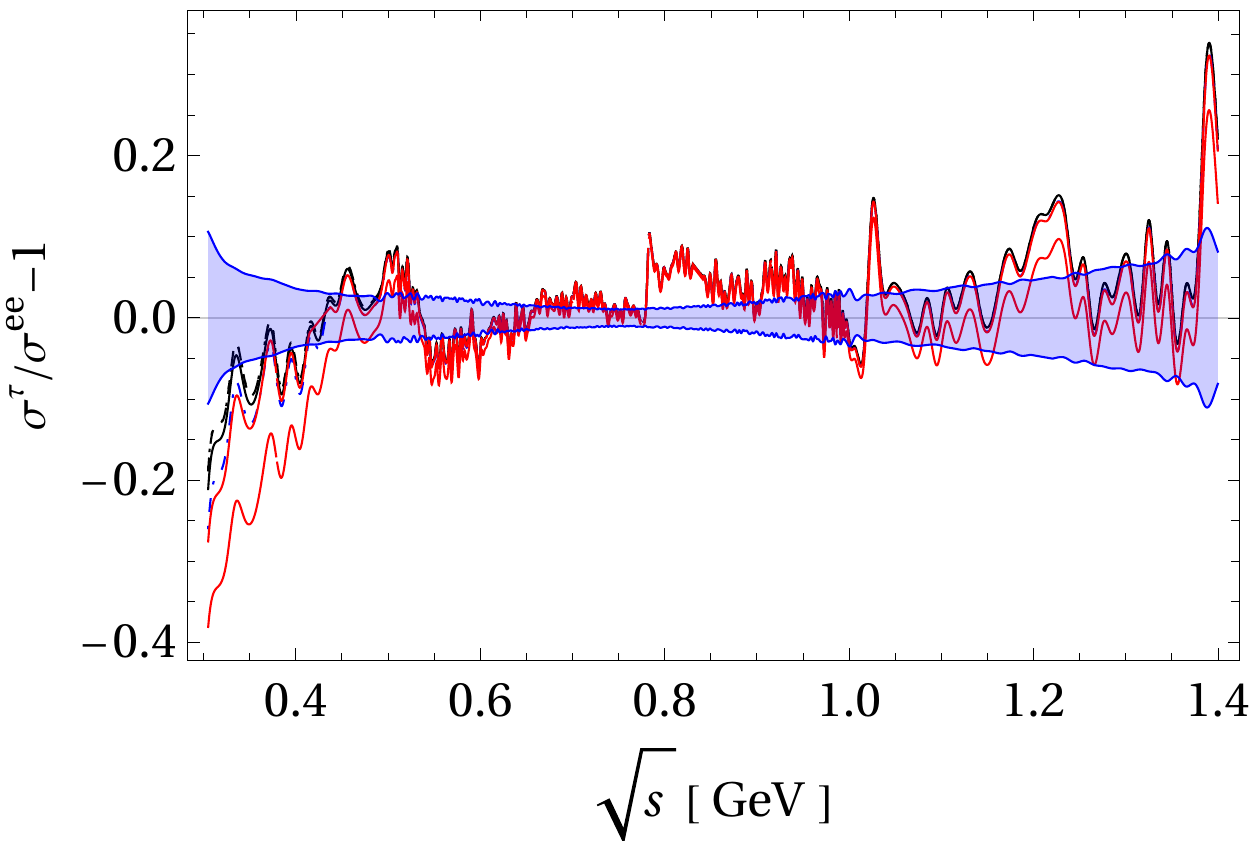}	
\includegraphics[width=7.4cm]{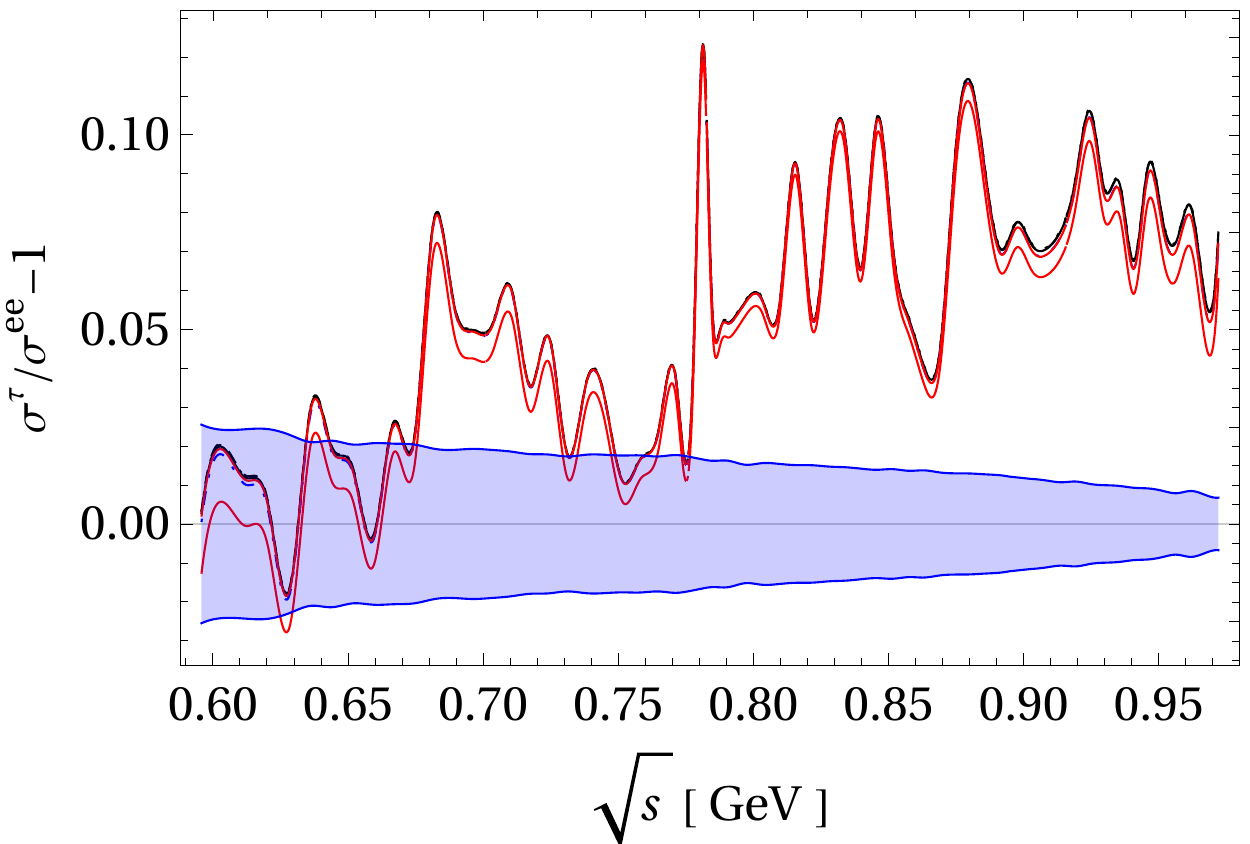}	
\includegraphics[width=7.4cm]{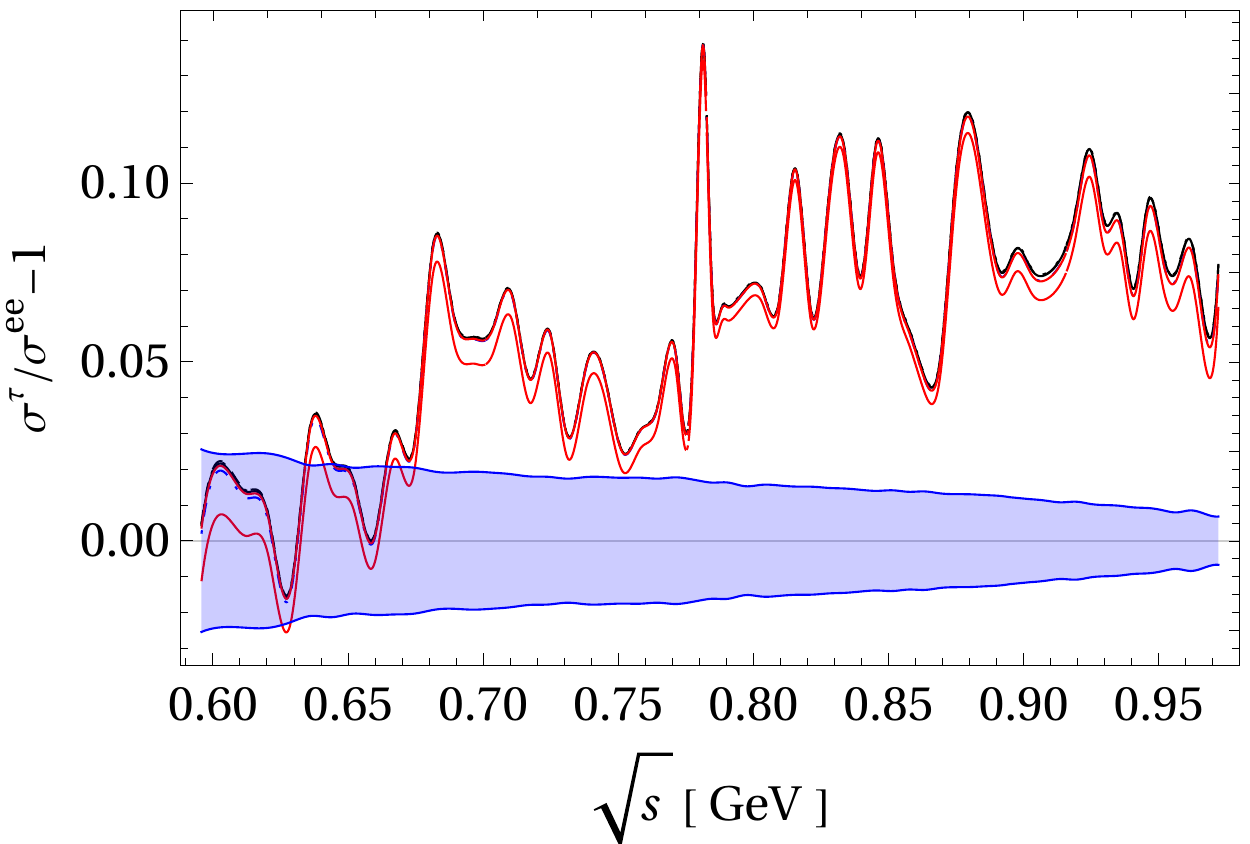}	
\centering			
\caption{Comparison between the different data sets from BaBar (above) and KLOE (below) with $\Delta\Gamma_{\pi\pi\gamma}=1.5\,\mathrm{MeV}$ (left-hand) and $\Delta\Gamma_{\pi\pi\gamma}=0.45\,\mathrm{MeV}$ (right-hand) for FF1 and FF2, respectively. The blue region corresponds to the experimental error on $\sigma_{\pi\pi(\gamma)}$. The solid and dashed lines represent the contributions with $F_V=\sqrt{3}F$ and $F_V=\sqrt{2}F$ at $\mathcal{O}(p^4)$, respectively. The dotted line is the SI contribution. The red line depicts the envelope of $G_{EM}(s)$ at $\mathcal{O}(p^6)$, that overestimates the uncertainty at this order. The blue dotdashed line is the $\mathcal{O}(p^6)$ contribution using only SD constraints.}\label{ee:fig1}
\end{figure}

 We show in fig. \ref{ee:fig1} the prediction for the $e^+e^-\to\pi^+\pi^-$ cross section using the data reported by Belle \cite{Fujikawa:2008ma} (as it is the most precise measurement of this spectrum) for the normalized spectrum $(1/N_{\pi\pi})(dN_{\pi\pi}/ds)$ compared to the last measurements from BaBar \cite{Lees:2012cj} and KLOE \cite{Babusci:2012rp}~\footnote{We have chosen to show in the comparison these two $e^+e^-$ data sets as the results from both Colls. are those deviating the most, and thus mainly responsible from the tension in $\sigma(e^+e^-\to\pi^+\pi^-)$.}.
 
We recall that the $e^+e^-\to\pi^+\pi^-$ cross section obtained using $\tau$ data is given by \cite{Fujikawa:2008ma}
\begin{equation}\label{eq:Belle}
\sigma_{\pi\pi}^0=\frac{1}{\mathcal{N}(s)}\times\left(\frac{B_{\pi\pi}}{B_e}\right)\times\left(\frac{1}{N_{\pi\pi}}\frac{dN_{\pi\pi}}{ds}\right)\left(\frac{R_{IB}(s)}{S_{EW}}\right)\,.
\end{equation}

In fig. \ref{ee:fig1} the $\tau$-based prediction is obtained using the $\mathcal{O}(p^4)$ result for $G_{EM}(s)$, with the estimated uncertainty from missing higher-order corrections given by the result at $\mathcal{O}(p^6)$ (employing only the SD constraints). The blue dotdashed line shown overestimates the error at $\mathcal{O}(p^6)$.

From fig. \ref{ee:fig1}, we observe good agreement between the BaBar data and the $\tau$ decays prediction (slightly better for FF1) 
\footnote{One can also check how important the $\rho^+-\rho^0$ width difference is around $s\simeq M_\rho^2$.}. The previous comparisons make us consider our evaluation with FF1 the reference one (so that its difference with FF2 will assess the size of the error induced by IB among the $\rho\to\pi\pi\gamma$ decay channels)~\footnote{We, nevertheless, recall that recent BESIII data \cite{Ablikim:2015orh, Ablikim:2020bah} and evaluations within the Hidden Local Symmetry model \cite{Benayoun:2011mm, Benayoun:2012wc, Benayoun:2015gxa, Benayoun:2019zwh} agree better with the KLOE data than with BaBar's.}.


Using eq. (\ref{eq:Belle}) we evaluate the IB-corrected $a_\mu^{HVP,LO}[\pi\pi,\tau]$ from the Belle mass spectrum. We use the PDG values \cite{Tanabashi:2018oca} for $m_\tau$, $V_{ud}$ and $B_e$. 

In tables \ref{HVP:tab4.01} (\ref{HVP:tab4.02})  we show IB-corrected $a_\mu^{HVP,LO}[\pi\pi,\tau]$ in units of $10^{-10}$ using the measured mass spectrum by Belle (ALEPH). For each dataset, results for the different approximations to $G_{EM}(s)$ are shown. We choose showing first the results with both Belle and ALEPH datasets as the first (second) one yields the most accurate spectral function (branching ratio) measurement. As in ref.~\cite{Davier:2009ag} (and later works by the Orsay group), the contributions are split in two intervals. In the first one, $\sqrt{s}\in[2m_{\pi^\pm},0.36\,\mathrm{GeV}]$, (the very scarce) data is not used, as this affects the precision of the integral. Instead, we use the results of the dispersive fits in ref. \cite{Gonzalez-Solis:2019iod}. We proceed analogously in tables \ref{HVP:tab4.03} and \ref{HVP:tab4.04} with the CLEO \cite{Anderson:1999ui} and OPAL \cite{Ackerstaff:1998yj}~\footnote{We thank to Jorge Portol\'es for providing us with the OPAL data set.} measurements.\\

\begin{table}[htbp]
\begin{center}
\small{\begin{tabular}{|c|c|c|c|c|c|c|c|}
\hline
 \multicolumn{8}{ |c| }{$\mathrm{FF1}$} \\
\hline
$\left[s_1,s_2\right]$ & $\mathrm{SI}$ & $F_V=\sqrt{2}F$ & $F_V=\sqrt{3}F$ & $\mathrm{SD}$ & $\mathrm{min}$ & $\mathrm{max}$ & $\mathrm{mean}$ \\
\hline
$\left[0.1296\,\mathrm{GeV}^2,1\,\mathrm{GeV}^2\right]$ & $499.43$ & $499.42$ & $499.05$ & $498.16$ & $492.18$ & $498.41$ & $495.30$ \\
$\left[0.1296\,\mathrm{GeV}^2,2\,\mathrm{GeV}^2\right]$ & $509.47$ & $509.46$ & $509.09$ & $508.14$ & $501.87$ & $508.40$ & $505.13$ \\
$\left[0.1296\,\mathrm{GeV}^2,3\,\mathrm{GeV}^2\right]$ & $509.68$ & $509.67$ & $509.30$ & $508.35$ & $502.08$ & $508.61$ & $505.34$ \\
$\left[0.1296\,\mathrm{GeV}^2,3.125\,\mathrm{GeV}^2\right]$ & $509.72$ & $509.71$ & $509.34$ & $508.40$ & $502.12$ & $508.65$ & $505.39$ \\
\hline\hline
 \multicolumn{8}{ |c| }{$\mathrm{FF2}$} \\
\hline
$\left[s_1,s_2\right]$ & $\mathrm{SI}$ & $F_V=\sqrt{2}F$ & $F_V=\sqrt{3}F$ & $\mathrm{SD}$ & $\mathrm{min}$ & $\mathrm{max}$ & $\mathrm{mean}$ \\
\hline
$\left[0.1296\,\mathrm{GeV}^2,1\,\mathrm{GeV}^2\right]$ & $503.03$ & $503.02$ & $502.65$ & $501.75$ & $495.76$ & $502.01$ & $498.88$ \\
$\left[0.1296\,\mathrm{GeV}^2,2\,\mathrm{GeV}^2\right]$ & $513.08$ & $513.06$ & $512.70$ & $511.75$ & $505.46$ & $512.00$ & $508.73$ \\
$\left[0.1296\,\mathrm{GeV}^2,3\,\mathrm{GeV}^2\right]$ & $513.29$ & $513.28$ & $512.91$ & $511.96$ & $505.66$ & $512.21$ & $508.94$ \\
$\left[0.1296\,\mathrm{GeV}^2,3.125\,\mathrm{GeV}^2\right]$ & $513.33$ & $513.32$ & $512.95$ & $512.01$ & $505.71$ & $512.26$ & $508.98$ \\
\hline
\end{tabular}}
\caption{IB-corrected $a_\mu^{HVP,LO}[\pi\pi,\tau]$ in units of $10^{-10}$ using the measured mass spectrum by Belle with $B_{\pi\pi}=(25.24\pm0.01\pm0.39)\%$. Different approximation to $G_{EM}(s)$ are displayed in the various columns. The last three of them show the results at $\mathcal{O}(p^6)$ and their differences overestimate the error at this order. The error of the $\mathcal{O}(p^4)$ prediction (obtained with $F_V=\sqrt{2}F$) can be quantified from its difference with the $\mathrm{SD}$ value (corresponding to the $\mathcal{O}(p^6)$ contribution using only SD constraints).}
\label{HVP:tab4.01}
\end{center}
\end{table}

\begin{table}[htbp]
\begin{center}
\small{\begin{tabular}{|c|c|c|c|c|c|c|c|}
\hline
 \multicolumn{8}{ |c| }{$\mathrm{FF1}$} \\
\hline
$\left[s_1,s_2\right]$ & $\mathrm{SI}$ & $F_V=\sqrt{2}F$ & $F_V=\sqrt{3}F$ & $\mathrm{SD}$ & $\mathrm{min}$ & $\mathrm{max}$ & $\mathrm{mean}$ \\
\hline
$\left[0.1296\,\mathrm{GeV}^2,1\,\mathrm{GeV}^2\right]$ & $495.28$ & $495.27$ & $494.92$ & $494.05$ & $488.25$ & $494.30$ & $491.27$ \\
$\left[0.1296\,\mathrm{GeV}^2,2\,\mathrm{GeV}^2\right]$ & $506.57$ & $506.56$ & $506.21$ & $505.29$ & $499.15$ & $505.53$ & $502.34$ \\
$\left[0.1296\,\mathrm{GeV}^2,3\,\mathrm{GeV}^2\right]$ & $506.82$ & $506.81$ & $506.45$ & $505.53$ & $499.38$ & $505.77$ & $502.58$ \\
$\left[0.1296\,\mathrm{GeV}^2,3.125\,\mathrm{GeV}^2\right]$ & $506.82$ & $506.81$ & $506.46$ & $505.53$ & $499.39$ & $505.78$ & $502.58$ \\
\hline\hline
 \multicolumn{8}{ |c| }{$\mathrm{FF2}$} \\
\hline
$\left[s_1,s_2\right]$ & $\mathrm{SI}$ & $F_V=\sqrt{2}F$ & $F_V=\sqrt{3}F$ & $\mathrm{SD}$ & $\mathrm{min}$ & $\mathrm{max}$ & $\mathrm{mean}$ \\
\hline
$\left[0.1296\,\mathrm{GeV}^2,1\,\mathrm{GeV}^2\right]$         & $498.86$ & $498.85$ & $498.50$ & $497.63$ & $491.81$ & $497.87$ & $494.84$ \\
$\left[0.1296\,\mathrm{GeV}^2,2\,\mathrm{GeV}^2\right]$         & $510.16$ & $510.15$ & $509.80$ & $508.87$ & $502.72$ & $509.12$ & $505.92$ \\
$\left[0.1296\,\mathrm{GeV}^2,3\,\mathrm{GeV}^2\right]$         & $510.41$ & $510.40$ & $510.04$ & $509.12$ & $502.95$ & $509.36$ & $506.16$ \\
$\left[0.1296\,\mathrm{GeV}^2,3.125\,\mathrm{GeV}^2\right]$ & $510.41$ & $510.40$ & $510.05$ & $509.12$ & $502.96$ & $509.36$ & $506.16$ \\
\hline
\end{tabular}}
\caption{IB-corrected $a_\mu^{HVP,LO}[\pi\pi,\tau]$ in units of $10^{-10}$ using the measured mass spectrum by ALEPH with $B_{\pi\pi}=(25.471\pm0.097\pm0.085)\%$. The rest is as in table \ref{HVP:tab4.01}.}
\label{HVP:tab4.02}
\end{center}
\end{table}

\begin{table}[htbp]
\begin{center}
\small{\begin{tabular}{|c|c|c|c|c|c|c|c|}
\hline
 \multicolumn{8}{ |c| }{$\mathrm{FF1}$} \\
\hline
$\left[s_1,s_2\right]$ & $\mathrm{SI}$ & $F_V=\sqrt{2}F$ & $F_V=\sqrt{3}F$ & $\mathrm{SD}$ & $\mathrm{min}$ & $\mathrm{max}$ & $\mathrm{mean}$ \\
\hline
$\left[0.1296\,\mathrm{GeV}^2,1\,\mathrm{GeV}^2\right]$ & $498.51$ & $498.50$ & $498.14$ & $497.27$ & $491.43$ & $497.52$ & $494.47$ \\
$\left[0.1296\,\mathrm{GeV}^2,2\,\mathrm{GeV}^2\right]$ & $508.98$ & $508.97$ & $508.61$ & $507.69$ & $501.54$ & $507.93$ & $504.74$ \\
$\left[0.1296\,\mathrm{GeV}^2,3\,\mathrm{GeV}^2\right]$ & $509.15$ & $509.14$ & $508.79$ & $507.86$ & $501.70$ & $508.11$ & $504.91$ \\
$\left[0.1296\,\mathrm{GeV}^2,3.125\,\mathrm{GeV}^2\right]$ & $509.20$ & $509.18$ & $508.83$ & $507.91$ & $501.75$ & $508.15$ & $504.95$ \\
\hline\hline
 \multicolumn{8}{ |c| }{$\mathrm{FF2}$} \\
\hline
$\left[s_1,s_2\right]$ & $\mathrm{SI}$ & $F_V=\sqrt{2}F$ & $F_V=\sqrt{3}F$ & $\mathrm{SD}$ & $\mathrm{min}$ & $\mathrm{max}$ & $\mathrm{mean}$ \\
\hline
$\left[0.1296\,\mathrm{GeV}^2,1\,\mathrm{GeV}^2\right]$         & $502.10$ & $502.09$ & $501.74$ & $500.86$ & $495.00$ & $501.11$ & $498.06$ \\
$\left[0.1296\,\mathrm{GeV}^2,2\,\mathrm{GeV}^2\right]$         & $512.58$ & $512.57$ & $512.22$ & $511.29$ & $505.12$ & $511.58$ & $508.33$ \\
$\left[0.1296\,\mathrm{GeV}^2,3\,\mathrm{GeV}^2\right]$         & $512.76$ & $512.75$ & $512.39$ & $511.47$ & $505.29$ & $511.71$ & $508.50$ \\
$\left[0.1296\,\mathrm{GeV}^2,3.125\,\mathrm{GeV}^2\right]$ & $512.80$ & $512.79$ & $512.43$ & $511.51$ & $505.33$ & $511.75$ & $508.54$ \\
\hline
\end{tabular}}
\caption{IB-corrected $a_\mu^{HVP,LO}[\pi\pi,\tau]$ in units of $10^{-10}$ using the measured mass spectrum by CLEO with $B_{\pi\pi}=(25.36\pm0.44)\%$. The rest is as in table \ref{HVP:tab4.01}.}
\label{HVP:tab4.03}
\end{center}
\end{table}

\begin{table}[htbp]
\begin{center}
\small{\begin{tabular}{|c|c|c|c|c|c|c|c|}
\hline
 \multicolumn{8}{ |c| }{$\mathrm{FF1}$} \\
\hline
$\left[s_1,s_2\right]$ & $\mathrm{SI}$ & $F_V=\sqrt{2}F$ & $F_V=\sqrt{3}F$ & $\mathrm{SD}$ & $\mathrm{min}$ & $\mathrm{max}$ & $\mathrm{mean}$ \\
\hline
$\left[0.1296\,\mathrm{GeV}^2,1\,\mathrm{GeV}^2\right]$         & $509.50$ & $509.51$ & $509.07$ & $508.04$ & $501.31$ & $508.34$ & $504.82$ \\
$\left[0.1296\,\mathrm{GeV}^2,2\,\mathrm{GeV}^2\right]$         & $521.29$ & $521.29$ & $520.86$ & $519.77$ & $512.69$ & $520.06$ & $516.34$ \\
$\left[0.1296\,\mathrm{GeV}^2,3\,\mathrm{GeV}^2\right]$         & $521.49$ & $521.49$ & $521.06$ & $519.96$ & $512.88$ & $520.25$ & $516.56$ \\
$\left[0.1296\,\mathrm{GeV}^2,3.125\,\mathrm{GeV}^2\right]$ & $521.49$ & $521.49$ & $521.06$ & $519.97$ & $512.88$ & $520.26$ & $516.57$ \\
\hline\hline
 \multicolumn{8}{ |c| }{$\mathrm{FF2}$} \\
\hline
$\left[s_1,s_2\right]$ & $\mathrm{SI}$ & $F_V=\sqrt{2}F$ & $F_V=\sqrt{3}F$ & $\mathrm{SD}$ & $\mathrm{min}$ & $\mathrm{max}$ & $\mathrm{mean}$ \\
\hline
$\left[0.1296\,\mathrm{GeV}^2,1\,\mathrm{GeV}^2\right]$         & $512.99$ & $512.99$ & $512.56$ & $511.53$ & $504.78$ & $511.82$ & $508.30$ \\
$\left[0.1296\,\mathrm{GeV}^2,2\,\mathrm{GeV}^2\right]$         & $524.79$ & $524.79$ & $524.36$ & $523.27$ & $516.17$ & $523.56$ & $519.86$ \\
$\left[0.1296\,\mathrm{GeV}^2,3\,\mathrm{GeV}^2\right]$         & $524.99$ & $524.99$ & $524.56$ & $523.46$ & $516.36$ & $523.76$ & $520.06$ \\
$\left[0.1296\,\mathrm{GeV}^2,3.125\,\mathrm{GeV}^2\right]$ & $524.99$ & $524.99$ & $524.56$ & $523.46$ & $516.36$ & $523.76$ & $520.06$ \\
\hline
\end{tabular}}
\caption{IB-corrected $a_\mu^{HVP,LO}[\pi\pi,\tau]$ in units of $10^{-10}$ using the measured mass spectrum by OPAL with $B_{\pi\pi}=(25.46\pm0.17\pm0.29)\%$. The rest is as in table \ref{HVP:tab4.01}.}
\label{HVP:tab4.04}
\end{center}
\end{table}

\begin{table}[htbp]
\begin{center}
\small{\begin{tabular}{|c|c|c|c|}
\hline
 \multicolumn{4}{ |c| }{$a_\mu^{HVP,LO}[\pi\pi,\tau]$} \\
\hline
Experiment & $2m_{\pi^\pm} - 0.36\,\mathrm{GeV}$ & $0.36 - 1.8\,\mathrm{GeV}$ & TOTAL \\
\hline
Belle & $8.81\pm0.00\pm0.14^{+0.16}_{-0.34}$ & $511.14\pm 1.94\pm 7.99^{+1.91}_{-2.09}$ & $519.95\pm 1.94 \pm 7.99^{+1.91}_{-2.12}$ \\
ALEPH & $8.89\pm0.00\pm0.05^{+0.16}_{-0.34}$ & $508.26\pm 4.48\pm 2.82^{+1.91}_{-2.09}$ & $517.15\pm 4.48\pm 2.82^{+1.91}_{-2.12}$ \\
CLEO & $8.85\pm0.00\pm0.15^{+0.16}_{-0.34}$ & $510.63\pm 3.40\pm 8.93^{+1.90}_{-2.08}$ & $519.48\pm 3.40\pm 8.93^{+1.90}_{-2.11}$ \\
OPAL & $8.89\pm0.00\pm0.12^{+0.15}_{-0.34}$ & $522.81\pm 10.04\pm 7.00^{+1.87}_{-2.12}$ & $531.70\pm 10.04 \pm 7.00^{+1.87}_{-2.15}$ \\
\hline
\end{tabular}}
\caption{IB-corrected $a_\mu^{HVP,LO}[\pi\pi,\tau]$ in units of $10^{-10}$ at $\mathcal{O}(p^4)$. The first error is related to the systematic uncertainties on the mass spectrum, and also include contributions from the $\tau$-mass and $V_{ud}$ uncertainties. The second error arises from $B_{\pi\pi^0}$ and $B_e$, and the third error from the isospin-breaking corrections.}
\label{HVP:tab4.4a}
\end{center}
\end{table}

\begin{table}[htbp]
\begin{center}
\small{\begin{tabular}{|c|c|c|c|}
\hline
 \multicolumn{4}{ |c| }{$a_\mu^{HVP,LO}[\pi\pi,\tau]$} \\
\hline
Experiment & $2m_{\pi^\pm} - 0.36\,\mathrm{GeV}$ & $0.36 - 1.8\,\mathrm{GeV}$ & TOTAL \\
\hline
Belle     & $7.77\pm0.00\pm0.12^{+1.20}_{-0.59}$ & $507.18\pm 1.91\pm 7.88^{+4.72}_{-3.76}$ & $514.95\pm 1.91 \pm 7.88^{+4.87}_{-3.81}$ \\
ALEPH  & $7.84\pm0.00\pm0.04^{+1.21}_{-0.60}$ & $504.37\pm 4.35\pm 2.79^{+4.63}_{-3.70}$ & $512.21\pm 4.35\pm 2.79^{+4.78}_{-3.75}$ \\
CLEO    & $7.80\pm0.00\pm0.14^{+1.21}_{-0.59}$ & $506.74\pm 3.28\pm 8.84^{+4.63}_{-3.71}$ & $514.54\pm 3.28\pm 8.84^{+4.78}_{-3.76}$ \\
OPAL    & $7.84\pm0.00\pm0.10^{+1.20}_{-0.60}$ & $518.32\pm 9.69\pm 6.92^{+5.25}_{-4.12}$ & $526.16\pm 9.69 \pm 6.92^{+5.39}_{-4.16}$ \\
\hline
\end{tabular}}
\caption{IB-corrected $a_\mu^{HVP,LO}[\pi\pi,\tau]$ in units of $10^{-10}$ at $\mathcal{O}(p^6)$. 
The rest is as in table \ref{HVP:tab4.4a}.}
\label{HVP:tab4.4b}
\end{center}
\end{table}

Taking into account all di-pion tau decay data from the ALEPH \cite{Schael:2005am}, Belle \cite{Fujikawa:2008ma}, CLEO \cite{Anderson:1999ui} and OPAL \cite{Ackerstaff:1998yj} Colls. (the latter yielding the largest contribution to $a_\mu^{HVP,LO|_{\pi\pi}}$ exceeding $\sim10.7\cdot10^{-10}$ the mean, although with the largest errors as well) in tables \ref{HVP:tab4.4a} and \ref{HVP:tab4.4b} at $\mathcal{O}(p^4)$ and $\mathcal{O}(p^6)$, respectively, we get the combined tau-data contribution
\begin{equation}\label{pipifromtau_p4}
 10^{10}\cdot a_\mu^{HVP,LO|_{\pi\pi,\tau\;\mathrm{data}}}\,=\,519.6\pm{2.8_{spectra+BRs}}{^{+1.9}_{-2.1}}_{IB}\,,
\end{equation}
at $\mathcal{O}(p^4)$ and 
\begin{equation}\label{pipifromtau_p6}
10^{10}\cdot a_\mu^{HVP,LO|_{\pi\pi,\tau\;\mathrm{data}}}\,=\,514.6\pm{2.8_{spectra+BRs}}{^{+5.0}_{-3.9}}_{IB}\,,
\end{equation}
at $\mathcal{O}(p^6)$.

The $IB$ errors come from the uncertainty on $\Gamma(\rho\to\pi\pi\gamma)$ (FF1 vs FF2) and either from the difference between the $F_V=\sqrt{2}F$ and $SD$ results (in eq.~(\ref{pipifromtau_p4})) or from the difference between the 'mean' and 'min'/'max' results (in eq.~(\ref{pipifromtau_p6})).

Contrary to previous estimates \cite{Cirigliano:2002pv, Davier:2009ag, Davier:2010nc, Jegerlehner:2011ti, Davier:2013sfa}, the errors in $a_\mu^{HVP,LO|_{\pi\pi,\tau\;\mathrm{data}}}$ happen to be dominated by the uncertainty on the IB contributions (but for the lower error on eq.~(\ref{pipifromtau_p4})).

When eqs.~(\ref{pipifromtau_p4}) and (\ref{pipifromtau_p6}) are supplemented with the four-pion tau decays measurements (up to $1.5$ GeV) and with $e^+e^-$ data at larger energies in these modes (and with $e^+e^-$ data in all other channels making up the hadronic cross section), we get \cite{Davier:2019can,Davier:2013sfa}

\begin{equation}\label{HVP,LO tau p4}
10^{10}\,\cdot\, a_\mu^{HVP,LO|_{\tau\;\mathrm{data}}}\,=\,705.7\pm{2.8_{spectra+BRs}}{^{+1.9}_{-2.1}}_{IB}\pm2.0_{e^+e^-}\pm0.1_{narrow\,res}\pm0.7_{QCD}\,,
\end{equation}
at $\mathcal{O}(p^4)$, and 

\begin{equation}\label{HVP,LO tau p6}
10^{10}\,\cdot\, a_\mu^{HVP,LO|_{\tau\;\mathrm{data}}}\,=\,700.7\pm{2.8_{spectra+BRs}}{^{+5.0}_{-3.9}}_{IB}\pm2.0_{e^+e^-}\pm0.1_{narrow\,res}\pm0.7_{QCD}\,,
\end{equation}

at $\mathcal{O}(p^6)$ and we have also included the uncertainties corresponding to using $e^+e^-$ data for those contributions not covered by tau decay measurements and to the inclusion of narrow resonances and the perturbative QCD part.

Adding errors in quadrature, an uncertainty of $^{+4.0}_{-4.1}$ ($^{+6.1}_{-5.2}$) is obtained at $\mathcal{O}(p^4)$ ($\mathcal{O}(p^6)$). These numbers (all in units of $10^{-10}$) have to be compared with the error of $4.0$ in ref. \cite{Aoyama:2020ynm}.

When all other (QED, EW and subleading hadronic) contributions are added to eqs. (\ref{HVP,LO tau p4}) and (\ref{HVP,LO tau p6}) according to ref. \cite{Aoyama:2020ynm}, the $3.7\sigma$ \cite{Aoyama:2020ynm} deficit of the SM prediction with respect to the $BNL$ measurement \cite{Bennett:2006fi} is reduced to

\begin{equation}
 \Delta a_\mu \equiv a_\mu^{exp}-a_\mu^{SM}=(15.3\pm 7.7)\cdot 10^{-10}\,,
\end{equation}
at $\mathcal{O}(p^4)$, and 

\begin{equation}
\Delta a_\mu \equiv a_\mu^{exp}-a_\mu^{SM}=(20.3^{+8.3}_{-8.9})\cdot 10^{-10}\,,
\end{equation}
at $\mathcal{O}(p^6)$, which are $2.0$ and $2.3\,\sigma$, respectively.

\section{Conclusions}\label{sec:Concl}

In this work we have revisited the resonance chiral Lagrangian computation of the isospin-breaking and radiative corrections to the $\tau^-\to\pi^-\pi^0\nu_\tau\gamma$ decays in ref. \cite{Cirigliano:2002pv}, by including the terms that start to contribute to the $\mathcal{O}(p^6)$ chiral LECs. Our main motivation for that was to revisit the determination of $a_\mu^{HVP,LO}$ using tau decay data, so that it could -when combined with the $e^+e^-$ measurements- reduce the Standard Model error on $a_\mu$, thus enhancing the sensitivity to new physics of the current BNL and future FNAL and J-PARC measurements.

Our isospin breaking corrections improve the agreement between $\tau$ and $e^+e^-$ di-pion data (both in the spectrum and its integral), which endorses our evaluation of $a_\mu^{HVP,LO|_{\tau\;\mathrm{data}}}$. Our main results are $a_\mu^{HVP,LO|_{\tau\;\mathrm{data}}}\,=\,(705.7^{+4.0}_{-4.1})\cdot10^{-10}$ (including the same contributions as in ref. \cite{Cirigliano:2002pv}), and $a_\mu^{HVP,LO|_{\tau\;\mathrm{data}}}\,=\,(700.7^{+6.1}_{-5.2})\cdot10^{-10}$ (when the operators starting to contribute to the $\mathcal{O}(p^6)$ LECs are also considered). These reduce the anomaly $\Delta a_\mu \equiv a_\mu^{exp}-a_\mu^{SM}$ to $2.0$ and $2.3\,\sigma$, respectively.

We also provide with a detailed study of the $\pi\pi$ spectrum, $E_\gamma$ distribution and branching ratio, for different cuts on the photon energy. These $\tau^-\to\pi^-\pi^0\nu_\tau\gamma$ decays observables have the potential to reduce drastically the error of our predictions, so we eagerly await their measurement at Belle-II.

\section*{Acknowledgements}
A.~M.~acknowledges Conacyt support through his Ph. D. scholarship. P.~R.~thanks the funding of Fondo SEP-Cinvestav 2018 (project number 142). A.~M.~ and P.~R.~ have benefitted from enriching discussions on this topic with Gabriel L\'opez Castro and Genaro Toledo S\'anchez. We thank Vincenzo Cirigliano, Antonio Pich and Jorge Portol\'es for helpful suggestions regarding the presentation and discussion of our results, and Antonio Rojas, Eduard de la Cruz Burelo and Iv\'an Heredia de la Cruz for their valuable help. We are indebted to Alex Keshavarzi, Bogdan Malaescu, Hisaki Hayashii and Jorge Portol\'es for providing us with the BaBar, Belle and OPAL data sets.

\appendix
\section{Fit results}\label{Fit}
Since the $\kappa_{i}^{V}$ couplings are related with the $\omega$ exchange which is known to give an important contribution to the $\tau\to\pi\pi\gamma\nu_\tau$ decays, we perform a global fit using the relations for the resonance saturation of the anomalous sector LECs at NLO \cite{Kampf:2011ty}, the eqs. (\ref{SD:eq01})-(\ref{SD:eq4}) in section \ref{subsec:SDC} and the estimation of the LECs in \cite{Jiang:2015dba}.\\

Neglecting all the other contributions, we get
\begin{subequations}\begin{align}
\kappa^{V}_{1}&=(-2.1\pm0.7)\cdot 10^{-2}\text{ GeV}^{-1},\\
\kappa^{V}_{2}&=(-8.8\pm9.1)\cdot 10^{-3}\text{ GeV}^{-1},\\
\kappa^{V}_{3}&=(2.2\pm5.8)\cdot 10^{-3}\text{ GeV}^{-1},\\
\kappa^{V}_{6}&=(-2.1\pm0.3)\cdot 10^{-2}\text{ GeV}^{-1},\\
\kappa^{V}_{7}&=(1.2\pm 0.5)\cdot 10^{-2}\text{ GeV}^{-1},\\
\kappa^{V}_{8}&=(3.1\pm 0.9)\cdot 10^{-2}\text{ GeV}^{-1},\\
\kappa^{V}_{9}&=(-0.1\pm 5.9)\cdot 10^{-3}\text{ GeV}^{-1},\\
\kappa^{V}_{10}&=(-5.9\pm 9.6)\cdot 10^{-3}\text{ GeV}^{-1},\\
\kappa^{V}_{11}&=(-3.0\pm 0.6)\cdot 10^{-2}\text{ GeV}^{-1},\\
\kappa^{V}_{12}&=(1.0\pm 0.8)\cdot 10^{-2}\text{ GeV}^{-1},\\
\kappa^{V}_{13}&=(-5.3\pm 1.1)\cdot 10^{-3}\text{ GeV}^{-1},\\
\kappa^{V}_{18}&=(4.7\pm 0.8)\cdot 10^{-3}\text{ GeV}^{-1}.
\end{align}\end{subequations}
These values are in good agreement with our earlier estimation in section \ref{subsec:SDC}, $\vert\kappa_i^V\vert \lesssim 0.025\, \mathrm{GeV}^{-1}$.

\section{Kinematics}\label{kin}
\subsection{$\tau^-(P)\to\pi^-(p_-)\pi^0(p_0)\gamma(k) \nu_\tau(q)$ kinematics}
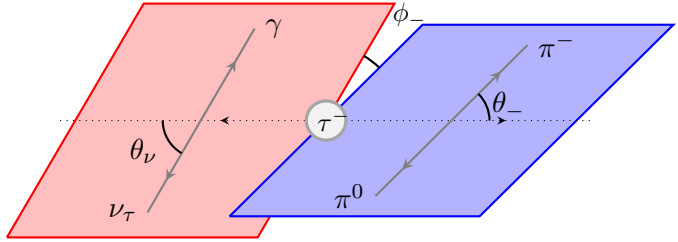
\begin{figure}[ht]
\centering
\begin{tikzpicture}
\filldraw[color=red,fill=pink, thick] (-4.2,-1.55)--(-2.4,1.55)--(0.9,1.55)--(-0.9,-1.55)--(-4.2,-1.55);
\filldraw[color=blue,fill=blue!30!white, thick] (-1.27,-1.27)--(1.27,1.27)--(4.57,1.27)--(2.02,-1.27)--(-1.27,-1.27);
\draw[black, dotted,reverse directed] (-3.5,0) -- (-0.25,0);
\draw[black, dotted, directed] (0.25,0) -- (3.5,0);
\draw[gray, thick, directed] (1.65,0) -- (2.65,1)node[below, right, black]{$\pi^-$};
\draw[gray, thick, directed] (1.65,0) -- (0.65,-1)node[above, left, black]{$\pi^0$};
\filldraw[color=gray!70, fill=black!5, very thick](0,0) circle (0.26);
\node at (0.1,0) {\small$\tau^-$};
\draw[thick] (2.15,0) arc (0:45:0.5);
\node at (2.4,0.2) {$\theta_-$};
\draw[gray, thick, directed] (-1.65,0)--(-0.94,1.22)node[right, black]{$\gamma$};
\draw[gray, thick, directed] (-1.65,0)--(-2.35,-1.22)node[left, black]{$\nu_\tau$};
\draw[thick] (-2.15,0) arc (180:240:0.5);
\node at (-2.4,-0.4) {$\theta_\nu$};
\draw[thick] (0.7,0.7) arc (45:60:0.99);
\node at (1.1,1.4) {\small$\phi_-$};
\end{tikzpicture}
\caption{The $\tau^-\to\pi^-\pi^0\gamma\nu_\tau$ decay in the $\tau$-lepton rest frame.}\label{Appx4:fig1}
\end{figure}

In order to describe this type of decays we need five independent variables. We choose $s=(p_-+p_0)^2$, $u=(P-p_-)^2$, $x=(k+q)^2$, $\theta_\nu$ which is the angle between the direction of the $\pi^-\pi^0$ CM frame in the $\tau$ lepton rest frame and the direction of $\vec{q}$ in the $\pi^-\pi^0$ CM frame (see fig. \ref{Appx4:fig1}) and $\phi_-$, which is angle between the plane of the $\pi^-\pi^0$ CM frame and the plane of the $\gamma\nu_\tau$ CM frame. \\ 
We can write the invariants in terms of these variables
\begin{subequations}\begin{align}
                     P\cdot p_0&=\frac{s+u-x-m_{\pi^0}^2}{2},\\
                     q\cdot k &=\frac{x-M_\gamma^2}{2},\\
                     p_-\cdot p_0&=\frac{s-m_{\pi^-}^2-m_{\pi^0}^2}{2},\\
                     p_-\cdot \left(q+k\right)&=\frac{u-x-m_{\pi^0}^2}{2},\\
                     P\cdot \left(q+k\right)&=\frac{x-s+m_\tau^2}{2},
                    \end{align}
\end{subequations}
\begin{equation}\begin{split}
                     P\cdot p_-&=\frac{(m_{\pi^-}^2-m_{\pi^0}^2+s)(m_\tau^2+s-x)}{4s}+\frac{\lambda^{1/2}(s,x,m_\tau^2)\lambda^{1/2}(m_\tau^2,m_{\pi^-}^2,m_{\pi^0}^2)}{4s}\cos \theta_-\\
					&=\frac{m_\tau^2+m_{\pi^-}^2-u}{2},
\end{split}\end{equation}
\begin{equation}
 P\cdot k=\frac{(m_\tau^2-s+x)(x+M_\gamma^2)}{4 x}-\frac{(x-M_\gamma^2)\lambda^{1/2}\left(s,x,m_\tau^2\right)}{4 x}\cos\theta_\nu,
\end{equation}
\begin{equation}\begin{split}
 p_-\cdot k =&\frac{\left(x+M_\gamma^2\right)\left(m_\tau^2-s-u+m_{\pi^0}^2\right)}{4x}-\frac{\left(x-M_\gamma^2\right)\cos\theta_\nu}{4x\,\lambda^{1/2}\left(s,x,m_\tau^2\right)}\,A\left(s,u,x\right)\\
 &-\frac{\left(x-M_\gamma^2\right)\lambda^{1/2}\left(s,m_{\pi^-}^2,m_{\pi^0}^2\right) }{4\sqrt{x}\sqrt{s}}\sin\theta_\nu \sin\theta_-\cos\phi_-,
\end{split}\end{equation}
\begin{equation}\begin{split}
\epsilon^{\mu\nu\alpha\beta}k_\mu P_\nu p_{-\alpha}q_\beta&=\frac{(x-M_\gamma^2)\lambda^{1/2}\left(s,m_{\pi^-}^2,m_{\pi^0}^2\right)\lambda^{1/2}\left(s,x,m_{\tau}^2\right)}{8\sqrt{s}\sqrt{x}}\times\\
&\quad\sin\theta_\nu\sin\theta_- \sin\phi_-,
\end{split}\end{equation}
\begin{equation}\begin{split}
\epsilon^{\mu\nu\alpha\beta}k_\mu P_\nu p_{-\alpha}p_{0\beta}&=\epsilon^{\mu\nu\alpha\beta}k_\mu P_\nu p_{0\alpha}q_\beta=\epsilon^{\mu\nu\alpha\beta}k_\mu p_{-\nu} p_{0\alpha}q_\beta=\epsilon^{\mu\nu\alpha\beta}P_\mu p_{-\nu} p_{0\alpha}q_\beta\\
&=-\epsilon^{\mu\nu\alpha\beta}k_\mu P_\nu p_{-\alpha}q_\beta,
\end{split}\end{equation}
where
\begin{equation}
 A\left(s,u,x\right)=m_\tau^4+s(s+u)+x(u-s-2m_{\pi^-}^2)+m_{\pi^0}^2(m_\tau^2-s+x)-m_\tau^2(2s+u+x).
\end{equation}

\begin{figure}[H]
\includegraphics[width=7cm]{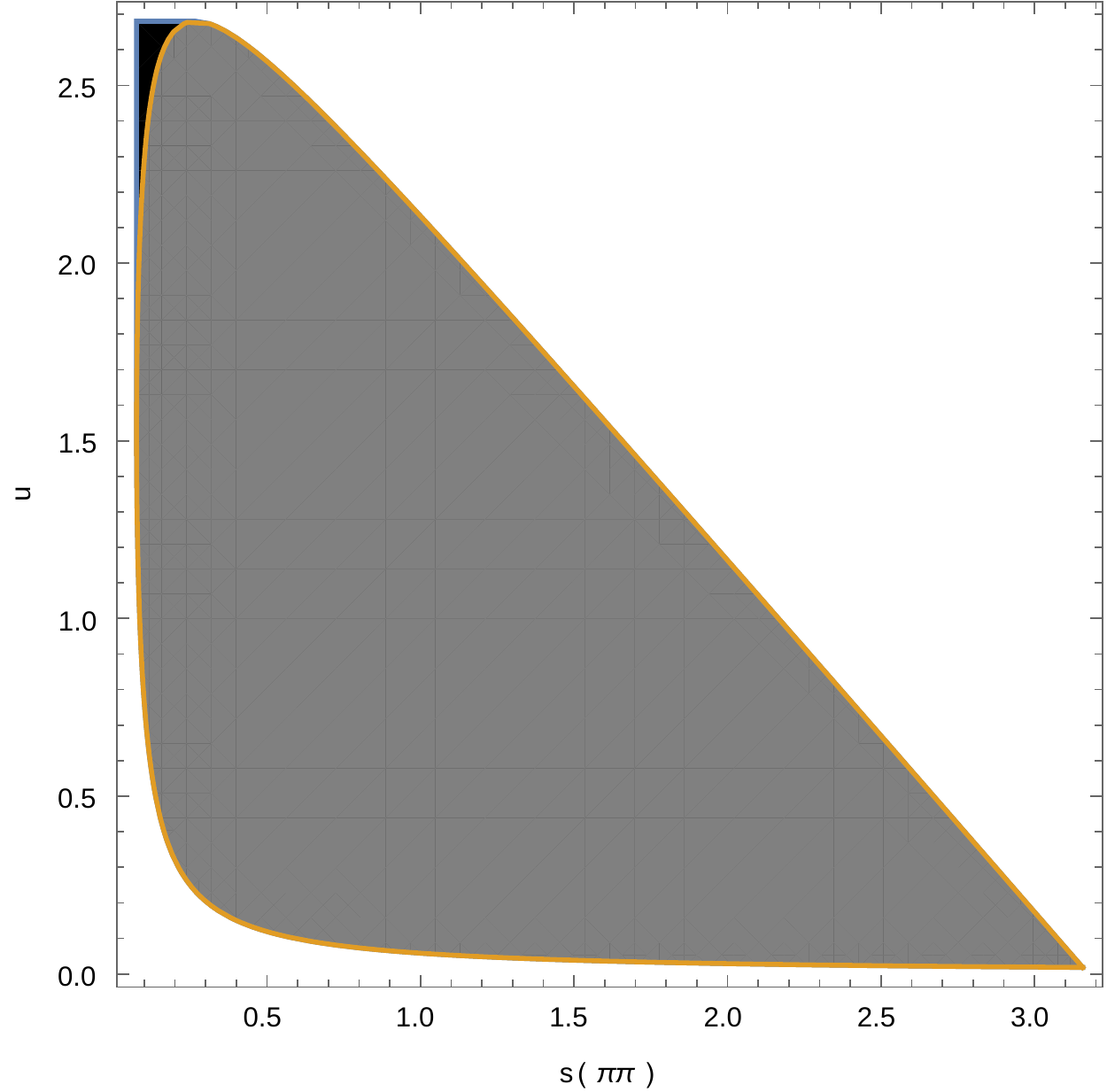}	
\centering			
\caption{Projection of the kinematic region for the non-radiative decay $\mathcal{R}^{III}$ (gray) and the radiative decay $\mathcal{R}^{IV}=\mathcal{R}^{IV/III}\cup \mathcal{R}^{III}$ (black and gray) onto the $su-$plane. $\mathcal{R}^{IV/III}$ (black) is the kinematic region which is only accessible to the radiative decay. }\label{Appx4:fig2}
\end{figure}

Working in the $\tau$-lepton rest frame, we have
\begin{equation}
 E_\gamma=\frac{(m_\tau^2-s+x)(x+M_\gamma^2)}{4m_\tau x}-\frac{(x-M_\gamma^2)\lambda^{1/2}\left(s,x,m_\tau^2\right)}{4m_\tau x}\cos\theta_\nu,
\end{equation}
\begin{equation}
 E_\nu= \vert\vec{q} \vert=\frac{(m_\tau^2-s+x)(x-M_\gamma^2)}{4m_\tau x}+\frac{(x-M_\gamma^2)\lambda^{1/2}\left(s,x,m_\tau^2\right)}{4m_\tau x}\cos\theta_\nu,
\end{equation}
\begin{equation}\begin{split}
 \vec{k}=&\left(-\frac{(x+M_\gamma^2)\lambda^{1/2}\left(s,x,m_\tau^2\right)}{4m_\tau x}+\frac{(m_\tau^2-s+x)(x-M_\gamma^2)}{4m_\tau x}\cos\theta_\nu\right)\hat{e}_z\\
&+\frac{x-M_\gamma^2}{2\sqrt{x}}\sin\theta_\nu \hat{e}_x,
\end{split}\end{equation}
\begin{equation}\begin{split}
 \vec{q}=&\left(-\frac{(x-M_\gamma^2)\lambda^{1/2}\left(s,x,m_\tau^2\right)}{4m_\tau x}-\frac{(m_\tau^2-s+x)(x-M_\gamma^2)}{4m_\tau x}\cos\theta_\nu\right)\hat{e}_z\\
&-\frac{x-M_\gamma^2}{2\sqrt{x}}\sin\theta_\nu \hat{e}_x,
\end{split}\end{equation}
\begin{equation}\begin{split}
 E_-&=\frac{(m_\tau^2+s-x)(s+m_{\pi^-}^2-m_{\pi^0}^2)}{4m_\tau s}+\frac{\lambda^{1/2}\left(s,x,m_\tau^2\right)\lambda^{1/2}\left(s,m_{\pi^-}^2,m_{\pi^0}^2\right)}{4m_\tau s}\cos\theta_-\\
&=\frac{m_\tau^2+m_{\pi^-}^2-u}{2m_\tau},
\end{split}\end{equation}
\begin{equation}\begin{split}
 E_0&=\frac{(m_\tau^2+s-x)(s-m_{\pi^-}^2+m_{\pi^0}^2)}{4m_\tau s}-\frac{\lambda^{1/2}\left(s,x,m_\tau^2\right)\lambda^{1/2}\left(s,m_{\pi^-}^2,m_{\pi^0}^2\right)}{4m_\tau s}\cos\theta_-\\
&=\frac{s+u-x-m_{\pi^-}^2}{2m_\tau},
\end{split}\end{equation}
\begin{equation}
 \vert \vec{p}_-\vert=\frac{\lambda^{1/2}\left(u,m_\tau^2,m_{\pi^-}^2\right)}{2m_\tau},
\end{equation}
\begin{equation}\begin{split}
 \vec{p}_-=&\left(\frac{(s+m_{\pi^-}^2-m_{\pi^0}^2)\lambda^{1/2}\left(s,x,m_\tau^2\right)}{4m_\tau s}+\frac{(m_\tau^2+s-x)\lambda^{1/2}\left(s,m_{\pi^-}^2,m_{\pi^0}^2\right)}{4m_\tau s}\cos\theta_-\right)\hat{e}_z\\
 &+\frac{\lambda^{1/2}\left(s,m_{\pi^-}^2,m_{\pi^0}^2\right)}{2\sqrt{s}}\sin\theta_-\hat{e}_\rho,
\end{split}\end{equation}
\begin{equation}\begin{split}
 \vec{p}_0=&\left(\frac{(s-m_{\pi^-}^2+m_{\pi^0}^2)\lambda^{1/2}\left(s,x,m_\tau^2\right)}{4m_\tau s}-\frac{(m_\tau^2+s-x)\lambda^{1/2}\left(s,m_{\pi^-}^2,m_{\pi^0}^2\right)}{4m_\tau s}\cos\theta_-\right)\hat{e}_z\\
 &-\frac{\lambda^{1/2}\left(s,m_{\pi^-}^2,m_{\pi^0}^2\right)}{2\sqrt{s}}\sin\theta_-\hat{e}_\rho,
\end{split}\end{equation}
\begin{equation}\label{Appx4:eq1}
 \cos\theta_-=\frac{2s(m_\tau^2+m_{\pi^-}^2-u)-(m_\tau^2+s-x)(s+m_{\pi^-}^2-m_{\pi^0}^2)}{\lambda^{1/2}\left(s,x,m_\tau^2\right)\lambda^{1/2}\left(s,m_{\pi^-}^2,m_{\pi^0}^2\right)},
\end{equation}
\begin{equation}
\cos\theta_\nu=\frac{(m_\tau^2-s+x)(x+M_\gamma^2)-4m_\tau E_\gamma x}{(x-M_\gamma^2)\lambda^{1/2}\left(s,x,m_\tau^2\right)},
\end{equation}
where $\lambda(x,y,z)=x^2+y^2+z^2-2xy-2xz-2yz$ is the Kallen function, and $\hat{e}_\rho=\cos\phi_-\hat{e}_x+\sin\phi_-\hat{e}_y$.\\
From eq. (\ref{Appx4:eq1}), we get
\begin{equation}\begin{split}\label{Appx4:eq18}
 x_\pm\left(s,u\right)=&\frac{-m_{\pi^-}^4+(m_{\pi^0}^2-s)(m_\tau^2-u)+m_{\pi^-}^2(m_\tau^2+m_{\pi^0}^2+s+u)}{2m_{\pi^-}^2}\\
 &\pm \frac{\lambda^{1/2}\left(u,m_\tau^2,m_{\pi^-}^2\right)\lambda^{1/2}\left(s,m_{\pi^-}^2,m_{\pi^0}^2\right)}{2m_{\pi^-}^2},
\end{split}\end{equation}
and 
\begin{equation}\begin{split}
 u_\pm\left(s,x\right)=&m_\tau^2+m_{\pi^-}^2-\frac{(m_\tau^2+s-x)(s+m_{\pi^-}^2-m_{\pi^0}^2)}{2s}\\
 &\pm\frac{\lambda^{1/2}\left(s,x,m_\tau^2\right)\lambda^{1/2}\left(s,m_{\pi^-}^2,m_{\pi^0}^2\right)}{2s},
\end{split}\end{equation}
these bounds on $u$ and $x$ correspond to the forward and backward direction, i.e. by taking $\theta_-=0,\,\pi$. \\
For the non-radiative decay, we have 
\begin{equation}
\mathcal{D}^{III}=\left\lbrace u_-\left(s,0\right)\leq u\leq u_+\left(s,0\right),\,\left(m_{\pi^-}+m_{\pi^0}\right)^2\leq s \leq m_\tau^2\right\rbrace,
\end{equation}
this region is plotted in fig. \ref{Appx4:fig2} which corresponds to the projection $\mathcal{R}^{III}$ onto the $su$-plane.\\
In the case of the radiative decay, we have
\begin{equation}\begin{split}
 \mathcal{D}^{IV}=&\left\lbrace x_{min}\left(s,u\right)\leq x\leq x_{max}\left(s,u\right),u_{min}\left(s\right)\leq u\leq u_{max}\left(s\right),\right.\\
 &\qquad\left.\left(m_{\pi^-}+m_{\pi^-}\right)^2\leq s \leq \left(m_\tau-M_\gamma\right)^2\right\rbrace,
\end{split}\end{equation}
with 
\begin{equation}
 x_{min}\left(s,u\right)=\left\lbrace \begin{array}{lll}
                                       x_-\left(s,u\right) & u_+\left(s,M_\gamma^2\right)\leq u\leq \left(m_\tau-m_{\pi^-}\right)^2, & \left(m_{\pi^-}+m_{\pi^0}\right)^2\leq s \leq s^*\\
                                       M_\gamma^2 & u_-\left(s,M_\gamma^2\right)\leq u \leq u_+\left(s,M_\gamma^2\right), & s^*\leq s \leq \left(m_\tau-M_\gamma\right)^2,\\
                                      \end{array}\right.
\end{equation}
\begin{equation}
 x_{max}\left(s,u\right)=x_+\left(s,u\right),
\end{equation}
\begin{equation}
 u_{min}\left(s\right)=u_-\left(s,M_\gamma^2\right),
\end{equation}

\begin{equation}
 u_{max}\left(s\right)=\left\lbrace \begin{array}{ll}
                                     \left(m_\tau-m_{\pi^-}\right)^2 & \left(m_{\pi^-}+m_{\pi^0}\right)^2\leq s \leq s^*,\\
                                     u_+\left(s,M_\gamma^2\right) & s^*\leq s \leq \left(m_\tau-M_\gamma\right)^2,\\
                                    \end{array}\right. 
\end{equation}
where $s^*=\frac{m_\tau\left(m_\tau m_{\pi^-}+m_{\pi^0}^2-m_{\pi^-}^2\right)-M_\gamma^2 m_{\pi^-}}{m_\tau-m_{\pi^-}}$ is the value that maximizes $u_+\left(s,M_\gamma^2\right)$. We will be working in the isospin-limit ($m_u=m_d$), i.e. $m_{\pi^-}^2=m_{\pi^0}^2$ and thus many of the last expressions will be simplified.\\
We use a non-vanishing $M_\gamma$ in order to deal with the IR divergences, at the end these divergences are canceled out by those divergences of the non-radiative decay so we can take the limit $M_\gamma\to 0$. The projection $\mathcal{R}^{IV}=\mathcal{R}^{IV/III}\cup \mathcal{R}^{III}$ of the $\mathcal{D}^{IV}$ is plotted in fig. \ref{Appx4:fig2} for $M_\gamma\to 0$.\\

\section{Vector Form Factors}\label{VF}
\begin{equation}\begin{split}
v_{1}^{R}&=\frac{1}{F^2} \Bigg(\frac{16 k\cdot p_{0}  (2 \kappa^{V}_{12}+\kappa^{V}_{16} )  (-(2 k\cdot p_{-}+s)( 2\kappa^{V}_{12}+\kappa^{V}_{16})+2 (P-q)^2  \kappa^{V}_{17} )}{D_{\omega}\left[(k+p_{0})^2\right]}\\
&+\frac{\sqrt{2} F_V }{M_{\rho}^2} \Big(2 s \lambda^{V}_{12}-(4 k\cdot p_{0}+s)\left( \lambda^{V}_{13}+\lambda^{V}_{14}-\lambda^{V}_{15}\right)+(2 k\cdot p_{-}+s) \lambda^{V}_{16}-4 k\cdot p_{-} \lambda^{V}_{17}\\
&+4 k\cdot p_{-} \lambda^{V}_{18}+s \lambda^{V}_{18}+4 k\cdot p_{-} \lambda^{V}_{19}+2 s \lambda^{V}_{19}+4 k\cdot p_{0} \lambda^{V}_{21}+2 k\cdot p_{0} \lambda^{V}_{22}-2 k\cdot p_{-} \lambda^{V}_{22} \Big)\\
&+\frac{1}{M_{a_{1}}^2 D_{a_{1}}\left[(k+p_{-})^2\right]}\Big(-8  (-2 k\cdot p_{0} M_{a_{1}}^2 s+ (k\cdot p_{-}-M_{a_{1}}^2 ) s^2+2 \left(k\cdot p_{0}\right) \left(k\cdot p_{-}\right)  \\
&(2 M_{a_{1}}^2+s ) ) (\lambda^{A}_{12})^2-8 k\cdot p_{-} (2 k\cdot p_{0}+s)  (2 M_{a_{1}}^2+s ) (\lambda^{A}_{13})^2-2 \sqrt{2} F_{A} \left(k\cdot p_{0}\right) s \lambda^{A}_{15}\\
&-2 \sqrt{2} F_{A} \left(k\cdot p_{-}\right) s \lambda^{A}_{15}-\sqrt{2} F_{A} s^2 \lambda^{A}_{15}-8 \sqrt{2} F_{A} k\cdot p_{-} M_{a_{1}}^2 \lambda^{A}_{17}-4 \sqrt{2} F_{A} \left(k\cdot p_{-}\right) s \lambda^{A}_{17}\\
&+16 \left(k\cdot p_{0}\right) \left(k\cdot p_{-}\right) s \lambda^{A}_{15} \lambda^{A}_{17}+16 \left(k\cdot p_{-}\right)^2 s \lambda^{A}_{15} \lambda^{A}_{17}+8 \left(k\cdot p_{-}\right) s^2 \lambda^{A}_{15} \lambda^{A}_{17}\\
&+32 \left(k\cdot p_{-}\right)^2 M_{a_{1}}^2 (\lambda^{A}_{17})^2+16 \left(k\cdot p_{-}\right)^2 s (\lambda^{A}_{17})^2+\lambda^{A}_{13}  (8 \left(k\cdot p_{-}\right) s (P-q)^2  \lambda^{A}_{15}\\
&+(2 k\cdot p_{0}-2 k\cdot p_{-}+s)  (2 M_{a_{1}}^2+s )  (\sqrt{2} F_{A}-8 k\cdot p_{-} \lambda^{A}_{17} ) )+\lambda^{A}_{12}  (-8  (s  (2 k\cdot p_{-}  (M_{a_{1}}^2-s )\\
&+M_{a_{1}}^2 s )+2 k\cdot p_{0}  (4 k\cdot p_{-} M_{a_{1}}^2-2 \left(k\cdot p_{-}\right) s+M_{a_{1}}^2 s ) ) \lambda^{A}_{13}-8  (k\cdot p_{-}-M_{a_{1}}^2 ) s (P-q)^2  \lambda^{A}_{15}\\
&+ (k\cdot p_{0}  (4 M_{a_{1}}^2-2 s )-s  (2 M_{a_{1}}^2+s )+k\cdot p_{-}  (-4 M_{a_{1}}^2+2 s ) )  (\sqrt{2} F_{A}-8 k\cdot p_{-} \lambda^{A}_{17} ) )\Big) \Bigg)
\end{split}\end{equation}

\begin{equation*}\begin{split}
v_{1}^{RR}&=\frac{1}{2 \sqrt{2}F^2} \Bigg(\frac{64 k\cdot p_{0} F_V  (-(2 k\cdot p_{-}+s)( 2\kappa^{V}_{12}+\kappa^{V}_{16})+2 (P-q)^2  \kappa^{V}_{17} ) \kappa^{VV}_3}{M_{\rho}^2 D_{\omega}\left[(k+p_{0})^2\right]}\\
&-\frac{64 k\cdot p_{0} (4 k\cdot p_{0}+2 k\cdot p_{-}+s)  (2 \kappa^{V}_{12}+\kappa^{V}_{16} ) \kappa^{VV}_3  (-F_V+2 \sqrt{2} (P-q)^2  \lambda^{V}_{22} )}{D_{\rho}\left[(P-q)^2\right] D_{\omega}\left[(k+p_{0})^2\right]}\\
&+\frac{2 F_V }{M_{a_{1}}^2 M_{\rho}^2 D_{a_{1}}\left[(k+p_{-})^2\right]} \Big(2  (2 \sqrt{2} F_{A} k\cdot p_{-} M_{a_{1}}^2-\sqrt{2} F_{A} \left(k\cdot p_{-}\right) s+\sqrt{2} F_{A} M_{a_{1}}^2 s\\
&+4  (-2 k\cdot p_{0} M_{a_{1}}^2 s+ (k\cdot p_{-}-M_{a_{1}}^2 ) s^2+2 \left(k\cdot p_{0}\right) \left(k\cdot p_{-}\right)  (2 M_{a_{1}}^2+s ) ) \lambda^{A}_{12}\\
&+4 (2 k\cdot p_{0}+s)  (2 k\cdot p_{-} M_{a_{1}}^2-\left(k\cdot p_{-}\right) s+M_{a_{1}}^2 s ) \lambda^{A}_{13}+8 \left(k\cdot p_{0}\right) \left(k\cdot p_{-}\right) s \lambda^{A}_{15}\\
&+8 \left(k\cdot p_{-}\right)^2 s \lambda^{A}_{15}-8 k\cdot p_{0} M_{a_{1}}^2 s \lambda^{A}_{15}-8 k\cdot p_{-} M_{a_{1}}^2 s \lambda^{A}_{15}+4 \left(k\cdot p_{-}\right) s^2 \lambda^{A}_{15}\\
&-4 M_{a_{1}}^2 s^2 \lambda^{A}_{15}-16 \left(k\cdot p_{-}\right)^2 M_{a_{1}}^2 \lambda^{A}_{17}+8 \left(k\cdot p_{-}\right)^2 s \lambda^{A}_{17}-8 k\cdot p_{-} M_{a_{1}}^2 s \lambda^{A}_{17} ) \lambda^{VA}_2\\
&-k\cdot p_{-}  (2 \sqrt{2} F_{A} M_{a_{1}}^2+\sqrt{2} F_{A} s+4  (4 k\cdot p_{0} M_{a_{1}}^2-2 \left(k\cdot p_{0}\right) s-s^2 ) \lambda^{A}_{12}\\
&+4 (2 k\cdot p_{0}+s)  (2 M_{a_{1}}^2+s ) \lambda^{A}_{13}-8 \left(k\cdot p_{0}\right) s \lambda^{A}_{15}-8 \left(k\cdot p_{-}\right) s \lambda^{A}_{15}-4 s^2 \lambda^{A}_{15}\\
&-16 k\cdot p_{-} M_{a_{1}}^2 \lambda^{A}_{17}-8 \left(k\cdot p_{-}\right) s \lambda^{A}_{17} )  (\lambda^{VA}_4+2 \lambda^{VA}_5 ) \Big)\\
&+\frac{ (\sqrt{2} F_V-4 (P-q)^2  \lambda^{V}_{22} )}{M_{a_{1}}^2 D_{a_{1}}\left[(k+p_{-})^2\right] D_{\rho}\left[(P-q)^2\right]}  \Big(-2  (-4 \sqrt{2}  (-2 k\cdot p_{0} M_{a_{1}}^2 s+ (k\cdot p_{-}-M_{a_{1}}^2 ) s^2\\
\end{split}\end{equation*}
\begin{equation}\begin{split}
\qquad&+2 \left(k\cdot p_{0}\right) \left(k\cdot p_{-}\right)  (2 M_{a_{1}}^2+s ) ) \lambda^{A}_{12}+ (4 k\cdot p_{0} M_{a_{1}}^2-2 \left(k\cdot p_{0}\right) s-s^2 )  \\
&(F_{A}-4 \sqrt{2} k\cdot p_{-} \lambda^{A}_{13}-4 \sqrt{2} k\cdot p_{-} \lambda^{A}_{17} ) ) \lambda^{VA}_2-2 s (P-q)^2   (F_{A}+4 \sqrt{2}  (k\cdot p_{-}-M_{a_{1}}^2 ) \lambda^{A}_{12}\\
&-4 \sqrt{2} k\cdot p_{-} \lambda^{A}_{13}-4 \sqrt{2} k\cdot p_{-} \lambda^{A}_{17} ) \lambda^{VA}_3+(2 k\cdot p_{0}+s)  (-4 \sqrt{2}  (2 k\cdot p_{-} M_{a_{1}}^2-\left(k\cdot p_{-}\right) s\\
&+M_{a_{1}}^2 s ) \lambda^{A}_{12}+ (2 M_{a_{1}}^2+s )  (F_{A}-4 \sqrt{2} k\cdot p_{-} \lambda^{A}_{13}-4 \sqrt{2} k\cdot p_{-} \lambda^{A}_{17} ) )  (\lambda^{VA}_4+2 \lambda^{VA}_5 ) \Big)\\
&+\frac{2 F_V}{M_{\rho}^2 D_{\rho}\left[(P-q)^2\right]}  \Big(-2 (P-q)^2  \lambda^{V}_{22}  (\left(P-q\right)^2-8 s \lambda^{VV}_2\\
&-2 (4 k\cdot p_{0}+s) \lambda^{VV}_3+8 k\cdot p_{0} \lambda^{VV}_4+2 s \lambda^{VV}_4-16 k\cdot p_{0} \lambda^{VV}_5-4 s \lambda^{VV}_5 )\\
&+\sqrt{2} F_V  (-4 s \lambda^{VV}_2-(4 k\cdot p_{0}+s)  (\lambda^{VV}_3-\lambda^{VV}_4+2 \lambda^{VV}_5 ) ) \Big)\\
&-\frac{4 F_V  (\sqrt{2} (4 k\cdot p_{0}+s) G_V \lambda^{VV}_7+s \lambda^{V}_{21}  (s-2 (4 k\cdot p_{0}+s) \lambda^{VV}_7 ) )}{M_{\rho}^2 D_{\rho}[s]} \Bigg)
\end{split}\end{equation}

\begin{equation}\begin{split}
v_{1}^{RRR}=&-\frac{F_V(-\sqrt{2} F_V+4 (P-q)^2  \lambda^{V}_{22} )}{\sqrt{2} F^2 M_{a_{1}}^2 M_{\rho}^2 D_{a_{1}}\left[(k+p_{-})^2\right]D_{\rho}\left[(P-q)^2\right]}  (-4  (-2 k\cdot p_{0} M_{a_{1}}^2 s\\
&+ (k\cdot p_{-}-M_{a_{1}}^2 ) s^2+2 \left(k\cdot p_{0}\right) \left(k\cdot p_{-}\right)  (2 M_{a_{1}}^2+s ) ) (\lambda^{VA}_2)^2\\
&-k\cdot p_{-}  (\lambda^{VA}_4+2 \lambda^{VA}_5 )  (-2 s (P-q)^2  \lambda^{VA}_3+(2 k\cdot p_{0}+s)  (2 M_{a_{1}}^2+s )  (\lambda^{VA}_4+2 \lambda^{VA}_5 ) )\\
&+2 \lambda^{VA}_2  (2  (k\cdot p_{-}-M_{a_{1}}^2 ) s (P-q)^2  \lambda^{VA}_3+ (s  (2 k\cdot p_{-}  (M_{a_{1}}^2-s )+M_{a_{1}}^2 s )\\
&+2 k\cdot p_{0}  (4 k\cdot p_{-} M_{a_{1}}^2-2 \left(k\cdot p_{-}\right) s+M_{a_{1}}^2 s ) )  (\lambda^{VA}_4+2 \lambda^{VA}_5 ) ) )
\end{split}\end{equation}

\begin{equation*}\begin{split}
v_{GI1}^{R+RR}=&\frac{1}{F^2 D_{\rho}\left[(P-q)^2\right] D_{\rho}[s]}\Bigg( \sqrt{2} F_V (2  (M_{\rho}^2-s ) s \lambda^{V}_{12}- (M_{\rho}^2-s ) (4 k\cdot p_{0}+s) \lambda^{V}_{13}\\
&-4 k\cdot p_{0} M_{\rho}^2 \lambda^{V}_{14}+4 \left(k\cdot p_{0}\right) s \lambda^{V}_{14}-M_{\rho}^2 s \lambda^{V}_{14}+s^2 \lambda^{V}_{14}+4 k\cdot p_{0} M_{\rho}^2 \lambda^{V}_{15}-4 \left(k\cdot p_{0}\right) s \lambda^{V}_{15}\\
&+M_{\rho}^2 s \lambda^{V}_{15}-s^2 \lambda^{V}_{15}-2 k\cdot p_{0} M_{\rho}^2 \lambda^{V}_{16}+2 \left(k\cdot p_{0}\right) s \lambda^{V}_{16}+4 k\cdot p_{0} M_{\rho}^2 \lambda^{V}_{17}-4 \left(k\cdot p_{0}\right) s \lambda^{V}_{17}\\
&+2 M_{\rho}^2 s \lambda^{V}_{17}-2 s^2 \lambda^{V}_{17}-4 k\cdot p_{0} M_{\rho}^2 \lambda^{V}_{18}+4 \left(k\cdot p_{0}\right) s \lambda^{V}_{18}-M_{\rho}^2 s \lambda^{V}_{18}+s^2 \lambda^{V}_{18}\\
&-2 k\cdot p_{0} M_{\rho}^2 \lambda^{V}_{19}+2 \left(k\cdot p_{0}\right) s \lambda^{V}_{19}-4 k\cdot p_{-} M_{\rho}^2 \lambda^{V}_{21}+4 \left(k\cdot p_{-}\right) s \lambda^{V}_{21}-2 M_{\rho}^2 s \lambda^{V}_{21}\\
&+s^2 \lambda^{V}_{21}+6 k\cdot p_{0} M_{\rho}^2 \lambda^{V}_{22}+2 k\cdot p_{-} M_{\rho}^2 \lambda^{V}_{22}-6 \left(k\cdot p_{0}\right) s \lambda^{V}_{22}-2 \left(k\cdot p_{-}\right) s \lambda^{V}_{22}\\
&+2 M_{\rho}^2 s \lambda^{V}_{22}-2 s^2 \lambda^{V}_{22}+8 \left(k\cdot p_{0}\right) s \lambda^{V}_{21} \lambda^{VV}_7+2 s^2 \lambda^{V}_{21} \lambda^{VV}_7 )+2 G_V  (\sqrt{2} (4 k\cdot p_{0}+s) \\
& (\left(P-q\right)^2-M_{\rho}^2)\lambda^{V}_7-(4 k\cdot p_{0}+s) F_V \lambda^{VV}_7+\sqrt{2} \lambda^{V}_{22}  (-16 \left(k\cdot p_{0}\right)^2- (2 M_{\rho}^2-s )\\
& (2 k\cdot p_{-}+s)-2 k\cdot p_{0}  (8 k\cdot p_{-}-2 M_{\rho}^2+3 s )+2 (4 k\cdot p_{0}+s) (P-q)^2  \lambda^{VV}_7 ) )\\
\end{split}\end{equation*}
\begin{equation}\begin{split}
&-4  (s (4 k\cdot p_{0}+s)  (\left(P-q\right)^2-M_{\rho}^2)\lambda^{V}_7 \lambda^{V}_{21}+\lambda^{V}_{22}  (2  (M_{\rho}^2-s ) s (P-q)^2  \lambda^{V}_{12}\\
&- (M_{\rho}^2-s ) (4 k\cdot p_{0}+s) (P-q)^2  \lambda^{V}_{13}-8 \left(k\cdot p_{0}\right)^2 M_{\rho}^2 \lambda^{V}_{14}-8 \left(k\cdot p_{0}\right) \left(k\cdot p_{-}\right) M_{\rho}^2 \lambda^{V}_{14}\\
\qquad&+8 \left(k\cdot p_{0}\right)^2 s \lambda^{V}_{14}+8 \left(k\cdot p_{0}\right) \left(k\cdot p_{-}\right) s \lambda^{V}_{14}-6 k\cdot p_{0} M_{\rho}^2 s \lambda^{V}_{14}-2 k\cdot p_{-} M_{\rho}^2 s \lambda^{V}_{14}\\
&+6 \left(k\cdot p_{0}\right) s^2 \lambda^{V}_{14}+2 \left(k\cdot p_{-}\right) s^2 \lambda^{V}_{14}-M_{\rho}^2 s^2 \lambda^{V}_{14}+s^3 \lambda^{V}_{14}+8 \left(k\cdot p_{0}\right)^2 M_{\rho}^2 \lambda^{V}_{15}\\
&+8 \left(k\cdot p_{0}\right) \left(k\cdot p_{-}\right) M_{\rho}^2 \lambda^{V}_{15}-8 \left(k\cdot p_{0}\right)^2 s \lambda^{V}_{15}-8 \left(k\cdot p_{0}\right) \left(k\cdot p_{-}\right) s \lambda^{V}_{15}\\
&+6 k\cdot p_{0} M_{\rho}^2 s \lambda^{V}_{15}+2 k\cdot p_{-} M_{\rho}^2 s \lambda^{V}_{15}-6 \left(k\cdot p_{0}\right) s^2 \lambda^{V}_{15}-2 \left(k\cdot p_{-}\right) s^2 \lambda^{V}_{15}+M_{\rho}^2 s^2 \lambda^{V}_{15}\\
&-s^3 \lambda^{V}_{15}-4 \left(k\cdot p_{0}\right)^2 M_{\rho}^2 \lambda^{V}_{16}-4 \left(k\cdot p_{0}\right) \left(k\cdot p_{-}\right) M_{\rho}^2 \lambda^{V}_{16}+4 \left(k\cdot p_{0}\right)^2 s \lambda^{V}_{16}\\
&+4 \left(k\cdot p_{0}\right) \left(k\cdot p_{-}\right) s \lambda^{V}_{16}-2 k\cdot p_{0} M_{\rho}^2 s \lambda^{V}_{16}+2 \left(k\cdot p_{0}\right) s^2 \lambda^{V}_{16}+8 \left(k\cdot p_{0}\right)^2 M_{\rho}^2 \lambda^{V}_{17}\\
&+8 \left(k\cdot p_{0}\right) \left(k\cdot p_{-}\right) M_{\rho}^2 \lambda^{V}_{17}-8 \left(k\cdot p_{0}\right)^2 s \lambda^{V}_{17}-8 \left(k\cdot p_{0}\right) \left(k\cdot p_{-}\right) s \lambda^{V}_{17}\\
&+8 k\cdot p_{0} M_{\rho}^2 s \lambda^{V}_{17}+4 k\cdot p_{-} M_{\rho}^2 s \lambda^{V}_{17}-8 \left(k\cdot p_{0}\right) s^2 \lambda^{V}_{17}-4 \left(k\cdot p_{-}\right) s^2 \lambda^{V}_{17}\\
&+2 M_{\rho}^2 s^2 \lambda^{V}_{17}-2 s^3 \lambda^{V}_{17}-8 \left(k\cdot p_{0}\right)^2 M_{\rho}^2 \lambda^{V}_{18}-8 \left(k\cdot p_{0}\right) \left(k\cdot p_{-}\right) M_{\rho}^2 \lambda^{V}_{18}\\
&+8 \left(k\cdot p_{0}\right)^2 s \lambda^{V}_{18}+8 \left(k\cdot p_{0}\right) \left(k\cdot p_{-}\right) s \lambda^{V}_{18}-6 k\cdot p_{0} M_{\rho}^2 s \lambda^{V}_{18}-2 k\cdot p_{-} M_{\rho}^2 s \lambda^{V}_{18}\\
&+6 \left(k\cdot p_{0}\right) s^2 \lambda^{V}_{18}+2 \left(k\cdot p_{-}\right) s^2 \lambda^{V}_{18}-M_{\rho}^2 s^2 \lambda^{V}_{18}+s^3 \lambda^{V}_{18}-4 \left(k\cdot p_{0}\right)^2 M_{\rho}^2 \lambda^{V}_{19}\\
&-4 \left(k\cdot p_{0}\right) \left(k\cdot p_{-}\right) M_{\rho}^2 \lambda^{V}_{19}+4 \left(k\cdot p_{0}\right)^2 s \lambda^{V}_{19}+4 \left(k\cdot p_{0}\right) \left(k\cdot p_{-}\right) s \lambda^{V}_{19}\\
&-2 k\cdot p_{0} M_{\rho}^2 s \lambda^{V}_{19}+2 \left(k\cdot p_{0}\right) s^2 \lambda^{V}_{19}-8 \left(k\cdot p_{0}\right) \left(k\cdot p_{-}\right) M_{\rho}^2 \lambda^{V}_{21}-8 \left(k\cdot p_{-}\right)^2 M_{\rho}^2 \lambda^{V}_{21}\\
&-16 \left(k\cdot p_{0}\right)^2 s \lambda^{V}_{21}-8 \left(k\cdot p_{0}\right) \left(k\cdot p_{-}\right) s \lambda^{V}_{21}+8 \left(k\cdot p_{-}\right)^2 s \lambda^{V}_{21}+4 k\cdot p_{0} M_{\rho}^2 s \lambda^{V}_{21}\\
&-8 k\cdot p_{-} M_{\rho}^2 s \lambda^{V}_{21}-6 \left(k\cdot p_{0}\right) s^2 \lambda^{V}_{21}+6 \left(k\cdot p_{-}\right) s^2 \lambda^{V}_{21}-2 M_{\rho}^2 s^2 \lambda^{V}_{21}+s^3 \lambda^{V}_{21}\\
&+8 \left(k\cdot p_{0}\right)^2 M_{\rho}^2 \lambda^{V}_{22}+8 \left(k\cdot p_{0}\right) \left(k\cdot p_{-}\right) M_{\rho}^2 \lambda^{V}_{22}-8 \left(k\cdot p_{0}\right)^2 s \lambda^{V}_{22}\\
&-8 \left(k\cdot p_{0}\right) \left(k\cdot p_{-}\right) s \lambda^{V}_{22}+6 k\cdot p_{0} M_{\rho}^2 s \lambda^{V}_{22}+2 k\cdot p_{-} M_{\rho}^2 s \lambda^{V}_{22}-6 \left(k\cdot p_{0}\right) s^2 \lambda^{V}_{22}\\
&-2 \left(k\cdot p_{-}\right) s^2 \lambda^{V}_{22}+M_{\rho}^2 s^2 \lambda^{V}_{22}-s^3 \lambda^{V}_{22}+16 \left(k\cdot p_{0}\right)^2 s \lambda^{V}_{21} \lambda^{VV}_7\\
&+16 \left(k\cdot p_{0}\right) \left(k\cdot p_{-}\right) s \lambda^{V}_{21} \lambda^{VV}_7+12 \left(k\cdot p_{0}\right) s^2 \lambda^{V}_{21} \lambda^{VV}_7+4 \left(k\cdot p_{-}\right) s^2 \lambda^{V}_{21} \lambda^{VV}_7\\
&+2 s^3 \lambda^{V}_{21} \lambda^{VV}_7 )) \Bigg)
\end{split}\end{equation}

\begin{equation}\begin{split}
v_{2}^{R}=&\frac{1}{F^2} \Bigg(\frac{8 (2 k\cdot p_{-}+s) }{M_{\omega}^2 D_{\omega}\left[(k+p_{0})^2\right]} \Big(-4  (k\cdot p_{0}-M_{\omega}^2 ) (2 k\cdot p_{-}+s) (\kappa^{V}_{12})^2+k\cdot p_{0} \kappa^{V}_{16} \\
& ( (4 k\cdot p_{0}+2 k\cdot p_{-}-2 M_{\omega}^2+s ) \kappa^{V}_{16}-2 (P-q)^2  \kappa^{V}_{17} )+\kappa^{V}_{12}  ( (-8 \left(k\cdot p_{0}\right)^2+4 k\cdot p_{0} M_{\omega}^2\\
&+2 M_{\omega}^2 (2 k\cdot p_{-}+s) ) \kappa^{V}_{16}+4  (k\cdot p_{0}-M_{\omega}^2 ) (P-q)^2  \kappa^{V}_{17} ) \Big)+\frac{\sqrt{2} F_V }{M_{\rho}^2} \Big(2 s \lambda^{V}_{12}-s \lambda^{V}_{13}\\
&+4 k\cdot p_{-} \lambda^{V}_{14}+s \lambda^{V}_{14}-4 k\cdot p_{-} \lambda^{V}_{15}-s \lambda^{V}_{15}-4 k\cdot p_{-} \lambda^{V}_{17}+2 k\cdot p_{-} \lambda^{V}_{18}-4 k\cdot p_{-} \lambda^{V}_{21}\\
&+2 k\cdot p_{0} \lambda^{V}_{22}-2 k\cdot p_{-} \lambda^{V}_{22} \Big)+\frac{2 \sqrt{2} }{M_{a_{1}}^2 D_{a_{1}}\left[(k+p_{-})^2\right]} \Big(4 \sqrt{2} \left(k\cdot p_{-}\right)^2  (2 k\cdot p_{0}+2 M_{a_{1}}^2+s ) \\
&(\lambda^{A}_{12})^2+\lambda^{A}_{12}  (4 \sqrt{2} k\cdot p_{-}  (-4 \left(k\cdot p_{0}\right) \left(k\cdot p_{-}\right)+2 k\cdot p_{0} M_{a_{1}}^2+2 k\cdot p_{-} M_{a_{1}}^2-2 \left(k\cdot p_{-}\right) s\\
&+M_{a_{1}}^2 s ) \lambda^{A}_{13}+4 \sqrt{2} \left(k\cdot p_{-}\right)^2 (P-q)^2  \lambda^{A}_{15}+ (-2 \left(k\cdot p_{0}\right) \left(k\cdot p_{-}\right)+2 \left(k\cdot p_{-}\right)^2\\
&+2 k\cdot p_{0} M_{a_{1}}^2+2 k\cdot p_{-} M_{a_{1}}^2-\left(k\cdot p_{-}\right) s+M_{a_{1}}^2 s )  (-F_{A}+4 \sqrt{2} k\cdot p_{-} \lambda^{A}_{17} ) )\\
&- (k\cdot p_{-}-M_{a_{1}}^2 )  (-4 \sqrt{2} k\cdot p_{-} (2 k\cdot p_{0}+s)( \lambda^{A}_{13})^2-4 k\cdot p_{-} \lambda^{A}_{17}  (F_{A}-2 \sqrt{2} k\cdot p_{-} \lambda^{A}_{17} )\\
&+(P-q)^2  \lambda^{A}_{15}  (-F_{A}+4 \sqrt{2} k\cdot p_{-} \lambda^{A}_{17} )+\lambda^{A}_{13}  (4 \sqrt{2} k\cdot p_{-} (P-q)^2  \lambda^{A}_{15}\\
&+(2 k\cdot p_{0}-2 k\cdot p_{-}+s)  (F_{A}-4 \sqrt{2} k\cdot p_{-} \lambda^{A}_{17} ) ) ) \Big) \Bigg)
\end{split}\end{equation}

\begin{equation*}\begin{split}
v_{2}^{RR}=&\frac{1}{\sqrt{2} F^2} \Bigg(\frac{8 F_V(2 k\cdot p_{-}+s) }{M_{\rho}^2 M_{\omega}^2 D_{\omega}\left[(k+p_{0})^2\right]}  \Big(2 (P-q)^2  \kappa^{V}_{17}  (-M_{\omega}^2 \kappa^{VV}_3+ (2 k\cdot p_{0}-M_{\omega}^2 ) \kappa^{VV}_4 )\\
&+\kappa^{V}_{16}  (M_{\omega}^2 (2 k\cdot p_{-}+s) \kappa^{VV}_3- (2 k\cdot p_{0}-M_{\omega}^2 ) (4 k\cdot p_{0}+2 k\cdot p_{-}+s) \kappa^{VV}_4 )\\
&+2 (2 k\cdot p_{-}+s) \kappa^{V}_{12}  (M_{\omega}^2 \kappa^{VV}_3+D_{\omega}\left[(k+p_{0})^2\right] \kappa^{VV}_4 ) \Big)\\
&+\frac{16 (2 k\cdot p_{-}+s) (-F_V+2 \sqrt{2} (P-q)^2  \lambda^{V}_{22} )}{M_{\omega}^2 D_{\rho}\left[(P-q)^2\right] D_{\omega}\left[(k+p_{0})^2\right]} \Big(k\cdot p_{0} \kappa^{V}_{16}  ( (\left(P-q\right)^2+M_{\omega}^2 ) \kappa^{VV}_3\\
&+ (2 k\cdot p_{0}-M_{\omega}^2 ) \kappa^{VV}_4 )+\kappa^{V}_{12}  ( (-4 \left(k\cdot p_{0}\right)^2+2 M_{\omega}^2 (2 k\cdot p_{-}+s)\\
&-2 k\cdot p_{0}  (2 k\cdot p_{-}-3 M_{\omega}^2+s ) ) \kappa^{VV}_3+2 k\cdot p_{0}  (-2 k\cdot p_{0}+M_{\omega}^2 ) \kappa^{VV}_4 ) \Big)\\
&  -\frac{2 k\cdot p_{-} F_V }{M_{a_{1}}^2 M_{\rho}^2 D_{a_{1}}\left[(k+p_{-})^2\right]} \Big(2 k\cdot p_{-}  (-\sqrt{2} F_{A}+4  (2 k\cdot p_{0}+2 M_{a_{1}}^2+s ) \lambda^{A}_{12}\\
&-4 (2 k\cdot p_{0}+s) \lambda^{A}_{13}+8 k\cdot p_{0} \lambda^{A}_{15}+8 k\cdot p_{-} \lambda^{A}_{15}+4 s \lambda^{A}_{15}+8 k\cdot p_{-} \lambda^{A}_{17} ) \lambda^{VA}_2\\
&- (-4  (2 k\cdot p_{0}  (k\cdot p_{-}-M_{a_{1}}^2 )-M_{a_{1}}^2 s+k\cdot p_{-}  (-2 M_{a_{1}}^2+s ) ) \lambda^{A}_{12}+ (k\cdot p_{-}-M_{a_{1}}^2 )\\
&  (\sqrt{2} F_{A}+4 (2 k\cdot p_{0}+s) \lambda^{A}_{13}-4 (P-q)^2  \lambda^{A}_{15}-8 k\cdot p_{-} \lambda^{A}_{17} ) )  (\lambda^{VA}_4+2 \lambda^{VA}_5 ) \Big)\\
\end{split}\end{equation*}
\begin{equation}\begin{split}
&-\frac{ (\sqrt{2} F_V-4 (P-q)^2  \lambda^{V}_{22} ) }{M_{a_{1}}^2 D_{a_{1}}\left[(k+p_{-})^2\right] D_{\rho}\left[(P-q)^2\right]} \Big(2  (4 \sqrt{2} \left(k\cdot p_{-}\right)^2  (2 k\cdot p_{0}+2 M_{a_{1}}^2+s ) \lambda^{A}_{12}\\
&+ (2 k\cdot p_{0}  (k\cdot p_{-}-M_{a_{1}}^2 )-M_{a_{1}}^2 s+k\cdot p_{-}  (-2 M_{a_{1}}^2+s ) )  (F_{A}-4 \sqrt{2} k\cdot p_{-} \lambda^{A}_{13}\\
&-4 \sqrt{2} k\cdot p_{-} \lambda^{A}_{17} ) ) \lambda^{VA}_2- (4 \sqrt{2} \left(k\cdot p_{-}\right)^2 \lambda^{A}_{12}+ (k\cdot p_{-}-M_{a_{1}}^2 )  (F_{A}-4 \sqrt{2} k\cdot p_{-} \lambda^{A}_{13}\\
&-4 \sqrt{2} k\cdot p_{-} \lambda^{A}_{17} ) )  (2 (P-q)^2  \lambda^{VA}_3-(2 k\cdot p_{0}+s)  (\lambda^{VA}_4+2 \lambda^{VA}_5 ) ) \Big)\\
&+\frac{F_V }{M_{\rho}^2 D_{\rho}\left[(P-q)^2\right]} \Big(\sqrt{2} F_V  (-4 s \lambda^{VV}_2+(4 k\cdot p_{-}+s)\left( \lambda^{VV}_3- \lambda^{VV}_4\right)-2 s \lambda^{VV}_5 )\\
&+2 (P-q)^2  \lambda^{V}_{22}  (\left(P-q\right)^2+8 s \lambda^{VV}_2-2 (4 k\cdot p_{-}+s) \left(\lambda^{VV}_3- \lambda^{VV}_4\right)+4 s \lambda^{VV}_5 ) \Big)\\
&+\frac{2 F_V  (\sqrt{2} (4 k\cdot p_{-}+s) G_V \lambda^{VV}_7+s \lambda^{V}_{21}  (s-2 (4 k\cdot p_{-}+s) \lambda^{VV}_7 ) )}{M_{\rho}^2 D_{\rho}[s]} \Bigg)
\end{split}\end{equation}

\begin{equation}\begin{split}
v_{2}^{RRR}=&-\frac{\sqrt{2} k\cdot p_{-} F_V(-\sqrt{2} F_V+4 (P-q)^{2} \lambda _{22} )}{F^2 M_{a_{1}}^2 M_{\rho}^2 D_{a_{1}}\left[(k+p_{-})^2\right] D_{\rho}\left[(P-q)^2\right]}   \Bigg(4 k\cdot p_{-}  (2 k\cdot p_{0}+2 M_{a_{1}}^2+s ) (\lambda^{VA}_2)^2\\
&- (k\cdot p_{-}-M_{a_{1}}^2 )  (\lambda^{VA}_4+2 \lambda^{VA}_5 )  (2 (P-q)^{2} \lambda^{VA}_3-(2 k\cdot p_{0}+s)  (\lambda^{VA}_4+2 \lambda^{VA}_5 ) )\\
&-2 \lambda^{VA}_2  (2 k\cdot p_{-} (P-q)^{2} \lambda^{VA}_3+ (-4 \left(k\cdot p_{0}\right) \left(k\cdot p_{-}\right)+2 k\cdot p_{0} M_{a_{1}}^2\\
&+2 k\cdot p_{-} M_{a_{1}}^2-2 \left(k\cdot p_{-}\right) s+M_{a_{1}}^2 s )  (\lambda^{VA}_4+2 \lambda^{VA}_5 ) ) \Bigg)
\end{split}\end{equation}

\begin{equation*}\begin{split}
v_{GI2}^{R+RR}=&\frac{\sqrt{2} F_V}{F^2 D_{\rho}\left[(P-q)^2\right] D_{\rho}[s]}  \Bigg(2  (M_{\rho}^2-s ) s \lambda^{V}_{12}+s  (-M_{\rho}^2+s ) \lambda^{V}_{13}+4 k\cdot p_{-} M_{\rho}^2 \lambda^{V}_{14}\\
&-4 \left(k\cdot p_{-}\right) s \lambda^{V}_{14}+M_{\rho}^2 s \lambda^{V}_{14}-s^2 \lambda^{V}_{14}-4 k\cdot p_{-} M_{\rho}^2 \lambda^{V}_{15}+4 \left(k\cdot p_{-}\right) s \lambda^{V}_{15}-M_{\rho}^2 s \lambda^{V}_{15}\\
&+s^2 \lambda^{V}_{15}-2 k\cdot p_{0} M_{\rho}^2 \lambda^{V}_{16}-2 k\cdot p_{-} M_{\rho}^2 \lambda^{V}_{16}+2 \left(k\cdot p_{0}\right) s \lambda^{V}_{16}+2 \left(k\cdot p_{-}\right) s \lambda^{V}_{16}\\
&-M_{\rho}^2 s \lambda^{V}_{16}+s^2 \lambda^{V}_{16}+4 k\cdot p_{0} M_{\rho}^2 \lambda^{V}_{17}-4 \left(k\cdot p_{0}\right) s \lambda^{V}_{17}+2 M_{\rho}^2 s \lambda^{V}_{17}-2 s^2 \lambda^{V}_{17}\\
&+2 k\cdot p_{-} M_{\rho}^2 \lambda^{V}_{18}-2 \left(k\cdot p_{-}\right) s \lambda^{V}_{18}+2 k\cdot p_{0} M_{\rho}^2 \lambda^{V}_{19}+2 k\cdot p_{-} M_{\rho}^2 \lambda^{V}_{19}-2 \left(k\cdot p_{0}\right) s \lambda^{V}_{19}\\
&-2 \left(k\cdot p_{-}\right) s \lambda^{V}_{19}+M_{\rho}^2 s \lambda^{V}_{19}-s^2 \lambda^{V}_{19}-4 k\cdot p_{-} M_{\rho}^2 \lambda^{V}_{21}+4 \left(k\cdot p_{-}\right) s \lambda^{V}_{21}+s^2 \lambda^{V}_{21}\\
&+6 k\cdot p_{0} M_{\rho}^2 \lambda^{V}_{22}+2 k\cdot p_{-} M_{\rho}^2 \lambda^{V}_{22}-6 \left(k\cdot p_{0}\right) s \lambda^{V}_{22}-2 \left(k\cdot p_{-}\right) s \lambda^{V}_{22}+2 M_{\rho}^2 s \lambda^{V}_{22}\\
&-2 s^2 \lambda^{V}_{22}-8 \left(k\cdot p_{-}\right) s \lambda^{V}_{21} \lambda^{VV}_7-2 s^2 \lambda^{V}_{21} \lambda^{VV}_7 )-2 G_V  (\sqrt{2} (4 k\cdot p_{-}+s) \\
& (\left(P-q\right)^2-M_{\rho}^2)\lambda^{V}_7-(4 k\cdot p_{-}+s) F_V \lambda^{VV}_7+\sqrt{2} \lambda^{V}_{22}  (-16 \left(k\cdot p_{-}\right)^2+8 k\cdot p_{-} M_{\rho}^2\\
&-10 \left(k\cdot p_{-}\right) s-s^2-2 k\cdot p_{0} (8 k\cdot p_{-}+s)+2 (P-q)^2  (4 k\cdot p_{-}+s) \lambda^{VV}_7 ) )\\
&+4  (s (4 k\cdot p_{-}+s)  (\left(P-q\right)^2-M_{\rho}^2)\lambda^{V}_7 \lambda^{V}_{21}+\lambda^{V}_{22}  (-2  (M_{\rho}^2-s ) s (P-q)^2  \lambda^{V}_{12}\\
&+ (M_{\rho}^2-s ) s (P-q)^2  \lambda^{V}_{13}-8 \left(k\cdot p_{0}\right) \left(k\cdot p_{-}\right) M_{\rho}^2 \lambda^{V}_{14}-8 \left(k\cdot p_{-}\right)^2 M_{\rho}^2 \lambda^{V}_{14}\\
\end{split}\end{equation*}
\begin{equation}\begin{split}
\qquad&+8 \left(k\cdot p_{0}\right) \left(k\cdot p_{-}\right) s \lambda^{V}_{14}+8 \left(k\cdot p_{-}\right)^2 s \lambda^{V}_{14}-2 k\cdot p_{0} M_{\rho}^2 s \lambda^{V}_{14}-6 k\cdot p_{-} M_{\rho}^2 s \lambda^{V}_{14}\\
&+2 \left(k\cdot p_{0}\right) s^2 \lambda^{V}_{14}+6 \left(k\cdot p_{-}\right) s^2 \lambda^{V}_{14}-M_{\rho}^2 s^2 \lambda^{V}_{14}+s^3 \lambda^{V}_{14}+8 \left(k\cdot p_{0}\right) \left(k\cdot p_{-}\right) M_{\rho}^2 \lambda^{V}_{15}\\
&+8 \left(k\cdot p_{-}\right)^2 M_{\rho}^2 \lambda^{V}_{15}-8 \left(k\cdot p_{0}\right) \left(k\cdot p_{-}\right) s \lambda^{V}_{15}-8 \left(k\cdot p_{-}\right)^2 s \lambda^{V}_{15}+2 k\cdot p_{0} M_{\rho}^2 s \lambda^{V}_{15}\\
&+6 k\cdot p_{-} M_{\rho}^2 s \lambda^{V}_{15}-2 \left(k\cdot p_{0}\right) s^2 \lambda^{V}_{15}-6 \left(k\cdot p_{-}\right) s^2 \lambda^{V}_{15}+M_{\rho}^2 s^2 \lambda^{V}_{15}-s^3 \lambda^{V}_{15}\\
&+4 \left(k\cdot p_{0}\right)^2 M_{\rho}^2 \lambda^{V}_{16}+8 \left(k\cdot p_{0}\right) \left(k\cdot p_{-}\right) M_{\rho}^2 \lambda^{V}_{16}+4 \left(k\cdot p_{-}\right)^2 M_{\rho}^2 \lambda^{V}_{16}-4 \left(k\cdot p_{0}\right)^2 s \lambda^{V}_{16}\\
&-8 \left(k\cdot p_{0}\right) \left(k\cdot p_{-}\right) s \lambda^{V}_{16}-4 \left(k\cdot p_{-}\right)^2 s \lambda^{V}_{16}+4 k\cdot p_{0} M_{\rho}^2 s \lambda^{V}_{16}+4 k\cdot p_{-} M_{\rho}^2 s \lambda^{V}_{16}\\
&-4 \left(k\cdot p_{0}\right) s^2 \lambda^{V}_{16}-4 \left(k\cdot p_{-}\right) s^2 \lambda^{V}_{16}+M_{\rho}^2 s^2 \lambda^{V}_{16}-s^3 \lambda^{V}_{16}-8 \left(k\cdot p_{0}\right)^2 M_{\rho}^2 \lambda^{V}_{17}\\
&-8 \left(k\cdot p_{0}\right) \left(k\cdot p_{-}\right) M_{\rho}^2 \lambda^{V}_{17}+8 \left(k\cdot p_{0}\right)^2 s \lambda^{V}_{17}+8 \left(k\cdot p_{0}\right) \left(k\cdot p_{-}\right) s \lambda^{V}_{17}\\
&-8 k\cdot p_{0} M_{\rho}^2 s \lambda^{V}_{17}-4 k\cdot p_{-} M_{\rho}^2 s \lambda^{V}_{17}+8 \left(k\cdot p_{0}\right) s^2 \lambda^{V}_{17}+4 \left(k\cdot p_{-}\right) s^2 \lambda^{V}_{17}\\
&-2 M_{\rho}^2 s^2 \lambda^{V}_{17}+2 s^3 \lambda^{V}_{17}-4 \left(k\cdot p_{0}\right) \left(k\cdot p_{-}\right) M_{\rho}^2 \lambda^{V}_{18}-4 \left(k\cdot p_{-}\right)^2 M_{\rho}^2 \lambda^{V}_{18}\\
&+4 \left(k\cdot p_{0}\right) \left(k\cdot p_{-}\right) s \lambda^{V}_{18}+4 \left(k\cdot p_{-}\right)^2 s \lambda^{V}_{18}-2 k\cdot p_{-} M_{\rho}^2 s \lambda^{V}_{18}+2 \left(k\cdot p_{-}\right) s^2 \lambda^{V}_{18}\\
&-4 \left(k\cdot p_{0}\right)^2 M_{\rho}^2 \lambda^{V}_{19}-8 \left(k\cdot p_{0}\right) \left(k\cdot p_{-}\right) M_{\rho}^2 \lambda^{V}_{19}-4 \left(k\cdot p_{-}\right)^2 M_{\rho}^2 \lambda^{V}_{19}+4 \left(k\cdot p_{0}\right)^2 s \lambda^{V}_{19}\\
&+8 \left(k\cdot p_{0}\right) \left(k\cdot p_{-}\right) s \lambda^{V}_{19}+4 \left(k\cdot p_{-}\right)^2 s \lambda^{V}_{19}-4 k\cdot p_{0} M_{\rho}^2 s \lambda^{V}_{19}-4 k\cdot p_{-} M_{\rho}^2 s \lambda^{V}_{19}\\
&+4 \left(k\cdot p_{0}\right) s^2 \lambda^{V}_{19}+4 \left(k\cdot p_{-}\right) s^2 \lambda^{V}_{19}-M_{\rho}^2 s^2 \lambda^{V}_{19}+s^3 \lambda^{V}_{19}+8 \left(k\cdot p_{0}\right) \left(k\cdot p_{-}\right) M_{\rho}^2 \lambda^{V}_{21}\\
&+8 \left(k\cdot p_{-}\right)^2 M_{\rho}^2 \lambda^{V}_{21}-24 \left(k\cdot p_{0}\right) \left(k\cdot p_{-}\right) s \lambda^{V}_{21}-24 \left(k\cdot p_{-}\right)^2 s \lambda^{V}_{21}+12 k\cdot p_{-} M_{\rho}^2 s \lambda^{V}_{21}\\
&-2 \left(k\cdot p_{0}\right) s^2 \lambda^{V}_{21}-14 \left(k\cdot p_{-}\right) s^2 \lambda^{V}_{21}-s^3 \lambda^{V}_{21}-8 \left(k\cdot p_{0}\right)^2 M_{\rho}^2 \lambda^{V}_{22}\\
&-8 \left(k\cdot p_{0}\right) \left(k\cdot p_{-}\right) M_{\rho}^2 \lambda^{V}_{22}+8 \left(k\cdot p_{0}\right)^2 s \lambda^{V}_{22}+8 \left(k\cdot p_{0}\right) \left(k\cdot p_{-}\right) s \lambda^{V}_{22}\\
&-6 k\cdot p_{0} M_{\rho}^2 s \lambda^{V}_{22}-2 k\cdot p_{-} M_{\rho}^2 s \lambda^{V}_{22}+6 \left(k\cdot p_{0}\right) s^2 \lambda^{V}_{22}+2 \left(k\cdot p_{-}\right) s^2 \lambda^{V}_{22}-M_{\rho}^2 s^2 \lambda^{V}_{22}\\
&+s^3 \lambda^{V}_{22}+16 \left(k\cdot p_{0}\right) \left(k\cdot p_{-}\right) s \lambda^{V}_{21} \lambda^{VV}_7+16 \left(k\cdot p_{-}\right)^2 s \lambda^{V}_{21} \lambda^{VV}_7+4 \left(k\cdot p_{0}\right) s^2 \lambda^{V}_{21} \lambda^{VV}_7\\
&+12 \left(k\cdot p_{-}\right) s^2 \lambda^{V}_{21} \lambda^{VV}_7+2 s^3 \lambda^{V}_{21} \lambda^{VV}_7 ) \Bigg)
\end{split}\end{equation}

\begin{equation}\begin{split}
v_{3}^{R}=&\frac{2}{F^2}  \Bigg(\frac{8 }{M_{\omega}^2 D_{\omega}\left[(k+p_{0})^2\right]} \Big(-4  (4 \left(k\cdot p_{0}\right)^2-M_{\omega}^2 (2 k\cdot p_{-}+s)+k\cdot p_{0}  \\
&(2 k\cdot p_{-}-2 M_{\omega}^2+s ) ) (\kappa^{V}_{12})^2+k\cdot p_{0} \kappa^{V}_{16}  ((2 k\cdot p_{-}+s) \kappa^{V}_{16}-2 (P-q)^2  \kappa^{V}_{17} )\\
&+2 \kappa^{V}_{12}  ( (4 \left(k\cdot p_{0}\right)^2-2 k\cdot p_{0} M_{\omega}^2+M_{\omega}^2 (2 k\cdot p_{-}+s) ) \kappa^{V}_{16}+2  (k\cdot p_{0}-M_{\omega}^2 ) \\
&(P-q)^2  \kappa^{V}_{17} ) \Big)+\frac{\sqrt{2} F_V  (2 \lambda^{V}_{13}+\lambda^{V}_{16}+\lambda^{V}_{18}+2 \lambda^{V}_{19} )}{M_{\rho}^2}\\
&+\frac{8 }{M_{a_{1}}^2 D_{a_{1}}\left[(k+p_{-})^2\right]}\Big( (-2 \left(k\cdot p_{0}\right) \left(k\cdot p_{-}\right)+2 k\cdot p_{0} M_{a_{1}}^2+2 k\cdot p_{-} M_{a_{1}}^2-\left(k\cdot p_{-}\right) s\\
&+M_{a_{1}}^2 s ) (\lambda^{A}_{12})^2-8 k\cdot p_{-} (2 k\cdot p_{0}+s) (\lambda^{A}_{13})^2-2 \sqrt{2} F_{A} k\cdot p_{0} \lambda^{A}_{15}-2 \sqrt{2} F_{A} k\cdot p_{-} \lambda^{A}_{15}\\
&-\sqrt{2} F_{A} s \lambda^{A}_{15}-4 \sqrt{2} F_{A} k\cdot p_{-} \lambda^{A}_{17}+16 \left(k\cdot p_{0}\right) \left(k\cdot p_{-}\right) \lambda^{A}_{15} \lambda^{A}_{17}+16 \left(k\cdot p_{-}\right)^2 \lambda^{A}_{15} \lambda^{A}_{17}\\
&+8 \left(k\cdot p_{-}\right) s \lambda^{A}_{15} \lambda^{A}_{17}+16 \left(k\cdot p_{-}\right)^2 (\lambda^{A}_{17})^2+\lambda^{A}_{13}  (8 k\cdot p_{-} (P-q)^2  \lambda^{A}_{15}+(2 k\cdot p_{0}\\
&-2 k\cdot p_{-}+s)  (\sqrt{2} F_{A}-8 k\cdot p_{-} \lambda^{A}_{17} ) )+\lambda^{A}_{12}  (8  (k\cdot p_{0}  (4 k\cdot p_{-}-2 M_{a_{1}}^2 )-M_{a_{1}}^2 s\\
&+2 k\cdot p_{-}  (M_{a_{1}}^2+s ) ) \lambda^{A}_{13}-8  (k\cdot p_{-}-M_{a_{1}}^2 ) (P-q)^2  \lambda^{A}_{15}+ (2 k\cdot p_{0}-2 k\cdot p_{-}+4 M_{a_{1}}^2\\
&+s )  (-\sqrt{2} F_{A}+8 k\cdot p_{-} \lambda^{A}_{17} ) )\Big) \Bigg)
\end{split}\end{equation}

\begin{equation}\begin{split}
v_{3}^{RR}=&\frac{1}{\sqrt{2} F^2} \Bigg(-\frac{16 F_V }{M_{\rho}^2 M_{\omega}^2 D_{\omega}\left[(k+p_{0})^2\right]} \Big(2 \kappa^{V}_{12}  (-M_{\omega}^2 (2 k\cdot p_{-}+s) \kappa^{VV}_3+ (2 k\cdot p_{0}-M_{\omega}^2 )\\
& (4 k\cdot p_{0}+2 k\cdot p_{-}+s) \kappa^{VV}_4 )- ((2 k\cdot p_{-}+s) \kappa^{V}_{16}-2 (P-q)^2  \kappa^{V}_{17} )  (M_{\omega}^2 \kappa^{VV}_3\\
&+D_{\omega}\left[(k+p_{0})^2\right] \kappa^{VV}_4 ) \Big)+\frac{32  (-F_V+2 \sqrt{2} (P-q)^2  \lambda^{V}_{22} )}{M_{\omega}^2 D_{\rho}\left[(P-q)^2\right] D_{\omega}\left[(k+p_{0})^2\right]} \Big(-2 \kappa^{V}_{12}  ( (2 \left(k\cdot p_{0}\right)^2\\
&-M_{\omega}^2 (2 k\cdot p_{-}+s)+k\cdot p_{0}  (2 k\cdot p_{-}-3 M_{\omega}^2+s ) ) \kappa^{VV}_3+k\cdot p_{0}  (-2 k\cdot p_{0}+M_{\omega}^2 ) \kappa^{VV}_4 )\\
&+k\cdot p_{0} \kappa^{V}_{16}  ( (\left(P-q\right)^2+M_{\omega}^2 ) \kappa^{VV}_3+D_{\omega}\left[(k+p_{0})^2\right] \kappa^{VV}_4 ) \Big)\\
& +\frac{2 F_V }{M_{a_{1}}^2 M_{\rho}^2 D_{a_{1}}\left[(k+p_{-})^2\right]} \Big(-2  (-4  (2 k\cdot p_{0}  (k\cdot p_{-}-M_{a_{1}}^2 )-M_{a_{1}}^2 s+k\cdot p_{-} \\
& (-2 M_{a_{1}}^2+s ) ) \lambda^{A}_{12}+ (k\cdot p_{-}-M_{a_{1}}^2 )  (\sqrt{2} F_{A}+4 (2 k\cdot p_{0}+s) \lambda^{A}_{13}-4 (P-q)^2  \lambda^{A}_{15}\\
&-8 k\cdot p_{-} \lambda^{A}_{17} ) ) \lambda^{VA}_2+k\cdot p_{-}  (-\sqrt{2} F_{A}+4  (2 k\cdot p_{0}+2 M_{a_{1}}^2+s ) \lambda^{A}_{12}-4 (2 k\cdot p_{0}+s) \lambda^{A}_{13}\\
&+8 k\cdot p_{0} \lambda^{A}_{15}+8 k\cdot p_{-} \lambda^{A}_{15}+4 s \lambda^{A}_{15}+8 k\cdot p_{-} \lambda^{A}_{17} )  (\lambda^{VA}_4+2 \lambda^{VA}_5 ) \Big)\\
&+\frac{ (-\sqrt{2} F_V+4 (P-q)^2  \lambda^{V}_{22} ) }{M_{a_{1}}^2 D_{a_{1}}\left[(k+p_{-})^2\right] D_{\rho}\left[(P-q)^2\right]} \Big(-2  (4 \sqrt{2}  (2 k\cdot p_{0}  (k\cdot p_{-}-M_{a_{1}}^2 )-M_{a_{1}}^2 s\\
&+k\cdot p_{-}  (-2 M_{a_{1}}^2+s ) ) \lambda^{A}_{12}+ (2 k\cdot p_{0}+2 M_{a_{1}}^2+s )  (F_{A}-4 \sqrt{2} k\cdot p_{-} \lambda^{A}_{13}\\
&-4 \sqrt{2} k\cdot p_{-} \lambda^{A}_{17} ) ) \lambda^{VA}_2+ (F_{A}+4 \sqrt{2}  (k\cdot p_{-}-M_{a_{1}}^2 ) \lambda^{A}_{12}-4 \sqrt{2} k\cdot p_{-} \lambda^{A}_{13}\\
&-4 \sqrt{2} k\cdot p_{-} \lambda^{A}_{17} )  (2 (P-q)^2  \lambda^{VA}_3-(2 k\cdot p_{0}+s)  (\lambda^{VA}_4+2 \lambda^{VA}_5 ) ) \Big)\\
&+\frac{8 F_V  (\sqrt{2} F_V-4 (P-q)^2  \lambda^{V}_{22} ) \lambda^{VV}_5}{M_{\rho}^2 D_{\rho}\left[(P-q)^2\right]} \Bigg)
\end{split}\end{equation}

\begin{equation}\begin{split}
v_{3}^{RRR}=&-\frac{\sqrt{2} F_V (-\sqrt{2} F_V+4 (P-q)^2  \lambda^{V}_{22} )}{F^2 M_{a_{1}}^2 M_{\rho}^2 D_{a_{1}}\left[(k+p_{-})^2\right] D_{\rho}\left[(P-q)^2\right]}  \Bigg(-4  (2 k\cdot p_{0}  (k\cdot p_{-}-M_{a_{1}}^2 )\\
&-M_{a_{1}}^2 s+k\cdot p_{-}  (-2 M_{a_{1}}^2+s ) ) (\lambda^{VA}_2)^2+k\cdot p_{-}  (\lambda^{VA}_4+2 \lambda^{VA}_5 )  (2 (P-q)^2  \lambda^{VA}_3\\
&-(2 k\cdot p_{0}+s)  (\lambda^{VA}_4+2 \lambda^{VA}_5 ) )+2 \lambda^{VA}_2  (2  (k\cdot p_{-}-M_{a_{1}}^2 ) (P-q)^2  \lambda^{VA}_3\\
&- (k\cdot p_{0}  (4 k\cdot p_{-}-2 M_{a_{1}}^2 )-M_{a_{1}}^2 s+2 k\cdot p_{-}  (M_{a_{1}}^2+s ) )  (\lambda^{VA}_4+2 \lambda^{VA}_5 ) ) \Bigg)
\end{split}\end{equation}

\begin{equation}
v_{GI3}^{R+RR}=\frac{2 \sqrt{2}}{F^2 D_{\rho}\left[(P-q)^2\right]}  (F_V-2 \sqrt{2} (P-q)^2  \lambda^{V}_{22} ) (2 \lambda^{V}_{13}+\lambda^{V}_{16}+\lambda^{V}_{18}+\lambda^{V}_{19} ) 
\end{equation}

\begin{equation}\begin{split}
v_{4}^{R}=&\frac{2}{F^2}  \Bigg(\frac{8 }{M_{\omega}^2 D_{\omega}\left[(k+p_{0})^2\right]} \Big(4  (k\cdot p_{0}-M_{\omega}^2 ) (2 k\cdot p_{-}+s) (\kappa^{V}_{12})^2+k\cdot p_{0} \kappa^{V}_{16} \\
& (-(2 k\cdot p_{-}+s) \kappa^{V}_{16}+2  (2 k\cdot p_{-}+M_{\omega}^2+s ) \kappa^{V}_{17} )-2 \kappa^{V}_{12}  (M_{\omega}^2 (2 k\cdot p_{-}+s) \kappa^{V}_{16}\\
&+2  (-M_{\omega}^2 (2 k\cdot p_{-}+s)+k\cdot p_{0}  (2 k\cdot p_{-}-M_{\omega}^2+s ) ) \kappa^{V}_{17} ) \Big)\\
&-\frac{\sqrt{2} F_V  (\lambda^{V}_{13}+\lambda^{V}_{14}-\lambda^{V}_{15}-\lambda^{V}_{21} )}{M_{\rho}^2}\\
&-\frac{\sqrt{2}  (2 \lambda^{A}_{12}+\lambda^{A}_{15} )(-F_{A}+4 \sqrt{2} k\cdot p_{-}\left( \lambda^{A}_{12}+\lambda^{A}_{13}+\lambda^{A}_{17}\right) ) }{D_{a_{1}}\left[(k+p_{-})^2\right]}  \Bigg)
\end{split}\end{equation}

\begin{equation}\begin{split}
v_{4}^{RR}=&\frac{\sqrt{2}}{F^2}  \Bigg(-\frac{8 F_V }{M_{\rho}^2 M_{\omega}^2 D_{\omega}\left[(k+p_{0})^2\right]} \Big((2 k\cdot p_{-}+s) \kappa^{V}_{16}  (M_{\omega}^2 \kappa^{VV}_3+ (-2 k\cdot p_{0}+M_{\omega}^2 ) \kappa^{VV}_4 )\\
&-2 \kappa^{V}_{17}  (M_{\omega}^2 (P-q)^2  \kappa^{VV}_3- (2 k\cdot p_{0}-M_{\omega}^2 ) (2 k\cdot p_{-}+s) \kappa^{VV}_4 )+2 (2 k\cdot p_{-}+s) \kappa^{V}_{12}\\
&  (M_{\omega}^2 \kappa^{VV}_3+D_{\omega}\left[(k+p_{0})^2\right] \kappa^{VV}_4 ) \Big)-\frac{16 \kappa^{VV}_3 (-F_V+2 \sqrt{2} (P-q)^2  \lambda^{V}_{22} )}{M_{\omega}^2 D_{\rho}\left[(P-q)^2\right] D_{\omega}\left[(k+p_{0})^2\right]}\\
&\Big( (2 M_{\omega}^2 (2 k\cdot p_{-}+s)-2 k\cdot p_{0}  (2 k\cdot p_{-}-2 M_{\omega}^2+s ) ) \kappa^{V}_{12}+k\cdot p_{0}  (2 k\cdot p_{-}+2 M_{\omega}^2+s ) \kappa^{V}_{16} \Big)\\
& +\frac{ (\sqrt{2} F_V-4 (P-q)^2  \lambda^{V}_{22} )  (-F_{A}+4 \sqrt{2} k\cdot p_{-}\left( \lambda^{A}_{12}+\lambda^{A}_{13}+\lambda^{A}_{17}\right) )  (2 \lambda^{VA}_2-\lambda^{VA}_3 )}{D_{a_{1}}\left[(k+p_{-})^2\right] D_{\rho}\left[(P-q)^2\right]}\\
&+\frac{4 k\cdot p_{-} F_V  (2 \lambda^{A}_{12}+\lambda^{A}_{15} )  (2 \lambda^{VA}_2-\lambda^{VA}_4-2 \lambda^{VA}_5 )}{M_{\rho}^2 D_{a_{1}}\left[(k+p_{-})^2\right]}-\frac{2 F_V  (\sqrt{2} G_V-2 s \lambda^{V}_{21} ) \lambda^{VV}_7}{M_{\rho}^2 D_{\rho}[s]}\\
&-\frac{F_V  (\sqrt{2} F_V  (\lambda^{VV}_3-\lambda^{VV}_4+2 \lambda^{VV}_5 )-2 (P-q)^2  \lambda^{V}_{22}  (-1+2 \lambda^{VV}_3-2 \lambda^{VV}_4+4 \lambda^{VV}_5 ) )}{M_{\rho}^2 D_{\rho}\left[(P-q)^2\right]}\Bigg)
\end{split}\end{equation}

\begin{equation}\begin{split}
v_{4}^{RRR}=-\frac{2 \sqrt{2} F_V k\cdot p_{-}(\sqrt{2} F_V-4 (P-q)^2  \lambda^{V}_{22} )  (2 \lambda^{VA}_2-\lambda^{VA}_3 ) (2 \lambda^{VA}_2-\lambda^{VA}_4-2 \lambda^{VA}_5 )}{F^2 M_{\rho}^2 D_{a_{1}}\left[(k+p_{-})^2\right] D_{\rho}\left[(P-q)^2\right]}  
\end{split}\end{equation}

\begin{equation}\begin{split}
v_{GI4}^{R+RR}=&\frac{2 \sqrt{2} F_V }{F^2 D_{\rho}\left[(P-q)^2\right] D_{\rho}[s]} \Bigg( (-M_{\rho}^2+s ) \left(\lambda^{V}_{13}+ \lambda^{V}_{14}- \lambda^{V}_{15}+ \lambda^{V}_{18}+ \lambda^{V}_{19}\right)+2 M_{\rho}^2 \lambda^{V}_{21}-s \lambda^{V}_{21}\\
&+2 s \lambda^{V}_{21} \lambda^{VV}_7 )+4 G_V  (\sqrt{2}  (\left(P-q\right)^2-M_{\rho}^2)\lambda^{V}_7-F_V \lambda^{VV}_7+\sqrt{2} \lambda^{V}_{22}  (-4 k\cdot p_{0}-4 k\cdot p_{-}\\
&+3 M_{\rho}^2-2 s+2 (P-q)^2  \lambda^{VV}_7 ) )-8  (s  (\left(P-q\right)^2-M_{\rho}^2)\lambda^{V}_7 \lambda^{V}_{21}+\lambda^{V}_{22}  (- (M_{\rho}^2-s )\\
& (P-q)^2  \lambda^{V}_{13}- (M_{\rho}^2-s ) (P-q)^2  \lambda^{V}_{14}+2 k\cdot p_{0} M_{\rho}^2 \lambda^{V}_{15}+2 k\cdot p_{-} M_{\rho}^2 \lambda^{V}_{15}-2 \left(k\cdot p_{0}\right) s \lambda^{V}_{15}\\
&-2 \left(k\cdot p_{-}\right) s \lambda^{V}_{15}+M_{\rho}^2 s \lambda^{V}_{15}-s^2 \lambda^{V}_{15}-2 k\cdot p_{0} M_{\rho}^2 \lambda^{V}_{18}-2 k\cdot p_{-} M_{\rho}^2 \lambda^{V}_{18}+2 \left(k\cdot p_{0}\right) s \lambda^{V}_{18}\\
&+2 \left(k\cdot p_{-}\right) s \lambda^{V}_{18}-M_{\rho}^2 s \lambda^{V}_{18}+s^2 \lambda^{V}_{18}-2 k\cdot p_{0} M_{\rho}^2 \lambda^{V}_{19}-2 k\cdot p_{-} M_{\rho}^2 \lambda^{V}_{19}+2 \left(k\cdot p_{0}\right) s \lambda^{V}_{19}\\
&+2 \left(k\cdot p_{-}\right) s \lambda^{V}_{19}-M_{\rho}^2 s \lambda^{V}_{19}+s^2 \lambda^{V}_{19}+4 k\cdot p_{0} M_{\rho}^2 \lambda^{V}_{21}+4 k\cdot p_{-} M_{\rho}^2 \lambda^{V}_{21}-8 \left(k\cdot p_{0}\right) s \lambda^{V}_{21}\\
&-8 \left(k\cdot p_{-}\right) s \lambda^{V}_{21}+5 M_{\rho}^2 s \lambda^{V}_{21}-4 s^2 \lambda^{V}_{21}+4 \left(k\cdot p_{0}\right) s \lambda^{V}_{21} \lambda^{VV}_7+4 \left(k\cdot p_{-}\right) s \lambda^{V}_{21} \lambda^{VV}_7\\
&+2 s^2 \lambda^{V}_{21} \lambda^{VV}_7 ) \Bigg)
\end{split}\end{equation}

\section{Axial Form Factors}\label{AF}
\begin{equation*}\begin{split}
a_{1}^{R}=&\frac{\sqrt{2} }{3 F^2} \Bigg(-\frac{2 (P-q)^2  F_V  (\kappa^{V}_1-\kappa^{V}_2+\kappa^{V}_3+\kappa^{V}_6+\kappa^{V}_7-\kappa^{V}_8-2 \kappa^{V}_{12}-\kappa^{V}_{16}+\kappa^{V}_{17} )}{M_{\omega}^2}\\
&+\frac{4 ((P-q)^2  \kappa^{V}_{11}+s \kappa^{V}_{12}-(k\cdot p_{0}+k\cdot p_{-}) \kappa^{V}_{16} )  (-G_V+\sqrt{2} s \lambda^{V}_{21} )}{D_{\rho}[s]}\\
&+\frac{1}{M_{\rho}^2 D_{\rho}\left[(k+p_{-})^2\right]}\Big(F_V (2 k\cdot p_{-}+s)  (2  (k\cdot p_{-}-M_{\rho}^2 ) \kappa^{V}_{12}-k\cdot p_{-}\kappa^{V}_{16} )\\
&+G_V  (-4  (2 \left(k\cdot p_{-}\right)^2+\left(k\cdot p_{-}\right) s-M_{\rho}^2 s ) \kappa^{V}_{12}+2 k\cdot p_{-}  (2k\cdot p_{-}-2 M_{\rho}^2+s ) \kappa^{V}_{16} )\\
&+\sqrt{2}  (-2 \kappa^{V}_{12}  ( (k\cdot p_{-}-M_{\rho}^2 ) (2 k\cdot p_{0}+s) (2 k\cdot p_{-}+s) \lambda^{V}_{16}-2  (k\cdot p_{-}-M_{\rho}^2 )\\
& (2 k\cdot p_{0}+s) (2 k\cdot p_{-}+s) \lambda^{V}_{17}+8 \left(k\cdot p_{0}\right) \left(k\cdot p_{-}\right)^2 \lambda^{V}_{18}+8 \left(k\cdot p_{-}\right)^3 \lambda^{V}_{18}\\
&+4 \left(k\cdot p_{0}\right) \left(k\cdot p_{-}\right) s \lambda^{V}_{18}+8 \left(k\cdot p_{-}\right)^2 s \lambda^{V}_{18}-4 k\cdot p_{0} M_{\rho}^2 s \lambda^{V}_{18}-4 k\cdot p_{-} M_{\rho}^2 s \lambda^{V}_{18}\\
&+2 \left(k\cdot p_{-}\right) s^2 \lambda^{V}_{18}-2 M_{\rho}^2 s^2 \lambda^{V}_{18}+8 \left(k\cdot p_{0}\right) \left(k\cdot p_{-}\right)^2 \lambda^{V}_{19}+8 \left(k\cdot p_{0}\right) \left(k\cdot p_{-}\right) M_{\rho}^2 \lambda^{V}_{19}\\
&+8 \left(k\cdot p_{-}\right)^2 M_{\rho}^2 \lambda_{19}+4 \left(k\cdot p_{0}\right) \left(k\cdot p_{-}\right) s \lambda^{V}_{19}+4 \left(k\cdot p_{-}\right)^2 s \lambda^{V}_{19}-4 k\cdot p_{0} M_{\rho}^2 s \lambda^{V}_{19}\\
&+2 \left(k\cdot p_{-}\right) s^2 \lambda^{V}_{19}-2 M_{\rho}^2 s^2 \lambda^{V}_{19}-8 \left(k\cdot p_{-}\right)^3 \lambda^{V}_{21}-4 \left(k\cdot p_{-}\right)^2 s \lambda^{V}_{21}+4 k\cdot p_{-} M_{\rho}^2 s \lambda^{V}_{21}\\
&+8 \left(k\cdot p_{-}\right)^3 \lambda^{V}_{22}-8 \left(k\cdot p_{-}\right)^2 M_{\rho}^2 \lambda^{V}_{22}+4 \left(k\cdot p_{-}\right)^2 s \lambda^{V}_{22}-4 k\cdot p_{-} M_{\rho}^2 s \lambda^{V}_{22} )\\
&+k\cdot p_{-} \kappa^{V}_{16}  ((2 k\cdot p_{0}+s) (2 k\cdot p_{-}+s) \lambda^{V}_{16}+2  (-(2 k\cdot p_{0}+s) (2 k\cdot p_{-}+s) \lambda^{V}_{17}\\
&+(P-q)^2   (2 k\cdot p_{-}-2 M_{\rho}^2+s ) \lambda^{V}_{18}+4 \left(k\cdot p_{0}\right) \left(k\cdot p_{-}\right) \lambda^{V}_{19}-8 k\cdot p_{0} M_{\rho}^2 \lambda^{V}_{19}\\
&-4 k\cdot p_{-} M_{\rho}^2 \lambda^{V}_{19}+2 \left(k\cdot p_{0}\right) s \lambda^{V}_{19}+2 \left(k\cdot p_{-}\right) s \lambda^{V}_{19}-2 M_{\rho}^2 s \lambda^{V}_{19}+s^2 \lambda^{V}_{19}-4 \left(k\cdot p_{-}\right)^2 \\
&\lambda^{V}_{21}+4 k\cdot p_{-} M_{\rho}^2 \lambda^{V}_{21}-2 \left(k\cdot p_{-}\right) s \lambda^{V}_{21}+4 \left(k\cdot p_{-}\right)^2 \lambda^{V}_{22}+2 \left(k\cdot p_{-}\right) s \lambda^{V}_{22} ) ) )\Big)\\
&+\frac{1}{M_{\rho}^2 D_{\rho}\left[(k+p_{0})^2\right]}\Big(F_V(4 k\cdot p_{0}+2 k\cdot p_{-}+s)   (2  (k\cdot p_{0}-M_{\rho}^2 ) \kappa^{V}_{12}-k\cdot p_{0} \kappa^{V}_{16} )\\
&+G_V  (-4  (4 \left(k\cdot p_{0}\right)^2-M_{\rho}^2 (2 k\cdot p_{-}+s)+k\cdot p_{0}  (2 k\cdot p_{-}-2 M_{\rho}^2+s ) ) \kappa^{V}_{12}+2 k\cdot p_{0} \\
& (4 k\cdot p_{0}+2 k\cdot p_{-}-2 M_{\rho}^2+s ) \kappa^{V}_{16} )+\sqrt{2}  (k\cdot p_{0} \kappa^{V}_{16}  ((2 k\cdot p_{-}+s) (4 k\cdot p_{0}+2 k\cdot p_{-}+s)\\
& \lambda^{V}_{16}+2  (-(2 k\cdot p_{-}+s) (4 k\cdot p_{0}+2 k\cdot p_{-}+s) \lambda^{V}_{17}+ (8 \left(k\cdot p_{0}\right)^2+12 \left(k\cdot p_{0}\right) \left(k\cdot p_{-}\right)\\
&+4 \left(k\cdot p_{-}\right)^2-4 k\cdot p_{0} M_{\rho}^2-4 k\cdot p_{-} M_{\rho}^2+6 \left(k\cdot p_{0}\right) s+4 \left(k\cdot p_{-}\right) s-2 M_{\rho}^2 s+s^2 ) \lambda^{V}_{18}\\
&+8 \left(k\cdot p_{0}\right) \left(k\cdot p_{-}\right) \lambda^{V}_{19}+4 \left(k\cdot p_{-}\right)^2 \lambda^{V}_{19}-4 k\cdot p_{-} M_{\rho}^2 \lambda^{V}_{19}+4 \left(k\cdot p_{0}\right) s \lambda^{V}_{19}\\
&+4 \left(k\cdot p_{-}\right) s \lambda^{V}_{19}-2 M_{\rho}^2 s \lambda^{V}_{19}+s^2 \lambda^{V}_{19}-8 \left(k\cdot p_{0}\right)^2 \lambda^{V}_{21}-4 \left(k\cdot p_{0}\right) \left(k\cdot p_{-}\right) \lambda^{V}_{21}\\
&+4 k\cdot p_{0} M_{\rho}^2 \lambda^{V}_{21}-2 \left(k\cdot p_{0}\right) s \lambda^{V}_{21}+8 \left(k\cdot p_{0}\right)^2 \lambda^{V}_{22}+4 \left(k\cdot p_{0}\right) \left(k\cdot p_{-}\right) \lambda^{V}_{22}+2 \left(k\cdot p_{0}\right) s \lambda^{V}_{22} ) )\\
&-2 \kappa^{V}_{12}  ( (k\cdot p_{0}-M_{\rho}^2 ) (2 k\cdot p_{-}+s) (4 k\cdot p_{0}+2 k\cdot p_{-}+s) \lambda^{V}_{16}+2  (- (k\cdot p_{0}-M_{\rho}^2 ) \\
&(2 k\cdot p_{-}+s) (4 k\cdot p_{0}+2 k\cdot p_{-}+s) \lambda^{V}_{17}+(P-q)^2   (4 \left(k\cdot p_{0}\right)^2-M_{\rho}^2 (2 k\cdot p_{-}+s)\\
&+k\cdot p_{0}  (2 k\cdot p_{-}-2 M_{\rho}^2+s ) ) \lambda^{V}_{18}+8 \left(k\cdot p_{0}\right)^2 \left(k\cdot p_{-}\right) \lambda^{V}_{19}+4 \left(k\cdot p_{0}\right) \left(k\cdot p_{-}\right)^2 \lambda^{V}_{19}\\
\end{split}\end{equation*}
\begin{equation}\begin{split}
\qquad&-4 \left(k\cdot p_{0}\right) \left(k\cdot p_{-}\right) M_{\rho}^2 \lambda^{V}_{19}-4 \left(k\cdot p_{-}\right)^2 M_{\rho}^2 \lambda^{V}_{19}+4 \left(k\cdot p_{0}\right)^2 s \lambda^{V}_{19}+4 \left(k\cdot p_{0}\right) \left(k\cdot p_{-}\right) s \lambda^{V}_{19}\\
&-2 k\cdot p_{0} M_{\rho}^2 s \lambda^{V}_{19}-4 k\cdot p_{-} M_{\rho}^2 s \lambda^{V}_{19}+\left(k\cdot p_{0}\right) s^2 \lambda^{V}_{19}-M_{\rho}^2 s^2 \lambda^{V}_{19}-8 \left(k\cdot p_{0}\right)^3 \lambda^{V}_{21}\\&-4 \left(k\cdot p_{0}\right)^2 \left(k\cdot p_{-}\right) \lambda^{V}_{21}+4 \left(k\cdot p_{0}\right)^2 M_{\rho}^2 \lambda^{V}_{21}+4 \left(k\cdot p_{0}\right) \left(k\cdot p_{-}\right) M_{\rho}^2 \lambda^{V}_{21}-2 \left(k\cdot p_{0}\right)^2 s \lambda^{V}_{21}\\
&+2 k\cdot p_{0} M_{\rho}^2 s \lambda^{V}_{21}+8 \left(k\cdot p_{0}\right)^3 \lambda^{V}_{22}+4 \left(k\cdot p_{0}\right)^2 \left(k\cdot p_{-}\right) \lambda^{V}_{22}-8 \left(k\cdot p_{0}\right)^2 M_{\rho}^2 \lambda^{V}_{22}\\
&-4 \left(k\cdot p_{0}\right) \left(k\cdot p_{-}\right) M_{\rho}^2 \lambda^{V}_{22}+2 \left(k\cdot p_{0}\right)^2 s \lambda^{V}_{22}-2 k\cdot p_{0} M_{\rho}^2 s \lambda^{V}_{22} ) ) )\Big)\\
&+\frac{2 (P-q)^2   (\kappa^{A}_5-\kappa^{A}_6+\kappa^{A}_7 )  (-F_{A}+2 \sqrt{2} (P-q)^2  \lambda^{A}_{17} )}{D_{a_{1}}\left[(P-q)^2\right]} \Bigg)
\end{split}\end{equation}

\begin{equation*}\begin{split}
a_{1}^{RR}=&\frac{1}{3 \sqrt{2} F^2} \Bigg(\frac{4 s F_V \kappa^{VV}_3  (-\sqrt{2} G_V+2 s \lambda^{V}_{21} )}{M_{\omega}^2 D_{\rho}[s]}\\
&-\frac{1}{M_{\rho}^2 M_{\omega}^2 D_{\rho}\left[(k+p_{-})^2\right]}\Big(F_V  (\sqrt{2} (2 k\cdot p_{-}+s) F_V  (M_{\rho}^2 \kappa^{VV}_3+D_{\rho}\left[(k+p_{-})^2\right] \kappa^{VV}_4 )\\
&+2 \sqrt{2} G_V  (M_{\rho}^2 (2 k\cdot p_{-}-s) \kappa^{VV}_3-(2 k\cdot p_{-}+s) D_{\rho}\left[(k+p_{-})^2\right] \kappa^{VV}_4 )\\
&+2  (2 k\cdot p_{-}-M_{\rho}^2 ) (2 k\cdot p_{-}+s) \kappa^{VV}_4  ((2 k\cdot p_{0}+s) \lambda^{V}_{16}-2 (2 k\cdot p_{0}+s) \lambda^{V}_{17}+4 k\cdot p_{0} \lambda^{V}_{18}\\
&+4 k\cdot p_{-} \lambda^{V}_{18}+2 s \lambda^{V}_{18}+4 k\cdot p_{0} \lambda^{V}_{19}+2 s \lambda^{V}_{19}-4 k\cdot p_{-} \lambda^{V}_{21}+4 k\cdot p_{-} \lambda^{V}_{22} )-2 M_{\rho}^2 \kappa^{VV}_3  \\
&((2 k\cdot p_{0}+s) (2 k\cdot p_{-}+s) \lambda^{V}_{16}-2 (2 k\cdot p_{0}+s) (2 k\cdot p_{-}+s) \lambda^{V}_{17}-2 (2 k\cdot p_{-}-s) \\
&(P-q)^2  \lambda^{V}_{18}-24 \left(k\cdot p_{0}\right) \left(k\cdot p_{-}\right) \lambda^{V}_{19}-16 \left(k\cdot p_{-}\right)^2 \lambda^{V}_{19}+4 \left(k\cdot p_{0}\right) s \lambda^{V}_{19}-4 \left(k\cdot p_{-}\right) s \lambda^{V}_{19}\\
&+2 s^2 \lambda^{V}_{19}+8 \left(k\cdot p_{-}\right)^2 \lambda^{V}_{21}-4 \left(k\cdot p_{-}\right) s \lambda^{V}_{21}+8 \left(k\cdot p_{-}\right)^2 \lambda^{V}_{22}+4 \left(k\cdot p_{-}\right) s \lambda^{V}_{22} ) )\Big)\\
&+\frac{1}{M_{\rho}^2 M_{\omega}^2 D_{\rho}\left[(k+p_{0})^2\right]}\Big(F_V  (-\sqrt{2} F_V (4 k\cdot p_{0}+2 k\cdot p_{-}+s)  (M_{\rho}^2 \kappa^{VV}_3+D_{\rho}\left[(k+p_{0})^2\right]\\
& \kappa^{VV}_4 )+2 \sqrt{2} G_V  (M_{\rho}^2 (2 k\cdot p_{-}+s) \kappa^{VV}_3+(4 k\cdot p_{0}+2 k\cdot p_{-}+s) D_{\rho}\left[(k+p_{0})^2\right] \kappa^{VV}_4 )\\
&-2  (2 k\cdot p_{0}-M_{\rho}^2 ) (4 k\cdot p_{0}+2 k\cdot p_{-}+s) \kappa^{VV}_4  ((2 k\cdot p_{-}+s) \lambda^{V}_{16}-2 (2 k\cdot p_{-}+s) \lambda^{V}_{17}\\
&+4 k\cdot p_{0} \lambda^{V}_{18}+4 k\cdot p_{-} \lambda^{V}_{18}+2 s \lambda^{V}_{18}+4 k\cdot p_{-} \lambda^{V}_{19}+2 s \lambda^{V}_{19}-4 k\cdot p_{0} \lambda^{V}_{21}+4 k\cdot p_{0} \lambda^{V}_{22} )\\
&+2 M_{\rho}^2 \kappa^{VV}_3  ((2 k\cdot p_{-}+s) (4 k\cdot p_{0}+2 k\cdot p_{-}+s) \lambda^{V}_{16}+2  (-(2 k\cdot p_{-}+s)\\
& (4 k\cdot p_{0}+2 k\cdot p_{-}+s) \lambda^{V}_{17}+(2 k\cdot p_{-}+s) (P-q)^2  \lambda^{V}_{18}+4 \left(k\cdot p_{-}\right)^2 \lambda^{V}_{19}\\
&+4 \left(k\cdot p_{-}\right) s \lambda^{V}_{19}+s^2 \lambda^{V}_{19}-4 \left(k\cdot p_{0}\right) \left(k\cdot p_{-}\right) \lambda^{V}_{21}-2 \left(k\cdot p_{0}\right) s \lambda^{V}_{21}+8 \left(k\cdot p_{0}\right)^2 \lambda^{V}_{22}\\
&+4 \left(k\cdot p_{0}\right) \left(k\cdot p_{-}\right) \lambda^{V}_{22}+2 \left(k\cdot p_{0}\right) s \lambda^{V}_{22} ) ) )\Big)\\
&+\frac{2 (P-q)^2  F_V  (\kappa^{VA}_2-\kappa^{VA}_3-\kappa^{VA}_4 )  (-\sqrt{2} F_{A}+4 (P-q)^2  \lambda^{A}_{17} )}{M_{\omega}^2 D_{a_{1}}\left[(P-q)^2\right]}\\
&-\frac{4 (P-q)^2  \kappa^{VA}_5  (-\sqrt{2}G_V +2 s \lambda^{V}_{21}  )(-F_{A}+2 \sqrt{2} (P-q)^2  \lambda^{A}_{17} )}{D_{a_{1}}\left[(P-q)^2\right] D_{\rho}[s]}\\
\end{split}\end{equation*}
\begin{equation}\begin{split}
\qquad&-\frac{2(F_{A}-2 \sqrt{2} (P-q)^2  \lambda^{A}_{17} ) }{M_{\rho}^2 D_{a_{1}}\left[(P-q)^2\right] D_{\rho}\left[(k+p_{-})^2\right]}   \Big(k\cdot p_{-} \kappa^{V}_{16}  (-2  (- (2 M_{\rho}^2-s ) (2 k\cdot p_{-}+s)\\
&+2 k\cdot p_{0}  (2 k\cdot p_{-}-4 M_{\rho}^2+s ) ) \lambda^{VA}_2-2 M_{\rho}^2 (4 k\cdot p_{0}+2 k\cdot p_{-}+s) \lambda^{VA}_3\\&+(2 k\cdot p_{-}+s)  ( (2 k\cdot p_{0}+4 k\cdot p_{-}-2 M_{\rho}^2+s ) \lambda^{VA}_4+2 (2 k\cdot p_{0}+s) \lambda^{VA}_5 ) )\\
&+2 \kappa^{V}_{12}  (2  (\left(k\cdot p_{-}\right) s^2-M_{\rho}^2 s^2+2 \left(k\cdot p_{-}\right)^2  (2 M_{\rho}^2+s )+2 k\cdot p_{0}  (2 \left(k\cdot p_{-}\right)^2\\
&-M_{\rho}^2 s+k\cdot p_{-}  (2 M_{\rho}^2+s ) ) ) \lambda^{VA}_2-2 k\cdot p_{-} M_{\rho}^2 (4 k\cdot p_{0}+2 k\cdot p_{-}+s) \lambda^{VA}_3\\&-(2 k\cdot p_{-}+s)  ( (4 \left(k\cdot p_{-}\right)^2-2 k\cdot p_{-} M_{\rho}^2+2 k\cdot p_{0}  (k\cdot p_{-}-M_{\rho}^2 )+\left(k\cdot p_{-}\right) s\\
&-M_{\rho}^2 s ) \lambda^{VA}_4+2  (k\cdot p_{-}-M_{\rho}^2 ) (2 k\cdot p_{0}+s) \lambda^{VA}_5 ) ) \Big)\\
&+\frac{2  (F_{A}-2 \sqrt{2} (P-q)^2  \lambda^{A}_{17} )}{M_{\rho}^2 D_{a_{1}}\left[(P-q)^2\right] D_{\rho}\left[(k+p_{0})^2\right]} \Big(-k\cdot p_{0} \kappa^{V}_{16}  (-2 (2 k\cdot p_{-}+s)  (4 k\cdot p_{0}+2 k\cdot p_{-}\\
&-2 M_{\rho}^2+s ) \lambda^{VA}_2-2 M_{\rho}^2 (2 k\cdot p_{-}+s) \lambda^{VA}_3+(4 k\cdot p_{0}+2 k\cdot p_{-}+s)  ( (4 k\cdot p_{0}+2 k\cdot p_{-}\\
&-2 M_{\rho}^2+s ) \lambda^{VA}_4+2 (2 k\cdot p_{-}+s) \lambda^{VA}_5 ) )+2 \kappa^{V}_{12}  (-2 (2 k\cdot p_{-}+s)  (4 \left(k\cdot p_{0}\right)^2\\
&-M_{\rho}^2 (2 k\cdot p_{-}+s)+k\cdot p_{0}  (2 k\cdot p_{-}-2 M_{\rho}^2+s ) ) \lambda^{VA}_2+2 k\cdot p_{0} M_{\rho}^2 (2 k\cdot p_{-}+s) \lambda^{VA}_3\\
&+(4 k\cdot p_{0}+2 k\cdot p_{-}+s)  ( (4 \left(k\cdot p_{0}\right)^2-M_{\rho}^2 (2 k\cdot p_{-}+s)+k\cdot p_{0}  (2 k\cdot p_{-}-2 M_{\rho}^2+s ) )\\
& \lambda^{VA}_4+2  (k\cdot p_{0}-M_{\rho}^2 ) (2 k\cdot p_{-}+s) \lambda^{VA}_5 ) ) \Big) \Bigg)
\end{split}\end{equation}

\begin{equation}\begin{split}
a_{1}^{RRR}=&-\frac{F_V (\sqrt{2} F_{A}-4 (P-q)^2  \lambda^{A}_{17} )}{3 \sqrt{2} F^2 M_{\rho}^2 M_{\omega}^2 D_{a_{1}}\left[(P-q)^2\right] D_{\rho}\left[(k+p_{0})^2\right] D_{\rho}\left[(k+p_{-})^2\right]}    \Bigg(-M_{\rho}^2 (2 k\cdot p_{-}+s)\\
& D_{\rho}\left[(k+p_{-})^2\right] \kappa^{VV}_3  (2 (2 k\cdot p_{-}+s) \lambda^{VA}_2+4 k\cdot p_{0} \lambda^{VA}_3-(4 k\cdot p_{0}+2 k\cdot p_{-}+s)\\
&  (\lambda^{VA}_4+2 \lambda^{VA}_5 ) )+D_{\rho}\left[(k+p_{0})^2\right]  (2 D_{\rho}\left[(k+p_{-})^2\right] \kappa^{VV}_4  (-2 (3 k\cdot p_{0}+k\cdot p_{-}+s) \\
&(2 k\cdot p_{-}+s) \lambda^{VA}_2+ (8 \left(k\cdot p_{0}\right)^2+6 \left(k\cdot p_{-}\right)^2+5 \left(k\cdot p_{-}\right) s+s^2+5 k\cdot p_{0} \\
&(2 k\cdot p_{-}+s) ) \lambda^{VA}_4+2 (3 k\cdot p_{0}+k\cdot p_{-}+s) (2 k\cdot p_{-}+s) \lambda^{VA}_5 )+M_{\rho}^2 \kappa^{VV}_3 \\
& (2  (12 \left(k\cdot p_{0}\right) \left(k\cdot p_{-}\right)+8 \left(k\cdot p_{-}\right)^2-2 \left(k\cdot p_{0}\right) s+2 \left(k\cdot p_{-}\right) s-s^2 ) \lambda^{VA}_2\\
&-4 k\cdot p_{-} (4 k\cdot p_{0}+2 k\cdot p_{-}+s) \lambda^{VA}_3+(2 k\cdot p_{0}+s) (2 k\cdot p_{-}+s)  (\lambda^{VA}_4+2 \lambda^{VA}_5 ) ) ) \Bigg)
\end{split}\end{equation}

\begin{equation*}\begin{split}
a_{2}^{R}&=\frac{4 \sqrt{2}}{3 F^2}  \Bigg(-\frac{2 (P-q)^2  F_V  (\kappa^{V}_1-\kappa^{V}_2+\kappa^{V}_3 )}{M_{\omega}^2 D_{\pi}\left[(P-q)^2\right]}\\
&+\frac{F_V  (3 \kappa^{V}_1-3 \kappa^{V}_2+3 \kappa^{V}_3+\kappa^{V}_6+\kappa^{V}_7-\kappa^{V}_8-2 \kappa^{V}_{12}-\kappa^{V}_{16}+\kappa^{V}_{17} )}{M_{\omega}^2}\\
&+\frac{2 \sqrt{2} k\cdot p_{0}  (2 \kappa^{V}_{12}+\kappa^{V}_{16} )  (\lambda^{V}_{18}+2 \lambda^{V}_{19} )}{D_{\rho}\left[(k+p_{0})^2\right]}-\frac{2 k\cdot p_{0}  (2 \kappa^{V}_{12}+\kappa^{V}_{16} )  (-G_V+2 \sqrt{2} k\cdot p_{0} \lambda^{V}_{21} )}{D_{\pi}\left[(P-q)^2\right] D_{\rho}\left[(k+p_{0})^2\right]}\\
&-\frac{2 k\cdot p_{-}  (2 \kappa^{V}_{12}+\kappa^{V}_{16} )  (-G_V+2 \sqrt{2} k\cdot p_{-} \lambda^{V}_{21} )}{D_{\pi}\left[(P-q)^2\right] D_{\rho}\left[(k+p_{-})^2\right]}+\frac{2  (\kappa^{V}_{11}+\kappa^{V}_{12} )  (G_V-\sqrt{2} s \lambda^{V}_{21} )}{D_{\rho}[s]}\\
&-\frac{2 (k\cdot p_{0}+k\cdot p_{-})  (2 \kappa^{V}_{12}+\kappa^{V}_{16} )  (G_V-\sqrt{2} s \lambda^{V}_{21} )}{D_{\pi}\left[(P-q)^2\right] D_{\rho}[s]}+\frac{1}{M_{\rho}^2 D_{\rho}\left[(k+p_{-})^2\right]}\\
&\Big(G_V  (4  (k\cdot p_{-}-M_{\rho}^2 ) \kappa^{V}_{12}-2 k\cdot p_{-} \kappa^{V}_{16} )+F_V  (-2  (k\cdot p_{-}-M_{\rho}^2 ) \kappa^{V}_{12}+k\cdot p_{-} \kappa^{V}_{16} )\\
&+\sqrt{2}  (-k\cdot p_{-} \kappa^{V}_{16}  ((2 k\cdot p_{0}+s) \lambda^{V}_{16}-2 (2 k\cdot p_{0}+s) \lambda^{V}_{17}+4 k\cdot p_{0} \lambda^{V}_{18}+4 k\cdot p_{-} \lambda^{V}_{18}\\
\end{split}\end{equation*}
\begin{equation}\begin{split}
&-2 M_{\rho}^2 \lambda^{V}_{18}+2 s \lambda^{V}_{18}+4 k\cdot p_{0} \lambda^{V}_{19}+2 s \lambda^{V}_{19}-4 k\cdot p_{-} \lambda^{V}_{21}+4 k\cdot p_{-} \lambda^{V}_{22} )+2 \kappa^{V}_{12}  \\
&( (k\cdot p_{-}-M_{\rho}^2 ) (2 k\cdot p_{0}+s) \lambda^{V}_{16}-2  (k\cdot p_{-}-M_{\rho}^2 ) (2 k\cdot p_{0}+s) \lambda^{V}_{17}+2  (2 \left(k\cdot p_{-}\right)^2\\
&+2 k\cdot p_{0}  (k\cdot p_{-}-M_{\rho}^2 )-M_{\rho}^2 s+k\cdot p_{-}  (-M_{\rho}^2+s ) ) \lambda^{V}_{18}+2  (k\cdot p_{-}-M_{\rho}^2 ) \\
& ((2 k\cdot p_{0}+s) \lambda^{V}_{19}+2 k\cdot p_{-}  (-\lambda^{V}_{21}+\lambda^{V}_{22} ) ) ) )\Big)+\frac{ (\kappa^{A}_5-\kappa^{A}_6+\kappa^{A}_7 )  (F_{A}-2 \sqrt{2} (P-q)^2  \lambda^{A}_{17} )}{D_{a_{1}}\left[(P-q)^2\right]} \Bigg)
\end{split}\end{equation}

\begin{equation}\begin{split}
a_{2}^{RR}=&\frac{2 \sqrt{2}}{3 F^2}  \Bigg(\frac{8 k\cdot p_{0} F_V \kappa^{VV}_3  (\lambda^{V}_{18}+2 \lambda^{V}_{19} )}{M_{\omega}^2 D_{\rho}\left[(k+p_{0})^2\right]}+\frac{4 k\cdot p_{0} F_V \kappa^{VV}_3  (\sqrt{2} G_V-4 k\cdot p_{0} \lambda^{V}_{21} )}{M_{\omega}^2 D_{\pi}\left[(P-q)^2\right] D_{\rho}\left[(k+p_{0})^2\right]}\\
&+\frac{4 k\cdot p_{-} F_V \kappa^{VV}_3  (\sqrt{2} G_V-4 k\cdot p_{-} \lambda^{V}_{21} )}{M_{\omega}^2 D_{\pi}\left[(P-q)^2\right] D_{\rho}\left[(k+p_{-})^2\right]}+\frac{2 F_V \kappa^{VV}_3  (\sqrt{2} G_V-2 s \lambda^{V}_{21} )}{M_{\omega}^2 D_{\rho}[s]}\\
&+\frac{F_V }{M_{\rho}^2 M_{\omega}^2 D_{\rho}\left[(k+p_{-})^2\right]} \Big(\sqrt{2} F_V  (M_{\rho}^2 \kappa^{VV}_3+D_{\rho}\left[(k+p_{-})^2\right] \kappa^{VV}_4 )\\
&-2 \sqrt{2} G_V  (M_{\rho}^2 \kappa^{VV}_3+D_{\rho}\left[(k+p_{-})^2\right] \kappa^{VV}_4 )-2 M_{\rho}^2 \kappa^{VV}_3  ((2 k\cdot p_{0}+s) \lambda^{V}_{16}\\
&-2 (2 k\cdot p_{0}+s) \lambda^{V}_{17}+4 k\cdot p_{0} \lambda^{V}_{18}+2 s \lambda^{V}_{18}+4 k\cdot p_{0} \lambda^{V}_{19}+2 s \lambda^{V}_{19}-4 k\cdot p_{-} \lambda^{V}_{21}\\
&+4 k\cdot p_{-} \lambda^{V}_{22} )+2  (2 k\cdot p_{-}-M_{\rho}^2 ) \kappa^{VV}_4  ((2 k\cdot p_{0}+s) \lambda^{V}_{16}-2 (2 k\cdot p_{0}+s) \lambda^{V}_{17}\\
&+4 k\cdot p_{0} \lambda^{V}_{18}+4 k\cdot p_{-} \lambda^{V}_{18}+2 s \lambda^{V}_{18}+4 k\cdot p_{0} \lambda^{V}_{19}+2 s \lambda^{V}_{19}-4 k\cdot p_{-} \lambda^{V}_{21}\\
&+4 k\cdot p_{-} \lambda^{V}_{22} ) \Big)+\frac{F_V  (\kappa^{VA}_2-\kappa^{VA}_3-\kappa^{VA}_4 )  (\sqrt{2} F_{A}-4 (P-q)^2  \lambda^{A}_{17} )}{M_{\omega}^2 D_{a_{1}}\left[(P-q)^2\right]}\\
&+\frac{2 \kappa^{VA}_5  (\sqrt{2}G_V  -2 s \lambda^{V}_{21} ) (F_{A}-2 \sqrt{2} (P-q)^2  \lambda^{A}_{17} )}{D_{a_{1}}\left[(P-q)^2\right] D_{\rho}[s]}\\
&+\frac{4 k\cdot p_{0}  (2 \kappa^{V}_{12}+\kappa^{V}_{16} )  (F_{A}-2 \sqrt{2} (P-q)^2  \lambda^{A}_{17} )  (2 \lambda^{VA}_2-\lambda^{VA}_3 )}{D_{a_{1}}\left[(P-q)^2\right] D_{\rho}\left[(k+p_{0})^2\right]}\\
&+\frac{2  (F_{A}-2 \sqrt{2} (P-q)^2  \lambda^{A}_{17} ) }{M_{\rho}^2 D_{a_{1}}\left[(P-q)^2\right] D_{\rho}\left[(k+p_{-})^2\right]} \Big(k\cdot p_{-} \kappa^{V}_{16}  (-2 (2 k\cdot p_{0}+s) \lambda^{VA}_2\\
&+ (2 k\cdot p_{0}+4 k\cdot p_{-}-2 M_{\rho}^2+s ) \lambda^{VA}_4+2 (2 k\cdot p_{0}+s) \lambda^{VA}_5 )+2 \kappa^{V}_{12}  \\
&(2  (k\cdot p_{-}-M_{\rho}^2 ) (2 k\cdot p_{0}+s) \lambda^{VA}_2+ (-2 \left(k\cdot p_{0}\right) \left(k\cdot p_{-}\right)-4 \left(k\cdot p_{-}\right)^2\\
&+2 k\cdot p_{0} M_{\rho}^2+2 k\cdot p_{-} M_{\rho}^2-\left(k\cdot p_{-}\right) s+M_{\rho}^2 s ) \lambda^{VA}_4-2  (k\cdot p_{-}-M_{\rho}^2 )\\
& (2 k\cdot p_{0}+s) \lambda^{VA}_5 ) \Big) \Bigg)
\end{split}\end{equation}

\begin{equation}\begin{split}
a_{2}^{RRR}&=-\frac{2 \sqrt{2} F_V (\sqrt{2} F_{A}-4 (P-q)^2  \lambda^{A}_{17} )}{3 F^2 M_{\rho}^2 M_{\omega}^2 D_{a_{1}}\left[(P-q)^2\right] D_{\rho}\left[(k+p_{0})^2\right] D_{\rho}\left[(k+p_{-})^2\right]}   \Bigg(4 k\cdot p_{0} M_{\rho}^2 \\
&D_{\rho}\left[(k+p_{-})^2\right] \kappa^{VV}_3  (-2 \lambda^{VA}_2+\lambda^{VA}_3 )+D_{\rho}\left[(k+p_{0})^2\right]  (M_{\rho}^2 (2 k\cdot p_{0}+s) \kappa^{VV}_3 \\
& (2 \lambda^{VA}_2-\lambda^{VA}_4-2 \lambda^{VA}_5 )+D_{\rho}\left[(k+p_{-})^2\right] \kappa^{VV}_4  (2 (2 k\cdot p_{0}+s) \lambda^{VA}_2\\
&-(2 k\cdot p_{0}+4 k\cdot p_{-}+s) \lambda^{VA}_4-2 (2 k\cdot p_{0}+s) \lambda^{VA}_5 ) ) \Bigg)
\end{split}\end{equation}

\begin{equation*}\begin{split}
a_{3}^{R}=&\frac{\sqrt{2}}{3 F^2}  \Bigg(-\frac{4 F_V (k\cdot p_{0}-k\cdot p_{-})  (\kappa^{V}_1-\kappa^{V}_2+\kappa^{V}_3+\kappa^{V}_6+\kappa^{V}_7-\kappa^{V}_8-\kappa^{V}_{17} )}{M_{\omega}^2}\\
&-\frac{4 (k\cdot p_{0}-k\cdot p_{-})  (2 \kappa^{V}_{11}-\kappa^{V}_{16} )  (G_V-\sqrt{2} s \lambda^{V}_{21} )}{D_{\rho}[s]}\\
&+\frac{1}{M_{\rho}^2 D_{\rho}\left[(k+p_{0})^2\right]}\Big(-2 G_V  (2  (4 \left(k\cdot p_{0}\right)^2+M_{\rho}^2 (2 k\cdot p_{-}+s)-k\cdot p_{0}  \\
&(2 k\cdot p_{-}+2 M_{\rho}^2+s ) ) \kappa^{V}_{12}+k\cdot p_{0}  (-4 k\cdot p_{0}+2 k\cdot p_{-}+2 M_{\rho}^2+s ) \kappa^{V}_{16} )\\
&+F_V  (2  (4 \left(k\cdot p_{0}\right)^2-k\cdot p_{0} (2 k\cdot p_{-}+s)+M_{\rho}^2 (2 k\cdot p_{-}+s) ) \kappa^{V}_{12}+k\cdot p_{0} \\
& (-4 k\cdot p_{0}+2 k\cdot p_{-}+4 M_{\rho}^2+s ) \kappa^{V}_{16} )-\sqrt{2}  (k\cdot p_{0} \kappa^{V}_{16}  (- (4 k\cdot p_{0}-2 k\cdot p_{-}\\
&-4 M_{\rho}^2-s ) (2 k\cdot p_{-}+s) \lambda^{V}_{16}+2  ( (4 k\cdot p_{0}-2 k\cdot p_{-}-4 M_{\rho}^2-s ) (2 k\cdot p_{-}+s) \lambda^{V}_{17}\\
&+ (-8 \left(k\cdot p_{0}\right)^2+4 \left(k\cdot p_{-}\right)^2-2 k\cdot p_{0}  (2 k\cdot p_{-}-2 M_{\rho}^2+s )+4 k\cdot p_{-}  (M_{\rho}^2+s )\\
&+s  (2 M_{\rho}^2+s ) ) \lambda^{V}_{18}-8 \left(k\cdot p_{0}\right) \left(k\cdot p_{-}\right) \lambda^{V}_{19}+4 \left(k\cdot p_{-}\right)^2 \lambda^{V}_{19}+4 k\cdot p_{-} M_{\rho}^2 \lambda^{V}_{19}\\
&-4 \left(k\cdot p_{0}\right) s \lambda^{V}_{19}+4 \left(k\cdot p_{-}\right) s \lambda^{V}_{19}+2 M_{\rho}^2 s \lambda^{V}_{19}+s^2 \lambda^{V}_{19}+8 \left(k\cdot p_{0}\right)^2 \lambda^{V}_{21}\\
&-4 \left(k\cdot p_{0}\right) \left(k\cdot p_{-}\right) \lambda^{V}_{21}-4 k\cdot p_{0} M_{\rho}^2 \lambda^{V}_{21}-2 \left(k\cdot p_{0}\right) s \lambda^{V}_{21}-8 \left(k\cdot p_{0}\right)^2 \lambda^{V}_{22}\\
&+4 \left(k\cdot p_{0}\right) \left(k\cdot p_{-}\right) \lambda^{V}_{22}+8 k\cdot p_{0} M_{\rho}^2 \lambda^{V}_{22}+2 \left(k\cdot p_{0}\right) s \lambda^{V}_{22} ) )+2 \kappa^{V}_{12}  ((2 k\cdot p_{-}+s) \\
& (4 \left(k\cdot p_{0}\right)^2-k\cdot p_{0} (2 k\cdot p_{-}+s)+M_{\rho}^2 (2 k\cdot p_{-}+s) ) \lambda^{V}_{16}+2  (-(2 k\cdot p_{-}+s) \\
& (4 \left(k\cdot p_{0}\right)^2-k\cdot p_{0} (2 k\cdot p_{-}+s)+M_{\rho}^2 (2 k\cdot p_{-}+s) ) \lambda^{V}_{17}+ (8 \left(k\cdot p_{0}\right)^3-k\cdot p_{0}\\
& (2 k\cdot p_{-}+s)^2+M_{\rho}^2 (2 k\cdot p_{-}+s)^2+2 \left(k\cdot p_{0}\right)^2  (2 k\cdot p_{-}-2 M_{\rho}^2+s ) ) \lambda^{V}_{18}\\
&+8 \left(k\cdot p_{0}\right)^2 \left(k\cdot p_{-}\right) \lambda^{V}_{19}-4 \left(k\cdot p_{0}\right) \left(k\cdot p_{-}\right)^2 \lambda^{V}_{19}-4 \left(k\cdot p_{0}\right) \left(k\cdot p_{-}\right) M_{\rho}^2 \lambda^{V}_{19}\\
&+4 \left(k\cdot p_{-}\right)^2 M_{\rho}^2 \lambda^{V}_{19}+4 \left(k\cdot p_{0}\right)^2 s \lambda^{V}_{19}-4 \left(k\cdot p_{0}\right) \left(k\cdot p_{-}\right) s \lambda^{V}_{19}-2 k\cdot p_{0} M_{\rho}^2 s \lambda^{V}_{19}\\
&+4 k\cdot p_{-} M_{\rho}^2 s \lambda^{V}_{19}-\left(k\cdot p_{0}\right) s^2 \lambda^{V}_{19}+M_{\rho}^2 s^2 \lambda^{V}_{19}-8 \left(k\cdot p_{0}\right)^3 \lambda^{V}_{21}+4 \left(k\cdot p_{0}\right)^2 \left(k\cdot p_{-}\right) \lambda^{V}_{21}\\
&+4 \left(k\cdot p_{0}\right)^2 M_{\rho}^2 \lambda^{V}_{21}-4 \left(k\cdot p_{0}\right) \left(k\cdot p_{-}\right) M_{\rho}^2 \lambda^{V}_{21}+2 \left(k\cdot p_{0}\right)^2 s \lambda^{V}_{21}-2 k\cdot p_{0} M_{\rho}^2 s \lambda^{V}_{21}\\
&+8 \left(k\cdot p_{0}\right)^3 \lambda^{V}_{22}-4 \left(k\cdot p_{0}\right)^2 \left(k\cdot p_{-}\right) \lambda^{V}_{22}+4 \left(k\cdot p_{0}\right) \left(k\cdot p_{-}\right) M_{\rho}^2 \lambda^{V}_{22}-2 \left(k\cdot p_{0}\right)^2 s \lambda^{V}_{22}\\
&+2 k\cdot p_{0} M_{\rho}^2 s \lambda^{V}_{22} ) ) )\Big)+\frac{1}{M_{\rho}^2 D_{\rho}\left[(k+p_{-})^2\right]}\Big(2 G_V  (2  (2 \left(k\cdot p_{-}\right)^2-\left(k\cdot p_{-}\right) s+M_{\rho}^2 s ) \kappa^{V}_{12}\\
&+k\cdot p_{-}  (-2 k\cdot p_{-}+2 M_{\rho}^2+s ) \kappa^{V}_{16} )-F_V  (2  (2 \left(k\cdot p_{-}\right)^2+2 k\cdot p_{-} M_{\rho}^2-\left(k\cdot p_{-}\right) s\\
&+M_{\rho}^2 s ) \kappa^{V}_{12}+k\cdot p_{-}  (-2 k\cdot p_{-}+4 M_{\rho}^2+s ) \kappa^{V}_{16} )+\sqrt{2}  (k\cdot p_{-} \kappa^{V}_{16}  (- (2 k\cdot p_{-}-4 M_{\rho}^2-s ) \\
&(2 k\cdot p_{0}+s) \lambda^{V}_{16}+2  ( (2 k\cdot p_{-}-4 M_{\rho}^2-s ) (2 k\cdot p_{0}+s) \lambda^{V}_{17}- (2 k\cdot p_{-}-2 M_{\rho}^2-s )\\
& (P-q)^2  \lambda^{V}_{18}-4 \left(k\cdot p_{0}\right) \left(k\cdot p_{-}\right) \lambda^{V}_{19}+4 k\cdot p_{-} M_{\rho}^2 \lambda^{V}_{19}+2 \left(k\cdot p_{0}\right) s \lambda^{V}_{19}-2 \left(k\cdot p_{-}\right) s \lambda^{V}_{19}\\
&+2 M_{\rho}^2 s \lambda^{V}_{19}+s^2 \lambda^{V}_{19}+4 \left(k\cdot p_{-}\right)^2 \lambda^{V}_{21}-4 k\cdot p_{-} M_{\rho}^2 \lambda^{V}_{21}-2 \left(k\cdot p_{-}\right) s \lambda^{V}_{21}-4 \left(k\cdot p_{-}\right)^2 \lambda^{V}_{22}\\
\end{split}\end{equation*}
\begin{equation}\begin{split}
\qquad&+8 k\cdot p_{-} M_{\rho}^2 \lambda^{V}_{22}+2 \left(k\cdot p_{-}\right) s \lambda^{V}_{22} ) )+2 \kappa^{V}_{12}  ((2 k\cdot p_{0}+s)  (2 \left(k\cdot p_{-}\right)^2+2 k\cdot p_{-} M_{\rho}^2\\
&-\left(k\cdot p_{-}\right) s+M_{\rho}^2 s ) \lambda^{V}_{16}+2  ((2 k\cdot p_{0}+s)  (-2 \left(k\cdot p_{-}\right)^2-M_{\rho}^2 s+k\cdot p_{-}  (-2 M_{\rho}^2+s ) ) \lambda^{V}_{17}\\
&+(P-q)^2   (2 \left(k\cdot p_{-}\right)^2-\left(k\cdot p_{-}\right) s+M_{\rho}^2 s ) \lambda^{V}_{18}+4 \left(k\cdot p_{0}\right) \left(k\cdot p_{-}\right)^2 \lambda^{V}_{19}-4 \left(k\cdot p_{0}\right) \\
&\left(k\cdot p_{-}\right) M_{\rho}^2 \lambda^{V}_{19}+4 \left(k\cdot p_{-}\right)^2 M_{\rho}^2 \lambda^{V}_{19}-2 \left(k\cdot p_{0}\right) \left(k\cdot p_{-}\right) s \lambda^{V}_{19}+2 \left(k\cdot p_{-}\right)^2 s \lambda^{V}_{19}\\&+2 k\cdot p_{0} M_{\rho}^2 s \lambda^{V}_{19}-\left(k\cdot p_{-}\right) s^2 \lambda^{V}_{19}+M_{\rho}^2 s^2 \lambda^{V}_{19}-4 \left(k\cdot p_{-}\right)^3 \lambda^{V}_{21}+2 \left(k\cdot p_{-}\right)^2 s \lambda^{V}_{21}\\
&-2 k\cdot p_{-} M_{\rho}^2 s \lambda^{V}_{21}+4 \left(k\cdot p_{-}\right)^3 \lambda^{V}_{22}+4 \left(k\cdot p_{-}\right)^2 M_{\rho}^2 \lambda^{V}_{22}-2 \left(k\cdot p_{-}\right)^2 s \lambda^{V}_{22}\\
&+2 k\cdot p_{-} M_{\rho}^2 s \lambda^{V}_{22} ) ) )\Big)+\frac{4 (-k\cdot p_{0}+k\cdot p_{-})  (\kappa^{A}_5-\kappa^{A}_6+\kappa^{A}_7 )  (F_{A}-2 \sqrt{2} (P-q)^2  \lambda^{A}_{17} )}{D_{a_{1}}\left[(P-q)^2\right]} \Bigg)
\end{split}\end{equation}

\begin{equation*}\begin{split}
a_{3}^{RR}=&\frac{1}{3 \sqrt{2} F^2} \Bigg(\frac{F_V }{M_{\rho}^2 M_{\omega}^2 D_{\rho}\left[(k+p_{0})^2\right]} \Big(\sqrt{2} F_V  (M_{\rho}^2 (4 k\cdot p_{0}+2 k\cdot p_{-}+s) \kappa^{VV}_3\\
&+(-4 k\cdot p_{0}+2 k\cdot p_{-}+s) D_{\rho}\left[(k+p_{0})^2\right] \kappa^{VV}_4 )-2  (\sqrt{2} G_V  (M_{\rho}^2 (2 k\cdot p_{-}+s) \kappa^{VV}_3\\
&+(-4 k\cdot p_{0}+2 k\cdot p_{-}+s) D_{\rho}\left[(k+p_{0})^2\right] \kappa^{VV}_4 )+ (2 k\cdot p_{0}-M_{\rho}^2 ) (4 k\cdot p_{0}\\
&-2 k\cdot p_{-}-s) \kappa^{VV}_4  ((2 k\cdot p_{-}+s) \lambda^{V}_{16}-2 (2 k\cdot p_{-}+s) \lambda^{V}_{17}+4 k\cdot p_{0} \lambda^{V}_{18}\\
&+4 k\cdot p_{-} \lambda^{V}_{18}+2 s \lambda^{V}_{18}+4 k\cdot p_{-} \lambda^{V}_{19}+2 s \lambda^{V}_{19}-4 k\cdot p_{0} \lambda^{V}_{21}+4 k\cdot p_{0} \lambda^{V}_{22} )\\
&+M_{\rho}^2 \kappa^{VV}_3  ((2 k\cdot p_{-}+s) (4 k\cdot p_{0}+2 k\cdot p_{-}+s) \lambda^{V}_{16}+2  (-(2 k\cdot p_{-}+s)\\
& (4 k\cdot p_{0}+2 k\cdot p_{-}+s) \lambda^{V}_{17}+(2 k\cdot p_{-}+s) (P-q)^2  \lambda^{V}_{18}+4 \left(k\cdot p_{-}\right)^2 \lambda^{V}_{19}\\
&+4 \left(k\cdot p_{-}\right) s \lambda^{V}_{19}+s^2 \lambda^{V}_{19}-4 \left(k\cdot p_{0}\right) \left(k\cdot p_{-}\right) \lambda^{V}_{21}-2 \left(k\cdot p_{0}\right) s \lambda^{V}_{21}+8 \left(k\cdot p_{0}\right)^2 \lambda^{V}_{22}\\
&+4 \left(k\cdot p_{0}\right) \left(k\cdot p_{-}\right) \lambda^{V}_{22}+2 \left(k\cdot p_{0}\right) s \lambda^{V}_{22} ) ) ) \Big)+\frac{F_V }{M_{\rho}^2 M_{\omega}^2 D_{\rho}\left[(k+p_{-})^2\right]} \\
&\Big(-\sqrt{2} F_V  (M_{\rho}^2 (6 k\cdot p_{-}+s) \kappa^{VV}_3+(-2 k\cdot p_{-}+s) D_{\rho}\left[(k+p_{-})^2\right] \kappa^{VV}_4 )\\
&+2  (\sqrt{2} G_V  (M_{\rho}^2 (2 k\cdot p_{-}+s) \kappa^{VV}_3+(-2 k\cdot p_{-}+s) D_{\rho}\left[(k+p_{-})^2\right] \kappa^{VV}_4 )\\
&+ (2 k\cdot p_{-}-M_{\rho}^2 ) (2 k\cdot p_{-}-s) \kappa^{VV}_4  ((2 k\cdot p_{0}+s) \lambda^{V}_{16}-2 (2 k\cdot p_{0}+s) \lambda^{V}_{17}\\
&+4 k\cdot p_{0} \lambda^{V}_{18}+4 k\cdot p_{-} \lambda^{V}_{18}+2 s \lambda^{V}_{18}+4 k\cdot p_{0} \lambda^{V}_{19}+2 s \lambda^{V}_{19}-4 k\cdot p_{-} \lambda^{V}_{21}\\
&+4 k\cdot p_{-} \lambda^{V}_{22} )+M_{\rho}^2 \kappa^{VV}_3  ((2 k\cdot p_{0}+s) (6 k\cdot p_{-}+s) \lambda^{V}_{16}+2  (-(2 k\cdot p_{0}+s)\\
& (6 k\cdot p_{-}+s) \lambda^{V}_{17}+(2 k\cdot p_{-}+s) (P-q)^2  \lambda^{V}_{18}-4 \left(k\cdot p_{0}\right) \left(k\cdot p_{-}\right) \lambda^{V}_{19}\\
&+8 \left(k\cdot p_{-}\right)^2 \lambda^{V}_{19}+2 \left(k\cdot p_{0}\right) s \lambda^{V}_{19}+2 \left(k\cdot p_{-}\right) s \lambda^{V}_{19}+s^2 \lambda^{V}_{19}-4 \left(k\cdot p_{-}\right)^2 \lambda^{V}_{21}\\
&-2 \left(k\cdot p_{-}\right) s \lambda^{V}_{21}+12 \left(k\cdot p_{-}\right)^2 \lambda^{V}_{22}+2 \left(k\cdot p_{-}\right) s \lambda^{V}_{22} ) ) ) \Big)+\frac{4 F_V }{M_{\omega}^2 D_{a_{1}}\left[(P-q)^2\right]}\\
\end{split}\end{equation*}
\begin{equation}\begin{split}
\qquad&\Big( (-k\cdot p_{0}+k\cdot p_{-})  (\kappa^{VA}_2-\kappa^{VA}_3-\kappa^{VA}_4 )  (\sqrt{2} F_{A}-4 (P-q)^2  \lambda^{A}_{17} )\Big)\\
&+\frac{8 (-k\cdot p_{0}+k\cdot p_{-}) \kappa^{VA}_5  (\sqrt{2}G_V -2 s \lambda^{V}_{21})  (F_{A}-2 \sqrt{2} (P-q)^2  \lambda^{A}_{17} ) }{D_{a_{1}}\left[(P-q)^2\right] D_{\rho}[s]}\\
&+\frac{2  (F_{A}-2 \sqrt{2} (P-q)^2  \lambda^{A}_{17} ) }{M_{\rho}^2 D_{a_{1}}\left[(P-q)^2\right] D_{\rho}\left[(k+p_{-})^2\right]} \Big(k\cdot p_{-} \kappa^{V}_{16}  ( (-8 \left(k\cdot p_{0}\right) \left(k\cdot p_{-}\right)+8 k\cdot p_{-} M_{\rho}^2\\&+4 \left(k\cdot p_{0}\right) s-4 \left(k\cdot p_{-}\right) s+4 M_{\rho}^2 s+2 s^2 ) \lambda^{VA}_2+2 M_{\rho}^2 (4 k\cdot p_{0}-2 k\cdot p_{-}+s) \lambda^{VA}_3\\
&+4 \left(k\cdot p_{0}\right) \left(k\cdot p_{-}\right) \lambda^{VA}_4+8 \left(k\cdot p_{-}\right)^2 \lambda^{VA}_4-8 k\cdot p_{0} M_{\rho}^2 \lambda^{VA}_4-4 k\cdot p_{-} M_{\rho}^2 \lambda^{VA}_4\\
&-2 \left(k\cdot p_{0}\right) s \lambda^{VA}_4-2 \left(k\cdot p_{-}\right) s \lambda^{VA}_4-2 M_{\rho}^2 s \lambda^{VA}_4-s^2 \lambda^{VA}_4+8 \left(k\cdot p_{0}\right) \left(k\cdot p_{-}\right) \lambda^{VA}_5\\
&-16 k\cdot p_{0} M_{\rho}^2 \lambda^{VA}_5-4 \left(k\cdot p_{0}\right) s \lambda^{VA}_5+4 \left(k\cdot p_{-}\right) s \lambda^{VA}_5-8 M_{\rho}^2 s \lambda^{VA}_5-2 s^2 \lambda^{VA}_5 )\\&+2 \kappa^{V}_{12}  (2  (2 k\cdot p_{0}  (k\cdot p_{-}-M_{\rho}^2 ) (2 k\cdot p_{-}-s)-\left(k\cdot p_{-}\right) s^2+M_{\rho}^2 s^2\\
&+2 \left(k\cdot p_{-}\right)^2  (2 M_{\rho}^2+s ) ) \lambda^{VA}_2+2 k\cdot p_{-} M_{\rho}^2 (4 k\cdot p_{0}-2 k\cdot p_{-}+s) \lambda^{VA}_3\\
&-4 \left(k\cdot p_{0}\right) \left(k\cdot p_{-}\right)^2 \lambda^{VA}_4-8 \left(k\cdot p_{-}\right)^3 \lambda^{VA}_4-4 \left(k\cdot p_{0}\right) \left(k\cdot p_{-}\right) M_{\rho}^2 \lambda^{VA}_4\\
&+4 \left(k\cdot p_{-}\right)^2 M_{\rho}^2 \lambda^{VA}_4+2 \left(k\cdot p_{0}\right) \left(k\cdot p_{-}\right) s \lambda^{VA}_4+2 \left(k\cdot p_{-}\right)^2 s \lambda^{VA}_4\\
&-2 k\cdot p_{0} M_{\rho}^2 s \lambda^{VA}_4-4 k\cdot p_{-} M_{\rho}^2 s \lambda^{VA}_4+\left(k\cdot p_{-}\right) s^2 \lambda^{VA}_4-M_{\rho}^2 s^2 \lambda^{VA}_4\\
&-8 \left(k\cdot p_{0}\right) \left(k\cdot p_{-}\right)^2 \lambda^{VA}_5-8 \left(k\cdot p_{0}\right) \left(k\cdot p_{-}\right) M_{\rho}^2 \lambda^{VA}_5+4 \left(k\cdot p_{0}\right) \left(k\cdot p_{-}\right) s \lambda^{VA}_5\\
&-4 \left(k\cdot p_{-}\right)^2 s \lambda^{VA}_5-4 k\cdot p_{0} M_{\rho}^2 s \lambda^{VA}_5-4 k\cdot p_{-} M_{\rho}^2 s \lambda^{VA}_5+2 \left(k\cdot p_{-}\right) s^2 \lambda^{VA}_5\\
\qquad&-2 M_{\rho}^2 s^2 \lambda^{VA}_5 ) \Big)+\frac{2  (F_{A}-2 \sqrt{2} (P-q)^2  \lambda^{A}_{17} )}{M_{\rho}^2 D_{a_{1}}\left[(P-q)^2\right] D_{\rho}\left[(k+p_{0})^2\right]}  \Big(k\cdot p_{0} \kappa^{V}_{16}  (2  (4 k\cdot p_{0}-2 k\cdot p_{-}\\
&-2 M_{\rho}^2-s ) (2 k\cdot p_{-}+s) \lambda^{VA}_2-2 M_{\rho}^2 (2 k\cdot p_{-}+s) \lambda^{VA}_3-16 \left(k\cdot p_{0}\right)^2 \lambda^{VA}_4\\
&+4 \left(k\cdot p_{-}\right)^2 \lambda^{VA}_4+8 k\cdot p_{0} M_{\rho}^2 \lambda^{VA}_4+4 k\cdot p_{-} M_{\rho}^2 \lambda^{VA}_4+4 \left(k\cdot p_{-}\right) s \lambda^{VA}_4\\
&+2 M_{\rho}^2 s \lambda^{VA}_4+s^2 \lambda^{VA}_4-16 \left(k\cdot p_{0}\right) \left(k\cdot p_{-}\right) \lambda^{VA}_5+8 \left(k\cdot p_{-}\right)^2 \lambda^{VA}_5\\
&+16 k\cdot p_{-} M_{\rho}^2 \lambda^{VA}_5-8 \left(k\cdot p_{0}\right) s \lambda^{VA}_5+8 \left(k\cdot p_{-}\right) s \lambda^{VA}_5+8 M_{\rho}^2 s \lambda^{VA}_5+2 s^2 \lambda^{VA}_5 )\\
&+2 \kappa^{V}_{12}  (-2 (2 k\cdot p_{-}+s)  (4 \left(k\cdot p_{0}\right)^2+M_{\rho}^2 (2 k\cdot p_{-}+s)-k\cdot p_{0}  \\
&(2 k\cdot p_{-}+2 M_{\rho}^2+s ) ) \lambda^{VA}_2-2 k\cdot p_{0} M_{\rho}^2 (2 k\cdot p_{-}+s) \lambda^{VA}_3+16 \left(k\cdot p_{0}\right)^3 \lambda^{VA}_4\\
&-4 \left(k\cdot p_{0}\right) \left(k\cdot p_{-}\right)^2 \lambda^{VA}_4-8 \left(k\cdot p_{0}\right)^2 M_{\rho}^2 \lambda^{VA}_4+4 \left(k\cdot p_{0}\right) \left(k\cdot p_{-}\right) M_{\rho}^2 \lambda^{VA}_4\\
&+4 \left(k\cdot p_{-}\right)^2 M_{\rho}^2 \lambda^{VA}_4-4 \left(k\cdot p_{0}\right) \left(k\cdot p_{-}\right) s \lambda^{VA}_4+2 k\cdot p_{0} M_{\rho}^2 s \lambda^{VA}_4\\
&+4 k\cdot p_{-} M_{\rho}^2 s \lambda^{VA}_4-\left(k\cdot p_{0}\right) s^2 \lambda^{VA}_4+M_{\rho}^2 s^2 \lambda^{VA}_4+16 \left(k\cdot p_{0}\right)^2 \left(k\cdot p_{-}\right) \lambda^{VA}_5\\
&-8 \left(k\cdot p_{0}\right) \left(k\cdot p_{-}\right)^2 \lambda^{VA}_5+8 \left(k\cdot p_{-}\right)^2 M_{\rho}^2 \lambda^{VA}_5+8 \left(k\cdot p_{0}\right)^2 s \lambda^{VA}_5\\
&-8 \left(k\cdot p_{0}\right) \left(k\cdot p_{-}\right) s \lambda^{VA}_5+8 k\cdot p_{-} M_{\rho}^2 s \lambda^{VA}_5-2 \left(k\cdot p_{0}\right) s^2 \lambda^{VA}_5+2 M_{\rho}^2 s^2 \lambda^{VA}_5 ) \Big) \Bigg)
\end{split}\end{equation}

\begin{equation}\begin{split}
a_3^{RRR}=&\frac{F_V(\sqrt{2} F_{A}-4 (P-q)^2  \lambda^{A}_{17} )}{3 \sqrt{2} F^2 M_{\rho}^2 M_{\omega}^2 D_{a_{1}}\left[(P-q)^2\right] D_{\rho}\left[(k+p_{0})^2\right] D_{\rho}\left[(k+p_{-})^2\right]}  \Bigg(-M_{\rho}^2 (2 k\cdot p_{-}+s)\\
& D_{\rho}\left[(k+p_{-})^2\right] \kappa^{VV}_3  (2 (2 k\cdot p_{-}+s) \lambda^{VA}_2+4 k\cdot p_{0} \lambda^{VA}_3-(4 k\cdot p_{0}+2 k\cdot p_{-}+s) \\
& (\lambda^{VA}_4+2 \lambda^{VA}_5 ) )-D_{\rho}\left[(k+p_{0})^2\right]  (2 (k\cdot p_{0}-k\cdot p_{-}) D_{\rho}\left[(k+p_{-})^2\right] \kappa^{VV}_4 \\
& (-2 (2 k\cdot p_{-}+3 s) \lambda^{VA}_2+(8 k\cdot p_{0}+6 k\cdot p_{-}+s) \lambda^{VA}_4+2 (2 k\cdot p_{-}+3 s) \lambda^{VA}_5 )\\
&+M_{\rho}^2 \kappa^{VV}_3  (2  (4 \left(k\cdot p_{0}\right) \left(k\cdot p_{-}\right)-8 \left(k\cdot p_{-}\right)^2-2 \left(k\cdot p_{0}\right) s-2 \left(k\cdot p_{-}\right) s-s^2 ) \lambda^{VA}_2\\
&-4 k\cdot p_{-} (4 k\cdot p_{0}-2 k\cdot p_{-}+s) \lambda^{VA}_3+(2 k\cdot p_{0}+s) (6 k\cdot p_{-}+s)  (\lambda^{VA}_4+2 \lambda^{VA}_5 ) ) ) \Bigg)
\end{split}\end{equation}

\begin{equation}\begin{split}
a_{4}^{R}&=\frac{4 \sqrt{2}}{3 F^2 M_{\rho}^2}  \Bigg(\frac{1}{D_{\rho}\left[(k+p_{0})^2\right]}\Big(G_V  (4  (k\cdot p_{0}-M_{\rho}^2 ) \kappa^{V}_{12}-2 k\cdot p_{0} \kappa^{V}_{16} )\\
&+F_V  (-2  (k\cdot p_{0}-M_{\rho}^2 ) \kappa^{V}_{12}+k\cdot p_{0} \kappa^{V}_{16} )+\sqrt{2}  (-k\cdot p_{0} \kappa^{V}_{16}  ((2 k\cdot p_{-}+s) \lambda^{V}_{16}\\
&-2 (2 k\cdot p_{-}+s) \lambda^{V}_{17}+4 k\cdot p_{0} \lambda^{V}_{18}+4 k\cdot p_{-} \lambda^{V}_{18}+2 s \lambda^{V}_{18}+4 k\cdot p_{-} \lambda^{V}_{19}+4 M_{\rho}^2 \lambda^{V}_{19}\\
&+2 s \lambda^{V}_{19}-4 k\cdot p_{0} \lambda^{V}_{21}+4 k\cdot p_{0} \lambda^{V}_{22} )+2 \kappa^{V}_{12}  ( (k\cdot p_{0}-M_{\rho}^2 ) (2 k\cdot p_{-}+s) \lambda^{V}_{16}\\
&-2  (k\cdot p_{0}-M_{\rho}^2 ) (2 k\cdot p_{-}+s) \lambda^{V}_{17}+4 \left(k\cdot p_{0}\right)^2 \lambda^{V}_{18}+4 \left(k\cdot p_{0}\right) \left(k\cdot p_{-}\right) \lambda^{V}_{18}\\
&-4 k\cdot p_{0} M_{\rho}^2 \lambda^{V}_{18}-4 k\cdot p_{-} M_{\rho}^2 \lambda^{V}_{18}+2 \left(k\cdot p_{0}\right) s \lambda^{V}_{18}-2 M_{\rho}^2 s \lambda^{V}_{18}+4 \left(k\cdot p_{0}\right) \left(k\cdot p_{-}\right) \lambda^{V}_{19}\\
&-4 k\cdot p_{0} M_{\rho}^2 \lambda^{V}_{19}-4 k\cdot p_{-} M_{\rho}^2 \lambda^{V}_{19}+2 \left(k\cdot p_{0}\right) s \lambda^{V}_{19}-2 M_{\rho}^2 s \lambda^{V}_{19}-4 \left(k\cdot p_{0}\right)^2 \lambda^{V}_{21}\\
&+4 k\cdot p_{0} M_{\rho}^2 \lambda^{V}_{21}+4 \left(k\cdot p_{0}\right)^2 \lambda^{V}_{22}-4 k\cdot p_{0} M_{\rho}^2 \lambda^{V}_{22} ) )\Big)+\frac{1}{D_{\rho}\left[(k+p_{-})^2\right]}\\
&\Big(F_V  (2  (k\cdot p_{-}-M_{\rho}^2 ) \kappa^{V}_{12}-k\cdot p_{-} \kappa^{V}_{16} )+G_V  (-4  (k\cdot p_{-}-M_{\rho}^2 ) \kappa^{V}_{12}+2 k\cdot p_{-} \kappa^{V}_{16} )\\
&+\sqrt{2}  (k\cdot p_{-} \kappa^{V}_{16}  ((2 k\cdot p_{0}+s) \lambda^{V}_{16}-2 (2 k\cdot p_{0}+s) \lambda^{V}_{17}+4 k\cdot p_{0} \lambda^{V}_{18}+4 k\cdot p_{-} \lambda^{V}_{18}\\
&+2 s \lambda^{V}_{18}+4 k\cdot p_{0} \lambda^{V}_{19}+4 M_{\rho}^2 \lambda^{V}_{19}+2 s \lambda^{V}_{19}-4 k\cdot p_{-} \lambda^{V}_{21}+4 k\cdot p_{-} \lambda^{V}_{22} )\\
&+\kappa^{V}_{12}  (-2  (k\cdot p_{-}-M_{\rho}^2 ) (2 k\cdot p_{0}+s) \lambda^{V}_{16}+4  (k\cdot p_{-}-M_{\rho}^2 ) (2 k\cdot p_{0}+s) \lambda^{V}_{17}\\
&-4  (k\cdot p_{-}-M_{\rho}^2 ) (P-q)^2  \lambda^{V}_{18}-8 \left(k\cdot p_{0}\right) \left(k\cdot p_{-}\right) \lambda^{V}_{19}+8 k\cdot p_{0} M_{\rho}^2 \lambda^{V}_{19}+8 k\cdot p_{-} M_{\rho}^2 \lambda^{V}_{19}\\
&-4 \left(k\cdot p_{-}\right) s \lambda^{V}_{19}+4 M_{\rho}^2 s \lambda^{V}_{19}+8 \left(k\cdot p_{-}\right)^2 \lambda^{V}_{21}-8 k\cdot p_{-} M_{\rho}^2 \lambda^{V}_{21}-8 \left(k\cdot p_{-}\right)^2 \lambda^{V}_{22}\\
&+8 k\cdot p_{-} M_{\rho}^2 \lambda^{V}_{22} ) )\Big) \Bigg)
\end{split}\end{equation}

\begin{equation*}\begin{split}
a_{4}^{RR}=&\frac{2 \sqrt{2}}{3 F^2 M_{\rho}^2}  \Bigg(\frac{F_V}{M_{\omega}^2 D_{\rho}\left[(k+p_{0})^2\right]}  \Big(\sqrt{2} F_V  (M_{\rho}^2 \kappa^{VV}_3+D_{\rho}\left[(k+p_{0})^2\right] \kappa^{VV}_4 )\\
&-2 \sqrt{2} G_V  (M_{\rho}^2 \kappa^{VV}_3+D_{\rho}\left[(k+p_{0})^2\right] \kappa^{VV}_4 )+2  (2 k\cdot p_{0}-M_{\rho}^2 ) \kappa^{VV}_4 \\
& ((2 k\cdot p_{-}+s) \lambda^{V}_{16}-2 (2 k\cdot p_{-}+s) \lambda^{V}_{17}+4 k\cdot p_{0} \lambda^{V}_{18}+4 k\cdot p_{-} \lambda^{V}_{18}+2 s \lambda^{V}_{18}\\
&+4 k\cdot p_{-} \lambda^{V}_{19}+2 s \lambda^{V}_{19}-4 k\cdot p_{0} \lambda^{V}_{21}+4 k\cdot p_{0} \lambda^{V}_{22} )-2 M_{\rho}^2 \kappa^{VV}_3  ((2 k\cdot p_{-}+s) \lambda^{V}_{16}\\
&-2 (2 k\cdot p_{-}+s) \lambda^{V}_{17}+4 k\cdot p_{0} \lambda^{V}_{18}+4 k\cdot p_{-} \lambda^{V}_{18}+2 s \lambda^{V}_{18}+8 k\cdot p_{0} \lambda^{V}_{19}+4 k\cdot p_{-} \lambda^{V}_{19}\\
&+2 s \lambda^{V}_{19}-4 k\cdot p_{0} \lambda^{V}_{21}+4 k\cdot p_{0} \lambda^{V}_{22} ) \Big)+\frac{F_V }{M_{\omega}^2 D_{\rho}\left[(k+p_{-})^2\right]} \Big(-\sqrt{2} F_V  (M_{\rho}^2 \kappa^{VV}_3\\
&+D_{\rho}\left[(k+p_{-})^2\right] \kappa^{VV}_4 )+2 \sqrt{2} G_V  (M_{\rho}^2 \kappa^{VV}_3+D_{\rho}\left[(k+p_{-})^2\right] \kappa^{VV}_4 )\\
&-2  (2 k\cdot p_{-}-M_{\rho}^2 ) \kappa^{VV}_4  ((2 k\cdot p_{0}+s) \lambda^{V}_{16}-2 (2 k\cdot p_{0}+s) \lambda^{V}_{17}+4 k\cdot p_{0} \lambda^{V}_{18}\\
&+4 k\cdot p_{-} \lambda^{V}_{18}+2 s \lambda^{V}_{18}+4 k\cdot p_{0} \lambda^{V}_{19}+2 s \lambda^{V}_{19}-4 k\cdot p_{-} \lambda^{V}_{21}+4 k\cdot p_{-} \lambda^{V}_{22} )\\
&+2 M_{\rho}^2 \kappa^{VV}_3  ((2 k\cdot p_{0}+s) \lambda^{V}_{16}-2 (2 k\cdot p_{0}+s) \lambda^{V}_{17}+4 k\cdot p_{0} \lambda^{V}_{18}+4 k\cdot p_{-} \lambda^{V}_{18}\\
\end{split}\end{equation*}
\begin{equation}\begin{split}
\qquad&+2 s \lambda^{V}_{18}+4 k\cdot p_{0} \lambda^{V}_{19}+8 k\cdot p_{-} \lambda^{V}_{19}+2 s \lambda^{V}_{19}-4 k\cdot p_{-} \lambda^{V}_{21}+4 k\cdot p_{-} \lambda^{V}_{22} ) \Big)\\
&-\frac{2  (F_{A}-2 \sqrt{2} (P-q)^2  \lambda^{A}_{17} )}{D_{a_{1}}\left[(P-q)^2\right] D_{\rho}\left[(k+p_{-})^2\right]}  \Big(k\cdot p_{-} \kappa^{V}_{16}  (-2  (2 k\cdot p_{0}+2 M_{\rho}^2+s ) \lambda^{VA}_2\\
&+2 M_{\rho}^2 \lambda^{VA}_3+2 k\cdot p_{0} \lambda^{VA}_4+4 k\cdot p_{-} \lambda^{VA}_4-2 M_{\rho}^2 \lambda^{VA}_4+s \lambda^{VA}_4+4 k\cdot p_{0} \lambda^{VA}_5+2 s \lambda^{VA}_5 )\\
&+2 \kappa^{V}_{12}  (2  (2 \left(k\cdot p_{0}\right) \left(k\cdot p_{-}\right)-2 k\cdot p_{0} M_{\rho}^2-2 k\cdot p_{-} M_{\rho}^2+\left(k\cdot p_{-}\right) s-M_{\rho}^2 s ) \lambda^{VA}_2\\
&+2 k\cdot p_{-} M_{\rho}^2 \lambda^{VA}_3-2 \left(k\cdot p_{0}\right) \left(k\cdot p_{-}\right) \lambda^{VA}_4-4 \left(k\cdot p_{-}\right)^2 \lambda^{VA}_4+2 k\cdot p_{0} M_{\rho}^2 \lambda^{VA}_4\\
&+2 k\cdot p_{-} M_{\rho}^2 \lambda^{VA}_4-\left(k\cdot p_{-}\right) s \lambda^{VA}_4+M_{\rho}^2 s \lambda^{VA}_4-4 \left(k\cdot p_{0}\right) \left(k\cdot p_{-}\right) \lambda^{VA}_5\\
&+4 k\cdot p_{0} M_{\rho}^2 \lambda^{VA}_5-2 \left(k\cdot p_{-}\right) s \lambda^{VA}_5+2 M_{\rho}^2 s \lambda^{VA}_5 ) \Big)+\frac{2  (F_{A}-2 \sqrt{2} (P-q)^2  \lambda^{A}_{17} )}{D_{a_{1}}\left[(P-q)^2\right] D_{\rho}\left[(k+p_{0})^2\right]} \\
& \Big(k\cdot p_{0} \kappa^{V}_{16}  (-2  (2 k\cdot p_{-}+2 M_{\rho}^2+s ) \lambda^{VA}_2+2 M_{\rho}^2 \lambda^{VA}_3+4 k\cdot p_{0} \lambda^{VA}_4+2 k\cdot p_{-} \lambda^{VA}_4\\
&-2 M_{\rho}^2 \lambda^{VA}_4+s \lambda^{VA}_4+4 k\cdot p_{-} \lambda^{VA}_5+2 s \lambda^{VA}_5 )+\kappa^{V}_{12}  (4  (-M_{\rho}^2 (2 k\cdot p_{-}+s)\\
&+k\cdot p_{0}  (2 k\cdot p_{-}-2 M_{\rho}^2+s ) ) \lambda^{VA}_2+4 k\cdot p_{0} M_{\rho}^2 \lambda^{VA}_3-8 \left(k\cdot p_{0}\right)^2 \lambda^{VA}_4\\
&-4 \left(k\cdot p_{0}\right) \left(k\cdot p_{-}\right) \lambda^{VA}_4+4 k\cdot p_{0} M_{\rho}^2 \lambda^{VA}_4+4 k\cdot p_{-} M_{\rho}^2 \lambda^{VA}_4-2 \left(k\cdot p_{0}\right) s \lambda^{VA}_4\\
&+2 M_{\rho}^2 s \lambda^{VA}_4-8 \left(k\cdot p_{0}\right) \left(k\cdot p_{-}\right) \lambda^{VA}_5+8 k\cdot p_{-} M_{\rho}^2 \lambda^{VA}_5-4 \left(k\cdot p_{0}\right) s \lambda^{VA}_5\\
&+4 M_{\rho}^2 s \lambda^{VA}_5 ) \Big) \Bigg)
\end{split}\end{equation}

\begin{equation}\begin{split}
a_{4}^{RRR}=&-\frac{4 \sqrt{2} F_V(k\cdot p_{0}-k\cdot p_{-})  (\sqrt{2} F_{A}-4 (P-q)^2  \lambda^{A}_{17} )  }{3 F^2 M_{\rho}^2 M_{\omega}^2 D_{a_{1}}\left[(P-q)^2\right] D_{\rho}\left[(k+p_{0})^2\right] D_{\rho}\left[(k+p_{-})^2\right]}\\
& \Bigg(-D_{\rho}\left[(k+p_{0})^2\right] D_{\rho}\left[(k+p_{-})^2\right] \kappa^{VV}_4  (2 \lambda^{VA}_2+\lambda^{VA}_4-2 \lambda^{VA}_5 )\\
&+M_{\rho}^2 \kappa^{VV}_3  (2  (M_{\rho}^2+\left(P-q\right)^2) \lambda^{VA}_2-2 M_{\rho}^2 \lambda^{VA}_3+D_{\rho}\left[(P-q)^2\right] (\lambda^{VA}_4+2 \lambda^{VA}_5 ) ) \Bigg)
\end{split}\end{equation}

\bibliographystyle{hunsrt}
\bibliography{bibl}
\addcontentsline{toc}{section}{References}

\end{document}